%% file: main_tension.tex
\documentclass[%
 reprint,
superscriptaddress,
nofootinbib,
 amsmath,amssymb,
 aps,
 prd,
floatfix,
]{revtex4-2}

\ifdefined\pdfminorversion
\fi

\usepackage{color}
\usepackage{amsmath}
\usepackage{amssymb}
\usepackage{capt-of}
\usepackage{graphicx}%
\usepackage{dcolumn}%
\usepackage{bm}%
\usepackage[pdfusetitle]{hyperref}
\hypersetup{hidelinks,pdfauthor={Yuan Feng, Elias R. Most, and Lorenzo Gavassino}}
\usepackage{cleveref}
\usepackage{pbox}
\usepackage[utf8]{inputenc}
\usepackage[T1,OT1]{fontenc}
\DeclareTextCommandDefault{\k}[1]{\UseTextAccent{T1}{\k}{#1}}
\usepackage[]{times}
\usepackage{tensor}
\usepackage{xcolor}
\usepackage{tabularx}
\usepackage[normalem]{ulem}
\usepackage{tikz}
\usetikzlibrary{arrows.meta,positioning,fit}
\usepackage{aas_macros}

\usepackage{etoolbox}
\makeatletter
\patchcmd{\@bibdataout@aps}{author="08"}{author="48"}{}{}
\patchcmd{\@bibdataout@aps}{author="08"}{author="48"}{}{}
\makeatother

\definecolor{criticalred}{RGB}{0,0,0}
\definecolor{proofgreen}{RGB}{0,0,0}
\definecolor{closuregreen}{RGB}{0,0,0}
\definecolor{closuremissing}{RGB}{0,0,0}
\definecolor{revisionblue}{RGB}{0,0,0}
\newcommand{\revadd}[1]{\textcolor{revisionblue}{#1}}
\definecolor{revisionpurple}{RGB}{0,0,0}

\definecolor{revisiongreen}{RGB}{0,0,0}

\definecolor{notationred}{RGB}{0,0,0}

\definecolor{physicsred}{RGB}{0,0,0}

\definecolor{derivationgreen}{RGB}{0,0,0}
\newcolumntype{Y}{>{\raggedright\arraybackslash}X}

\begin{document}

\interfootnotelinepenalty=10000
\title{{Nonlinearly Causal General-Relativistic Superfluid Hydrodynamics:\texorpdfstring{\\}{ }Simulating Vortex Dynamics in Neutron-Star Interiors}}

\author{Yuan Feng}\email{yfeng2@caltech.edu}
\affiliation{TAPIR, Mailcode 350-17, California Institute of Technology, Pasadena, CA 91106, USA}
\author{Elias R. Most}\email{emost@caltech.edu}
\affiliation{TAPIR, Mailcode 350-17, California Institute of Technology, Pasadena, CA 91106, USA}
\affiliation{Walter Burke Institute for Theoretical Physics, California Institute of Technology, Pasadena, CA 91106, USA}
\author{Lorenzo Gavassino}
\affiliation{Department of Applied Mathematics and Theoretical Physics,
University of Cambridge, Cambridge CB3 0WA, UK}

\date{November 19, 2026}%

\begin{abstract}
{Neutron stars offer a natural laboratory for superfluid matter under extreme conditions. Understanding its nonlinear dynamics is important for interpreting pulsar glitches, including their possible connection to fast radio bursts, as well as for modeling tides in inspiralling neutron star binaries. To enable their full nonlinear numerical study, }
we formulate general-relativistic superfluid hydrodynamics with
entrainment, vortex forces and line energy, and derive necessary and sufficient conditions for
nonlinear causality and strong hyperbolicity for such a system for the first time. As a side outcome, we also show that commonly used dissipative force-balanced approaches without line tension are mathematically ill-posed due to the generic appearance of the short-wavelength limit of the Donnelly-Glaberson instability.
Leveraging the mathematical similarities between the vortex array subsystem and force-free electrodynamics when tension is included, we formulate superfluid vortex dynamics in terms of effective electric and magnetic fields with drift speeds. This allows us to develop a novel finite-volume scheme with (superfluid) vorticity constrained
transport and implicit electric field evolution capable of stably solving the general-relativistic superfluid system numerically. We then assess the correctness of the scheme and implementation, among other tests, by presenting an exact solution of the relativistic superfluid Riemann problem, as well as by computing Kelvin-wave dispersion relations for this system.
Finally, we perform dynamical global simulations of general-relativistic two-fluid neutron star equilibria, validate their global oscillation modes, and perform the first (simplified) calculation of general-relativistic superfluid interior spin-down, focusing on vortex-array dynamics, {prescribed release and repinning}. We also demonstrate that the resulting stationary state retains a net twist in the vortex array, as well as a residual degree of differential rotation of the superfluid component. In short, our work makes it now possible to numerically study general-relativistic superfluid dynamics in neutron star interiors.
\end{abstract}
\maketitle

\section{Introduction}
\label{sec:introduction}

Neutron stars are expected to contain superfluid matter in their
interiors. Below the relevant pairing temperatures, neutrons
can form a superfluid and protons a superconductor, with the extent of
pairing remaining sensitive to the nuclear interaction and medium effects
\cite{Elgaroy:1996mg,Baldo:1998ca,Schwenk:2003bc}.
The paired state has a coherent macroscopic phase
\cite{Bardeen:1957mv,Yang:1962zz}, allowing the neutron condensate to move
independently of the charged matter. This relative motion introduces an
internal reservoir of angular momentum and additional degrees of freedom
that an ordinary single-fluid description cannot retain
(see, e.g., Refs.~\cite{Haskell:2017lkl,Andersson:2021zrx} for recent reviews).

Observations provide several ways to constrain this picture. The measured
cooling of the Cassiopeia A neutron star~\cite{Heinke:2010cr} has been
interpreted as the onset of neutron pairing in a core with already paired
protons~\cite{Page:2010aw,Shternin:2010qi}, although the inferred cooling
rate remains sensitive to detector systematics~\cite{Zhao:2025koi}.
Post-outburst observations trace crustal thermal relaxation
\cite{Cackett:2008tq}. Combined with the deposited heat and an inferred
core temperature, they place lower bounds on the core heat capacity
\cite{Cumming:2016weq}. 
Quasi-periodic oscillations following magnetar flares
\cite{Israel:2005av} offer a complementary dynamical probe, since
superfluid decoupling changes the magnetoelastic spectrum
\cite{Gabler:2013ova}. Interpreting these signals requires specifying how
the superfluid couples to the thermal, magnetic, and elastic response.

Pulsar glitches provide a particularly direct probe of angular-momentum
exchange. Vela observations revealed an abrupt spin-up together
with a change in the spin-down rate
\cite{1969Natur.222..228R,1969Natur.222..229R}.
{Ref.~\cite{1969Natur.224..872B} modeled postglitch relaxation using a
weakly coupled neutron superfluid, with a starquake triggering the
event.} {Ref. ~\cite{Anderson:1975zze} proposed
instead that vortices pinned to the crust sustain differential rotation
and release angular momentum when they unpin.} Vortex-creep models relate
the recovery to thermally activated motion through the pinning barriers
\cite{1984ApJ...276..325A}.

Population studies constrain the angular momentum transferred in glitches
\cite{Espinoza:2011pq}, while resolved timing of the 2016 Vela glitch
\cite{2018Natur.556..219P} constrains rapid transfer. The inferred rise
time has a $90\%$ upper limit of $12.6\,\mathrm{s}$, and the rotation
frequency overshoots before relaxing~\cite{Ashton:2019hqd}.
SGR~1935+2154 also exhibited a spin-down glitch in 2020, followed by
renewed radio activity~\cite{Younes:2022yet}. In a separate event in 2022,
two spin-up glitches bracketed a fast radio burst~\cite{Hu:2024fdt}.
These temporal associations motivate a joint study of rotational and
magnetospheric activity.

The material composition supplies a natural multifluid description of
these processes. In the inner crust, unbound neutrons coexist with
{{neutron-rich nuclear clusters and electrons. The clusters dissolve into}}
uniform matter toward the core
\cite{Negele:1971vb,Douchin:2001sv}.
On macroscopic scales, the lattice or core charged particles can be
grouped into an electromagnetically coupled component, while the neutron
condensate retains an independent velocity
\cite{Carter:2004he,Sourie:2016hrk}.
The two components are nevertheless coupled by entrainment: the momentum
of either constituent depends on both currents
\cite{1975JETP...42..164A}. Crustal band-structure calculations predict
strong entrainment in some layers~\cite{Chamel:2012zn}, which can increase
the reservoir required by glitch activity
\cite{Andersson:2012iu,Chamel:2012ae}. Pairing can reduce this effect
\cite{Watanabe:2017nzj}, so the available angular momentum depends
on both the microphysics and the components coupled during the event.

The superfluid's rotational response is governed by quantized vortices.
Away from vortex
cores, the superfluid momentum is locally proportional to the condensate
phase gradient, while loops
enclosing vortices carry quantized circulation
\cite{ANDERSON:1966bps,Gavassino:2020zfk}.
On scales larger than the intervortex
spacing, the array can be represented by a continuous vorticity field,
as in Hall--Vinen--Bekarevich--Khalatnikov (HVBK) hydrodynamics
\cite{1956RSPSA.238..215H,publisher:BekarevichKhalatnikov1961} and its relativistic
extensions~\cite{Gusakov:2016eom,Andersson:2016asl}.
The velocity transporting this circulation need not equal either fluid
velocity. {Electron scattering from magnetized neutron vortices provides
drag in the core~\cite{Alpar:1984zz,Andersson:2005rs},
while crustal nuclei can pin the array~\cite{2016MNRAS.455.3952S}.
Where protons form a type-II superconductor, magnetic flux can penetrate
as quantized tubes, or fluxoids. Pinning between neutron vortices and
fluxoids couples the two condensates
\cite{Sourie:2020drw,Sourie:2021vsj}.
{{Vortices cutting through proton fluxoids can dissipate energy by exciting
Kelvin waves~\cite{Link:2003hq,Haskell:2013hja}.}}}
The balance of lift, drag, and any holding force therefore determines how
the superfluid exchanges angular momentum with the charged component.
Relativistic vortex theories also account for the energy and anisotropic
stress of the array~\cite{Carter:1995mj,Gusakov:2016eom}. Lattice shear
elasticity describes changes in their relative arrangement and requires
additional strain variables~\cite{Andersson:2019fdj}. 

Existing calculations address this problem at several scales. Reservoir
and creep models connect pinning to glitch amplitudes and recovery
\cite{1984ApJ...276..325A,Pizzochero:2011dd}, while spatial two-fluid and
three-dimensional HVBK simulations follow differential rotation and
mutual friction
\cite{Haskell:2011xe,Antonelli:2016nka,Fuentes:2024gvv}.
Gross--Pitaevskii simulations resolve collective vortex avalanches
\cite{Warszawski:2011vy}, and filament calculations capture bending
and history-dependent pinning~\cite{Link:2022vum}.
Microscopic vortex--nucleus calculations~\cite{Wlazlowski:2016yoe}
and mesoscopic line averaging~\cite{2016MNRAS.455.3952S} constrain the forces
entering these models. A deformable continuum vortex array has also
been evolved through accumulation, reconnection and superfluid river formation \cite{Levin:2023ewn}. These approaches establish important parts of the
transfer mechanism. Connecting local release and repinning to the
response of a relativistic star additionally requires evolving the fluid
motion and circulation transport together.

General relativity affects both the stellar structure and the coupling
between the fluids. Relativistic two-fluid equilibria have been
constructed in slow and rapid rotation
\cite{Andersson:2000hy,Prix:2004wq,Aranguren:2022ted}, including realistic equations of
state and entrainment~\cite{Sourie:2016hrk}.
{{Ref.~\cite{Sourie:2016tgd} calculated mutual-friction-driven glitch
rises through quasi-stationary sequences of uniformly rotating two-fluid
equilibria. Relativistic corrections changed the rise time by approximately
$20$--$40\%$ for the models studied.}}
Relativistic calculations also relate vortex geometry to coupling times
\cite{Gavassino:2020sry}.
{{The coupled fluid and vortex evolution considered here allows the fluid
response and spatial
redistribution of circulation to develop without imposing uniform
rotation or a fixed vortex geometry.}}

Carter's covariant multifluid theory provides a consistent thermodynamic
basis for such an evolution~\cite{Carter:1992gmy,1992AnPhy.219..243C}.
A master function of the current scalar products determines the
constituent momenta, entrainment, and material stress--energy tensor.
The nondissipative theory can be expressed in relativistic Landau-type
variables~\cite{Son:2000ht,Gusakov:2007px}, with the thermodynamic
mapping developed in Ref.~\cite{Gavassino:2019wzd}.
The choice of independent currents specifies the physical degrees of
freedom. In Landau--Tisza theory, the normal component carries entropy
\cite{Tisza:1938ydz,1941ZhETF..11..592L,1982ZhETF..83.1601L}, whereas neutron and charged particle currents can remain independent even at negligible temperature.
Multifluid neutron-star theories account for this material distinction
and its electromagnetic and vortex-mediated couplings
\cite{1991ApJ...380..515M,Langlois:1997bz,Carter:1998rn,Rau:2020wnv}.

The vortex dynamics requires a force law in addition to the material
thermodynamics. In the commonly used force-equilibrium approximation, vortex
acceleration is neglected and Magnus and generalized Iordanskii forces
balance drag~\cite{Gavassino:2020zfk}. 
Pinning additionally requires a holding reaction, while
dissipation must enter the internal work and entropy production
\cite{Haskell:2012vp,Gavassino:2020zfk}.

{{For a well-posed evolution, the formulation must also be strongly
hyperbolic and nonlinearly causal.}} Relative motion changes the collective modes
and can lead to two-stream instabilities
\cite{2004MNRAS.354..101A,Samuelsson:2009up,Hawke:2013haa},
whose onset depends on constitutive coupling and dissipation
\cite{Haber:2015exa,Andersson:2019ezz}.
Near equilibrium, thermodynamic stability constrains multifluid
entrainment and causality~\cite{Gavassino:2022zzz}, while {{Onsager--Casimir symmetry yields symmetric-hyperbolic linear equations
for thermodynamically stable equilibrium states under the assumptions
of Ref.~\cite{Gavassino:2022nff}, including the absence of macroscopic
magnetic fields and spin degrees of freedom.}}
{For viscous superfluids, suitable hydrodynamic frames admit
linearly stable and causal regimes at nonzero superfluid
velocity~\cite{Hoult:2024cyx}.}
Finite-amplitude evolution requires a separate characteristic analysis,
as in dissipative relativistic fluids~\cite{Bemfica:2020xym} and recent
two-fluid plasma formulations~\cite{Most:2026uga}, which has been lacking for 
relativistic superfluids, especially with full vortex line dynamics.

In this work, we formulate general-relativistic superfluid hydrodynamics with
entrainment, vortex forces and line energy, formulate the first numerical scheme capable of integrating these equations in general situations.
To enable this, we derive necessary and sufficient conditions for
nonlinear causality and strong hyperbolicity, and show that commonly used dissipative force-balanced approaches without line tension are mathematically ill-posed due to the generic appearance of the short-wavelength limit of the Donnelly-Glaberson instability.
 To assess correctness of the scheme, among others we present an exact solution of the relativistic superfluid Riemann problem, as well as compute Kelvin-wave dispersion relations for this system.
Finally, we perform dynamical global simulations of general-relativistic two-fluid neutron star equilibria \cite{Prix:2004wq}, incl. simulating neutron star interior spin down with vortex unpinning.

This work is structured as follows.
Section~\ref{sec:theory} presents the formulation and characteristic
conditions, and Sec.~\ref{sec:numerical_methods} describes the numerical
scheme. Sections~\ref{sec:numerical-results} and
\ref{sec:stellar_applications} give the flat-spacetime and stellar tests.
Section~\ref{sec:results-summary} discusses the results. Supporting
derivations and numerical details are collected in the appendices.
We use signature $(-,+,+,+)$ and units $G=c=k_{\rm B}=1$, with Greek
indices denoting spacetime components and Latin indices spatial
components.

We use the $3+1$ metric \cite{Gourgoulhon:2007ue}
\begin{align}
 ds^2&=-\alpha^2dt^2+
       \gamma_{ij}(dx^i+\beta^idt)(dx^j+\beta^jdt),\notag\\
 n^\mu&=\alpha^{-1}(1,-\beta^i),\qquad
 \sqrt{-g}=\alpha\sqrt\gamma,\notag\\
 K_{ij}&=-\frac{\partial_t\gamma_{ij}-\mathcal L_\beta\gamma_{ij}}{2\alpha}.
 \label{eq:adm_metric_unified}
\end{align}
\textcolor{black}{Here $\alpha$ is the lapse, $\beta^i$ the shift, $\gamma_{ij}$
the spatial metric, and $n^\mu$ the future-directed unit slice normal.
We write $g=\det g_{\mu\nu}$ and use $\mathcal L_\beta$ for the Lie
derivative along the shift. The spacetime and spatial Levi-Civita
connections are denoted by $\nabla_\mu$ and $\nabla_i$, respectively.}
Here $\gamma=\det\gamma_{ij}$,
\textcolor{black}{$K_{ij}$ is the extrinsic curvature,} $K=\gamma^{ij}K_{ij}$ and
$\epsilon^{ijk}=[ijk]/\sqrt\gamma$\textcolor{black}{, where $[123]=+1$ fixes
the alternating symbol}. Spatial indices are raised and
lowered with $\gamma_{ij}$.

\section{Relativistic superfluid matter and vortex motion}
\label{sec:theory}
\label{sec:methods}

We briefly review relativistic hydrodynamics in 
Sec.~\ref{sec:moment_starting_point}. Section~\ref{sec:moment_common_frame}
then develops the two-component superfluid description and {vortex force balance, circulation transport and internal heating.}
Carter's master function supplies the thermodynamic closure in
Sec.~\ref{sec:moment_carter_closure}. 
\textcolor{black}{We then add vortex tension in Sec.~\ref{sec:model_b_vortex_tension},
before describing the characteristics and their causal, strongly hyperbolic
domain in Secs.~\ref{sec:primitive_hyperbolicity_summary}
and~\ref{sec:hyperbolicity}.}

\subsection{14-moment description of relativistic fluids}
\label{sec:moment_starting_point}

A single relativistic fluid can be described in terms of its particle number current, $N^\mu$, and its energy--momentum tensor $T^{\mu\nu}$. While these can be prescribed phenomenologically, they are commonly derived from kinetic theory~\cite{Israel:1979wp,Denicol:2012cn} or a variational principle~\cite{Carter:1992gmy,Andersson:2006nr}. Here, we defer the constitutive closure to Sec.~\ref{sec:moment_carter_closure}, and first discuss the structure of the available degrees of freedom, as well as implications of basic conservation.
The combined system contains 14 independent degrees of freedom, which we can see as follows. We can pick a timelike observer, {$o^\mu$}\textcolor{black}{, normalized by $o_\mu o^\mu=-1$}, usually the fluid velocity, relative to which we can decompose the tensors into orthogonal components,
\begin{align}
 N^\mu&=n {o^\mu}+V^\mu,\notag\\
 T^{\mu\nu}&=e {o^\mu o^\nu}+p_{\rm iso} {\Delta_o^{\mu\nu}}
             +2 {o^{(\mu}}Q^{\nu)}+\pi^{\mu\nu},\notag\\
 {\Delta_o^{\mu\nu}}&=g^{\mu\nu}+{o^\mu o^\nu}.
 \label{eq:moments_observer_decomposition}
\end{align}
{Here $n=-o_\mu N^\mu$ is the observer-measured particle density,
$e=o_\mu o_\nu T^{\mu\nu}$ its energy density, and
$p_{\rm iso}=\Delta_{o\mu\nu}T^{\mu\nu}/3$ the mean isotropic pressure relative to the frame, $o^\mu$.}
The particle diffusion vector $V^\mu$ and heat flux vector $Q^\mu$ are orthogonal to {$o^\mu$} by construction. The anisotropic pressure $\pi^{\mu\nu}$ is symmetric, trace-free and also orthogonal. It should be noted that special hydrodynamics frames $o^\mu$ can be chosen such that, e.g., $Q^\mu=0$ in the Landau energy frame~\cite{Denicol:2012cn}.
The details of the microphysical theory, especially the presence of other fields or components, usually determine the precise conservation laws, which we will state in
{Secs.~\ref{sec:moment_carter_closure}
and~\ref{sec:equilibrium_3plus1}}. After introducing the phase dynamics
and choosing a common frame, we obtain the reversible moment decomposition from Carter's master function and determine its scalar coefficients in Gusakov--Son variables in Sec.~\ref{sec:moment_gusakov_matching}.

\label{sec:canonical_motivation}

Superfluidity constrains canonical circulation.
To define circulation in relativity, we first recall this concept for a perfect fluid and consider
an ordinary perfect fluid with $N^\mu=n o^\mu$ and
$T^{\mu\nu}=(e+P)o^\mu o^\nu+Pg^{\mu\nu}$. \textcolor{black}{Here $P$ is the
pressure, $s$ the rest-frame entropy density, and $T$ the temperature.
The fluid acceleration is $a_\nu=o^\rho\nabla_\rho o_\nu$.} Define entropy per particle
$\bar s=s/n$, specific enthalpy $h=(e+P)/(m_pn)$, and the enthalpy
one-form $\pi_\mu=h o_\mu$, where $m_p$ is the particle rest mass. Conservation of stress--energy for a perfect fluid implies the Carter--Lichnerowicz equation for
$\Omega^{(h)}_{\mu\nu}=2\nabla_{[\mu}\pi_{\nu]}$
\cite{Gourgoulhon:2006bn},
\begin{align}
 o^\rho\Omega^{(h)}_{\rho\nu}
 &=h a_\nu+\Delta_{o\nu}{}^\alpha\nabla_\alpha h\notag\\
 &=\frac{T}{m_p}\Delta_{o\nu}{}^\alpha\nabla_\alpha\bar s
 =\frac{T}{m_p}\nabla_\nu\bar s.
 \label{eq:moments_carter_lichnerowicz}
\end{align}
The final equality uses ideal entropy advection. The mass factor follows
from the definition of $h$; it is unity in the unit-rest-mass convention.
Homentropic flow satisfies $o^\rho\Omega^{(h)}_{\rho\nu}=0$.

Vorticity is tightly coupled to circulation and it is instructive to briefly introduce the concept here for later use. Formally, general-relativistic circulation is the integral of momentum per particle around a closed spacelike contour $C$. For the perfect fluid,
$p_\mu=m_p\pi_\mu=(e+P)o_\mu/n$, and
\begin{align}
 \mathcal C_p[C]&=\oint_C p_\mu\,dx^\mu,\notag\\
 \frac{d\mathcal C_p[C_\lambda]}{d\lambda}
 &=\oint_{C_\lambda}V^\rho
      (2\nabla_{[\rho}p_{\mu]})\,dx^\mu.
 \label{eq:relativistic_circulation_transport}
\end{align}
Choose an oriented closed loop $C_0$ on a spacelike slice, and let
$C_\lambda$ be its image under the flow of a smooth vector $V^\mu$,
with all loops remaining spacelike. The choice specifies which
circulation is followed: a material loop uses $V^\mu\parallel o^\mu$,
whereas a fixed spatial loop or a loop transported with vortices need
not follow the matter. On slices labeled by coordinate time $t$, the
material choice is $V^\mu=o^\mu/o^t$, its normalization only fixes the
transport parameter. Cartan's identity gives the
second relation because the integral of the exact part of
$\mathcal L_Vp$ vanishes on a closed contour. For a homentropic perfect
fluid and $V^\mu\parallel o^\mu$, it yields the relativistic Kelvin
theorem~\cite[Sec.~7.5]{Gourgoulhon:2006bn}. In the weak-gravity, low-speed
limit with $e-m_pn,P\ll m_pn$, $h\to1$ and $p_i/m_p\to v_i$.
On a constant-time contour,
\begin{equation}
 \frac{\mathcal C_p[C]}{m_p}
 \longrightarrow\oint_C\boldsymbol v\cdot d\boldsymbol\ell,
 \label{eq:newtonian_circulation_limit}
\end{equation}
recovering ordinary velocity circulation and its Kelvin theorem.

\subsection{Relativistic superfluids}
\label{sec:moment_common_frame}

We model the relativistic superfluid phase of a neutron star interior as 
a multi-component hydrodynamical flow. Neutron-star matter contains neutrons, protons, and electrons. We detail in the following how these degrees of freedom are modeled in the system, which are retained dynamically and what simplifications have been made.

The constituent split depends on the chemical gauge
(Sec.~\ref{sec:theory_force_specialization}), while the current--momentum
relations follow from Carter's construction
(Sec.~\ref{sec:moment_carter_closure}). We use the physical
neutron--proton basis: $N_s^\mu=n_n^\mu\textcolor{black}{=\nu_nu_n^\mu}$ counts all neutrons, including
those associated with nuclei, and $N_c^\mu=n_pu^\mu$ is the proton
current. \textcolor{black}{The proper neutron and proton number densities are
$\nu_n$ and $n_p$, and their unit four-velocities are $u_n^\mu$ and
$u^\mu$, respectively.} Electrons are included through local charge neutrality and
share the carrier velocity. Quantized neutron vortices transport
canonical circulation and mediate momentum and angular-momentum exchange
with the carrier, as described in Sec.~\ref{sec:moment_vortex_closure}.

We make the following approximations, which we detail in subsequent subsections:
\begin{itemize}
 \setlength{\itemsep}{1pt}
 \item \emph{Local charge neutrality:} We neglect electromagnetic dynamics and fix the electrons by assuming local charge neutrality. This means we only need to model a proton current as a carrier fluid.
 \item {\emph{Averaged vortex dynamics:} 
{We average the hydrodynamic fields over scales larger than the
 intervortex spacing and follow the collective array. This choice does
 not resolve individual lines or their quantum cores. Classical filament
 models can instead follow individual quantized vortices with prescribed
 circulation and core parameters~\cite{Link:2022vum}.}
 The force-equilibrium
 approximation balances Magnus lift, Iordanskii lift and drag to
 determine the transverse drift, as described in
 Sec.~\ref{sec:moment_vortex_closure}. It neglects line momentum and
{stress. Generic dissipative counterflow then fails strong
 hyperbolicity in three dimensions (see also Sec. \ref{sec:donnelly_glaberson}).
 \textcolor{black}{Section~\ref{sec:model_b_vortex_tension} introduces independent
 line inertia and tension. Its causal and strongly hyperbolic domain
 is given in Sec.~\ref{sec:vortex_tension_hyperbolicity}.}}}
 \item \emph{Two-current approximation:} {We keep two material currents and the canonical superfluid
momentum. Particle number is conserved, while an entropy-carrying
component also accounts for heating. The vortex energy supplies the
additional stress needed to evolve the vortex lines independently.} The physical interpretation of the currents is fixed by a chemical gauge choice \cite{Gavassino:2020zfk}.
\end{itemize}

\subsubsection{Canonical superfluid momentum and circulation}
{
In the rotating neutron star configurations we wish to model, an array
of quantized vortices supports the superfluid circulation and associated
angular momentum.\footnote{For an azimuthal circle, axisymmetry gives
$\mathcal C_\xi=2\pi\xi_\phi$. In the chosen neutron--proton basis,
the neutron canonical angular momentum on an axisymmetric spatial
slice is
$J_n=\int D_n\xi_\phi\,dV
=(2\pi)^{-1}\int D_n\mathcal C_\xi\,dV$,
where $D_n=-n_\mu N_s^\mu$, $n^\mu$ is the future-directed unit
slice normal, and $dV=\sqrt{\gamma}\,d^3x$
\cite[Eq.~(86)]{Andersson:2000hy}.
Thus angular momentum is a density-weighted volume integral of
the circulation.}
}
{Away from vortex cores, the condensate momentum per particle is
locally a phase gradient, $\xi_\mu=\nabla_\mu\varphi$, on a smooth
phase patch} \cite{Carter:1992gmy,Son:2000ht}. \textcolor{black}{The phase
$\varphi$ defines the canonical vorticity
$\Omega_{\mu\nu}=2\nabla_{[\mu}\xi_{\nu]}$.
Let $\Sigma$ be an oriented spacelike surface with boundary $C$ and
area element $d\Sigma^{\mu\nu}$.} Its circulation is
\begin{align}
 \mathcal C_\xi[C]&=\oint_C\xi_\mu\,dx^\mu
 =\frac12\int_\Sigma\Omega_{\mu\nu}\,d\Sigma^{\mu\nu},\notag\\
 \mathcal C_\xi[C]&=\kappa_vN_v,\qquad \partial\Sigma=C,
 \label{eq:superfluid_circulation_quantization}
\end{align}
where $N_v$ is the signed number of vortex intersections and
$\kappa_v=\pi\hbar$ for paired neutrons with phase normalized per neutron
\cite{Gavassino:2020zfk}. A contractible contour in a defect-free
phase patch has zero circulation. A contour enclosing a vortex can have
nonzero circulation despite local irrotationality, which we capture via a coarse-grained vorticity through the cores.
\textcolor{black}{For neutron rest mass $m_n$,} the Newtonian limit has the intuitive interpretation,
\begingroup\color{black}
\begin{equation}
 \frac{\mathcal C_\xi[C]}{m_n}
 =\oint_C\boldsymbol v_{\rm phase}\cdot d\boldsymbol\ell
 =\kappa_{\rm vel} N_v,\qquad \kappa_{\rm vel}=\frac{\kappa_v}{m_n},
 \label{eq:superfluid_newtonian_circulation}
\end{equation}
\endgroup
Here $\boldsymbol v_{\rm phase}=\boldsymbol\xi/m_n$ is the phase
velocity at leading Newtonian order.

\textcolor{black}{In the hydrodynamical mean-field approach, the coarse-grained
canonical vorticity obeys}
\begin{equation}
 \nabla_{[\lambda}\Omega_{\mu\nu]}=0,\qquad
 N_s^\rho\Omega_{\rho\nu}=f_{s\nu}.
 \label{eq:moments_superfluid_vorticity}
\end{equation}
Here $f_{s\nu}$ is the force density. 
Thermodynamics supplies the current--momentum relation
in Sec.~\ref{sec:moment_carter_closure}, whereas the vortex-force prescription is presented
in Sec.~\ref{sec:moment_vortex_closure}. \textcolor{black}{Let $u_L^\mu$ be the
unit vortex-line four-velocity.} For an ordered array, a loop transported with
$V^\mu\parallel u_L^\mu$ preserves its circulation because
$\Omega_{\mu\nu}u_L^\nu=0$. Vortices crossing a fixed spatial loop
instead change its enclosed winding number.

\subsubsection{Two-component description and common frame}

In the inner crust, the assignment of neutrons to free or confined
baryons additionally depends on the chemical gauge~\cite{Carter:2004he,Gavassino:2020zfk}.
Neutrons carried with the nuclei can be assigned to the carrier current
without introducing another independent velocity. 
{{This change of constituent basis preserves the total baryon
current, the carrier velocity, the condensate momentum $\xi_\mu$ and
the total stress--energy tensor when the constitutive relations are
transformed consistently~\cite{Gavassino:2020zfk}.
The observer frame independently specifies the temporal and spatial
projections of these currents.}}
For separately conserved particle constituents, we restrict these basis
changes to constant chemical-gauge parameters, see also
Sec.~\ref{sec:theory_force_specialization}.
To resolve both currents in one frame, construct
\begin{equation}
 K^\mu=a_sN_s^\mu+a_cN_c^\mu=\mathcal N_KU^\mu,
 \qquad \mathcal N_K=\sqrt{-K_\mu K^\mu}\,,
 \label{eq:moments_weighted_frame}
\end{equation}
{The weights $a_s$ and $a_c$ select a future-directed timelike
current combination in a fixed constituent basis. Its normalized
four-velocity $U^\mu$, with $U^\mu U_\mu=-1$, defines the common observer.}
Choosing $o^\mu = U^\mu$ in the language of the 14-moment fluid gives, 
\begin{equation}
 N_x^\mu=N_x^{(U)}U^\mu+j_x^\mu,\qquad
 U\cdot j_x=0,\qquad a_sj_s^\mu+a_cj_c^\mu=0.
 \label{eq:moments_common_currents}
\end{equation}

The canonical momentum is resolved in the same frame,
\begin{equation}
 \xi_\mu=\mu_UU_\mu+w_\mu,\qquad
 U^\mu w_\mu=0,\qquad \mu_U=-U^\mu\xi_\mu\,,
 \label{eq:moments_josephson_matching}
\end{equation}
where, as a constitutive relation, thermodynamic matching identifies
$\mu_U$ with the EOS chemical potential. The resulting Josephson condition, $U^\mu\xi_\mu=-\mu_U$, fixes the phase evolution along that
flow~\cite{Son:2000ht,Gusakov:2007px}.

{{We adopt the physical neutron--proton basis and choose
$a_s=0$ and $a_c=1$ in Eq.~\eqref{eq:moments_weighted_frame}.
The common observer is then the proton four-velocity,
$U^\mu=u_c^\mu\equiv u^\mu$.
We write $\mu_U=\mu$ and $w^\mu=q^\mu$.}} Then
$\Delta_u^{\mu\nu}=g^{\mu\nu}+u^\mu u^\nu$ projects
orthogonally to the carrier, and
$q^\mu=\Delta_u^{\mu\nu}\xi_\nu$ is the reversible counterflow (i.e., not aligned with $u^\mu$)
momentum used in the subsequent equations, not to be confused with $Q^\mu$ in
Eq.~\eqref{eq:moments_observer_decomposition}, which denotes energy flux.

\textcolor{black}{For Eulerian projections, write
$u^\mu=W(n^\mu+v_c^\mu)$ and
$u_n^\mu=W_n(n^\mu+v_n^\mu)$, with $n\cdot v_x=0$.
The spatial four-velocities are $z^i=Wv_c^i$ and
$z_n^i=W_nv_n^i$, so $W=\sqrt{1+\gamma_{ij}z^iz^j}$ and
$W_n=\sqrt{1+\gamma_{ij}z_n^iz_n^j}$. We also use $z_c=z$ and
$W_c=W$. On the timelike phase branch, the invariant norm is
$\varepsilon_\xi=\sqrt{-\xi_\mu\xi^\mu}$ and the Eulerian phase
energy is $\Xi=-n^\mu\xi_\mu
=\sqrt{\varepsilon_\xi^2+\gamma^{ij}\xi_i\xi_j}$.}

The material stress has the same moment form and may be written
schematically as
\begin{equation*}
 T^{\mu\nu}=T_c^{\mu\nu}+T_s^{\mu\nu}+\Delta T^{\mu\nu},
\end{equation*}
with $\Delta T^{\mu\nu}$ collecting interaction contributions.
Its exact combined form, including entrainment, must follow from
thermodynamic consistency, which we impose using Carter's construction
in Sec.~\ref{sec:moment_carter_closure}, and will in general have the 14-moment form of Eq.~\eqref{eq:moments_observer_decomposition}\footnote{Even a carrier that is perfect in its own rest frame acquires energy
flux and anisotropic stress in a generic common frame. Take
$T_c^{\mu\nu}=(e_c+P_c)u_c^\mu u_c^\nu+P_cg^{\mu\nu}$ and write
$u_c^\mu=\Gamma_c(U^\mu+v_c^\mu)$, with $U\cdot v_c=0$ and
$\Gamma_c=-U\cdot u_c$. Defining
$\mathcal W_c=(e_c+P_c)\Gamma_c^2$, its contributions to
Eq.~\eqref{eq:moments_observer_decomposition} are
\begin{align*}
 e_U^{(c)}&=\mathcal W_c-P_c,
 &p_{\rm iso}^{(c)}&=P_c+\frac{\mathcal W_cv_c^2}{3},\\
 Q_{(c)}^\mu&=\mathcal W_cv_c^\mu,\\
 \pi_{(c)}^{\mu\nu}&=\mathcal W_c
 \left(v_c^\mu v_c^\nu-\frac{v_c^2}{3}\Delta_U^{\mu\nu}\right),
\end{align*}
where $\Delta_U^{\mu\nu}=g^{\mu\nu}+U^\mu U^\nu$.
These are reversible frame effects. $Q_{(c)}^\mu$ and
$\pi_{(c)}^{\mu\nu}$ vanish for $U^\mu=u_c^\mu$.}.

\subsubsection{Vortex interactions and angular momentum transport}
\label{sec:moment_vortex_closure}

{We model the dynamics of the averaged vortex lines through
force balance with the generalized Iordanskii force
\cite{Gavassino:2020zfk}. In this section we discuss primarily forces acting on vortex lines, while {{Sec.~\ref{sec:model_b_vortex_tension}}} introduces vortex tension and line energy.}
The force has three principal channels
\cite{Andersson:2005rs}.
The Magnus force is the transverse hydrodynamic reaction to constituent
flow across quantized circulation, analogous to aerodynamic lift for
relative line motion through the superfluid
\cite{Carter:2000ty,Gavassino:2020zfk}.
{Generalized Iordanskii lift denotes an additional transverse
carrier-current force whose coefficient depends on the constituent
basis and circulation prescription~\cite{Gavassino:2020zfk}.
In thermal-excitation models, an Iordanskii contribution can arise from
asymmetric phonon scattering, with its relation to the total lift
depending on the dynamical regime~\cite{Stone:2000zz,Gavassino:2020zfk}.
For a cold neutron vortex with attached proton fluxoids, the additional
carrier lift instead includes the proton Magnus reaction associated
with their circulation~\cite{Sourie:2020drw,Sourie:2021vsj}.}
Longitudinal drag comes from momentum transfer to microscopic degrees of
freedom: in the core, electrons scatter from the electromagnetic field
of neutron vortices, generated by entrained proton currents and, for
appropriate neutron pairing states, intrinsic vortex magnetization
\cite{Alpar:1984zz,Andersson:2005rs,Haskell:2017lkl}.
Additional crustal channels, not modeled explicitly here, include nuclear
encounters that excite lattice phonons and short-wave
Kelvin oscillations of the vortex~\cite{Haskell:2017lkl,1992ApJ...387..276E}.
In a type-II proton superconductor, cutting magnetic flux tubes likewise
excites Kelvin waves and produces nonlinear drag
\cite{Haskell:2013hja}.
{Static pinning to nuclei or flux tubes, thermally or
quantum-activated creep, and lattice shear elasticity require additional
holding, transport, or elastic prescriptions}
\cite{Wlazlowski:2016yoe,Haskell:2015jra,Gusakov:2016eom}.

These processes enter through averaged force prescriptions for the
resolved canonical vorticity. The ordered-array description below fixes
the local line geometry and its transverse motion.

A vortex worldsheet is the two-dimensional timelike surface swept out
by a moving vortex line. Its tangent plane is spanned by a unit line
velocity $u_L^\mu$ and an oriented unit tangent $\ell^\mu$ along the
line. \textcolor{black}{The spacetime volume form is oriented by
$\epsilon_{0123}=+\sqrt{-g}$.} For a regular ordered array, the coarse-grained canonical
vorticity is its normal area form~\cite{Gavassino:2020zfk}:
\begin{align}
 \Omega_{\mu\nu}&=b\epsilon_{\mu\nu\rho\sigma}u_L^\rho\ell^\sigma,
 &b&=\kappa_v\mathcal N>0,\notag\\
 u_L^2&=-1,\qquad\ell^2=1,
 &u_L\cdot\ell&=0,
 \label{eq:moments_ordered_worldsheet}
\end{align}
Here $\mathcal N$ is the proper line density.
{The force law selects the transverse line motion and hence
the transport of quantized circulation.}

\begin{figure*}[tbp]
 \centering
 \includegraphics[width=\textwidth]{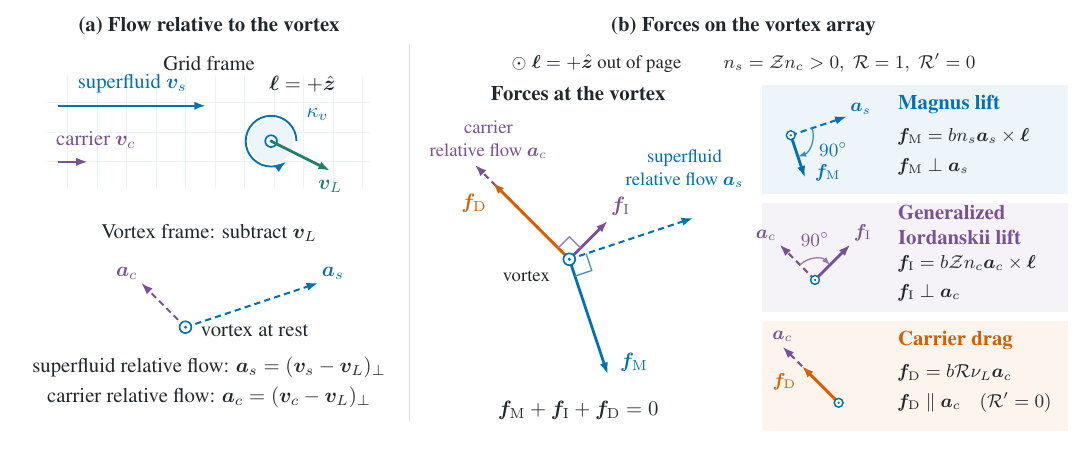}
 \caption{{Hydrodynamic vortex-force geometry in the local low-speed
 limit, with the line direction $\boldsymbol\ell=+\hat{\boldsymbol z}$ out of the page.
 (a)~Subtracting $\boldsymbol v_L$ from the grid-frame velocities
 (upper view) gives the flows past the moving vortex (lower view),
 $\boldsymbol a_s$ and $\boldsymbol a_c$; $\perp$ retains motion across
 the line, with the loop denoting canonical circulation. (b)~The left part shows all three forces
 on the vortex array together with their incident relative flows. The
 individual diagrams (right) separate Magnus lift, {generalized Iordanskii lift}, and
 carrier drag. For the positive lift weights shown, each lift points
 $90^\circ$ clockwise from its own relative flow, as indicated by the
 right-angle marks and curved arrows. Drag follows the carrier relative
 flow when $\mathcal R'=0$. \textcolor{black}{Here
 $\boldsymbol a_x=(\boldsymbol v_x-\boldsymbol v_L)_\perp$,
 $n_s$ and $n_c$ are the constituent number densities,
 $\mathcal Z$ is the carrier lift weight, and $\mathcal R$ and
 $\mathcal R'$ are the dimensionless drag coefficients.}
 The illustrated velocities, with
 $n_s=\mathcal Z n_c>0$, $\mathcal R=1$, and $\mathcal R'=0$, give
 $\boldsymbol f_{\rm M}+\boldsymbol f_{\rm I}+\boldsymbol f_{\rm D}=0$.
 This balance selects $\boldsymbol v_L$ through
 Eq.~\eqref{eq:moments_newtonian_forces}.}}
 \label{fig:vortex_microphysics}
\end{figure*}

We introduce the total lift current, $L^\mu$, and the projector, $\perp^\mu{}_{\nu}$, normal to the worldsheet,
\begin{align}
 L^\mu&=N_s^\mu+\mathcal ZN_c^\mu,
 &{\nu_L}& {=-u_\mu L^\mu},\notag\\
 \perp^\mu{}_{\nu}
 &=\delta^\mu{}_{\nu}+u_L^\mu u_{L\nu}-\ell^\mu\ell_\nu.
 \label{eq:moments_lift_current}
\end{align}
{The force-balanced branch below takes $\nu_L>0$.}
{Neglecting line inertia, tension, and external holding forces, the
covariant force balance is~\cite{Gavassino:2020zfk}}
\begin{align}
 \underbrace{N_s^\rho\Omega_{\rho\nu}}_{\rm Magnus}
 +\underbrace{\mathcal ZN_c^\rho\Omega_{\rho\nu}}_{\text{{generalized Iordanskii lift}}}
 &= \underbrace{b\perp^\rho{}_{\nu}d_\rho}_{Drag} ,\notag\\
 d_\rho&=\mathcal R\nu_Lu_\rho + \mathcal R'N_{s\rho}.
 \label{eq:moments_vortex_force}
\end{align} 
{Here $\mathcal Z$ specifies the generalized Iordanskii carrier lift}, while
$\mathcal R$ and $\mathcal R'$ normalize longitudinal carrier and
superfluid-constituent drag.

{
In the local low-speed limit, these dimensionless coefficients compare
vortex drag relaxation with the transverse lift response:
\begin{equation}
 \mathcal R=\frac{\tau_{\rm lift}}{\tau_c},\qquad
 \mathcal R'=\frac{\nu_L}{n_s}\frac{\tau_{\rm lift}}{\tau_s}.
 \label{eq:moments_drag_time_ratios}
\end{equation}
Here $n_s$ is the superfluid-constituent number density, and the local
vortex response times $\tau_c$, $\tau_s$, and $\tau_{\rm lift}$ are
defined below in Eq.~\eqref{eq:moments_vortex_response_times}.
}

{It is instructive to connect these expressions to their Newtonian
limit.} At low speeds in a local inertial frame, let $n_s,n_c$ be constituent
number densities, $n_v\simeq\mathcal N$ the vortex density, and
$\boldsymbol a_x=(\boldsymbol v_x-\boldsymbol v_L)_\perp$ the slip
normal to the line. Then $b=\kappa_vn_v$,
$\nu_L=n_s+\mathcal Z n_c$, and the forces per volume on the array are
\begin{align}
 \boldsymbol f_{\rm M}&=bn_s\boldsymbol a_s\times\boldsymbol\ell,\notag\\
 \boldsymbol f_{\rm I}&=b\mathcal Z n_c\boldsymbol a_c\times\boldsymbol\ell,\notag\\
 \boldsymbol f_{\rm D}
 &=b(\mathcal R'n_s\boldsymbol a_s+\mathcal R\nu_L\boldsymbol a_c),\notag\\
 \boldsymbol f_{\rm M}+\boldsymbol f_{\rm I}
                          +\boldsymbol f_{\rm D}&=0\,,
 \label{eq:moments_newtonian_forces}
\end{align}
where $\boldsymbol f_{\rm M}$ is the Magnus, $\boldsymbol f_{\rm I}$ the Iordanskii, and $\boldsymbol f_{\rm D}$ the drag force density.
{Figure~\ref{fig:vortex_microphysics}(b) shows these forces on the
array at a common origin (left) and separates the three force channels
(right), in the same transverse view as panel~(a).
Each lift is perpendicular to its own
incident slip, whereas each drag contribution is parallel to that slip.
Changing $\boldsymbol v_L$ changes both slips, so zero total force selects
the velocity rather than requiring the individual forces to vanish.
In this local limit, the plotted lift forces are minus the spatial
current--vorticity contractions on the left of
Eq.~\eqref{eq:moments_vortex_force}.}
{Here} $\boldsymbol v_s$ is the constituent-current velocity, rather than
$\boldsymbol v_{\rm phase}$. Restoring the neutron
mass $m_n$, the Iordanskii term is
$\textcolor{black}{n_v\kappa_{\rm vel}\rho_I\boldsymbol a_c\times\boldsymbol\ell}$, with
$\rho_I=m_n\mathcal Z n_c$ and $\textcolor{black}{\kappa_{\rm vel}=\kappa_v/m_n}$.
The quantity $\rho_I$ is a basis-dependent lift weight, which can have
either sign. The drag coefficients per unit line length are
\begingroup\color{black}
\begin{equation}
 D_s=\kappa_{\rm vel}(m_nn_s)\mathcal R',\qquad
 D_c=\kappa_{\rm vel}(m_n\nu_L)\mathcal R.
 \label{eq:moments_drag_normalization}
\end{equation}
\endgroup

{
To define these response times, consider transverse motion of a straight
vortex with fixed, uniform background fluid velocities and introduce an
auxiliary positive inertial mass per unit line length $\mu_v$. The
per-line drag coefficients in Eq.~\eqref{eq:moments_drag_normalization}
give
\begingroup\color{black}
\begin{equation}
 \tau_c=\frac{\mu_v}{D_c},\qquad
 \tau_s=\frac{\mu_v}{D_s},\qquad
 \tau_{\rm lift}=\frac{\mu_v}{\kappa_{\rm vel} m_n\nu_L}.
 \label{eq:moments_vortex_response_times}
\end{equation}
\endgroup
The corresponding auxiliary Newtonian evolution equation is
$\dot{\boldsymbol v}_{L\perp}
=-(\boldsymbol v_L-\boldsymbol v_c)_\perp/\tau_c
-(\boldsymbol v_L-\boldsymbol v_s)_\perp/\tau_s
+\boldsymbol\ell\times(\boldsymbol v_L-\boldsymbol v_{\rm lift})_\perp/\tau_{\rm lift}$,
where the dot denotes a time derivative along the vortex trajectory.
A vanishing drag
coefficient corresponds to an infinite drag time. Lift alone instead
turns the vortex velocity relative to the lift-weighted background
\begin{equation}
 \boldsymbol v_{{\rm lift}\perp}
 =\frac{n_s\boldsymbol v_{s\perp}
       +\mathcal Z n_c\boldsymbol v_{c\perp}}{\nu_L}
 \label{eq:moments_lift_weighted_velocity}
\end{equation}
at angular frequency $1/\tau_{\rm lift}$, without damping that relative
velocity. Thus $\tau_{\rm lift}$ is a turning time, with full period
$2\pi\tau_{\rm lift}$, rather than a relaxation time. These are local
vortex response times; bulk counterflow equilibration also depends on
the fluid inertia and vortex density. The algebraic limit
retains only the ratios in Eq.~\eqref{eq:moments_drag_time_ratios},
which are independent of the auxiliary $\mu_v$.
}

\subsubsection{Circulation transport and electromagnetic analogy}
\label{sec:moment_electromagnetic_duality}

It is useful both for physical interpretation and for our later numerical implementation to connect circulation transport to relativistic electrodynamics (see Sec.~{\ref{sec:numerical_ct}}).
As an antisymmetric tensor constructed from a vector potential, $\Omega_{\mu\nu}$ naturally admits an electromagnetic decomposition.
For an observer $o^\mu$ we can define
\begin{align}
 E_\mu&=\Omega_{\mu\nu}o^\nu,\notag\\
 B^\mu&=-\frac12\epsilon^{\mu\nu\rho\sigma}o_\nu\Omega_{\rho\sigma},\notag\\
 \Omega_{\mu\nu}
 &=o_\mu E_\nu-o_\nu E_\mu
   +\epsilon_{\mu\nu\rho\sigma}o^\rho B^\sigma.
 \label{eq:moments_vorticity_fields}
\end{align}
{In 3+1 coordinates,
$\Omega_{ij}=\epsilon_{ijk}B^k$ and
$\Omega_{0i}=-\alpha E_i-\beta^j\Omega_{ij}$.
Here $\epsilon_{ijk}=\sqrt\gamma[ijk]$ and the homogeneous identity gives}
\begin{align}
 \partial_i(\sqrt\gamma B^i)&=0,\notag\\
 \partial_t(\sqrt\gamma B^i)
 &+\partial_j\!\left[\sqrt\gamma\{
 \alpha\epsilon^{ijk}E_k+\beta^iB^j-\beta^jB^i\}\right]=0.
 \label{eq:moments_faraday_identity}
\end{align}
{The vorticity of an ordered array vanishes along its timelike
worldsheet,}
\begin{align}
 \Omega_{\mu\nu}u_L^\nu&=0,
 &\boldsymbol E&=-\boldsymbol v_L\times\boldsymbol B,\notag\\
 \boldsymbol E\cdot\boldsymbol B&=0,
 &B^2-E^2&=b^2.
 \label{eq:moments_line_transport}
\end{align}
Here $\boldsymbol B$ represents vortex flux and $\boldsymbol E$
transports circulation. In a right-handed local orthonormal frame of
$o^\mu$, Eq.~\eqref{eq:moments_line_transport} fixes the vortex velocity
orthogonal to $\boldsymbol B$ exactly:
\begin{equation}
 \boldsymbol v_{L\perp}
 =\frac{\boldsymbol E\times\boldsymbol B}{B^2},\qquad
 v_{L\perp}^2=\frac{E^2}{B^2}<1.
 \label{eq:moments_vortex_drift_velocity}
\end{equation}
This is the same drift speed expression as in relativistic force-free electrodynamics ~\cite{Gralla:2014yja}.
{At force equilibrium,
Eq.~\eqref{eq:moments_vortex_force} imposes zero net force on the
array.} The Magnus, Iordanskii, and drag terms can
be individually finite, as can the constituent force
$N_s^\rho\Omega_{\rho\nu}$ and the \textcolor{black}{dissipative work}.
When projected drag vanishes, the balance reduces to
$L^\rho\Omega_{\rho\nu}=0$, formally resembling the electromagnetic
force-free condition $F^{\rm em}_{\mu\nu}j_{\rm em}^\nu=0$
\cite{Gralla:2014yja}\textcolor{black}{, where $F^{\rm em}_{\mu\nu}$ is the
electromagnetic field tensor and $j_{\rm em}^\mu$ the electric current}.
Here $L^\mu$ is a material lift
current, and the bulk matter retains its pressure and inertia.
Thus the force-free analogy concerns the superfluid vorticity and
drift geometry only, while the dynamics of the whole system remain those of two coupled fluids. In these variables, the actual force balance reads
\begin{equation}
 -L^0E_i-[ijk]L^jB^k=b\perp^\rho{}_{i}d_\rho.
 \label{eq:moments_electric_balance}
\end{equation}
For $L^0\ne0$ its contributions are
\begin{align}
 E_i&=E_i^{\rm M}+E_i^{\rm I}+E_i^{\rm D},\notag\\
 E_i^{\rm M}&=-\frac{[ijk]N_s^jB^k}{L^0},\notag\\
 E_i^{\rm I}&=-\frac{\mathcal Z[ijk]N_c^jB^k}{L^0},\notag\\
 E_i^{\rm D}&=-\frac b{L^0}
   (d_i+u_{Li}u_L^\rho d_\rho-\ell_i\ell^\rho d_\rho).
 \label{eq:moments_explicit_electric_field}
\end{align}
The total lift enters the denominator and the drag projector depends
on $u_L^\mu$. The force terms must therefore be solved together with
Eq.~\eqref{eq:moments_line_transport} to obtain a single line velocity.

The force balance and kernel condition together determine $u_L^\mu$
algebraically for prescribed local drag coefficients and lift weight.
For $\mathcal R\geq0$ and $\nu_L>0$, define
\begin{align}
 \epsilon_R&=\frac{1}{1+\mathcal R},\qquad0\leq\epsilon_R\leq1,\notag\\
 \widetilde L^\mu&=\frac{\epsilon_R}{\nu_L}L^\mu,\notag\\
 \widetilde d^\mu
 &=(1-\epsilon_R)u^\mu
   +\epsilon_R\mathcal R'\frac{N_s^\mu}{\nu_L}.
 \label{eq:moments_bounded_drag_inputs}
\end{align}
Choose the line parametrization $\boldsymbol v_L\cdot\boldsymbol B=0$,
set $\boldsymbol n=\boldsymbol B/B$, so $\ell^\mu=(0,\boldsymbol n)$,
and project spatial vectors orthogonally to $\boldsymbol n$. Write
$\widetilde K^\mu=(\widetilde L^0,\widetilde{\boldsymbol L}_\perp)$ and
$\widetilde D^\mu=(\widetilde d^0,\widetilde{\boldsymbol d}_\perp)$, with
$C_v=\widetilde D^2-\widetilde K^2$ and
$p_v=-\widetilde K\cdot\widetilde D$.
For $\epsilon_R>0$, retain the future-timelike lift branch
$L^0>|\boldsymbol L_\perp|$. In every evaluation use the paired roots
\begin{align}
 H_v&=\operatorname{hypot}(C_v,2p_v),\qquad
 h_\pm=\sqrt{\frac{H_v\pm C_v}{2}},\notag\\
 (\alpha_v,\lambda_v)
 &=\begin{cases}
 (h_+,p_v/h_+),&C_v\geq0,\\
 (|p_v|/h_-,\operatorname{sgn}_+(p_v)h_-),&C_v<0.
 \end{cases}
 \label{eq:moments_relativistic_line_scalars}
\end{align}
Here $\operatorname{hypot}(x,y)=\sqrt{x^2+y^2}$ denotes the stable norm
operation, and $\operatorname{sgn}_+(p)$ equals $+1$ for $p\geq0$ and
$-1$ otherwise. Only the root used by the selected branch is evaluated.
The pair satisfies $\alpha_v\lambda_v=p_v$ and
$\alpha_v^2-\lambda_v^2=C_v$ without subtractive cancellation.
The general exact four-velocity is
\begin{align}
 \widetilde{\mathcal U}_v^0
 &=\alpha_v\widetilde L^0+\lambda_v\widetilde d^0
   -\boldsymbol n\cdot
       (\widetilde{\boldsymbol L}_\perp\times
        \widetilde{\boldsymbol d}_\perp),\notag\\
 \widetilde{\boldsymbol{\mathcal U}}_v
 &=\alpha_v\widetilde{\boldsymbol L}_\perp
   +\lambda_v\widetilde{\boldsymbol d}_\perp\notag\\
 &\quad+\boldsymbol n\times
       (\widetilde L^0\widetilde{\boldsymbol d}_\perp
        -\widetilde d^0\widetilde{\boldsymbol L}_\perp),\notag\\
 u_L^\mu&=\frac{\widetilde{\mathcal U}_v^\mu}
                    {\lambda_v^2-\widetilde K^2},\qquad
 \boldsymbol v_{L\perp}
 =\frac{\widetilde{\boldsymbol{\mathcal U}}_v}
       {\widetilde{\mathcal U}_v^0}.
 \label{eq:moments_relativistic_line_velocity}
\end{align}
Here $\alpha_v=-u_L\cdot\widetilde K$ and
$\lambda_v=-u_L\cdot\widetilde D$. For $\epsilon_R>0$, the common
rescaling by $\epsilon_R/\nu_L$ cancels between the degree-two numerator
and denominator. This normalized expression therefore solves the full
force law with both drag channels and Iordanskii lift.
Zero drag means $\epsilon_R=1$ and $\mathcal R'=0$,
giving $\boldsymbol v_{L\perp}=\boldsymbol L_\perp/L^0$ with the
selected lift prescription.

The small-velocity expansion recovers the resolved transport fields in the
Newtonian transverse plane. Define
$\beta_s=n_s/\nu_L$, $\boldsymbol w_\perp=(\boldsymbol v_s-\boldsymbol v_c)_\perp$,
and $\boldsymbol j_\perp=n_s\boldsymbol w_\perp$. The two drag
combinations are
\begin{equation}
 \mathcal R_{\rm tot}=\mathcal R+\beta_s\mathcal R',\qquad
 \mathcal R_{\rm heat}=\mathcal R-\frac{\mathcal Z n_c}{\nu_L}\mathcal R'.
 \label{eq:moments_two_channel_drag}
\end{equation}
Solving Eq.~\eqref{eq:moments_newtonian_forces} gives
\begin{align}
 (\boldsymbol v_L-\boldsymbol v_c)_\perp
 &=\frac{\beta_s}{1+\mathcal R_{\rm tot}^2}
 \bigl[(1+\mathcal R_{\rm tot}\mathcal R')\boldsymbol w_\perp\notag\\
 &\hspace{28mm}-\mathcal R_{\rm heat}\boldsymbol\ell\times\boldsymbol w_\perp\bigr].
 \label{eq:moments_newtonian_line_velocity}
\end{align}
At this order $\boldsymbol B=b\boldsymbol\ell$, and the resolved fields are
\begin{align}
 \boldsymbol E^{\rm M}&=-\beta_s\boldsymbol v_s\times\boldsymbol B,\notag\\
 \boldsymbol E^{\rm I}&=-(1-\beta_s)\boldsymbol v_c\times\boldsymbol B,\notag\\
 \boldsymbol E^{\rm D}&=\eta_D^v
  (\boldsymbol j_\perp-\mathcal R_{\rm tot}\boldsymbol\ell\times\boldsymbol j_\perp),\notag\\
 \eta_D^v&=\frac{b\mathcal R_{\rm heat}}
                  {\nu_L(1+\mathcal R_{\rm tot}^2)}.
 \label{eq:moments_solved_electric_contributions}
\end{align}
Combining them gives the non-ideal electric field relative to the carrier,
\begin{align}
 \boldsymbol E_c&=\boldsymbol E+\boldsymbol v_c\times\boldsymbol B\notag\\
 &=\eta_D^v\boldsymbol j_\perp
      +\eta_H^v\boldsymbol\ell\times\boldsymbol j_\perp,\notag\\
 \eta_H^v&=\frac{b(1+\mathcal R_{\rm tot}\mathcal R')}
                    {\nu_L(1+\mathcal R_{\rm tot}^2)},\notag\\
 W_v&=\boldsymbol j_\perp\cdot\boldsymbol E_c
      =\eta_D^v|\boldsymbol j_\perp|^2.
 \label{eq:moments_vortex_ohm_law}
\end{align}
This has the resistive and Hall terms of an MHD Ohm law.
The resistive part converts counterflow energy into heat, while the
Hall part redirects circulation without work. Here $\boldsymbol j_\perp$ is a
relative particle flux. Iordanskii lift enters both responses
through $\nu_L$. For $\mathcal R'=0$, the coefficients become $\eta_D^v=b\mathcal R/[\nu_L(1+\mathcal R^2)]$ and $\eta_H^v=b/[\nu_L(1+\mathcal R^2)]$.
Without drag, vortices follow the lift-weighted flow. Strong carrier
drag instead approaches ideal-MHD-like flux freezing, with the lines
advected by the charged fluid.

{The strong carrier-drag limit reproduces the kinematic condition
of perfect pinning, with no transverse line motion relative to the
carrier~\cite{Andersson:2005rs}. Microscopic binding requires a separate
holding interaction and can occur at finite drag~\cite{Link:2022vum}.
For the drag law in Eq.~\eqref{eq:moments_vortex_force}, the locked
endpoint is $\mathcal R\to\infty$, or $\tau_c\to0$ at fixed auxiliary
line mass and lift response,}
with $\nu_L>0$ and fixed $\mathcal Z$ and $\mathcal R'$. Dividing
Eq.~\eqref{eq:moments_vortex_force} by $b\mathcal R\nu_L$ gives
\begin{align}
 \perp^\rho{}_{\nu}u_\rho
 &=\frac{L^\rho\Omega_{\rho\nu}/b
             -\mathcal R'\perp^\rho{}_{\nu}N_{s\rho}}
            {\mathcal R\nu_L}\longrightarrow0,\notag\\
 \Omega_{\mu\nu}^{\rm pin}u^\nu&=0.
 \label{eq:moments_pinning_limit}
\end{align}
The carrier therefore lies in the vortex worldsheet, so circulation is
advected with the carrier even when the particle currents differ. In a
carrier-adapted line parametrization the limiting representative is
$u_L^\mu=u^\mu$.

{This transport limit does not imply increasing dissipation.
For $\mathcal Z=\mathcal R'=0$, the dissipative and reactive
mutual-friction coefficients are
$\mathcal B=\mathcal R/(1+\mathcal R^2)$ and
$\mathcal B'=\mathcal R^2/(1+\mathcal R^2)$, respectively.
Thus $\mathcal B\to0$ and $\mathcal B'\to1$ as
$\mathcal R\to\infty$: the carrier locks the lines while a reactive
force remains~\cite{Andersson:2005rs}. A holding threshold and release
rule must be supplied separately, as in
Sec.~\ref{sec:unpinning_forces}.}

Setting $\epsilon_R=0$ in the general solution,
Eqs.~\eqref{eq:moments_bounded_drag_inputs}--
\eqref{eq:moments_relativistic_line_velocity}, gives
$\widetilde K=0$ and $\widetilde D=u-(u\cdot\ell)\ell$.
The paired root returns $\alpha_v=0$ and
$\lambda_v=\sqrt{1+(u\cdot\ell)^2}$, hence
\begin{equation}
 u_{L,{\rm pin}}^\mu
 =\frac{u^\mu-(u\cdot\ell)\ell^\mu}
        {\sqrt{1+(u\cdot\ell)^2}}.
 \label{eq:moments_regular_pinned_endpoint}
\end{equation}
This is the normalized transverse representative in the chosen observer
frame and places $u^\mu$ in the same worldsheet.

{Setting $\mathcal Z=\mathcal R=\mathcal R'=0$ gives the
free-line limit, leaving only the Magnus balance
$N_s^\rho\Omega_{\rho\nu}=0$. The analytic solution then has
$\widetilde d^\mu=0$, $\lambda_v=0$, and
$u_L^\mu=\widetilde K^\mu/\sqrt{-\widetilde K^2}$, so
$\boldsymbol v_{L\perp}=\boldsymbol N_{s\perp}/N_s^0
=\boldsymbol v_{n\perp}$. Vortices follow the neutron flow across
the lines.
The resulting field $\boldsymbol E=-\boldsymbol v_n\times\boldsymbol B$
transports canonical circulation, while the resolved fluid dynamics
supplies the Kelvin bending response tested in
Sec.~\ref{sec:results-kelvin}.}

In analogy with the resistivity in MHD, the dissipation of counterflow energy leads to heat that needs to be absorbed by the fluid.
Let $T_c$ and $s_c$ denote the local carrier temperature and entropy
density. \textcolor{black}{The partial mobility $f^2$ relates the carrier-frame
particle flux to the counterflow momentum through
$\Delta_u^\mu{}_\nu N_s^\nu=f^2q^\mu$. Its thermodynamic closure is
derived in Sec.~\ref{sec:moment_carter_closure}.}
The constituent forces exchange momentum internally and cancel
in the total matter balance. In analogy to MHD \cite{1957PhRv..108.1357H}, the released heat changes the entropy by
\begin{align}
 \nabla_\mu(s_cu^\mu)&=\frac{1}{T_c}f^2q^\mu\Omega_{\mu\nu}u^\nu,
 \label{eq:force_balanced_internal_vortex_work}
\end{align}

\subsection{Carter multifluid and thermodynamic consistency}
\label{sec:moment_carter_closure}

The preceding current and vortex equations specify kinematics and
momentum exchange. Their thermodynamic closure requires a common
generator for the momenta, stress, and equation of state.
{The following relations define the material master function
$\Lambda_0$ and its stress $T_0^{\mu\nu}$. The vortex line contribution is introduced separately in
Sec.~\ref{sec:model_b_vortex_tension}.} We adopt
Carter--Khalatnikov theory~\cite{1992AnPhy.219..243C}, see also
Ref.~\cite{Andersson:2006nr}.

Following Refs.~\cite{Carter:1992gmy,Gavassino:2019wzd}, we introduce a master function
$\Lambda=\Lambda(I_{ss},I_{cc},I_{sc})$, where
$I_{xy}=-g_{\mu\nu}N_x^\mu N_y^\nu$ and
{$\nabla_\mu N_x^\mu=0$ conserves the number of particles of each
constituent. When one current carries entropy instead, its production
is governed by {Eq.~\eqref{eq:adm_entropy_work}}.}
The three $I_{xy}$ exhaust the algebraic scalar products of the two
currents. At fixed metric, the differential
$d\Lambda=\xi_\mu dN_s^\mu+\mu^c_\mu dN_c^\mu$ defines their canonical
momenta, $\xi_\mu$, including the carrier momentum $\mu^c_\mu$:

\begin{align}
 \xi_\mu&=\mathcal B_sN_{s\mu}+\mathcal A N_{c\mu},
 &\mu^c_\mu&=\mathcal A N_{s\mu}+\mathcal B_cN_{c\mu},\notag\\
 \mathcal B_x&=-2\frac{\partial\Lambda}{\partial I_{xx}},
 &\mathcal A&=-\frac{\partial\Lambda}{\partial I_{sc}}.
 \label{eq:moments_carter_map}
\end{align}
By construction of $\Lambda$, each momentum is a function of both number currents. We can then define a generalized pressure $\Psi$\footnote{{The generalized pressure is the thermodynamic pressure, $\Psi=P$. This can be seen by considering a homogeneous equilibrium state, for which the grand potential is $\Omega_{\rm grand}=-PV=-\Psi V$, see Ref.~\cite[Eqs.~(14) and (17)]{Gavassino:2025ldj}.}}
\begin{align}
 \Psi&=\Lambda-N_s^\mu\xi_\mu-N_c^\mu\mu^c_\mu,\notag\\
 d\Psi&=-N_s^\mu d\xi_\mu-N_c^\mu d\mu^c_\mu,\notag\\
 T^{\mu\nu}
 &=\Psi g^{\mu\nu}+N_s^\mu\xi^\nu+N_c^\mu\mu_c^\nu.
 \label{eq:moments_carter_pressure_stress}
\end{align}
The common cross coefficient makes the material stress symmetric.
{This stress contains both fluid contributions.}
Conserved-current variations yield $N_x^\rho(2\nabla_{[\rho}\mu^x_{\nu]})=f_{x\nu}$,
recovering a Carter--Lichnerowicz-type relation, Eq.~\eqref{eq:moments_superfluid_vorticity}. Internal forces {exchange momentum with $\sum_xf_{x\nu}=0$ in the
force-equilibrium approximation and are not added again to total
stress--energy conservation.}

The closure can now be projected onto the original moment variables.
In any common frame, write
$\mu^x_\nu=\mu_x^{(U)}U_\nu+p^x_\nu$ with $U\cdot p^x=0$.
In line with relativistic two-fluid reductions \cite{Most:2021uck}, Equation~\eqref{eq:moments_carter_pressure_stress} can be related to a 14-moment system by
\begin{align}
 e_U&=-\Psi+\sum_x\mu_x^{(U)}N_x^{(U)},\notag\\
 Q^\mu&=\sum_x\mu_x^{(U)}j_x^\mu
                 =\sum_xN_x^{(U)}p_x^\mu,\notag\\
 p_{{\rm iso}}&=\Psi+\frac13\sum_xj_x^\mu p^x_\mu,\notag\\
 \pi^{\mu\nu}&=\sum_xj_x^{\langle\mu}p_x^{\nu\rangle},\notag\\
 de_U&=\sum_x\mu_x^{(U)}dN_x^{(U)}+\sum_xj_x^\mu dp^x_\mu.
 \label{eq:moments_carter_projections}
\end{align}
The differential holds at fixed observer and metric, and angle brackets
denote the spatial symmetric trace-free projection when applied to two
indices. {Thus the thermodynamic pressure $\Psi=P$ need not
equal the observer-dependent spatial mean stress $p_{\rm iso}$}, and
$e_U=-\Lambda+\sum_xj_x\cdot p_x$ need not equal $-\Lambda$.
For example, equal weights in Eq.~\eqref{eq:moments_weighted_frame}
give $j_s^\mu=-j_c^\mu=j^\mu$, so
$Q_{\rm rev}^\mu=(\mu_s^{(U)}-\mu_c^{(U)})j^\mu$ and
$\pi_{\rm rev}^{\mu\nu}=(\mathcal B_s+\mathcal B_c-2\mathcal A)
j^{\langle\mu}j^{\nu\rangle}$. We stress that these non-ideal degrees of freedom are just a statement about the two-fluid character of the theory expressed in a single-fluid frame \cite{Most:2021uck}.

In the carrier/proton frame $N_c^\mu=n_cu^\mu$, let
$f^2=\mathcal B_s^{-1}$ on a regular constitutive branch. Inverting
Eq.~\eqref{eq:moments_carter_map} gives
\begin{align}
 N_s^\mu&=f^2(\xi^\mu-\mathcal A n_cu^\mu),\notag\\
 T^{\mu\nu}
 &=Cu^\mu u^\nu+\Psi g^{\mu\nu}+f^2\xi^\mu\xi^\nu,\notag\\
 C&=(\mathcal B_c-\mathcal A^2f^2)n_c^2.
 \label{eq:moments_carrier_stress}
\end{align}
Its energy density and spatial moments are therefore
\begin{align}
 e_u&=C-\Psi+f^2\mu^2,
 &Q^\mu&=\mu f^2q^\mu,\notag\\
 p_{{\rm iso}}&=\Psi+\frac13f^2q_\mu q^\mu,
 &\pi^{\mu\nu}&=f^2q^{\langle\mu}q^{\nu\rangle}.
 \label{eq:moments_reversible_counterflow}
\end{align}
These are the reversible contributions to the moment decomposition of
Sec.~\ref{sec:moment_starting_point}. To use them as a closure, we must
express the currents and their response coefficients in a common set of
thermodynamic variables. The Gusakov--Son representation below supplies this scalar matching.

\subsubsection{Entrainment matrix and relativistic inertia}
\label{sec:entrainment_mass_properties}

\begin{figure}[tbp]
 \centering
 \includegraphics[width=\columnwidth]{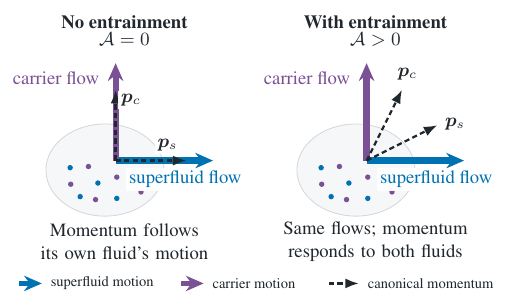}
 \caption{The same two particle flows with and without entrainment.
 Blue and {purple} arrows show the spatial particle currents measured
 by one local orthonormal observer; black dashed arrows show the
 corresponding canonical momenta $\boldsymbol p_s$ and
 $\boldsymbol p_c$. For $\mathcal A=0$, each momentum follows its own
 current. With entrainment, each also responds to the other current
 through the off-diagonal terms in Eq.~\eqref{eq:entrainment_matrix_dictionary}.
 The right panel illustrates $\mathcal A>0$ with nonparallel flows.}
 \label{fig:entrainment_matrix_sketch}
\end{figure}

Entrainment specifies the inertia of relative motion. As we can see in Eq.~\eqref{eq:moments_carter_map}, the canonical momenta can depend linearly on both
conserved currents, i.e.,
\begin{align}
 \mu^x_\nu&=\mathsf K_{xy}N_{y\nu},\qquad \mu^s_\nu=\xi_\nu,\notag\\
 \mathsf K&=\begin{pmatrix}\mathcal B_s&\mathcal A\\
                           \mathcal A&\mathcal B_c\end{pmatrix},
 &\mathsf Y&=\mathsf K^{-1},\notag\\
 \Delta_K&=\mathcal B_s\mathcal B_c-\mathcal A^2.
 \label{eq:entrainment_matrix_dictionary}
\end{align}
Figure~\ref{fig:entrainment_matrix_sketch} illustrates this mixing of
current directions into canonical momenta.
Here $x,y\in\{s,c\}$ label constituents. Symmetry follows from the
common master function, while the coefficients depend on its density,
composition, and relative-flow arguments~\cite{Gavassino:2019wzd,Andersson:2021vad}.
At a specified state, $\mathsf Y$ is the algebraic mobility obtained
when both momenta replace both currents. The scalar $Y=f^2$ in the preceding subsection instead
comes from the partial inversion retaining $N_c^\mu$:
\begin{equation}
 f^2=\frac1{\mathcal B_s}
 =\mathsf Y^{ss}-\frac{(\mathsf Y^{sc})^2}{\mathsf Y^{cc}},
 \qquad C=\frac{n_c^2\Delta_K}{\mathcal B_s}.
 \label{eq:entrainment_partial_mobility}
\end{equation}

The relativistic mass interpretation follows directly from the
current--momentum relation. Write $N_s^\mu=\nu_su_s^\mu$ and
$N_c^\mu=n_cu^\mu$, where $\nu_s,n_c$ are proper densities, and define
$\Gamma_r=-u_s\cdot u$ and the constituent-rest-frame chemical
potentials $\widehat\mu_s=-u_s\cdot\xi$,
$\widehat\mu_c=-u\cdot\mu^c$. The momenta take the covariant form
\begin{align}
 \xi_\mu&=m_s^*u_{s\mu}+\mathcal A n_cu_\mu,\notag\\
 \mu^c_\mu&=m_c^*u_\mu+\mathcal A\nu_su_{s\mu},\notag\\
 m_s^*&=\nu_s\mathcal B_s
       =\widehat\mu_s-\mathcal A n_c\Gamma_r,\notag\\
 m_c^*&=n_c\mathcal B_c
       =\widehat\mu_c-\mathcal A\nu_s\Gamma_r.
 \label{eq:entrainment_effective_masses}
\end{align}
These effective masses are state-dependent Lorentz scalars measuring
each momentum's coefficient along its own constituent velocity. They
have units of energy per particle. Even without entrainment,
$m_x^*=\widehat\mu_x$ includes the interaction and enthalpy contribution.

In the carrier frame, let
$u_s^\mu=\Gamma_r(u^\mu+v_s^\mu)$, with $u\cdot v_s=0$ and
$\Gamma_r=(1-v_s^2)^{-1/2}$. Projection gives the exact relations
\begin{align}
 \mu&=\Gamma_rm_s^*+\mathcal A n_c,\notag\\
 q^\mu&=\Gamma_rm_s^*v_s^\mu,
 &\Delta_u^\mu{}_{\nu}N_s^\nu&=f^2q^\mu,\notag\\
 f^2&=\frac{\nu_s}{m_s^*}.
 \label{eq:entrainment_carrier_relative_mass}
\end{align}
Thus $f^2$ is the relativistic mobility relative to the carrier: a larger
$m_s^*$ at fixed assigned density gives less relative particle flux
for a specified spatial phase momentum.

Finally, in the cold, rest-mass-dominated Newtonian limit,
$\widehat\mu_x\to m_x$ and $\Gamma_r\to1$. The relativistic relations
then recover
\begin{align}
 m_s^*&\longrightarrow m_s-\mathcal A n_c,\notag\\
 C&\longrightarrow\rho_{\rm tot}-m_s^2f^2,\notag\\
 \rho_{\rm tot}&=\sum_xm_xn_x\,.
 \label{eq:entrainment_newtonian_inertia}
\end{align}

\subsubsection{Gusakov--Son variables and scalar matching}
\label{sec:moment_gusakov_matching}

We now need to determine the scalar coefficients in the Carter formalism as functions of the generalized pressure, $\Psi$. This can most naturally be done by using Gusakov-Son variables ~\cite{Gusakov:2007px,Son:2000ht,Gavassino:2019wzd}. Choose a current $J^\mu$
conjugate to $\xi_\mu$ and a carrier-aligned current $R^\mu=ru^\mu$
conjugate to $\varpi_\mu$, so that
$d\Psi=-J^\mu d\xi_\mu-R^\mu d\varpi_\mu$ at fixed metric.
The scalar $\vartheta=-u^\mu\varpi_\mu$ is the intensive variable
conjugate to $r$. Its physical meaning depends on the chosen carrier
current. The constitutive models in the next subsection specify those currents and conjugates exactly.

For isotropic local equilibrium, write
$\mathcal P(\vartheta,\mu,X)=\Psi$, with $\mu=-u^\mu\xi_\mu$ and
$X=\xi_\mu\xi^\mu$. The pressure differential and responses are
\begin{align}
 d\mathcal P&=r\,d\vartheta+a\,d\mu-\frac{Y}{2}\,dX,\notag\\
 r&=\left.\frac{\partial\mathcal P}{\partial\vartheta}\right|_{\mu,X},
 &a&=\left.\frac{\partial\mathcal P}{\partial\mu}\right|_{\vartheta,X},\notag\\
 Y&=-2\left.\frac{\partial\mathcal P}{\partial X}\right|_{\vartheta,\mu}=f^2.
 \label{eq:moments_gusakov_thermodynamics}
\end{align}
The scalar dictionary, without any further moment construction, is
\begin{align}
 J^\mu&=au^\mu+Y\xi^\mu,
 &R^\mu&=ru^\mu,\notag\\
 n&=-u_\mu J^\mu=a+\mu Y,
 &C&=\vartheta r+\mu a,\notag\\
 e_u&=-\mathcal P+\vartheta r+\mu n.
 \label{eq:moments_son_form}
\end{align}
Thus $n$ is the carrier-frame density of the selected $J^\mu$, whereas
$a$ is its aligned coefficient in the momentum representation.
Equivalently, using $Q=q_\mu q^\mu=X+\mu^2$ and
$\overline{\mathcal P}(\vartheta,\mu,Q)
=\mathcal P(\vartheta,\mu,Q-\mu^2)$, the scalar first laws read
\begin{align}
 d\overline{\mathcal P}
 &=r\,d\vartheta+n\,d\mu-\frac{Y}{2}\,dQ,\notag\\
 de_u&=\vartheta\,dr+\mu\,dn+\frac{Y}{2}\,dQ.
 \label{eq:moments_gusakov_scalar_matching}
\end{align}
In summary, the pressure and thermodynamic relations are sufficient to fully specify all remaining (scalar) parameters of the Carter multifluid construction, Eq.~\eqref{eq:moments_reversible_counterflow}.

\subsubsection{Explicit models for a two-component superfluid}
\label{sec:theory_model_layers}

We use two constitutive models. The condensate--phonon (CP) model couples a condensate
to thermal excitations and provides a numerical benchmark.
{It captures thermal counterflow and second sound familiar from
laboratory superfluids in a simplified, phonon-dominated
setting~\cite{Carter:1995if}.}
The neutron--charged-fluid (NC) model describes neutron--charged counterflow for neutron-star
applications, with a carrier thermal completion.{}
{Its two material components allow relative motion and entrainment
even in the cold limit, providing a simplified description of
neutron-star differential rotation~\cite{Prix:2004wq}.}
The subscripts A and B on constitutive quantities denote CP and NC,
respectively.

The CP model describes one superfluid particle species with a normal component
of thermal excitations. The carrier follows the entropy current
$R_{\rm A}^\mu=S_{\rm A}^\mu=su^\mu$, while $J_{\rm A}^\mu$ counts
particles. Thus $\vartheta=T$ and $r=s$: the normal component carries
entropy, not a separately conserved excitation number. We use a cold
polytropic condensate pressure and a phenomenological phonon-inspired thermal term~\cite{Carter:1995if,Mannarelli:2008jq}. With particle rest mass set to unity, $\varepsilon_\xi=\sqrt{-X}$, polytropic exponent $\Gamma$,
and positive pressure constant $\kappa_p$, with $\Gamma>1$,
the complete generating pressure is
\begin{align}
 \Psi_{\rm A}&=P_{\rm A}=P_{\rm cold}+P_{\rm ph},\notag\\
 P_{\rm cold}
 &=\left[\frac{\Gamma-1}{\Gamma\kappa_{p}^{1/\Gamma}}
       (\varepsilon_\xi-1)\right]^{\Gamma/(\Gamma-1)},\notag\\
 P_{\rm ph}
 &=\frac{27T^4}{256c_s^3}
   \left(1-\frac{v_{\rm rel}^2}{c_s^2}\right)^{-2},\notag\\
 c_s^2&=\frac{(\Gamma-1)(\varepsilon_\xi-1)}{\varepsilon_\xi},
 &v_{\rm rel}^2&=1-\frac{\varepsilon_\xi^2}{\mu^2}.
 \label{eq:main_model_a_pressure}
\end{align}
Its thermodynamic responses and scalar dictionary are
\begin{align}
 s&=\frac{\partial P_{\rm A}}{\partial T},
 &\rho&=\frac{\partial P_{\rm A}}{\partial\mu},\notag\\
 f^2&=\frac{1}{\varepsilon_\xi}
       \frac{\partial P_{\rm A}}{\partial\varepsilon_\xi},
 &C_{\rm A}&=Ts+\mu\rho,\notag\\
 n_{\rm A}&=\rho+\mu f^2,
 &e_{u,{\rm A}}&=-P_{\rm A}+Ts+\mu n_{\rm A}.
 \label{eq:main_model_a_responses}
\end{align}
Here the derivatives hold the other coordinates in
$(T,\mu,\varepsilon_\xi)$ fixed, and
$(\vartheta,r,a,Y)=(T,s,\rho,f^2)$.
The particle current is $J_{\rm A}^\mu=N_{\rm A}^\mu
=\rho u^\mu+f^2\xi^\mu$, while $R_{\rm A}^\mu=S_{\rm A}^\mu$.
The intended regime is a low-temperature,
phonon-dominated condensate with local equilibration of the excitations
on the resolved scales. The physical branch requires $T>0$,
$\varepsilon_\xi>1$, $\mu>0$, $0\leq v_{\rm rel}^2<c_s^2$, and positive,
finite constitutive responses, including $f^2>0$.

The NC model describes independently moving superfluid neutrons and a charged
carrier. Local charge neutrality fixes the electron density, we additionally assume that protons and electrons share one velocity.
Proton number and total baryon number are conserved when weak reactions
are neglected. In the physical neutron--proton basis,
$n_n^\mu=\nu_nu_n^\mu$ and $n_p^\mu=n_pu^\mu$. We prescribe the
constant-entrainment master function, with{} $K_n,K_p>0$ and
$\Gamma_n,\Gamma_p>1$ \cite{Prix:2004wq},
\begin{align}
 \Lambda_{\rm B}&=-e_0(\nu_n,n_p)-\mathcal A\Delta,\notag\\
 \Delta&=\nu_n n_p(\Gamma_r-1),\qquad
 \Gamma_r=-u_n^\mu u_\mu,\notag\\
 e_0&=m_n\nu_n+m_pn_p\notag\\
 &\quad+\frac{K_n}{\Gamma_n-1}\nu_n^{\Gamma_n}
      +\frac{K_p}{\Gamma_p-1}n_p^{\Gamma_p}.
 \label{eq:main_model_b_master}
\end{align}
Here $m_n,m_p$ are the particle rest masses. This separable
degenerate-matter EOS is linear in the relative-flow invariant
$\Delta=I_{np}-\sqrt{I_{nn}I_{pp}}$, a Lorentz scalar that vanishes at
coflow. In the Newtonian limit,
$\Delta\simeq\nu_n n_p|\boldsymbol v_n-\boldsymbol v_p|^2/2$;
$\mathcal A$ therefore fixes the leading counterflow energy response
\cite{Andersson:2002jd}. Its density-conjugate chemical potentials
$\mu_n=\partial e_0/\partial\nu_n$ and
$\mu_p=\partial e_0/\partial n_p$, and its cold generating pressure, are
\begin{align}
 \mu_n&=m_n+\frac{\Gamma_nK_n}{\Gamma_n-1}
                   \nu_n^{\Gamma_n-1},\notag\\
 \mu_p&=m_p+\frac{\Gamma_pK_p}{\Gamma_p-1}
                   n_p^{\Gamma_p-1},\notag\\
 \Psi_{{\rm B},0}
 &=K_n\nu_n^{\Gamma_n}+K_pn_p^{\Gamma_p}+\mathcal A\Delta.
 \label{eq:main_model_b_pressure}
\end{align}
The exact scalar relations for this chosen master function are
\begin{align}
 H&=\mu_n-\mathcal A n_p,\qquad f^2=\frac{\nu_n}{H},\notag\\
 \mu&=H\Gamma_r+\mathcal A n_p,\qquad Q=H^2(\Gamma_r^2-1),\notag\\
 \chi&=\mu_p+\mathcal A\nu_n(\Gamma_r-1)-\mu,\notag\\
 n_{\rm eff}&=n_p(1-\mathcal A f^2),\qquad
 C_{{\rm B},0}=n_p\chi+\mu n_{\rm eff},\notag\\
 n_{\rm B}&=n_{\rm eff}+\mu f^2=n_p+\nu_n\Gamma_r,\notag\\
 e_{u,{{\rm B},0}}&=-\Psi_{{\rm B},0}+n_p\chi+\mu n_{\rm B}.
 \label{eq:main_model_b_responses}
\end{align}
The current choice is $J_{\rm B}^\mu=n_n^\mu+n_p^\mu$ and
$R_{\rm B}^\mu=n_p^\mu$. \textcolor{black}{The proton momentum is
$\mu^p_\mu\equiv\mu^c_\mu$, giving}
$\varpi_\mu=\mu^p_\mu-\xi_\mu$ and
$\vartheta=\chi=-u^\mu(\mu^p_\mu-\xi_\mu)$.
Consequently $(r,a,Y)=(n_p,n_{\rm eff},f^2)$ and $n_{\rm B}$ is the
carrier-frame total baryon density. Here $X=Q-\mu^2$.
To obtain the pressure at specified $(\chi,\mu,X)$, the inverse map is
explicitly reduced to a scalar root for $n_p$:
\begin{align}
 H(n_p)&=\sqrt{-X-2\mathcal A n_p\mu+\mathcal A^2n_p^2},\notag\\
 \nu_n(n_p)&=\left[\frac{\Gamma_n-1}{\Gamma_nK_n}
           \bigl(H(n_p)+\mathcal A n_p-m_n\bigr)\right]^{1/(\Gamma_n-1)},\notag\\
 \Gamma_r(n_p)&=\frac{\mu-\mathcal A n_p}{H(n_p)},\notag\\
 0&=\mu_p(n_p)+\mathcal A\nu_n(n_p)
                      [\Gamma_r(n_p)-1]-\mu-\chi.
 \label{eq:main_model_b_inverse}
\end{align}
Substituting this root into Eq.~\eqref{eq:main_model_b_pressure} gives
$\Psi_{{\rm B},0}(\chi,\mu,X)$. We select positive densities, $H>0$,
$\Gamma_r\geq1$, and a nonzero Jacobian
$\det[\partial(\chi,\mu,X)/\partial(\nu_n,n_p,\Gamma_r)]$. Its pressure differential is
\begin{equation}
 d\Psi_{{\rm B},0}
 =n_p\,d\chi+n_{\rm eff}\,d\mu-\frac{f^2}{2}\,dX.
 \label{eq:main_model_b_first_law}
\end{equation}
The physical currents are $n_n^\mu=f^2(\xi^\mu-\mathcal A n_pu^\mu)$
and $n_p^\mu=n_pu^\mu$, with $J_{\rm B}^\mu=n_n^\mu+n_p^\mu$ and
$R_{\rm B}^\mu=n_p^\mu$.

To retain the heat generated by vortex-mediated exchange, the thermal NC model
adds an independent entropy density $s_c$ carried by the charged fluid,
$S_c^\mu=s_cu^\mu$. Its production follows
{Eq.~\eqref{eq:adm_entropy_work}}, with the spatial evolution
given in Eq.~\eqref{eq:model_b_thermal_entropy_adm}. Thus entropy adds a
thermodynamic variable within the same two-velocity description. We
assume local thermal equilibrium and omit heat flow relative to $u^\mu$.
The additive thermal completion has $C_{\rm th}>0$ and obeys
\begin{align}
 \Lambda_{{\rm B},T}&=\Lambda_{\rm B}-e_{\rm th},\qquad
 e_{\rm th}=C_{\rm th}s_c^{4/3},\notag\\
 T_c&=\frac43C_{\rm th}s_c^{1/3},\notag\\
 P_{\rm th}&=\frac13C_{\rm th}s_c^{4/3}
             =\frac{27T_c^4}{256C_{\rm th}^3},\notag\\
 \Psi_{{\rm B},T}&=\Psi_{{\rm B},0}+P_{\rm th},\notag\\
 C_{{\rm B},T}&=C_{{\rm B},0}+T_cs_c,\notag\\
 e_{u,{{\rm B},T}}&=e_{u,{{\rm B},0}}+e_{\rm th},\notag\\
 d\Psi_{{\rm B},T}
 &=s_c\,dT_c+n_p\,d\chi+n_{\rm eff}\,d\mu-\frac{f^2}{2}\,dX.
 \label{eq:main_model_b_thermal}
\end{align}
The cold coefficients $(f^2,n_{\rm eff},\chi)$ are unchanged by this
entropy-only thermal term.

The cold model applies to paired,
degenerate matter when thermal pressure is negligible and reactions are
slow compared with the evolution, and the relative-flow expansion applies
while higher-order entrainment terms remain negligible. Its thermal
completion admits carrier
heating while retaining local equilibrium and the same two material
velocities. The{} admissible branch has $\nu_n>0$, $n_p>0$, $H>0$,
$\Gamma_r\geq1$ and the future-timelike phase condition in
Eq.~\eqref{eq:main_model_b_admission_compact}. The pressure-coordinate
map additionally requires $J_S\ne0$ in that equation. 

\subsubsection{Constituent basis and the Thouless prescription}
\label{sec:theory_force_specialization}

One of the remaining parameters of the system concerns the inclusion of the Iordanskii force \cite{Thouless:1996mt}. One option is to use the thermodynamically consistent Thouless form appropriate for slow vortex motion \cite{Thouless:1996mt,Gavassino:2020zfk}. A constant chemical gauge transformation of the form
$N_s'=N_s+a_{\rm ch}N_c$, $N_c'=(1-a_{\rm ch})N_c$ preserves $L^\mu$
when $\mathcal Z'=(\mathcal Z-a_{\rm ch})/(1-a_{\rm ch})$, with
$a_{\rm ch}\ne1$ \cite{Gavassino:2020zfk}.
This reassigns baryons between chemical constituents while keeping the
physical circulation.

For the cold two-particle closure derived in
Sec.~\ref{sec:moment_carter_closure}, $\mathcal B_s$, $\mathcal B_c$,
and $\mathcal A$ are the master-function response coefficients in
Eq.~\eqref{eq:moments_carter_map}, and $f^2=\mathcal B_s^{-1}$ on a
regular branch. Using the current--momentum map and its inversion in
Eq.~\eqref{eq:moments_carrier_stress}, the carrier momentum is
\begin{equation}
 \mu^c_\mu=\mathcal A f^2\xi_\mu
       +(\mathcal B_c-\mathcal A^2f^2)n_cu_\mu.
 \label{eq:moments_carrier_circulation}
\end{equation}
If a locally uniform, thermally relaxed rigid carrier contributes no
singular velocity circulation, the Thouless prescription gives
\cite{Carter:2000ty,Thouless:1996mt,Gavassino:2020zfk}
\begin{equation}
 \mathcal Z_{\rm Th}=\mathcal A f^2,\qquad
 L_{\rm Th}^\mu=f^2\xi^\mu,\qquad
 \nu_{L,\rm Th}=f^2\mu.
 \label{eq:moments_thouless_lift}
\end{equation}
The drag-free lines then follow the phase momentum, with
$\boldsymbol E=-(\boldsymbol\xi/\xi^0)\times\boldsymbol B$.

Their Newtonian lift is
${n_v\kappa_{\rm vel}\rho_s^{\rm phase}(\boldsymbol v_{\rm phase}
-\boldsymbol v_L)\times\boldsymbol\ell}$, where
$\rho_s^{\rm phase}=m_n^2f^2$. This is a circulation-dependent
specialization: taking $\mathcal Z=0$ in an entrained particle basis
would select a different force law, and a driven carrier need not
satisfy the thermalized circulation assumption.

\begingroup
\subsubsection{Vortex forces inside a neutron star}
\label{sec:theory_force_parameterization}

For superconducting, superfluid neutron star matter, explicit vortex force prescriptions have been derived that we can directly connect to our force prescription. A concrete
example is provided by neutron vortices attached to proton fluxoids
in a cold neutron-star core
\cite{Sourie:2020drw,Sourie:2021vsj}.
Fluxoids are the quantized magnetic structures of a type-II proton
superconductor. Each is associated with a winding of the charged
condensate phase and a magnetic flux quantum $\Phi_0=\textcolor{black}{\pi\hbar c/e}$ in
Gaussian units, where $e>0$ is the proton charge. We represent an
unresolved bundle of a neutron vortex and its attached fluxoids by
the timelike worldsheet of Eq.~\eqref{eq:moments_ordered_worldsheet}.
Its motion can differ from the carrier four-velocity $u^\mu$ even
when the bound quantized lines share a common worldsheet motion.

We retain the neutron--proton particle basis and assign a scalar
$N_\Phi\geq0$ to the mean number of attached proton fluxoids per
neutron vortex. The microscopic matching assumes a locally straight,
rigid bundle whose transverse extent is small compared with the
intervortex separation~\cite{Sourie:2020drw}.
Write $N_s^\mu=\nu_su_s^\mu$ and $N_c^\mu=n_cu^\mu$,
with proper densities $\nu_s,n_c$ and relative Lorentz factor
$\Gamma_r=-u_s\cdot u$ as in
Sec.~\ref{sec:entrainment_mass_properties}. The projector
$\perp^\mu{}_{\nu}$ removes both worldsheet-tangent directions.
With $u_{\perp\nu}=\perp^\rho{}_{\nu}u_\rho$, the covariant
force density is~\cite{Sourie:2021vsj}
\begin{align}
 f_{s\nu}&=N_s^\rho\Omega_{\rho\nu}\notag\\
 &=b\nu_s{\mathcal R_n}u_{\perp\nu}
       -N_\Phi N_c^\rho\Omega_{\rho\nu}.
 \label{eq:force_fluxoid_local}
\end{align}
Here {$\mathcal R_n\geq0$}. Both terms are covectors normal to the
vortex worldsheet. The
carrier-vorticity contraction describes the proton Magnus reaction
associated with the attached canonical circulation. The projected
carrier term describes electron scattering from the bundle's magnetic
field. Its coefficient {$\mathcal R_n$} is a dimensionless Lorentz
scalar, with corresponding drag coefficient per unit proper line
length {$D_e=\kappa_v\nu_s\mathcal R_n$}.

Comparison with Eq.~\eqref{eq:moments_vortex_force} fixes
\begin{align}
 \mathcal Z&=N_\Phi,\qquad \mathcal R'=0,\notag\\
 \mathcal R&={\frac{\nu_s\mathcal R_n}{\nu_L}},\qquad
 \nu_L=\Gamma_r\nu_s+N_\Phi n_c.
 \label{eq:force_fluxoid_parameters}
\end{align}
This relation retains the full relative Lorentz factor. The
normalization converts the microscopic drag ratio, measured against
neutron lift, to our drag relative to the total lift current
$L^\mu=N_s^\mu+N_\Phi N_c^\mu$.

The distinction between drag and circulation is explicit in the
carrier-frame work entering
{Eq.~\eqref{eq:force_balanced_internal_vortex_work}}:
\begin{equation}
 u^\nu f_{s\nu}
 =b\nu_s {\mathcal R_n}
 u_{\perp\mu}u_\perp^\mu\geq0.
 \label{eq:force_fluxoid_work}
\end{equation}
The normal space is spacelike, and antisymmetry of $\Omega_{\mu\nu}$
eliminates the proton-circulation term from this contraction
\cite{Sourie:2021vsj}. Thus positive drag supplies heat,
while proton lift changes circulation transport without contributing
to the carrier-frame work. When $\mathcal R_n=0$, the force balance
places $L^\mu$ in the worldsheet kernel. 
When the circulation of attached proton fluxoids dominates the force balance, the vortex worldsheet approaches carrier advection. This reactive suppression of transverse slip can occur even at weak drag and should be distinguished from the dissipative strong-drag pinning limit. Even at $N_\Phi=0$, entrainment magnetizes a neutron vortex and
allows electron drag~\cite{Sourie:2020drw}.

A microscopic closure must supply $\mathcal R_n$ as a function of
local scalar state variables, including the proper densities,
$\Gamma_r$, carrier temperature, vorticity magnitude $b$, and
bundle structure~\cite{Sourie:2021vsj}.
For a cold reference state with $\Gamma_r-1\ll1$, the
phenomenological
calibration of Ref.~\cite{Sourie:2020ima}
is 
\begin{equation}
 {\mathcal R_n}=
 \begin{cases}
  {\mathcal R_{n,0}},&N_\Phi=0,\\
  {\mathcal R_{n,0}}%
  \varepsilon_p^{-2}N_\Phi^{\alpha_\Phi},&N_\Phi>0,
 \end{cases}
 \qquad \alpha_\Phi=1\ \text{or}\ 2,
 \label{eq:force_fluxoid_drag_model}
\end{equation}
with
\begin{equation}
 {\mathcal R_{n,0}}=4\times10^{-4}
 \frac{\varepsilon_p^2}{\sqrt{1-\varepsilon_p}}
 \left(\frac{x_p}{0.05}\right)^{7/6}
 \frac{\rho_{14}^{1/6}}{1-x_p}.
 \label{eq:force_fluxoid_unpinned_drag}
\end{equation}
Here the reference rest-mass density
$\textcolor{black}{\rho_b=m_n\nu_s+m_pn_c}$ defines the scalars
$\textcolor{black}{x_p=m_pn_c/\rho_b}$ and
$\textcolor{black}{\rho_{14}=\rho_b/(10^{14}\,\mathrm{g\,cm^{-3}})}$.
The parameter $0<\varepsilon_p<1$ retains the microscopic proton
entrainment convention.
The two exponents explore the dependence on bundle structure.
For the NC model, entrainment is fixed by the master function of
Eq.~\eqref{eq:main_model_b_master}. To state the matching explicitly,
define {$\mu_p^{(u)}=-u^\mu\mu^p_\mu$} and use the relativistic
momentum convention~\cite[Appendix~A]{Antonelli:2017ytn}
\begin{equation}
 \frac{\mu^p_\mu}{{\mu_p^{(u)}}}
 =(1-\varepsilon_p)u_\mu
   +\frac{\varepsilon_p}{\Gamma_r}u^n_\mu.
 \label{eq:force_fluxoid_entrainment_convention}
\end{equation}
Differentiating the constant-entrainment master function gives
$\mu^p_\mu=(\mu_p-\mathcal A\nu_n)u_\mu
+\mathcal A\nu_nu^n_\mu$, and hence the exact relations
\begin{align}
 {\mu_p^{(u)}}&=\mu_p+\mathcal A\nu_n(\Gamma_r-1),\notag\\
 \varepsilon_p
 &=\frac{\mathcal A\nu_n\Gamma_r}
         {\mu_p+\mathcal A\nu_n(\Gamma_r-1)}.
 \label{eq:force_fluxoid_model_b_entrainment}
\end{align}
Here $\nu_n=\nu_s$ and
$\mu_p=m_p+\Gamma_pK_pn_p^{\Gamma_p-1}/(\Gamma_p-1)$ is the
density-conjugate chemical potential in
Eq.~\eqref{eq:main_model_b_pressure}.
At coflow, $\varepsilon_p=\mathcal A\nu_n/\mu_p$.

\input{section_vortex_tension_model_b}

\input{section_characteristics_tension}

\section{Numerical method}
\label{sec:numerical_methods}

We evolve the matter and canonical vortex fields with a shared
Implicit-Explicit (ImEx) Runge--Kutta update. The method combines conservative matter and
electric balances, constrained transport of circulation, and a
local implicit solve when the electric vortex force is stiff.
We first specify the evolved system and primitive recovery, then
describe its spatial discretization and stage coupling.

\input{section_evolution_tension}

\input{section_tension_imex}

\input{section_numerics_tension}

\section{Validation in flat spacetime}
\label{sec:numerical-results}

We first assess the implementation in flat spacetime. The sound-wave
and shock-tube tests use the CP model. The
coincident-sound, Kelvin-wave, and shear tests use the NC model. We use finite-volume SSPRK3 evolution with Courant
factor $0.25$, except for the Kelvin test, which uses
$\Delta t=0.15\Delta x$ and stationary-reference subtraction as described
in Sec.~\ref{sec:results-kelvin}.

\subsection{Linear propagation of first and second sound}
\label{sec:results-sound}

We test the propagation and polarization of both sound families using
the EOS in Eq.~\eqref{eq:main_model_a_pressure}, with $\Gamma=2$ and
$\textcolor{black}{\kappa_p=58}$.
We take the rotation-free branch, $F=0$, and recover the matter state
from particle number, energy, and momentum.

\textcolor{black}{For the CP sound and Riemann tests,
$e_{\rm S}\equiv\varepsilon_\xi$ is the invariant phase norm,
$w_x=z^x$ is the spatial carrier four-velocity, and $T$ is the
carrier temperature. In one dimension, $W=\sqrt{1+w_x^2}$ and
$\Xi=\sqrt{e_{\rm S}^2+\xi_x^2}$ give the carrier velocity
$v_c=w_x/W$ and the canonical velocity $v_{\rm phase}=\xi_x/\Xi$.
We denote the perturbation amplitude by $\epsilon$, the box length
by $L$, and the number of cells along the propagation direction by
$N_x$.}

\begin{figure}
  \centering
  \includegraphics[width=\columnwidth]{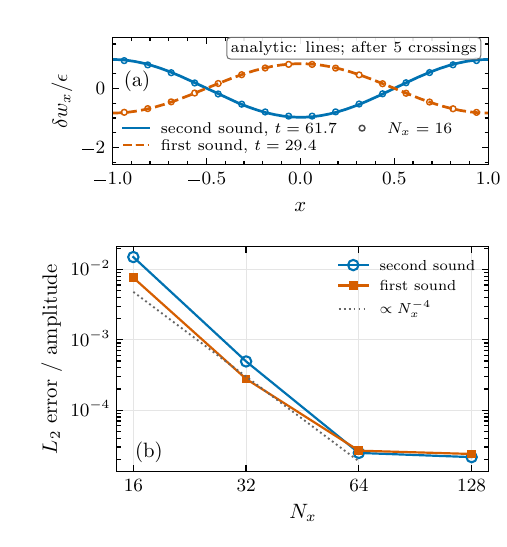}
  \caption{
  Linear propagation of first and second sound in the CP model.
  \textit{Top (a):} Velocity amplitude $\delta w_x/\epsilon$ for second sound (solid,
  $t=61.7$) and first sound (dashed, $t=29.4$), after five box crossings
  of each family.  Lines show the linear traveling waves and markers show
  every cell.
  \textit{Bottom (b):} normalized four-component error
  $\mathcal E_2$ from Eq.~\eqref{eq:sound-normalized-error} after five
  crossings. Blue circles and orange squares denote second and first
  sound. The dotted line gives an $N_x^{-4}$ reference slope over the
  coarse-grid interval; the finest-grid errors have reached a floor.}
  \label{fig:sound-linear}
\end{figure}

\begin{figure*}
  \centering
  \includegraphics[width=0.98\textwidth]{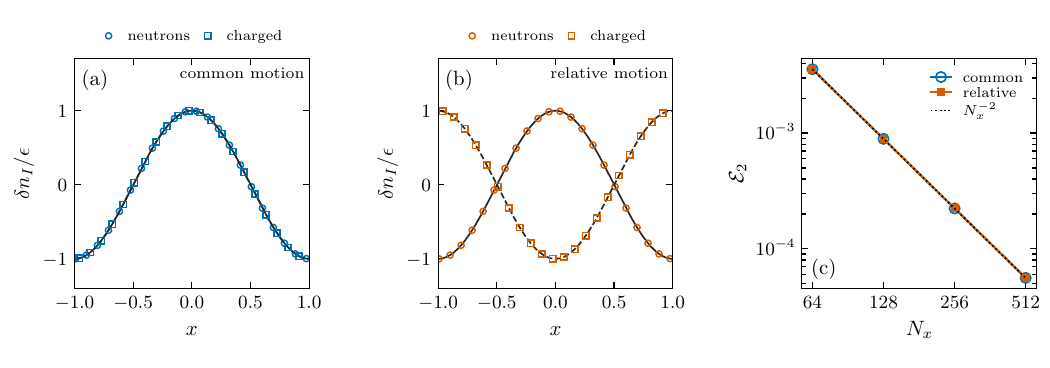}
  \caption{Coincident sound modes in the cold neutron--charged
  superfluid (NC model), both propagating at
  $c_{\rm ac}=1/\sqrt{2}$, and with identical unentrained constituents.
  (a,b)~Density perturbations after two crossings on the $N_x=128$ grid
  for common and relative motion. Here $n_I$ denotes $\nu_n$ or $n_p$.
  Circles and squares denote neutrons and
  charged particles; black curves are the linear traveling-wave reference.
  Here $\epsilon=10^{-6}$ is the fractional density amplitude.
  (c)~The four-component error $\mathcal E_2$, with an $N_x^{-2}$
  reference slope. Common and relative errors nearly coincide. The
  text explains the spatial-operator difference from
  Fig.~\ref{fig:sound-linear}.}
  \label{fig:identical-sound}
\end{figure*}

This is a one-dimensional test carried out on a two-dimensional grid, with extent $[-1,1]\times[-0.0625,0.0625]$,
and a fixed transverse resolution of 16 cells in the periodic direction.

We linearize the evolved conservative flux about
{$(e_{\rm S},T,w_x,\xi_x)=(1.12,0.03,0,0)$} and construct the
right-going first- and second-sound eigenvectors, $\boldsymbol r_1$ and
$\boldsymbol r_2$. In the primitive order
$(e_{\rm S},T,w_x,\xi_x)$, they are
\begin{equation}
 \boldsymbol r_1\simeq\begin{pmatrix}
 0.17616406\\0.04422219\\0.83624252\\0.51740610
 \end{pmatrix},\qquad
 \boldsymbol r_2\simeq\begin{pmatrix}
 0.03121046\\-0.05696534\\-0.97910771\\0.19268876
 \end{pmatrix}.
 \label{eq:sound-primitive-eigenvectors}
\end{equation}
The initial state for each mode is
\begin{equation}
  \boldsymbol p_{\rm A}(x,0)
  = \boldsymbol p_{{\rm A},0}
  + \epsilon\,\boldsymbol r_a\cos(kx),
  \qquad
  \boldsymbol p_{\rm A}=(e_{\rm S},T,w_x,\xi_x),
  \label{eq:sound-eigenmode-ic}
\end{equation}
with $\epsilon=10^{-6}$ and $k=\pi$, giving one wavelength across the box.
We normalize the eigenvectors by $\sum_{A=1}^4(r_a^A)^2=1$
and take $(r_a)_{e_{\rm S}}>0$. Thus $\epsilon$ is the norm of the
peak perturbation in these dimensionless primitives.
The sound speeds (characteristic eigenvalues) are $\lambda_1=0.34048$ and $\lambda_2=0.16197$,
and the linear reference is
$\boldsymbol p_{{\rm A},0}+\epsilon\,\boldsymbol r_a\cos[k(x-\lambda_at)]$.
We use resolutions $N_x=16$,
32, 64, and 128. The waves are evolved for five crossing times, $5L/\lambda_a$ with $L=2$, after which they are compared to the analytic solution. Figure~\ref{fig:sound-linear} shows the profiles and the error\textcolor{black}{, where $p_h^A,p_{\rm lin}^A$ are numerical and
linear-reference primitive components and $\langle\cdot\rangle_x$
is the spatial mean,}
\begin{equation}
 \mathcal E_2=\left[\frac14\sum_{A=1}^4
 \left\langle\left(\frac{p_h^A-p_{\rm lin}^A}
 {\epsilon|r_a^A|}\right)^2\right\rangle_x\right]^{1/2}.
 \label{eq:sound-normalized-error}
\end{equation}
The matched fourth-order spatial difference is applied to both the
material fluxes and the canonical-potential gradient. The fitted orders
over $N_x=16$--$64$ are $4.08$ and $4.62$ for first and second sound,
respectively. At $N_x=128$, the errors approach $2\times10^{-5}$ and
are dominated by harmonics absent from the linear reference. These
measurements describe the resolved spatial-error regime; SSPRK3 remains
third order in time.

\par
\noindent\begin{minipage}{\columnwidth}
  \centering
  \includegraphics[width=0.76\columnwidth]{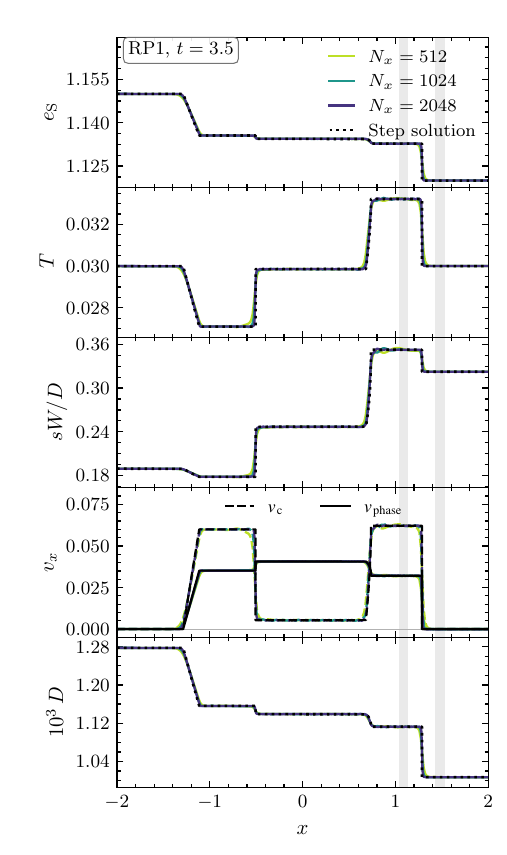}
  \captionof{figure}{{{Riemann problem RP1 for the CP model at $t=3.5$.
  From top to bottom, the panels show the invariant phase-momentum norm
  $e_{\rm S}=\sqrt{-\xi_\mu\xi^\mu}$, the normal-component temperature
  $T$, the {Eulerian entropy-to-particle density ratio} $sW/D$, the normal and phase velocities
  $\textcolor{black}{v_c=w_x/W}$ (dashed) and $\textcolor{black}{v_{\rm phase}=\xi_x/\Xi}$ (solid), and the Eulerian
  particle number density $D=N_{\rm A}^0$.
  Here $s$ is the proper entropy density, $W=\sqrt{1+w_x^2}$ and
  $\Xi=\sqrt{e_{\rm S}^2+\xi_x^2}$. The density panel plots $10^3D$.
  The colored curves distinguish the number $N_x$ of cells in the
  $x$-direction, and the dotted curves give the step solution.}}}
  \label{fig:riemann-first}
\end{minipage}\par

\noindent\begin{minipage}{\columnwidth}
  \centering
  \includegraphics[width=0.76\columnwidth]{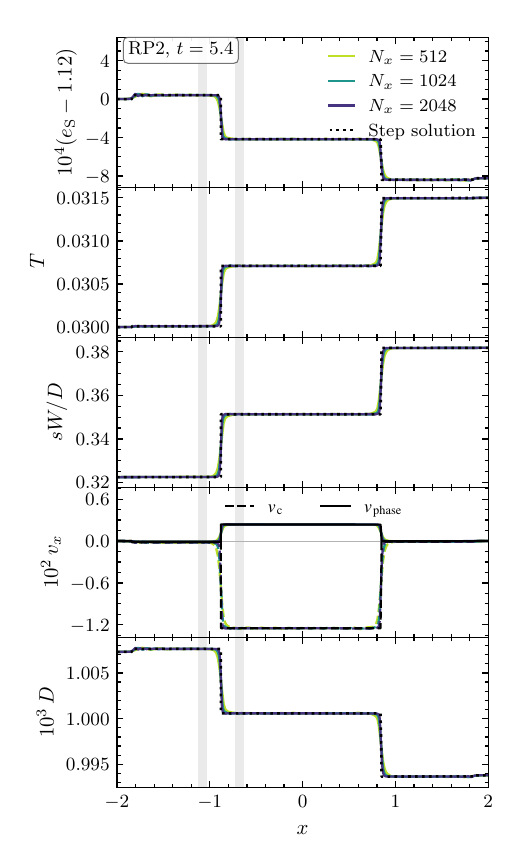}
  \captionof{figure}{{{Riemann problem RP2 at $t=5.4$, with quantities and
  notation as in Fig.~\ref{fig:riemann-first}. The top panel plots
  $10^4(e_{\rm S}-1.12)$, and the velocity panel plots $10^2v_x$.}}}
  \label{fig:riemann-second}
\end{minipage}\par

\subsubsection{Coincident sound speeds of identical constituents}
\label{sec:results-identical-sound}

We next test two independently excited modes at coincident sound speeds in the NC model.
The cold neutron--charged system has identical separable energy densities
$e_n(\nu_n)=\nu_n+\nu_n^2/2$ and $e_p(n_p)=n_p+n_p^2/2$,
background densities $\nu_n=n_p=1$, and zero entrainment. Both chemical
potentials are $2$, and Eq.~\eqref{eq:singlefluid_identical_speeds} gives
$c_+=c_-=c_{\rm ac}=1/\sqrt{2}$, with $c_{\rm ac}$ the common
acoustic speed. The right-going eigenvectors describe common and
relative motion. We initialize them separately with
$\delta\nu_n=\epsilon\cos(\pi x)$,
$\delta n_p=\pm\epsilon\cos(\pi x)$,
$v_n=c_{\rm ac}\epsilon\cos(\pi x)$, and
$v_p=\pm c_{\rm ac}\epsilon\cos(\pi x)$, where $v_I\equiv v_I^x$
for $I\in\{n,p\}$. {The upper and lower signs give
common and relative motion, respectively. Here $\epsilon=10^{-6}$ is
the fractional density amplitude about $\nu_{n,0}=n_{p,0}=1$, the
peak velocity perturbation is $c_{\rm ac}\epsilon$.}
We use the same domain and setup as in the previous test, and evolve $N_x=64$, 128, 256, and 512 through two crossings,
$t=2L/c_{\rm ac}=4\sqrt{2}$.
Both polarizations propagate at the common speed
(Fig.~\ref{fig:identical-sound}). The density perturbations remain
in phase for common motion and in opposition for relative motion; their relative phase changes by less than $2.1\times10^{-7}$ radians
on every grid.
We evaluate Eq.~\eqref{eq:sound-normalized-error} with
$\boldsymbol p=(\nu_n,n_p,v_n,v_p)$ and
$\boldsymbol r_\pm=(1,\pm1,c_{\rm ac},\pm c_{\rm ac})$. Both
polarizations converge at order $1.999$.

The spatial updates differ between the two sound tests. These NC
runs use the ordinary difference of point-reconstructed face fluxes,
whose leading smooth-wave phase error is second order. The preceding
CP test uses a matched fourth-order difference for both matter
and canonical transport. A comparison at the same spatial order requires
that difference in both NC updates, together with a timestep study
that makes the SSPRK3 error subdominant. Coincident acoustic speeds
retain independent eigenvectors and do not reduce the spatial order.

\subsection{Shocktube tests}
\label{sec:results-riemann}

Shock-tube tests assess characteristic propagation and the capture of
shocks and rarefactions in a conservative scheme~\cite{Amiram:2006zjz}. We
use Riemann initial data, consisting of two uniform states separated by
an interface. Matching the nonlinear wave curves gives an independent
reference solution, as in exact relativistic Riemann solvers
\cite{Giacomazzo:2005jy}. Appendix~\ref{app:riemann_analytic}
constructs those curves for the two sound families of our superfluid EOS.

Finite jumps distinguish compressive and counterflow propagation beyond
the linear regime. We use a strip
$[-4,4]\times[-0.0625,0.0625]$, with 16 transverse cells and
$N_x=512$, 1024, and 2048.  
{We evolve two Riemann problems, one aimed at each acoustic
family. For the background used here, first sound is predominantly
compressional, with normal and phase velocities of the same sign.
Second sound has opposing velocities and a thermal counterflow
polarization. Its particle flux and pressure perturbation need not
vanish~\cite{Alford:2014sounds}.
RP1 jumps the invariant phase-momentum norm at uniform temperature,
exciting mainly first sound. RP2 combines a temperature jump with
a small change in $e_{\rm S}$ along the second-sound eigenvector
tangent. To leading order, this thermodynamic jump is a superposition
of oppositely propagating second-sound modes with zero initial
velocities. The finite-amplitude solution has a predominantly
counterflow plateau and weak first-sound waves.}
The initial conditions are as follows.
{For RP1, we take}
\begin{equation}
  \begin{aligned}
    (e_{\rm S},T,w_x,\xi_x)_{L} &= (1.15,0.03,0,0),\\
    (e_{\rm S},T,w_x,\xi_x)_{R} &= (1.12,0.03,0,0).
  \end{aligned}
  \label{eq:rp1-ic}
\end{equation}
{For RP2, the complete initial states are
\begin{equation}
  \begin{aligned}
    (e_{\rm S},T,w_x,\xi_x)_{L} &= (1.12,\,0.03,\,0,\,0),\\
    (e_{\rm S},T,w_x,\xi_x)_{R} &= ({1.119178172379455},\,0.0315,\,0,\,0),
  \end{aligned}
  \label{eq:rp2-ic}
\end{equation}
i.e.\ a $5\%$ right-state temperature raise accompanied by the
$e_{\rm S}$ jump prescribed by the second-sound eigenvector tangent,
$\delta e_{\rm S} = ({r_{2,e_{\rm S}}}/{r_{2,T}})\,\delta T$, which
suppresses first sound up to the finite-amplitude departure from the
tangent.}

RP1 produces a right-going shock and a left-going rarefaction, separated
by the plateaus in Fig.~\ref{fig:riemann-first}. RP2 gives {a left-going second-sound shock --- a captured
discontinuity that carries the temperature jump at the second-sound
speed ---} and a right-going rarefaction
with a counterflow plateau (Fig.~\ref{fig:riemann-second}).
Both solutions agree well with their exact references obtained in Appendix \ref{app:riemann_analytic}.
\begin{figure}[t]
  \centering
  \includegraphics[width=\columnwidth]{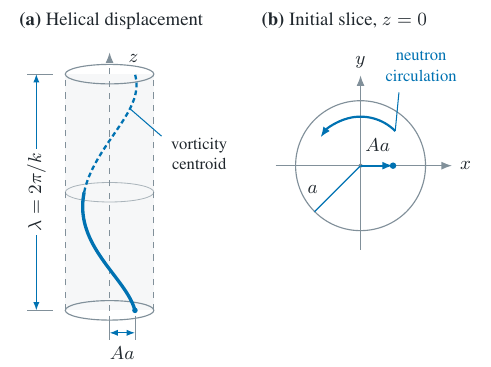}
  \caption{Initial helical displacement of the vortex column.
  (a) The blue curve traces the canonical-vorticity centroid,
  $X+iY=Aa e^{-ikz}$. Gray curves mark $r=a$ in the unperturbed
  state, and dashed blue segments lie behind the axis.
  \textcolor{black}{Here $X,Y$ are transverse centroid coordinates, $a$ is the
  column radius scale, $A$ the dimensionless displacement amplitude,
  and $k$ the axial wave number.}
  (b) At $z=0$, the centroid lies a distance $Aa$ from the axis.
  The displacement is enlarged for visibility.}
  \label{fig:kelvin-schematic}
\end{figure}

\subsection{{Kelvin waves along a vortex column}}
\label{sec:results-kelvin}

A transverse displacement of a vortex column produces a helical
$m=1$ perturbation\textcolor{black}{, where $m$ is the azimuthal mode number,}
that precesses about the unperturbed axis.
We study this motion in Minkowski spacetime using two columns
with smoothly distributed canonical vorticity
(Fig.~\ref{fig:kelvin-two-tests}). \textcolor{black}{We denote the central
angular velocity by $\Omega_c$, the axial wave number by $k$, and
the column radius scale by $a$.}
The first has no line energy, $F=0$, and vortex lines move with
the neutron flow. The displaced column induces a restoring flow,
with a frequency of order $\Omega_c$ at $ka\sim1$ that depends
on the vorticity profile. This tests canonical-vorticity transport
and the constraint-preserving update.

The second column includes the line energy $F(n,\mathcal N)$ and
its stress. The derivative $\mathcal T=F_{\mathcal N}$ supplies
both tension and inertia per line, allowing the lines to move
relative to the matter. For prescribed resting currents, tension
balanced by lift gives
$|\omega|\simeq\mathcal T k^2/(\kappa_v\nu)$ at long wavelength,
where $\nu=\nu_n+\mathcal Z n_p$
[Eqs.~\eqref{eq:kt_displacement} and \eqref{eq:kt_nondissipative}].
The evolved column also includes the radial fluid response, so
its frequency follows from the coupled mode calculation in
Appendix~\ref{app:kelvin_tension_prediction}.
This test includes line inertia, reactive coupling to the thermal
carrier, and the implicit evolution of the weighted electric field.

\begin{figure}[tbp]
  \centering
  \includegraphics[width=\columnwidth]{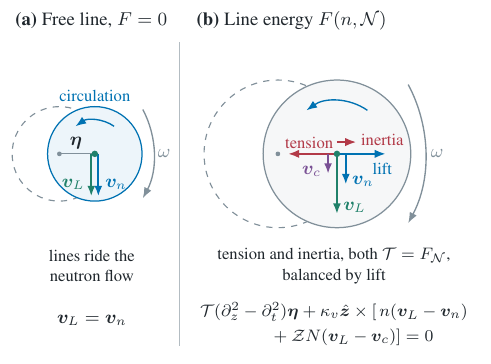}
  \caption{Restoring forces for the two Kelvin-wave tests, viewed
  along the column axis. Dashed and solid circles mark the unperturbed
  and displaced columns, separated by $\boldsymbol\eta$.
  The arrow labeled $\omega$ indicates retrograde precession.
  (a) With $F=0$, vortex lines follow the neutron flow.
  (b) With line energy, tension and line inertia balance lift from
  the neutrons and carrier. The line velocity $\boldsymbol v_L$
  can then differ from $\boldsymbol v_n$ and $\boldsymbol v_c$.
  The balance shown is Eq.~\eqref{eq:kt_displacement} at zero drag.}
  \label{fig:kelvin-two-tests}
\end{figure}

\subsubsection{Column without line stress}

We use the cold, unentrained NC model with
$\mathcal Z=\mathcal R=\mathcal R'=0$.
The unperturbed neutrons circulate about $z$ in a uniform proton
fluid at rest. Figure~\ref{fig:kelvin-schematic} illustrates the
bending perturbation, and Appendix~\ref{app:kelvin_native} gives
its dispersion relation and radial eigenfunction.

\begin{figure*}[tbp]
  \centering
  \includegraphics[width=\textwidth]{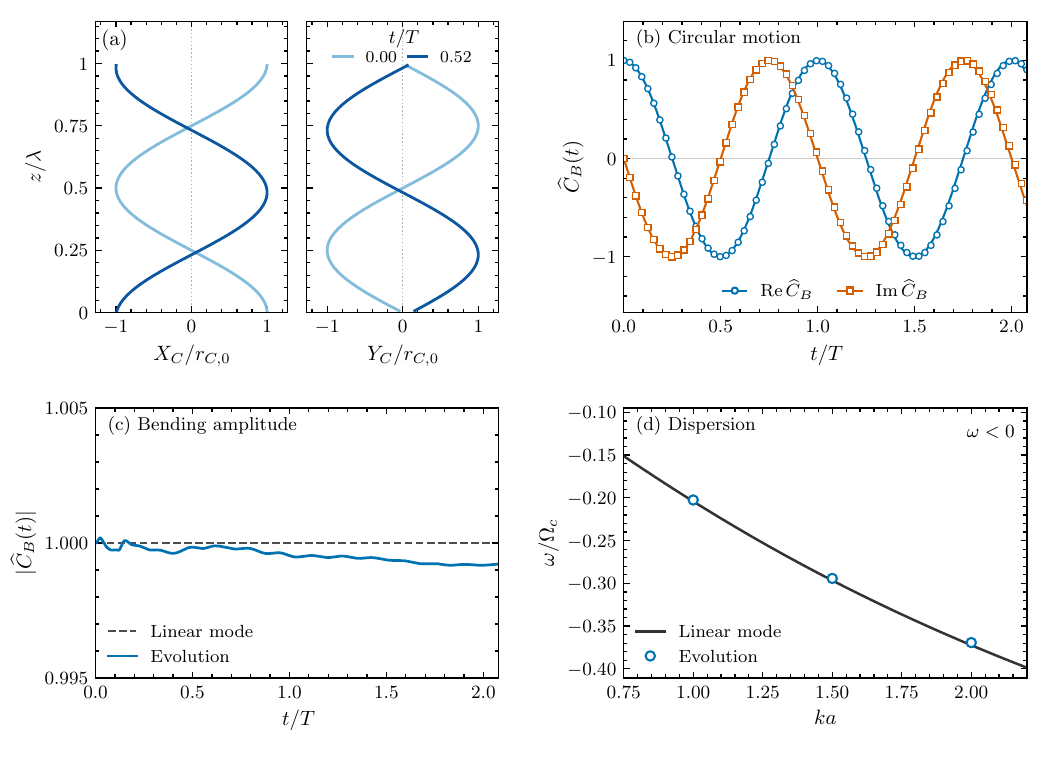}
  \caption{Kelvin wave in the column without line stress.
  (a) Gaussian-weighted vorticity centroids at $t=0$ and $0.520T$,
  normalized by the initial centroid radius $r_{C,0}$.
  (b) Real and imaginary parts of
  $\widehat C_B=C_B(t)/C_B(0)$.
  (c) Displacement magnitude compared with the constant linear
  prediction. Panels (a)--(c) use $ka=1$.
  (d) Measured dispersion compared with the radial eigenvalues.
  Here $a$ is the column scale, $k$ is the axial wave number,
  $T=2\pi/|\omega|$ is the period, and $\Omega_c$ is the central
  angular velocity.}
  \label{fig:kelvin-box}
\end{figure*}

\begin{figure*}[t]
  \centering
  \includegraphics[width=\textwidth]{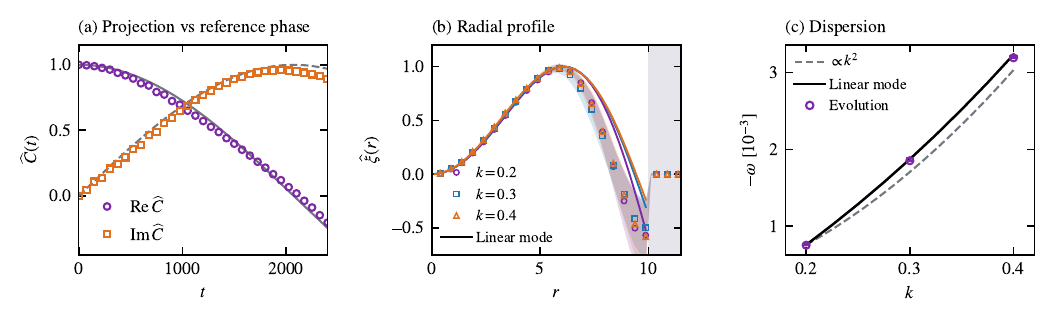}
  \caption{Collective column bending with line stress.
  (a) Real and imaginary parts of the mode projection
  $\widehat C(t)=C(t)/C(0)$ for $k=0.2$ on $96^2\times48$ cells,
  using $R_{\rm cut}=5$\textcolor{black}{ as the radial cutoff of the mode
  projection}.
  Gray curves give the reference phase
  $e^{-i\omega t}$ with $\omega=-7.5818\times10^{-4}$
  [Eq.~\eqref{eq:kt_warm_reference_frequency}].
  (b) Transverse neutron canonical-momentum profiles, divided by
  the projection and averaged over the run (symbols), compared
  with the linear eigenfunctions (lines). Each profile is normalized
  by its eigenfunction's peak magnitude. Shading gives the temporal
  standard deviation, and the gray strip marks the frozen exterior
  $r>R_b$\textcolor{black}{, where $R_b=10$ is the cylindrical wall radius}.
  (c) Linear dispersion (solid), $k^2$ scaling normalized at $k=0.2$
  (dashed), and measured frequencies. The bands span full-run fits
  over $3\leq R_{\rm cut}\leq6$.}
  \label{fig:kelvin-tension-results}
\end{figure*}

The azimuthal speed is the Kaufmann--Scully profile
$V(r)=\Omega_ca^2r/(r^2+a^2)$~\cite{2018arXiv180505064G},
where $r=(x^2+y^2)^{1/2}$ and $\Omega_c$ is the central angular
velocity. For $\mu_x=m_x+2K_xn_x$, $x\in\{n,p\}$ and
$n_n=\nu_n$, radial force balance fixes the initial state:
\begin{align}
 \boldsymbol v_{n0}&=\frac{\Omega_ca^2}{r^2+a^2}(-y,x,0),
 \quad \boldsymbol v_{p0}=0,\notag\\
 \mu_{n0}(r)&=\mu_{n,\infty}
 \exp\!\left[-\int_r^\infty\frac{V(s)^2}{s[1-V(s)^2]}\,ds\right],
 \notag\\
 n_{n0}(r)&=\frac{\mu_{n0}(r)-m_n}{2K_n},\notag\\
 n_{p0}&=\frac{\mu_p-m_p}{2K_p},\qquad W=(1-V^2)^{-1/2},\notag\\
 \xi_{\mu0}&=\mu_{n0}W(-1,\boldsymbol v_{n0}).
 \label{eq:kelvin_equilibrium_primitives}
\end{align}
The chemical-potential profile satisfies
$d\ln\mu_{n0}/dr=W^2V^2/r$.
We set $a=1$, $\Omega_c=0.4$, $m_n=m_p=1$, $K_n=0.024$,
$K_p=0.216$, $\mu_{n,\infty}=2$, and $\mu_p=3$.
The maximum speed is $V(a)=0.2$.
With perturbations proportional to $e^{i(\theta+kz-\omega t)}$,
negative $\omega$ denotes retrograde precession.
At $ka=1$ the radial eigenvalue is $\omega=-0.08165805313$,
corresponding to $T=76.94508$, and the sound speed on the axis
is $0.67614$. Increasing the outer matching radius from $16a$
to $32a$ leaves this frequency unchanged at the quoted precision.
The radial calculation gives $\omega=-0.11864207$ and
$-0.14893842$ at $ka=1.5$ and $2$.

The real eigenfunctions $R_n$, $P_n$ and $Q_n$ give the radial,
azimuthal and axial canonical-momentum perturbations, and $WH_n$
gives the chemical-potential perturbation. We initialize the
mode of Eq.~\eqref{eq:native_kelvin_radial_system} with $A=0.01$:
\begin{align}
 \delta\mu_n&=AW H_n\cos(\theta+kz),&
 \delta\xi_r&=-AR_n\sin(\theta+kz),\notag\\
 \delta\xi_\theta&=AP_n\cos(\theta+kz),&
 \delta\xi_z&=AQ_n\cos(\theta+kz).
 \label{eq:native_kelvin_seed}
\end{align}
The normalization
$\int_0^\infty r(R_n-P_n)\,dr=2\mu_{n,\infty}\Omega_ca^3$
sets the canonical-vorticity centroid displacement to $Aa$.
We sample the Cartesian canonical covector on the constrained-transport
edges and use its edge-to-cell averages to construct the conserved
initial fields. The protons remain unperturbed.

The domain is $[-6a,6a]^2\times[0,2\pi/k]$, with periodic axial
boundaries and equilibrium exterior ghost cells.
For $ka=1$, we use $64^2\times32$, $96^2\times48$ and
$128^2\times64$ cells, CFL factor $0.15$, and final time $t=160$.
The $ka=1.5$ and $2$ runs use $96^2\times48$ cells and end at
$t=110$ and $90$, respectively.

We extract the phase and amplitude with the momentum projection
\begin{equation}
 C_\xi(t)=\frac{\int w(r)(\xi_x+i\xi_y)e^{ikz}\,d^3x}
                  {\int w(r)\,d^3x},
 \qquad w(r)=e^{-r^2/(2a^2)}.
 \label{eq:native_kelvin_observable}
\end{equation}
For a mode with $\omega=\omega_r+ig$, this projection obeys
$C_\xi\propto e^{i\omega^*t}=e^{(g+i\omega_r)t}$.
Its phase slope gives the signed frequency, and its logarithmic
amplitude slope gives $g$.
An independent displacement diagnostic uses the canonical vorticity
$\mathcal B^z=\partial_x\xi_y-\partial_y\xi_x$:
\begin{equation}
 C_B(t)=\frac{\int w(r)(x+iy)\mathcal B^ze^{ikz}\,d^3x}
                  {\int w(r)\mathcal B^z\,d^3x}.
 \label{eq:kelvin_rel_vorticity_observable}
\end{equation}
A single mode gives both projections the same phase and envelope.
At each height, the weighted centroid is
$X_C+iY_C=\int w(r)(x+iy)\mathcal B^z\,dx\,dy/
\int w(r)\mathcal B^z\,dx\,dy$.
Figure~\ref{fig:kelvin-box} uses $\widehat C_B=C_B(t)/C_B(0)$
and $r_{C,0}=\langle X_C(z,0)^2+Y_C(z,0)^2\rangle_z^{1/2}$.

The evolution reproduces the helical precession and phase
quadrature in Fig.~\ref{fig:kelvin-box}. The plotted displacement
magnitude stays within $0.1\%$ of its initial value.
We fit $C_\xi=C_*\exp[(g+i\omega_h)(t-t_*)]$ over
$10\leq t\leq160$, with free complex amplitude, frequency and
envelope rate. The frequency error decreases from $1.7\%$ to
$0.4\%$ as the grid increases from $64^2\times32$ to
$128^2\times64$ (Table~\ref{tab:kelvin-native}).
The continuum mode has $g=0$, so the fitted growth and damping
rates measure numerical amplitude error. The measured dispersion
also follows the radial reference at the higher wave numbers
[Fig.~\ref{fig:kelvin-box}(d)].

\begin{table}[tbp]
 \centering
 {
 \begin{tabular}{cccc}
 \hline\hline
 Grid & $\omega_h$ & Error (\%) & $g$ \\
 \hline
 $64^2\times32$ & $-0.08024406$ & $1.732$ & $+3.002\times10^{-4}$ \\
 $96^2\times48$ & $-0.08098752$ & $0.821$ & $+7.292\times10^{-6}$ \\
 $128^2\times64$ & $-0.08128092$ & $0.462$ & $-5.333\times10^{-6}$ \\
 \hline\hline
 \end{tabular}}
 \caption{Frequency and numerical envelope rate for the
 $ka=1$ column without line stress. The frequency error is relative
 to the radial eigenvalue $\omega=-0.08165805313$.
 The continuum envelope rate is $g=0$, and positive $g$ denotes
 numerical amplification.}
 \label{tab:kelvin-native}
\end{table}

\subsubsection{Column with line stress}
\label{sec:results-kelvin-tension}

We next evolve a collective bending mode of the warm rotating
column in Sec.~\ref{sec:kt_warm_benchmark}.
The regulated line energy $F(n,\mathcal N)$ uses the BCS core
prescription, and reactive lift couples the array to the thermal
carrier. Appendix~\ref{app:kelvin_tension_prediction} constructs
the equilibrium and radial mode.
The parameters are $\mathcal Z=0.5$,
$\mathcal A=\mathcal R=\mathcal R'=0$, $C_{\rm th}=0.1$,
$\textcolor{black}{\varsigma=s/N=0.2}$, $\kappa_v=0.02$, $\mathcal N_0=0.1$,
$\textcolor{black}{\alpha_{\rm v}=2}$ and $R_b=10$.
We perturb Eq.~\eqref{eq:kt_warm_equilibrium} with the
eleven-component bending seed of
Eq.~\eqref{eq:kt_warm_initial_data}, using $\epsilon=10^{-3}$.

The domain is $[-12,12]^2\times[0,10\pi]$, with periodic axial
boundaries, CFL factor $0.2$, and stationary-reference subtraction.
Cells at $r>R_b$ hold the equilibrium state to impose the reflecting
boundary. At $k=0.2$ and $0.4$ we use $96^2\times48$ cells.
At $k=0.3$ the axial length is $20\pi/3$ and the grid is
$96^2\times32$.
We obtain $C(t)$ by projecting the transverse neutron canonical
momentum onto the radial eigenfunction within $r\leq R_{\rm cut}$.
A fit $C(t)=C_*e^{-i\omega_h t}$ over the full run gives the
signed frequency. The default projection radius is $R_{\rm cut}=5$.

The column precesses in the retrograde direction
[Fig.~\ref{fig:kelvin-tension-results}(a)]. Within $r\leq5$,
the radial profile agrees with the linear mode to $1$--$5\%$
[Fig.~\ref{fig:kelvin-tension-results}(b)], and all three fitted
frequencies are within $1.3\%$ of the radial reference values
(Table~\ref{tab:kelvin-tension}).
The leading uncertainty comes from the projection radius and
fitting window. The frozen exterior holds the unperturbed state,
but the radial eigenfunction does not vanish at $r=R_b$.
This mismatch launches a disturbance that reaches $r\approx7$
by $t\approx1500$ and changes the profile at $r\gtrsim6.5$.
It also excites prograde radial modes at the few-percent level
within $r\leq5$. The band in Fig.~\ref{fig:kelvin-tension-results} and
Table~\ref{tab:kelvin-tension} spans the full-run fits over
$3\leq R_{\rm cut}\leq6$.

The radial dispersion predicts
$\omega(0.4)/\omega(0.2)=4.264$, compared with $4$ for $k^2$
scaling [Fig.~\ref{fig:kt_warm_theory}(b)].
The measured ratio at $R_{\rm cut}=5$ is $4.24$, about $6\%$
above quadratic scaling. It ranges from $4.16$ to $4.24$ over
$3\leq R_{\rm cut}\leq6$
[Fig.~\ref{fig:kelvin-tension-results}(c)].
Thus the full-run ratios exceed the $k^2$ prediction across the
tested projection radii, with the fitting-window dependence
limiting the comparison. At $k=0.3$, the predicted excess over
$k^2$ scaling is $9.8\%$, and the measured frequency agrees with
the radial reference to $1.0\%$.
In short, we find good agreement between the numerical and the exact modes.

\begin{table}[tbp]
 \centering
 {
 \begin{tabular}{ccccc}
 \hline\hline
 $k$ & Grid & $\omega_h$ & Error (\%) & $R_{\rm cut}$ band (\%) \\
 \hline
 $0.2$ & $96^2\times48$ & $-7.531\times10^{-4}$ & $-0.68$ & $[-0.7,+1.5]$ \\
 $0.3$ & $96^2\times32$ & $-1.853\times10^{-3}$ & $-1.04$ & $[-1.0,+0.8]$ \\
 $0.4$ & $96^2\times48$ & $-3.192\times10^{-3}$ & $-1.27$ & $[-1.3,-0.7]$ \\
 \hline\hline
 \end{tabular}}
 \caption{Column-bending frequencies with line stress, fitted
 over the full runs at $R_{\rm cut}=5$. Errors are relative to the
 radial reference frequencies $-7.5818344185\times10^{-4}$
 ($k=0.2$), $-1.8726769\times10^{-3}$ ($k=0.3$), and
 $-3.2328881\times10^{-3}$ ($k=0.4$). The final column gives the
 range of relative frequency errors over $3\leq R_{\rm cut}\leq6$.}
 \label{tab:kelvin-tension}
\end{table}

\subsection{Kelvin--Helmholtz instability}
\label{sec:results-kh}

We study the Kelvin--Helmholtz instability of an entrained
neutron--charged mixture in Minkowski spacetime. We evolve the same
initial shear layers with {four} vortex-line prescriptions: free
transport with the neutron current, Thouless transport with the
condensate phase, {transport with the total baryon current,}
and pinning to the charged fluid{. In terms of the lift
coefficient and drag of Eq.~\eqref{eq:moments_lift_current}, these are
labelled $\mathcal Z=0$, $\mathcal Z=\mathcal Af^2$, and $\mathcal Z=1$
at zero drag, and $\mathcal R\to\infty$}. We follow the
counterflow and neutron canonical vorticity,
$\mathcal B^z=\partial_x\xi_y-\partial_y\xi_x$,
during shear-layer roll-up.

We use the equilibrium-thermal NC model, with the $(2,4)$ cold
EOS of Eq.~\eqref{eq:results_star_eos} and thermal energy
$e_{\rm th}=C_{\rm th}s_c^{4/3}$. The proper densities are uniform,
$(\nu_n,n_p)=(4,0.4)$, giving
$Y_p=n_p/(\nu_n+n_p)=0.091$. We set $\mathcal A=-0.6$ and
$e_{\rm th}=4.4\times10^{-8}$; dissipative sources vanish.
The cold pressure $\Psi=K_n\nu_n^2+K_pn_p^4$ is uniform, with
$K_p$ defined by the convention in
Eq.~\eqref{eq:results_star_eos}. The corresponding unentrained
neutron and proton sound speeds are $0.40$ and $0.42$.

On the periodic square $[-1,1]^2$, we initialize
two pressure-balanced shear layers,
\begingroup
\begin{align}
  v^x_p(y) &= V_0\,[h_+(y)-h_-(y)-1],
  \label{eq:kh-shear}\\
  v^y_p(x,y) &= A\sin(kx)\,[g_+(y)+g_-(y)],
  \label{eq:kh-perturbation}
  \\
  h_\pm(y) &= \tanh\!\left(\frac{y\pm0.5}{a}\right),
  \notag\\
  g_\pm(y) &= \exp\!\left[-\frac{(y\mp0.5)^2}{4a^2}\right],
  \notag
\end{align}
\endgroup
with $a=0.1$, $V_0=0.1$, $A=0.002$, and $k=2\pi$.
Both constituents have this analytic velocity field, so
$q_\mu=0$ initially. We initialize
$\xi_\mu=Hu^n_\mu+\mathcal A n_pu^p_\mu$, with
$H=\mu_n-\mathcal A n_p$, at cells and staggered edges.
Its curl supplies the initial shear-layer vorticity.
The prescriptions therefore start from the same analytic
line velocity, $\bm v_n=\bm v_{\rm ph}=\bm v_p$.
We evolve to $t=45$ with $128^2$ {, $256^2$, and $512^2$} cells in $(x,y)$
and four cells along the invariant $z$ direction.

The transport choices separate once relative motion develops.
Writing the counterflow as $\bm w=\bm v_n-\bm v_p$ and the line
slip as $\bm s=\bm v_L-\bm v_n$, free transport has
$\bm s=0$, whereas pinning gives $\bm s=-\bm w$.
For Thouless transport, the momentum map gives, to leading order
in the counterflow,
\begin{equation}
 \bm v_{\rm ph}-\bm v_n
 =-\frac{{\cal A}n_p}{\mu_n}\,\bm w
 \simeq+0.2013\,\bm w,
 \label{eq:kh_thouless_slip}
\end{equation}
Here $\mathcal A n_{p,0}/\mu_{n,0}=-0.20134$ and initially
$\xi_\mu/\mu_{n,0}\simeq1.20134u^n_\mu-0.20134u^p_\mu$.
The negative carrier contribution makes Thouless lines move ahead
of the neutron flow wherever neutrons lead the charge carrier. Pinned lines instead follow the carrier.

The fourth, baryon-current run keeps zero drag but fixes $\mathcal Z=1$ in
Eq.~\eqref{eq:moments_lift_current}, so that the lift current
$L^\mu=N_s^\mu+N_c^\mu=n_n^\mu+n_pu^\mu$ is the total baryon current. By
Eq.~\eqref{eq:moments_lift_weighted_velocity}, the lines then move with
the baryon-weighted mean velocity, and to leading order in the
counterflow
\begin{equation}
 \bm s=-\frac{n_p}{\nu_n+n_p}\,\bm w=-Y_p\bm w\simeq-0.091\,\bm w.
 \label{eq:kh_sonin_slip}
\end{equation}
This is the analogue of the superfluid-helium result in which the
Iordanskii force from the normal component adds to the superfluid
Magnus force, so that a vortex without friction moves with the
center-of-mass velocity~\cite{1997PhRvB..55..485S}. \revadd{The baryon-current lines} lag toward
the charged fluid, as pinned lines do, but with one eleventh of the
pinned slip. The four transports therefore bracket the slip from
$+0.20\,\bm w$ to $-\bm w$: Thouless lines lead the neutron flow,
whereas baryon-current and pinned lines lag behind it.

We distinguish these motions using averages weighted by
$|\mathcal B^z|$. After counterflow develops, the slip ratio
$c=\langle\bm s\cdot\bm w\rangle_{\mathcal B}/
\langle|\bm w|^2\rangle_{\mathcal B}$ is approximately
$+0.200$, $0$, {$-0.0909$,} and $-1.000$ for the Thouless, free, {baryon-current,} and pinned
runs. These values agree within one
percent with the background prediction
$\{-\mathcal A n_p/\mu_n,0,{-Y_p,}-1\}$.
This ratio checks the prescribed slip. The counterflow and
vorticity measure the evolving response.

\begin{figure}[tbp]
 \centering
 \includegraphics[width=\columnwidth]{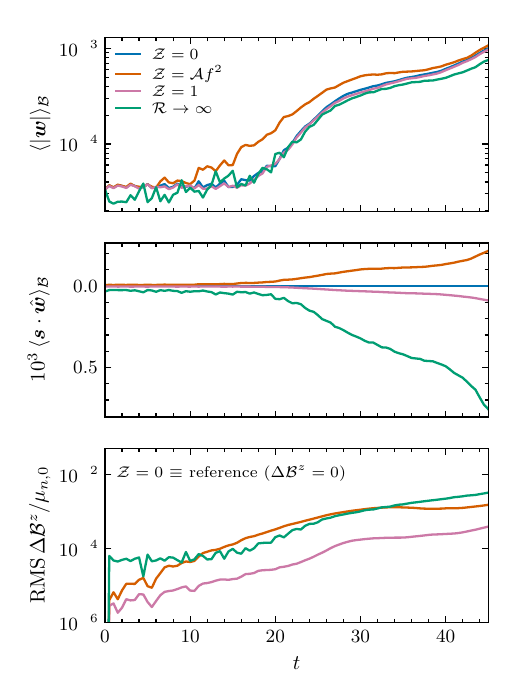}
 \caption{Effect of vortex-line transport in the
 entrained Kelvin--Helmholtz problem. Different colors denote different lift transport expressed in terms of $\mathcal Z$ of
 Eq.~\eqref{eq:moments_lift_current} at zero drag
 ($\mathcal R=\mathcal R'=0$), or by the drag limit
 $\mathcal R\to\infty$. The lines move with the neutron current for
 $\mathcal Z=0$ (free transport), with the condensate phase for the
 Thouless prescription $\mathcal Z=\mathcal Af^2$ (initially
 $\simeq-1.68$), with the total baryon current for $\mathcal Z=1$, and
 with the charged fluid for $\mathcal R\to\infty$ (pinning).
 Averages are weighted by $|\mathcal B^z|$.
 \textit{Top:} counterflow $\langle|\bm w|\rangle_{\mathcal B}$.
 \textit{Middle:} signed line slip
 $\langle\bm s\cdot\hat{\bm w}\rangle_{\mathcal B}$;
 positive values indicate motion ahead of the neutron flow along
 the counterflow. To leading order $\bm s=+0.201\,\bm w$, $0$,
 $-Y_p\bm w$, and $-\bm w$ for $\mathcal Z=\mathcal Af^2$, $0$, $1$, and
 $\mathcal R\to\infty$; the $\mathcal Z=0$ and $\mathcal R\to\infty$
 values are exact.
 \textit{Bottom:} root-mean-square vorticity difference
 $\mathrm{RMS}\,\Delta\mathcal B^z/\mu_{n,0}$ from the $\mathcal Z=0$ run,
 which serves as the reference.}
 \label{fig:kh-thouless-discriminator}
\end{figure}

\begin{figure*}[tbp]
 \centering
 \includegraphics[width=0.98\textwidth]{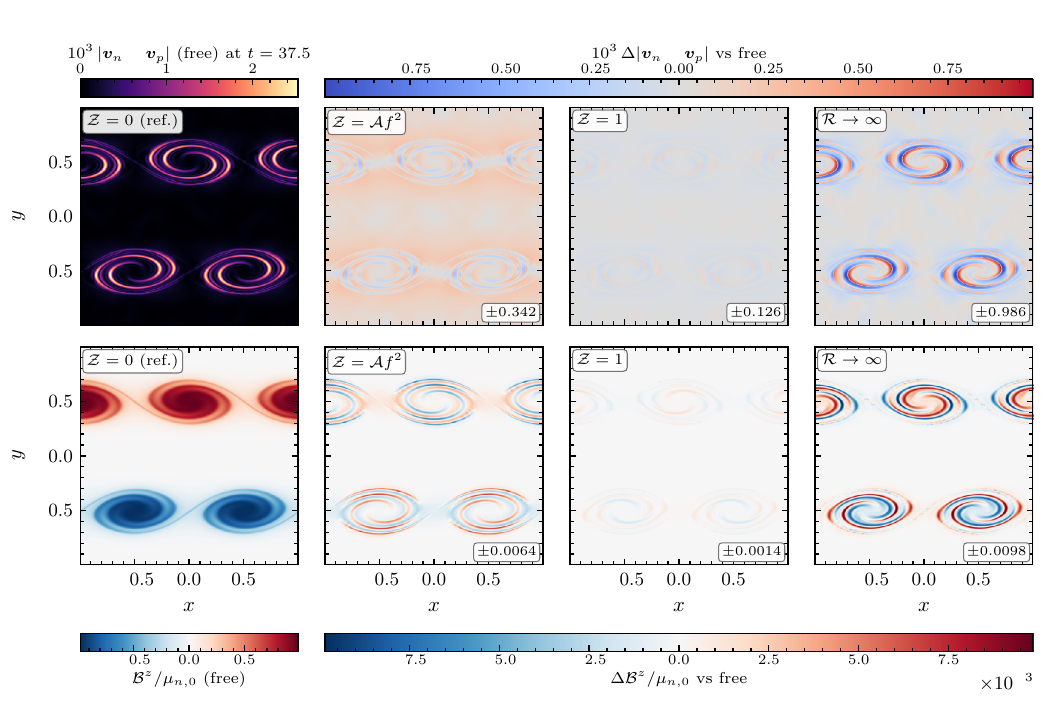}
 \caption{Spatial differences between the four
 vortex-line transports at $t=37.5$, labeled as
 in Fig.~\ref{fig:kh-thouless-discriminator}: out of plane vortex lines move with the
 neutron current ($\mathcal Z=0$), the condensate phase
 ($\mathcal Z=\mathcal Af^2$), the total baryon current
 ($\mathcal Z=1$), or the charged fluid ($\mathcal R\to\infty$).
  The left column shows the $\mathcal Z=0$ run:
 counterflow magnitude $|\bm v_n-\bm v_p|$ (\textit{top}) and
 canonical vorticity $\mathcal B^z/\mu_{n,0}$ (\textit{bottom}).
 The remaining columns show pointwise differences of the labeled
 run from the $\mathcal Z=0$ run. \revadd{The difference panels in each row
 share a symmetric color scale and show their 99.9th-percentile absolute amplitudes.}
 \textit{Top:} counterflow difference
 $\Delta|\bm v_n-\bm v_p|$. Lines that lead the neutron flow
 ($\mathcal Z=\mathcal Af^2$) enhance the counterflow, and lines
 that lag toward the charged fluid reduce it, weakly for
 $\mathcal Z=1$ and strongly for $\mathcal R\to\infty$.
 \textit{Bottom:} vorticity difference
 $\Delta\mathcal B^z/\mu_{n,0}$. \revadd{The annotated amplitude for
 $\mathcal Z=1$ is about $0.14$ times that for $\mathcal R\to\infty$.}}
 \label{fig:kh-thouless-maps}
\end{figure*}

Figure~\ref{fig:kh-thouless-discriminator} shows the stronger
response of the counterflow to the transport law. We find that pinning reduces the
counterflow on the vortex layers, while Thouless transport
retains more relative motion than free transport.

Figure~\ref{fig:kh-thouless-maps} shows these differences at
$t=37.5$. Counterflow forms spirals
through the billow cores in the free and Thouless runs, with
larger amplitudes for Thouless transport. Pinning weakens these
spirals{, and baryon-current transport weakens them
slightly}.

\section{Neutron star interiors}
\label{sec:stellar_applications}

As the main application of our general-relativistic superfluid code framework, we now present full nonlinear simulations of neutron star interiors.
Specifically, we construct self-gravitating two-fluid neutron stars in full general
relativity and evolve their oscillations and nonlinear spin-down response.

\subsection{Equilibrium construction and initial-state matching}
\label{sec:stellar_initial_data}

We construct general-relativistic superfluid neutron star equilibria following the
stationary, axisymmetric formulation of Ref.
\cite{Prix:2004wq}. Each constituent rotates uniformly about the same
axis, with angular velocities $\Omega_n$ and $\Omega_p$ specified
independently. In quasi-isotropic
coordinates, written on the meridian as $R=r\sin\theta$ and
$z=r\cos\theta$, the geometry and circular four-velocities are
\begin{align}
 ds^2={}&-\alpha^2dt^2+A_g^2(dR^2+dz^2)
       +B_g^2R^2(d\phi-\omega_{\rm fd}dt)^2,\notag\\
 u_x^\mu={}&\frac{W_x}{\alpha}(1,0,0,\Omega_x),
 \qquad x\in\{n,p\},\notag\\
 v_x^{\hat\phi}={}&\frac{B_gR}{\alpha}
                  (\Omega_x-\omega_{\rm fd}),\qquad
 W_x=[1-(v_x^{\hat\phi})^2]^{-1/2}.
 \label{eq:stellar_prix_geometry}
\end{align}
Here $\omega_{\rm fd}$ is the frame-dragging angular velocity, so
$\beta^\phi=-\omega_{\rm fd}$. The metric amplitudes $A_g$ and $B_g$
are distinct from the entrainment coefficient $\mathcal A$.
The neutron star consists of two currents $n_n^\mu=\nu_nu_n^\mu$ and
$n_p^\mu=n_pu_p^\mu$, for neutrons and protons respectively.

\begingroup
\begin{figure*}[tbp]
  \centering
  \includegraphics[width=0.98\textwidth]{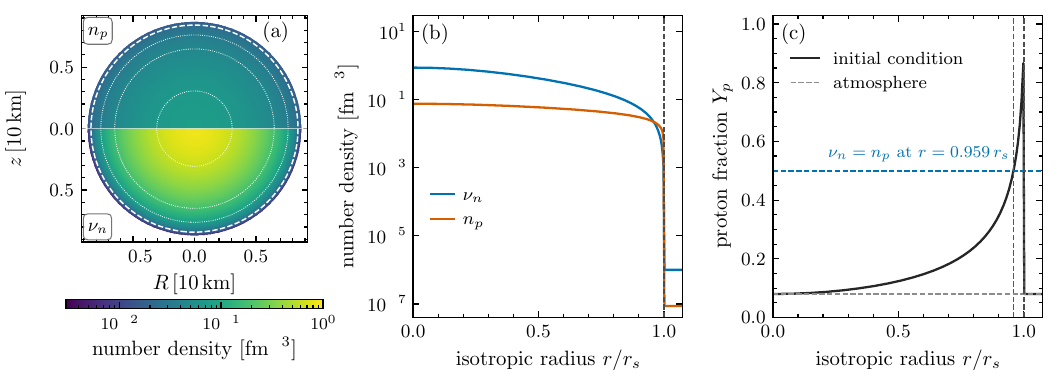}
  \caption{{Initial condition of the nonrotating two-fluid
  star.
  \textit{Left:} meridional rest-frame number density distribution, with proton density $n_p$
  above the equator and neutron density $\nu_n$ below it, on a common
  logarithmic scale.  The solid white curve marks the stellar surface and
  \revadd{the dashed curve marks proton fraction $Y_p=0.5$. Dotted curves
  mark $Y_p=0.1$, $0.2$, and $0.3$.}
  \textit{Middle:} initial number density profiles through the first grid row
  adjacent to the equator. The dashed vertical line marks the isotropic
  surface radius $r_s$, beyond which the dilute two-fluid atmosphere begins.
  \textit{Right:} the corresponding proton fraction $Y_p(r)$, with
  $\nu_n=n_p$ near $0.959r_s$.}
  }
  \label{fig:star-transfer}
\end{figure*}
\endgroup

The stationary Euler equations admit one first integral for each
uniformly rotating constituent \cite{Prix:2004wq},
\begin{equation}
 p^x_t+\Omega_xp^x_\phi=-C_x,\qquad
 p^n_\mu=\xi_\mu,\quad p^p_\mu=\mu^p_\mu .
 \label{eq:stellar_canonical_first_integrals}
\end{equation}
The constants $C_x>0$ fix the two matter distributions. Both currents
source the same metric, so their equilibria must be solved together.
For the cold, unentrained EOS used in both calculations,
$m_n=m_p=1$, $\mathcal A=0$, and $(\Gamma_n,\Gamma_p)=(2,4)$,
\begingroup
\begin{equation}
 e_0=\nu_n+n_p+0.024\,\nu_n^2
             +0.2394358512\,n_p^4 .
 \label{eq:results_star_eos}
\end{equation}
\endgroup
{Equation~\eqref{eq:results_star_eos} uses
$n_{\rm nuc}=0.1\,\mathrm{fm}^{-3}$ as the number-density unit and
$\rho_{\rm nuc}c^2$ as the energy-density unit, where
$\rho_{\rm nuc}=1.66\times10^{14}\,\mathrm{g\,cm^{-3}}$.
The mass unit is $m_B=\rho_{\rm nuc}/n_{\rm nuc}$.}
\textcolor{black}{The pressure coefficients are $K_n=0.024$ and $K_p=0.7183075536$,
in units of $m_Bc^2/n_{\rm nuc}$ and $m_Bc^2/n_{\rm nuc}^3$,
respectively. The charged contribution to $e_0$ is therefore
$K_p n_p^4/3$, as in Eq.~\eqref{eq:main_model_b_master}.} The momenta are
$p^x_\mu=\mu_xu^x_\mu$, and the first integrals give the explicit
matter recovery
\begingroup\color{black}
\begin{align}
 \mu_x&=\frac{C_xW_x}{\alpha},\notag\\
 \nu_n&=\frac{[\mu_n-1]_+}{2K_n},\qquad
 n_p=\left(\frac{3[\mu_p-1]_+}{4K_p}\right)^{1/3},
 \label{eq:stellar_equilibrium_recovery}
\end{align}
\endgroup
where $[a]_+=\max(a,0)$. The zero-density surfaces therefore satisfy
$\alpha/W_x=C_x/m_x$. Their relative extent follows from $C_n,C_p$
and the two rotation rates. Where one constituent vanishes, only the
remaining current sources the matter stress. The vanishing constituent
has zero pressure at its \revadd{surface. The} metric extends continuously to
the common vacuum exterior. For the initial equilibria used in
the stellar tests below, we choose $\Omega_n=\Omega_p$ and $C_n=C_p$,
giving chemical equilibrium and a common surface for the adopted EOS.
In the rotating-star experiment, subsequent spin-down of the charged
constituent generates a rotational lag.
Their different polytropic indices still produce different density
profiles.

For a trial metric, Eq.~\eqref{eq:stellar_equilibrium_recovery} 
determines the cold stress tensor
$T_0^{\mu\nu}=\sum_x n_x\mu_xu_x^\mu u_x^\nu+\Psi_0g^{\mu\nu}$,
where $n_n=\nu_n$ and
$\textcolor{black}{\Psi_0=\sum_xn_x\mu_x-e_0=K_n\nu_n^2+K_p n_p^4}$.
Define its Eulerian projections
$E=T_0^{\mu\nu}n_\mu n_\nu$,
$j^i=-\gamma^i{}_\mu T_0^{\mu\nu}n_\nu$, and
$S^i{}_j=\gamma^i{}_\mu T_0^{\mu\nu}\gamma_{\nu j}$,
with $S=S^i{}_i$ and $S_m=S^R{}_R+S^z{}_z$.
Writing $\nu_g=\ln\alpha$, $a_g=\ln A_g$, and $b_g=\ln B_g$,
the four elliptic metric equations are \cite{Prix:2004wq}
\begin{align}
 \Delta_3\nu_g={}&4\pi A_g^2(E+S)+A_g^2\mathcal K\notag\\
 &-\partial\nu_g\cdot\partial(\nu_g+b_g),\notag\\
 \widetilde\Delta_3(R\omega_{\rm fd})={}&
 -16\pi\alpha A_g^2Rj^\phi\notag\\
 &-R\,\partial\omega_{\rm fd}\cdot\partial(3b_g-\nu_g),
 \notag\\
 \Delta_2[(\alpha B_g-1)R]={}&
 8\pi\alpha A_g^2B_gR S_m,\notag\\
 \Delta_2(\nu_g+a_g)={}&8\pi A_g^2S^\phi{}_\phi
 +\frac32A_g^2\mathcal K\notag\\
 &-(\partial\nu_g)^2 ,
 \label{eq:stellar_metric_elliptic}
\end{align}
where $\Delta_2=\partial_R^2+\partial_z^2$,
$\Delta_3=\Delta_2+R^{-1}\partial_R$,
$\widetilde\Delta_3=\Delta_3-R^{-2}$, and
$\partial f\cdot\partial g=f_{,R}g_{,R}+f_{,z}g_{,z}$. Here
\begin{equation}
 \mathcal K=K_{ij}K^{ij}
 =\frac{B_g^2R^2}{2\alpha^2A_g^2}
   \left[(\partial_R\omega_{\rm fd})^2+
         (\partial_z\omega_{\rm fd})^2\right].
 \label{eq:stellar_extrinsic_curvature}
\end{equation}
Equation~\eqref{eq:stellar_metric_elliptic} uses geometrized
matter sources. With lengths in $L_0=10\,\mathrm{km}$ and matter
in the units of Eq.~\eqref{eq:results_star_eos}, its matter factors
$4\pi$, $8\pi$, and $16\pi$ become $\kappa_g$, $2\kappa_g$,
and $4\kappa_g$, respectively, with
$\kappa_g=4\pi G\rho_{\rm nuc}L_0^2/c^2$.
We impose asymptotic flatness,
$\alpha,A_g,B_g\to1$ and $\omega_{\rm fd}\to0$,
regularity on the rotation axis, including $A_g=B_g$ there,
and equatorial reflection symmetry.
Numerically, we use the publicly available \textsc{Sfluide} code, which is part of the \textsc{LORENE} library.

Both equilibria have central densities
$(\nu_{n,c},n_{p,c})=(0.873057,0.075918)\,\mathrm{fm}^{-3}$,
or central logarithmic enthalpies
$\ln(\mu_{x,c}/m_x)\simeq0.35$, and central proton fraction
$Y_p=n_p/(\nu_n+n_p)=0.08$.
Together with the EOS and boundary conditions above, these central
data and the rotation rates specify the two models.
Table~\ref{tab:stellar_equilibria} gives their first-integral
constants and basic properties. The nonrotating star also has
baryon mass $M_b=1.62783\,M_\odot$, circumferential radius
$R_s=11.0866\,\mathrm{km}$, and compactness
$GM/(R_sc^2)=0.19597$. The rotating model has an oblate common surface.

\begin{table}[ht]
 \caption{{Cold equilibria used for the neutron star
 calculations. The radii (equatorial: $r_{\rm eq}$, and polar: $r_{\rm pol}$) are quasi-isotropic coordinate radii.
 $J$ is given in geometrized units with $L_0=10\,\mathrm{km}$. $\Omega _n$ and $\Omega_p$ are the angular frequencies of the neutron and proton components, $M$ is the total gravitational mass, $C_n$ and $C_p$ are first-integral constants, and $\alpha_c$ is the central lapse value. }}
 \label{tab:stellar_equilibria}
 \begin{ruledtabular}
 \begin{tabular}{lcc}
  & Nonrotating & Rotating\\
 \hline
 $\Omega_n/(2\pi)=\Omega_p/(2\pi)\ [\mathrm{Hz}]$ & $0$ & $500$\\
 $C_n=C_p$ & $0.779786275$ & $0.772456963$\\
 $\alpha_c$ & $0.549506$ & $0.544341$\\
 $M/M_\odot$ & $1.47137$ & $1.4997$\\
 $r_{\rm eq}\ [\mathrm{km}]$ & $8.7796$ & $8.95671$\\
 $r_{\rm pol}\ [\mathrm{km}]$ & $8.7796$ & $8.54079$\\
 $J/L_0^2$ & $0$ & $0.0103343$\\
 \end{tabular}
 \end{ruledtabular}
\end{table}

Figure~\ref{fig:star-transfer} shows the nonrotating density and
composition profiles. The proton fraction increases from $0.08$
at the center to $0.5$ near $0.959r_s$, with
$r_s=r_{\rm eq}=r_{\rm pol}$. The still more proton-rich terminal
skin contains negligible matter. This stratification follows from
chemical equilibrium with the two different polytropic responses.

The equilibrium solution is specified in the physical neutron--proton
basis. For our Carter master function, the entrainment coefficients
are $K^{nn}=(\mu_n-\mathcal A n_p)/\nu_n$,
$K^{np}=K^{pn}=\mathcal A$, and
$K^{pp}=(\mu_p-\mathcal A\nu_n)/n_p$ on the positive-density branch.
For our simulations, we choose the charged fluid as the common frame,
$u^\mu=u_p^\mu$, and reconstruct the neutron canonical covector before
forming the Gusakov--Son scalars. The map follows
Eqs.~\eqref{eq:moments_carter_map} and
\eqref{eq:main_model_b_responses}:
\begin{align}
 \xi_\mu&=H u^n_\mu+\mathcal A n_pu_\mu,
 &H&=\mu_n-\mathcal A n_p,\notag\\
 \Gamma_r&=-u_n^\mu u_\mu,
 &\mu&=H\Gamma_r+\mathcal A n_p,\notag\\
 q_\mu&=\xi_\mu-\mu u_\mu,
 &X&=\xi_\mu\xi^\mu,\notag\\
 Q&=X+\mu^2=H^2(\Gamma_r^2-1),
 &f^2&=\frac{\nu_n}{H},\notag\\
 \chi&=\mu_p+\mathcal A\nu_n(\Gamma_r-1)-\mu,
 &n_{\rm eff}&=n_p(1-\mathcal A f^2).
 \label{eq:stellar_prix_gs_dictionary}
\end{align}
Thus the neutron chemical potential $\mu_n$, the proton-frame
projection $\mu=-u^\mu\xi_\mu$, and the Eulerian projection
$\Xi=-n^\mu\xi_\mu$ are different quantities away from coflow.
The selected currents are $J_{\rm B}^\mu=n_n^\mu+n_p^\mu$ and
$R_{\rm B}^\mu=n_pu^\mu$. In the pressure representation,
$(\vartheta,r,a,Y)=(\chi,n_p,n_{\rm eff},f^2)$, with
$n_{\rm B}=n_p+\nu_n\Gamma_r$. These relations supply the initial
pressure coordinates and response coefficients directly from the
imported physical state.

For the thermal completion{}, the carrier also carries
$S_c^\mu=s_cu^\mu$. Equation~\eqref{eq:main_model_b_thermal} {} gives
$\mathcal P=\Psi_{{\rm B},0}+P_{\rm th}$ and
$C=n_p\chi+\mu n_{\rm eff}+T_cs_c$, with
$P_{\rm th}=e_{\rm th}/3$ and $e_{\rm th}=C_{\rm th}s_c^{4/3}$.
The thermal term changes the pressure, energy and carrier enthalpy.

The numerical state is then constructed in the Eulerian frame of the
imported metric. Writing $W=W_p$, $z^i=Wv_p^i$, and
$\varepsilon_\xi=\sqrt{-X}$, the required projections are
\begin{align}
 \Xi&=\sqrt{\varepsilon_\xi^2+\gamma^{ij}\xi_i\xi_j},
 &\xi_0&=-\alpha\Xi+\beta^i\xi_i,\notag\\
 D_{\rm B}&=\nu_nW_n+n_pW,
 &D_p&=n_pW,\notag\\
 S_i&=CWz_i+f^2\Xi\xi_i,\notag\\
 \tau&=CW^2-\mathcal P+f^2\Xi^2-D_{\rm B}.
 \label{eq:stellar_initial_conserved}
\end{align}
The entropy density is $s_cW$. Multiplication by
$\sqrt\gamma=A_g^2B_gR$ gives the densitized conserved fields. \revadd{The axisymmetric storage convention}
divides the densitized cell variables, fluxes and sources by $R$.
This volume factor is
restored in physical particle-number and angular-momentum integrals.
Canonical momentum is transferred as a spatial covector. Matter
variables are sampled at cell centers and canonical components at their
staggered locations. The discrete curl supplies the initial
canonical-vorticity field.

For the actual initial models, $\mathcal A=0$ and the constituents
comove: $\Omega_n=\Omega_p=0$ in the oscillation model and
$\Omega_n=\Omega_p=\Omega_0$ in the rotating model. Consequently
\begin{align}
 \Gamma_r&=1,\quad \mu=\mu_n,\quad q_\mu=0,\quad
 f^2=\frac{\nu_n}{\mu_n},\notag\\
 n_{\rm eff}&=n_p,\quad \chi=\mu_p-\mu_n,\quad
 C=n_p\mu_p+T_cs_c.
 \label{eq:stellar_initial_coflow}
\end{align}

The charged-frame flux and trace-free stress initially vanish.
In the nonrotating model $\xi_i=0$, whereas rotation gives
$\xi_\phi=\mu_nWB_g^2R^2(\Omega_0-\omega_{\rm fd})/\alpha$.
The rotating star therefore carries canonical circulation
$\mathcal C_\xi=2\pi\xi_\phi$ even though its initial counterflow
vanishes. Spin-down subsequently creates the lag.

\subsection{Radial oscillations of a nonrotating superfluid neutron star}
\label{sec:results-star-breathing}

As a quantitative correctness test of the framework, we consider oscillations of a perturbed nonrotating superfluid neutron star. This is a natural {extension} of similar setups used to test relativistic hydro {\cite{Font:1999wh,Font:2001ew}} and GRMHD codes {\cite{Giacomazzo:2007ti,Cerda-Duran:2008qfl}}.
The spacetime is fixed in the relativistic
Cowling approximation
\cite{Cowling:1941nqk,1983ApJ...268..837M}.
This cold NC calculation has $\mathcal A=0$,
separately conserved neutron and charged particle numbers, and free
neutron-current transport, with pinning and mutual friction disabled.
We compare the evolution with an independent free-surface
eigenproblem on the same equilibrium (see Appendix \ref{app:model_b_radial_cowling}).

After importing the nonrotating initial condition given in Tab. \ref{tab:stellar_equilibria}, we add a controlled radial perturbation to the velocity, 
$\dot V_{\rm kick}^{\,r}=(A_k/\Delta t_k)\sin(\pi r/r_s)$
for $0<r<r_s$ over a short timespan $t_k\leq t<t_k+\Delta t_k$, with
$A_k=10^{-2}$, $t_k=0.05t_u$ and $\Delta t_k=5t_u$.
We evolve the system in axisymmetry using $256^2$ cells with
$\Delta x=0.0625\,\mathrm{km}$ to
$t=2000t_u=66.7\,\mathrm{ms}$, where
$t_u=L_0/c=33.3564\,\mu\mathrm{s}$ and $L_0=10\,\mathrm{km}$.
In order to avoid contamination from numerical errors at the surface, we evolve the system in a simple well-balanced form, by subtracting out the discrete $t=0$ right-hand-side update from the update \cite{2016arXiv160202711P}.

Using the approach in Appendix ~\ref{app:model_b_radial_cowling}, we can exactly calculate neutron and
charged-fluid fundamental (F) mode frequencies, and the first neutron acoustic overtone,
\begingroup
\begin{align}
 \omega_{n,F}t_u&=0.74439,
 &\omega_{p,F}t_u&=1.13242,
 \notag\\
 \omega_{n,p_1}t_u&=1.25677 .&&
 \label{eq:results_star_eigenvalues}
\end{align}
\endgroup
After unit conversion, the neutron fundamental is $3.552\,\mathrm{kHz}$.
The two acoustic families reflect the additional fluid degree of
freedom \cite{Prix:2002fk,Andersson:2002jd}.

\begin{figure}[tbp]
  \centering
  \includegraphics[width=\columnwidth]{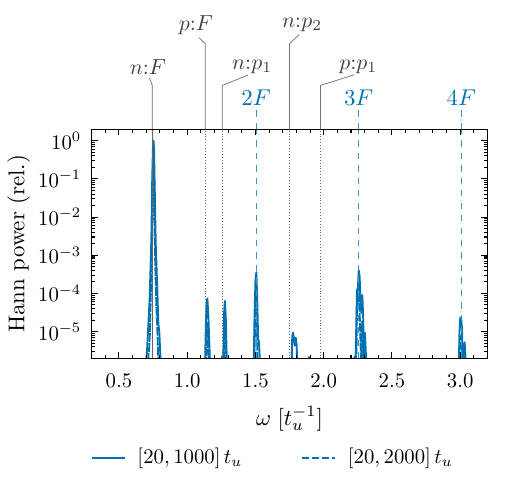}
  \caption{Central-density oscillation spectra of the two-fluid neutron star. These are separately computed over half the evolution and the full evolution time, $20\leq t/t_u\leq1000$ (solid) and $20\leq t/t_u\leq2000$ (dashed).
Gray guides mark the exact radial mode Cowling frequencies in Table~\ref{tab:app_cowling_modes}: $n$ and $p$ label the constituent
families, $F$ the fundamental and $p_j$ the acoustic overtones.
Blue dashed guides mark $2F$, $3F$ and $4F$ of the measured
dominant frequency. Here $t_u = 10\rm km/c$ is a characteristic time unit.}
 \label{fig:star-breathing}
\end{figure}

Figure~\ref{fig:star-breathing} shows a fundamental-dominated
response over approximately $237$ reference periods.
We can find excellent agreement with the exact F- and p-mode frequencies for both neutrons and protons, as well as with their overtones. 
This validates the correctness of the GR aspect of our relativistic superfluid implementation.

\subsection{{Unpinning, vortex transport and neutron star spin-down}}
\label{sec:results-model-b-star-unpinning}

We now want to assess vortex dynamics during neutron star spin-down in a fully relativistic neutron star model.
To this end, we next spin down the charged constituent of an initially
corotating star. The resulting lag loads the pinned vortex array,
leading to release, outward circulation transport and partial
repinning as the load falls
\cite{Anderson:1975zze,Haskell:2015jra,2016MNRAS.455.3952S}.
We follow this process in a dynamical toy calculation on a fixed
general-relativistic stellar spacetime, retaining line energy and tension.
Figure~\ref{fig:unpinning-schematic} illustrates the force balance
and the prescribed release and repinning sequence.
\textcolor{black}{We use $\ell_{\rm num}$ for the dimensionless numerical
Magnus load, $\bar f_c$ for the normalized holding strength, and
$\Lambda=\ell_{\rm num}/\bar f_c$ for their ratio. Their
normalization and switching proxy are specified in
Eqs.~\eqref{eq:results_model_b_pinning_units} and
\eqref{eq:results_model_b_star_load_code}.}

\begin{figure*}[tbp]
 \centering
 \includegraphics[width=0.98\textwidth]{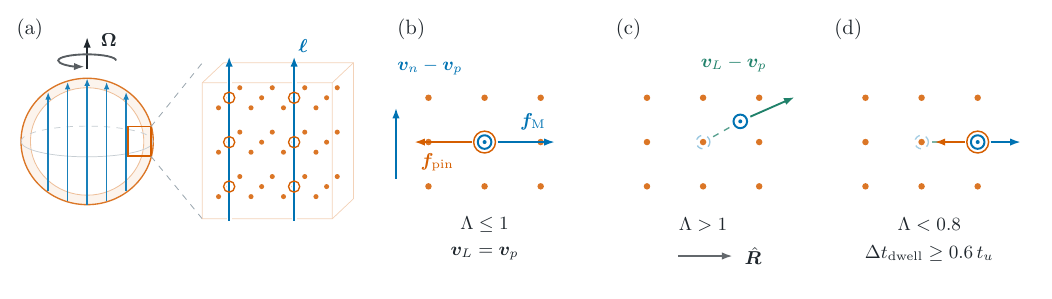}
 \caption{Forces on pinned and released neutron vortices
 during stellar spin-down. \textit{(a)} Axial vortices (blue) pass
 through an illustrative crustal lattice ({orange}), with rings marking
 pinning contacts and $\boldsymbol\ell$ the line direction.
 \textit{(b)--(d)} Local views in the carrier frame, with vortices
 pointing toward the reader, outward motion to the right and
 positive azimuthal motion upward.
 \textit{(b)} Faster neutron rotation, $\Omega_n>\Omega_p$,
 produces an outward Magnus load balanced by an inward pinning
 reaction. \textit{(c)} Release permits outward line motion when
 the numerical load ratio $\Lambda=\ell_{\rm num}/\bar f_c$ exceeds
 unity. \textit{(d)} A released cell can repin after accumulating
 at least $0.6t_u$ of eligible low-load time with $\Lambda<0.8$.
 Here $t_u=L_0/c$, with $L_0=10\,\mathrm{km}$.}
 \label{fig:unpinning-schematic}
\end{figure*}

\subsubsection{{Spin down prescription}}
\label{sec:unpinning_prescription}
\label{app:unpinning_prescription}

We use the rotating equilibrium in Table~\ref{tab:stellar_equilibria},
with $\Omega_0/(2\pi)=500\,\mathrm{Hz}$ and
$\Omega_0t_u=0.1047922511$. The EOS is
Eq.~\eqref{eq:results_star_eos}, with $\mathcal A=0$ and the thermal
completion of Eq.~\eqref{eq:main_model_b_thermal}. We initialize uniform
proper entropy $s_c=5.700665\times10^{-5}$ and set
$C_{\rm th}=K_p\lambda_{\rm th}^{4/3}=0.0432534728127$, with
$\lambda_{\rm th}=0.1215578225$. These values give
$T_c=2.220\times10^{-3}$ and $e_{\rm th}=9.490\times10^{-8}$ initially.
Here $\lambda_{\rm th}=(C_{\rm th}/K_p)^{3/4}$ is an EOS normalization.
We use a $256^2$ grid with resolution $\Delta x=62.5\,\mathrm{m}$.

The cold stellar equilibrium supplies a reference configuration.
Adding line energy changes the momenta and stress, adding a small,
but for our purposes negligible perturbation.
The length and time units are $L_0=10\,\mathrm{km}$ and $t_u=L_0/c$.

We model crustal forcing from magnetospheric spin-down as a prescribed change in rotation rate, via a proxy $\Omega_{\rm drv}$ through
\begin{align}
 \dot\Omega_{\rm drv}&=-10^{-3}t_u^{-2}\,r(t),\notag\\
 r(t)&={\cal S}(\bar t-0.01)\,[1-{\cal S}(\bar t-15)],
 \qquad \bar t=t/t_u,
 \label{eq:results_model_b_star_drive}
\end{align}
where ${\cal S}(x)=s^2(3-2s)$ with $s=\min[1,\max(0,x)]$.
The ramp rises over $0.01<\bar t<1.01$, remains unity through
$\bar t=15$, and switches off at $\bar t=16$.
This is an absolute angular acceleration, giving
$\Omega_{\rm drv}=0.0898022511t_u^{-1}=0.856955\Omega_0$ at the end of spin-down, i.e., a $14.3\%$ spin reduction.

The driving enters through an azimuthal momentum correction at each
accepted endpoint,
\begin{align}
 \Delta\widetilde S_\phi
 &={\cal I}_\phi(\Omega_{\rm drv}-\Omega_p),\notag\\
 {\cal I}_\phi&=\frac{\sqrt\gamma\,C_TW^3\gamma_{\phi\phi}}{\alpha},
 \label{eq:results_model_b_star_rigid_projection}
\end{align}
with $C_T=n_p\mu_p+T_cs_c$ for this unentrained model.

For a numerical evolution of the relativistic system, the intrinsic challenge is that the spin-down time can be very long.
In order to demonstrate the physical mechanism in sequence in a computationally efficient way for this first toy spin-down scenario, we impose the spin-down unrealistically rapidly to resolve release and circulation transport within a short calculation.
For a straight, uniformly rotating vortex array in Newtonian
force equilibrium \cite{Sourie:2020ima}, spatially uniform mobility gives
\begin{align}
 \dot\Omega_n&=-2\mathcal B_{\rm eff}\Omega_n\Delta\Omega,\notag\\
 \tau_{\rm SD}&\equiv\frac{\Omega_n}{|\dot\Omega_n|}
 =\frac{1}{2\mathcal B_{\rm eff}|\Delta\Omega|}.
 \label{eq:results_model_b_neutron_spindown}
\end{align}
Here $\mathcal B_{\rm eff}=\beta_s\mathcal R_{\rm heat}/(1+\mathcal R_{\rm tot}^2)$
from Eq.~\eqref{eq:moments_newtonian_line_velocity} and
$\textcolor{black}{n_v=2\Omega_n/\kappa_{\rm vel}}$.
This means that small $\Delta\Omega\propto f_c$, where $f_c$ is the holding strength, in the small-speed limit
of Eq.~\eqref{eq:results_model_b_star_critical_lag} can mean very long integration times.
For a first qualitative test of release and repinning, rescaling the
holding strengths permits larger resolved lags and shortens
$\tau_{\rm SD}$ at fixed mobility near the release threshold.
More realistic values relevant for glitch scenarios will be considered in future work.

\subsubsection{{Pinning force model}}
\label{sec:unpinning_forces}
\label{sec:unpinning_diagnostics}
\label{app:unpinning_diagnostics}

{We prescribe holding strengths in the crust and core, motivated
by vortex--nucleus pinning~\cite{2016MNRAS.455.3952S} and
vortex--fluxoid interaction estimates~\cite{Link:2003hq,Glampedakis:2011px}.
Below the release threshold, the lines are constrained to follow the
charged carrier. Above it, they move under the full line-force law.
In the core, this switch releases an effective carrier-locking
constraint on a still-bound neutron-vortex--fluxoid bundle. }

We use $\textcolor{black}{\kappa_v=\pi\hbar}$ for the physical momentum circulation
and $\textcolor{black}{\kappa_{\rm vel}=\kappa_v/m_n}$ for its Newtonian
velocity counterpart. The carrier-frame vorticity defines
\begin{align}
 {\cal B}_{(u)}^\mu
 &=-\frac12\epsilon^{\mu\nu\rho\sigma}u_\nu\Omega_{\rho\sigma},
 \notag\\
 H_u&=\sqrt{{\cal B}_{(u)}^\mu{\cal B}^{(u)}_\mu},\qquad
 \widehat\ell^\mu={\cal B}_{(u)}^\mu/H_u,
 \label{eq:results_model_b_star_carrier_vorticity}
\end{align}
The carrier measures a line density $\textcolor{black}{H_u/\kappa_v}$.
It is a proper density for pinned lines. For independently moving
lines the invariant density is $\textcolor{black}{\sqrt{B^2-E^2}/\kappa_v}$.
For a Magnus-only pinned reaction, the covariant holding load per
unit proper line length is
\begingroup\color{black}
\begin{align}
 {\cal L}_{\rm cov}&=\mathcal C_n\kappa_v
 \left(P_\perp^{\mu\nu}q_\mu q_\nu\right)^{1/2},\notag\\
 P_\perp^{\mu\nu}&=g^{\mu\nu}+u^\mu u^\nu-
 \widehat\ell^\mu\widehat\ell^\nu\,,
 \label{eq:results_model_b_star_load}
\end{align}
\endgroup
With line energy at $\mathcal A=0$, the dressed neutron enthalpy
is $h_n=\mu_n+F_{,\nu_n}$, and $\xi_\mu=h_nu^n_\mu$.
The corresponding current response is $\mathcal C_n=\nu_n/h_n$.
For $F=0$ these reduce to $\mu_n$ and $f^2=\nu_n/\mu_n$.
Circular flow with meridional lines orthogonal to both currents gives
\begingroup\color{black}
\begin{align}
 \Gamma_r&=W_nW_p(1-v_n^{\hat\phi}v_p^{\hat\phi}),\notag\\
 {\cal L}_{\rm cov}
 &=\kappa_v\nu_n\sqrt{\Gamma_r^2-1}\notag\\
 &=\kappa_{\rm vel}\rho_{n,\rm rest}W_nW_p\frac{\varpi}{\alpha}
       |\Omega_n-\Omega_p|\,,
 \label{eq:results_model_b_star_load_primitives}
\end{align}
\endgroup
where \textcolor{black}{$\rho_{n,\rm rest}=m_n\nu_n$ and}
$\varpi=\sqrt{\gamma_{\phi\phi}}=B_gR$.

The holding strength per unit proper line length is prescribed separately in the crust and core.
Using $\rho_b=m_n\nu_n+m_pn_p$, we set
\begin{equation}
 f_c(\rho_b,B_{\rm mag})=
 \begin{cases}
  f_\Phi(B_{\rm mag}),&\rho_b>\rho_{\rm cc},\\{\eta_N}\,f_N(\rho_b),&\rho_1\leq\rho_b\leq\rho_{\rm cc},\\
  0,&\rho_b<\rho_1.
 \end{cases}
 \label{eq:results_model_b_star_threshold}
\end{equation}
with $\rho_1=1.5\times10^{12}\,\mathrm{g\,cm^{-3}}$ and
$\rho_{\rm cc}=1.3\times10^{14}\,\mathrm{g\,cm^{-3}}$.
The dimensionless factors $\eta_N$ and $\eta_\Phi$ independently
scale the crustal and core holding strengths. Their reference
values are unity.
{In the core we assume coexisting neutron superfluidity and type-II
proton superconductivity. A pinned bundle follows the charged fluid.
After release, its attached fluxoids move with the neutron vortex and
may slip relative to that fluid~\cite{Sourie:2020drw,Sourie:2021vsj}.
We use the vortex--fluxoid intersection-force estimate as an effective
holding-strength scale~\cite{Link:2003hq,Glampedakis:2011px},}
\begin{equation}
 f_\Phi=3\times10^{15}\eta_\Phi
 \left(\frac{B_{\rm mag}}{10^{12}\,\mathrm G}\right)^{1/2}
 \,\mathrm{dyn\,cm^{-1}} .
 \label{eq:results_model_b_core_pinning}
\end{equation}
The square-root scaling follows from an intersection force divided
by the fluxoid spacing, $d_\Phi\propto B_{\rm mag}^{-1/2}$.

For nuclei we use the lattice-averaged force per unit vortex length {$f_N$} of Ref. ~\cite[Tables~1 and~5]{2016MNRAS.455.3952S}, including their
estimate of the rigidity length from vortex tension. We select their
weak-pairing case $\beta=3$ with variable vortex tension and interpolate their calculation (see Fig. \ref{fig:unpinning-two-zone-strength}).
The five density and force pairs are
\begin{align}
 \frac{\rho_i}{10^{12}\,\mathrm{g\,cm^{-3}}}
 &=(1.5,9.6,34,78,130),\notag\\
 \frac{f_i}{10^{15}\,\mathrm{dyn\,cm^{-1}}}
 &=(0.32,0.31,8.55,1.84,0.06).
 \label{eq:results_model_b_crust_pinning_nodes}
\end{align}
Between adjacent nodes, logarithmic interpolation gives
\begin{align}
 f_N(\rho_b)&=f_i\left(\frac{f_{i+1}}{f_i}\right)^{t_i},\notag\\
 t_i&=\frac{\ln(\rho_b/\rho_i)}{\ln(\rho_{i+1}/\rho_i)},
 \qquad \rho_i\leq\rho_b\leq\rho_{i+1}.
 \label{eq:results_model_b_crust_pinning}
\end{align}
The core and crustal maps are separate prescriptions and need not
match at $\rho_{\rm cc}$.

For the reference field in Fig.~\ref{fig:unpinning-two-zone-strength},
we use the regular Euclidean dipole flux function
\begin{equation}
 \psi_{\rm mag}=
 \begin{cases}
 \dfrac{B_{\rm pole}}4 R^2(5-3r^2/R_\star^2),&r\leq R_\star,\\[2pt]
 \dfrac{B_{\rm pole}R_\star^3R^2}{2r^3},&r>R_\star,
 \end{cases}
 \label{eq:results_model_b_pinning_dipole}
\end{equation}
where $r^2=R^2+z^2$, $R_\star=0.89375L_0$\textcolor{black}{ is the stellar
radius and $B_{\rm pole}$ the polar surface magnetic field},
$B_{{\rm mag},R}=-R^{-1}\partial_z\psi_{\rm mag}$ and
$B_{{\rm mag},z}=R^{-1}\partial_R\psi_{\rm mag}$.
This prescribed test field fixes the normalized holding strength as
\begingroup\color{black}
\begin{equation}
 \bar f_c=\frac{f_c}{\rho_{\rm nuc}\kappa_{\rm vel}c},
 \qquad
 \rho_{\rm nuc}\kappa_{\rm vel}c
 \simeq9.84\times10^{21}\,\mathrm{dyn\,cm^{-1}},
 \label{eq:results_model_b_pinning_units}
\end{equation}
\endgroup
with $\rho_{\rm nuc}=1.66\times10^{14}\,\mathrm{g\,cm^{-3}}$.
In the corresponding matter units, the prescribed numerical load is
\begin{align}
 \ell_{\rm num}&=\mathcal C_{\rm num}\sqrt{\max\{0,q_\mu q^\mu-(q_i\widehat B_E^i)^2\}},
 \notag\\
 B_E^2&=\gamma_{ij}B^iB^j,\qquad
 \widehat B_E^i=B^i/B_E,\qquad
 B^i=\mathfrak B^i/\sqrt\gamma,\notag\\
 \Lambda&=\ell_{\rm num}/\bar f_c.
 \label{eq:results_model_b_star_load_code}
\end{align}
For the unentrained numerical recovery closure,
$\mathcal C_{\rm num}=\nu_n^{\rm rec}/\varepsilon_\xi$, where
$\nu_n^{\rm rec}$ is the recovered neutron density.
The projection uses the Eulerian line direction. For circular flow orthogonal to meridional vorticity, it gives
the normalized carrier-frame Magnus load when the numerical and
covariant current responses agree. In a general moving state we use
$\ell_{\rm num}$ as the specified switching proxy and do not
identify it with an exact radial Magnus-force component.
The field $B^i$ here is canonical vorticity, distinct from the
prescribed magnetic field $B_{\rm mag}$.

\begin{figure}[tbp]
 \centering
 \includegraphics[width=\columnwidth]{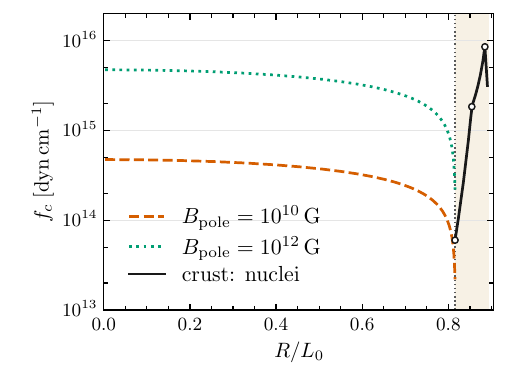}
 \caption{Holding force per unit vortex length in the
 neutron-star core and inner crust, sampled on the equator.
 Fluxoid pinning in the core is shown for
 $B_{\rm pole}=10^{10}\,\mathrm G$ (orange dashed) and
 $10^{12}\,\mathrm G$ (green dotted).
 The black curve gives nuclear pinning for the $\beta=3$
 variable-tension prescription of Ref.~\cite[Tables~1 and~5]{2016MNRAS.455.3952S}.
 The force reaches a maximum within the shaded inner crust.
 The vertical dotted line marks its interface with the core.
 These reference curves use $\eta_\Phi=\eta_N=1$, before the
 rescaling used in the evolution. Here $R$ is the cylindrical
 coordinate radius and $L_0=10\,\mathrm{km}$.}
 \label{fig:unpinning-two-zone-strength}
\end{figure}

\begin{figure*}[tbp]
 \centering
 \includegraphics[width=0.98\textwidth,height=0.81\textheight,keepaspectratio]{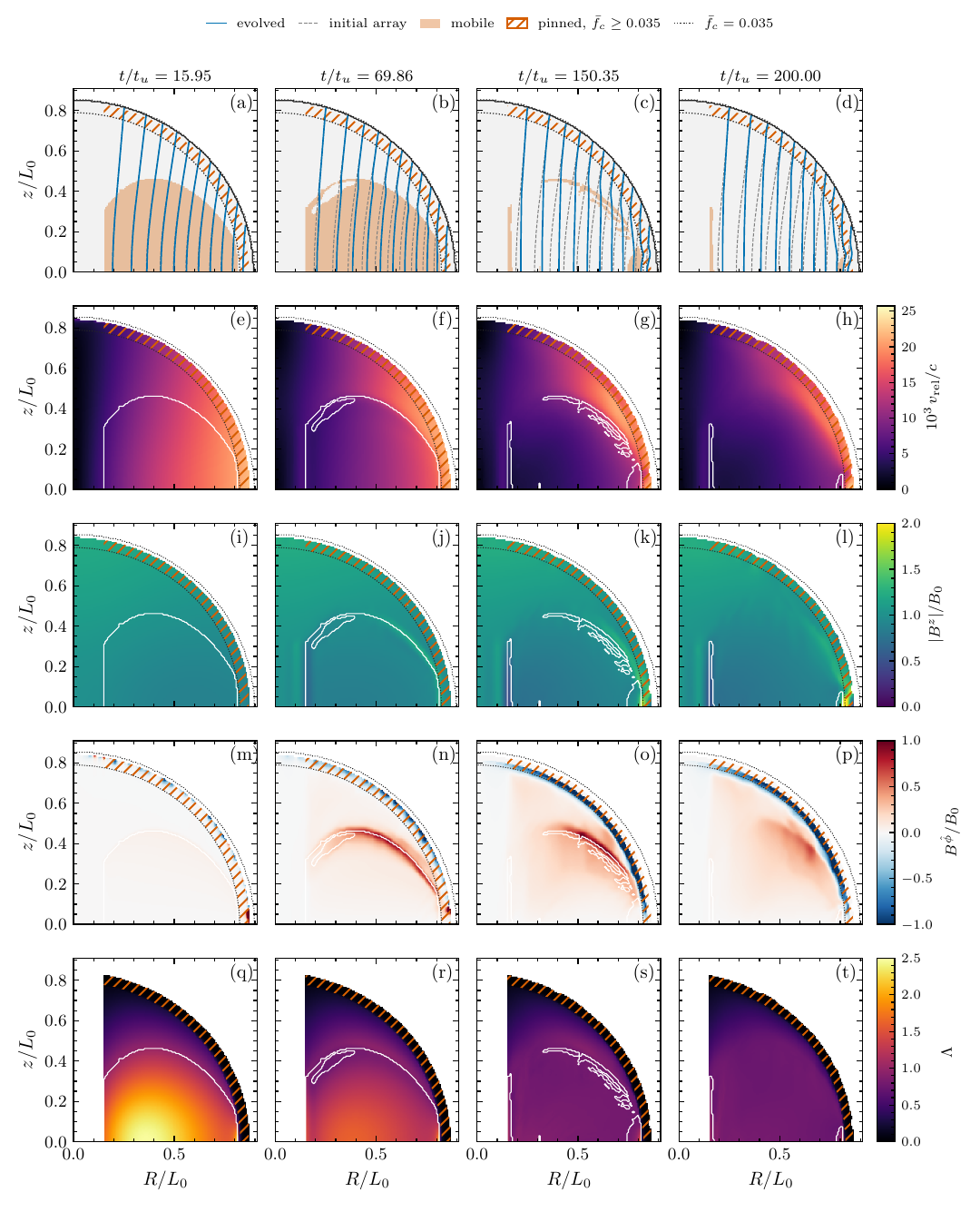}
 \caption{Outward transport and winding of the neutron
 vortex array during spin-down, at the four times shown above the
 columns. \textit{(a)--(d)} Poloidal circulation contours in the
 evolved state (blue) and initially (gray dashed).
 Orange shading marks mobile cells. Hatching selects pinned cells
 with normalized holding strength $\bar f_c\geq0.035$, and the dotted contour marks that
 strength threshold. White contours locate mobile regions in
 the lower panels.
 \textit{(e)--(h)} Neutron--carrier relative speed, $10^3v_{\rm rel}/c$,
 where $v_{\rm rel}/c=\sqrt{1-\Gamma_r^{-2}}$ and
 $\Gamma_r=-u_n^\mu u_{p\mu}$.
 \textit{(i)--(l)} Axial canonical vorticity,
 $|B^z|/B_0=|\mathfrak B^z|/(\sqrt\gamma B_0)$, with $B_0$ its
 initial central-cell magnitude.
 \textit{(m)--(p)} Signed orthonormal toroidal component,
 $B^{\hat\phi}/B_0=\sqrt{\gamma_{\phi\phi}}\,
 \mathfrak B^\phi/(\sqrt\gamma B_0)$.
 Its growth records winding of the initially poloidal array.
 \textit{(q)--(t)} Numerical load ratio
 $\Lambda=\ell_{\rm num}/\bar f_c$, as defined in
 Eq.~\eqref{eq:results_model_b_star_load_code}.
 The inert outer layer is blank in the field maps.
 Radii and times are measured in $L_0=10\,\mathrm{km}$ and
 $t_u=L_0/c$.}
 \label{fig:unpinning-lines}
\end{figure*}

For the evolution we take $B_{\rm pole}=10^{12}\,\mathrm G$ and
rescale the strengths by
\begin{equation}
 \eta_\Phi=6.28\times10^4,\qquad \eta_N=3\times10^6,
 \label{eq:results_model_b_pinning_rescale}
\end{equation}
These artificial strengths permit larger resolved lags in our
short \revadd{evolution, as discussed around Eq.~\eqref{eq:results_model_b_neutron_spindown}.}

For a prograde spin-down lag $\Delta\Omega=\Omega_n-\Omega_p>0$,
the circular-flow physical threshold obeys
\endgroup
\begingroup\color{black}
\begin{align}
 \Delta\Omega_{\rm crit}
 &=\frac{\alpha f_c}{\rho_{n,\rm rest}\kappa_{\rm vel} W_{n,\rm crit}W_p\varpi},\notag\\
 \Omega_{n,\rm crit}&=\Omega_p+\Delta\Omega_{\rm crit},
 \label{eq:results_model_b_star_critical_lag}
\end{align}
\endgroup
with $\Delta\Omega=\Omega_n-\Omega_p>0$ and
$W_{n,\rm crit}=W_n(\Omega_p+\Delta\Omega_{\rm crit})$.
This is implicit at fixed density and metric.
For small local speeds,
$\textcolor{black}{\Delta\Omega_{\rm crit}\simeq\alpha f_c/(\rho_{n,\rm rest}\kappa_{\rm vel}\varpi)}$.
A hypothetical uniform initial lag could release a chosen core
region $\mathcal K$ and leave the eligible crust $\mathcal C$
pinned if
\begin{equation}
 \max_{\mathcal K}\Delta\Omega_{\rm crit}
 <\Delta\Omega_0<\min_{\mathcal C}\Delta\Omega_{\rm crit}.
 \label{eq:results_model_b_star_critical_lag_window}
\end{equation}
The actual calculation starts in corotation and uses the numerical
load proxy during evolution, so this window is a physical
comparison and not an initial condition or an evaluated bound
for the trajectory.

The independent numerical circulation scale in the line-energy
model is denoted $\kappa_{\rm line}=2\times10^{-3}$ here.
It is the parameter called $\kappa_v$ in
Sec.~\ref{sec:model_b_vortex_tension}, with
$\mathcal N=\sqrt{B^2-E^2}/\kappa_{\rm line}$.
Dividing the pinning load by physical circulation does not set
$\kappa_{\rm line}$ to unity.
The other line parameters are $\mathcal N_0=20$ and
$\alpha_{\rm v}=0.05$.
The gap in Eq.~\eqref{eq:tension_bcs_core_law} is
$\Delta_{\rm pair}/\Delta_0=10^{-3}+(1-10^{-3})u/(1+u)$,
where $u=(\nu_n/0.6)^2$.
For $\nu_n<0.1$, the numerical prescription holds the line fields
inert and applies no vortex force.

{Mobile core bundles retain their attached fluxoids and use the
Sourie--Chamel lift and drag prescription~\cite{Sourie:2020drw,Sourie:2021vsj}, with}
\begin{align}
 N_\Phi&=30,\qquad \alpha_\Phi=2,\notag\\
 \mathcal Z&=30,\qquad \mathcal R'=0,\notag\\
 \mathcal R&=\frac{\mathcal R_n}{\Gamma_r+30n_p/\nu_n}.
 \label{eq:results_model_b_sc_transport_choice}
\end{align}
The unentrained EOS has $\varepsilon_p=0$.
Cancelling $\varepsilon_p^2$ before taking the attached-fluxoid
limit of Eqs.~\eqref{eq:force_fluxoid_drag_model}--\eqref{eq:force_fluxoid_unpinned_drag}
gives
\begin{equation}
 \mathcal R_n=0.36
 \left(\frac{x_p}{0.05}\right)^{7/6}
 \frac{\rho_{14}^{1/6}}{1-x_p}.
 \label{eq:results_model_b_sc_drag_choice}
\end{equation}
The local primitive state determines these coefficients.
They specify lift and drag sources in the line evolution.
In particular,
$\mathcal R\nu_L-\mathcal Z n_p\mathcal R'=\nu_n\mathcal R_n>0$
has the dissipative sign. It does not impose an algebraic
instantaneous slip velocity on the independently evolved fields. The line velocity is reconstructed from the independently evolved
fields,
$\boldsymbol v_L=\boldsymbol E\times\boldsymbol B/B^2$
(Sec.~\ref{sec:tension_electric_and_drift}).
Thus the prescribed lift and drag act together with the line
inertia and tension in the field evolution.

We prescribe distinct thresholds for unpinning and repinning,
together with a finite repinning delay, to suppress artificial
oscillations at the pinning interface. Pinned vortices
are released when the numerical load exceeds the local holding
strength,
\begin{equation}
 \Lambda=\frac{\ell_{\rm num}}{\bar f_c}>1\,.
 \label{eq:results_model_b_two_zone_release}
\end{equation}
Mobile vortices can repin only after spending at least $0.6t_u$
below $\Lambda=0.8$ following release, counting only time for which
the fluid state satisfies the numerical state requirements. If the load returns
to $\Lambda\geq0.8$ during such a step, this waiting period starts
again. The lower repinning threshold and the finite delay prevent
immediate repinning after release.

The prescription applies to regular fluid states outside the
numerical floor. A cell is treated as pinned when it first meets
these requirements and is released in that update if $\Lambda>1$.
If the state temporarily fails the requirements, its pinned or
mobile condition is preserved and that time does not count toward
the repinning delay.

\subsubsection{Vortex dynamics and angular momentum transport}

We then simulate the spin-down of the neutron star to $200\,t_u$. The evolution is shown in Fig. \ref{fig:unpinning-lines}, which we describe in the following. The top row shows the poloidal circulation contours, with orange shading marking mobile cells inside the neutron star. We can see that initially vortex arrays vertically thread the neutron star and remain pinned in the strong-pinning selection $\bar f_c\geq0.035$. As the charge carrier (proton) part of the crust is experiencing spin-down, a radial Magnus force begins to build up inside the star due to the relative drift of vortex arrays (anchored in the slowing crust) to the uniformly spinning neutron superfluid. This situation is highlighted especially well by comparing the numerical load ratio $\Lambda$ with unity. We can see that it exceeds unity in the released region, unpinning a maximum fraction of $50\%$ of the star. At that point, we can see that the vortex arrays begin to move outwards and a net relative flow difference, $v_{\rm rel}$, begins to build up, from the initial coflow. As the star spins down, more and more vorticity is transported toward the crust, as we can clearly see by the buildup of axial canonical vorticity, $|B^z|$, near the crust-core interface. Once the star spins down further, the vortex arrays begin to repin, until $\sim1.6\%$ of the eligible cells remain unpinned at the final time shown. At that point, the neutron fluid retains a lag relative to the carrier.

{The array also retains a net twist at the unpinning interface,
visible in the toroidal vorticity wound from the initially poloidal
configuration. Circulation transport allows differential rotation to
wind the array, while the line energy supplies the stress associated
with bending.} {Whether this twist develops into the vortex instabilities and tangling discussed in Ref.~\cite{Sidery:2007ge} requires a three-dimensional calculation we reserve for future work.}

We quantify the neutron angular-momentum change using the axial-current
balance of Refs.~\cite{Croft:2022gks,Xin:2025ymm}, following the radial
construction used for merger remnants in Ref.~\cite{Ng:2026jev}.

For the unentrained neutron fluid, $\xi_\mu=\varepsilon_\xi u_{n\mu}$
and $T^{a(n)}{}_{\phi}=N_s^a\xi_\phi$, with $a\in\{t,R,z\}$.
Particle conservation and Eq.~\eqref{eq:moments_superfluid_vorticity}
then give, in axisymmetry,
\begin{equation}
 \partial_a\!\left(\alpha\sqrt\gamma\,N_s^a\xi_\phi\right)
 =\alpha\sqrt\gamma\,f_{s\phi}.
 \label{eq:results_star_neutron_angular_balance}
\end{equation}
The neutron angular momentum is~\cite{Prix:2004wq}
\begin{align}
 J_n
 &=\int_{\Sigma_t}\alpha\sqrt\gamma\,T^{t(n)}{}_{\phi}\,d^3x,
 \notag\\
 &=4\pi\int_{\cal Q}\sqrt\gamma\,\nu_nW_n\xi_\phi\,dR\,dz,
 \label{eq:results_model_b_star_angular_momentum}
\end{align}
where ${\cal Q}$ is the quarter meridian and
$W_n^2=1+\gamma^{ij}\xi_i\xi_j/\varepsilon_\xi^2$.
Define the radial density
$\mathcal L_n=4\pi\int_0^\infty\sqrt\gamma\,\nu_nW_n\xi_\phi\,dz$,
so that $J_n(R,t)=\int_0^R\mathcal L_n(R',t)\,dR'$.
The outward flux through the cylinder at $R$ is
\begin{align}
 \mathcal F_n(R,t)
 &=4\pi\int_0^\infty\alpha\sqrt\gamma\,N_s^R\xi_\phi\,dz,
 \notag\\
 &=4\pi\int_0^\infty\sqrt\gamma\,\nu_nW_n\xi_\phi
 \left(\frac{\alpha\gamma^{Rj}\xi_j}{\varepsilon_\xi W_n}
       -\beta^R\right)dz.
 \label{eq:results_star_neutron_radial_flux}
\end{align}
Equatorial reflection and vanishing vertical boundary flux reduce
Eq.~\eqref{eq:results_star_neutron_angular_balance} to
\begin{equation}
 \partial_t\mathcal L_n+\partial_R\mathcal F_n=\mathcal T_n,
 \qquad
 \mathcal T_n=4\pi\int_0^\infty\alpha\sqrt\gamma\,f_{s\phi}\,dz.
 \label{eq:results_star_neutron_shell_balance}
\end{equation}

We can then express the change in neutron angular momentum relative to the initial state,
$\Delta J_n(R,t)=J_n(R,t)-J_n(R,0)$.
Integrating the shell balance in time gives
\begin{align}
 \frac{d\Delta J_n}{dR}
 &=\mathcal L_n(R,t)-\mathcal L_n(R,0),\notag\\
 &=-\partial_R\int_0^t\mathcal F_n(R,t')\,dt'
   +\int_0^t\mathcal T_n(R,t')\,dt'.
 \label{eq:results_star_neutron_shell_change}
\end{align}
The outward radial flux $\mathcal F_n$ and local force torque per
unit radius $\mathcal T_n$ both contribute to a shell's
angular-momentum change. A shell-density difference alone does
not separate these contributions.

The coordinate angular velocities $\Omega_I=n_I^\phi/n_I^t$ are
\begin{align}
 \Omega_n&=-\beta^\phi
       +\frac{\alpha\xi_\phi}{\varepsilon_\xi W_n\gamma_{\phi\phi}},
 \notag\\
 \Omega_p&=-\beta^\phi+\frac{\alpha z^\phi}{W},
 \label{eq:results_model_b_star_omega}
\end{align}
where $-\beta^\phi=\omega_{\rm fd}$ in
Eq.~\eqref{eq:stellar_prix_geometry}.

Panel~(a) shows that the neutron component spins down but retains
differential rotation, with the outer core rotating faster than
the inner core at late times. This is consistent with our observation that superfluid circulation, in terms of vortex arrays, starts to pile up near the crust-core interface (see Fig. \ref{fig:unpinning-lines}).

\begin{figure}[tbp]
 \centering
 \includegraphics[width=\columnwidth]{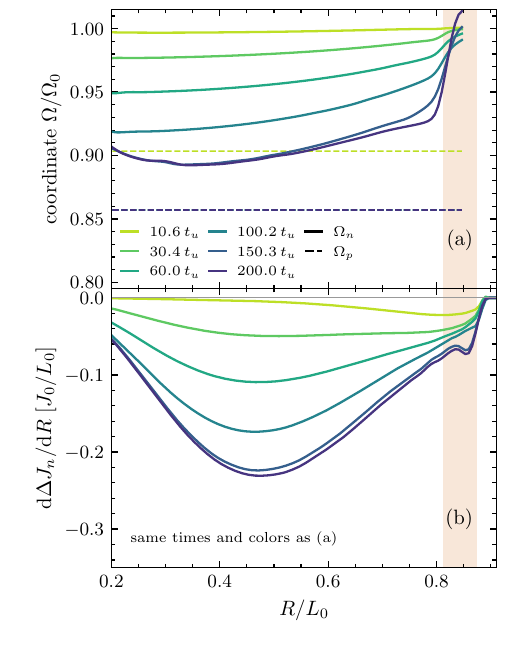}
 \caption{Neutron spin-down and interior angular-momentum
 depletion during vortex release and repinning.
 \textit{(a)} Coordinate angular velocities $\Omega_n/\Omega_0$
 (solid) and $\Omega_p/\Omega_0$ (dashed), sampled on the equator at different times indicated in the figure legend.
 Shading marks the initial outward rise in holding strength.
 The neutron fluid slows down but reaches a differentially rotating
 steady state above the imposed uniform carrier rate.
 \textit{(b)} Sampled neutron shell-density change from the initial
 state, $(L_0/J_0)d\Delta J_n/dR$, at the corresponding times
 and colors. Negative values denote local depletion, and
 $J_0=J_n(0)+J_p(0)$ is the initial material angular momentum.
  Here $L_0=10\,\mathrm{km}$.}
 \label{fig:model-b-star-unpinning-cycle}
\end{figure}

Figure~\ref{fig:model-b-star-unpinning-cycle}(b) evaluates the sampled
change directly from the saved angular-momentum densities and plots
$(L_0/J_0)d\Delta J_n/dR$, with $J_0=J_n(0)+J_p(0)$.
We can see that the neutron fluid indeed spins down by losing angular momentum in the interior of the star.
The strongest local neutron depletion lies near $R\simeq0.47L_0$.
Most importantly, the final steady state does not have uniform rotation of the neutron superfluid. As the vortices repin, the frozen-in rotation profile of the neutron component is mildly differential, with the region near the crust rotating slightly faster than the inside.

\section{Conclusions}
\label{sec:results-summary}

We have developed a general-relativistic formulation and numerical
scheme for the coupled evolution of superfluid matter and an averaged
vortex array. The formulation combines multifluid thermodynamics,
entrainment and vortex forces, and includes independent line motion
with the energy and stress of the array in neutron--charged matter.
Evolving the material currents together with canonical circulation via {constrained transport}
allows local vortex forces to change the fluid motion and the spatial
distribution of circulation within a relativistic star.
{To the best of our knowledge, our work represents the first nonlinear
numerical solution of relativistic superfluid dynamics in \revadd{multiple spatial dimensions with vortex dynamics included}.}

Earlier covariant mutual-friction and relativistic HVBK formulations
established vortex transport and vortex-energy stresses
\cite{Andersson:2016asl,Gusakov:2016eom}.
As we show, the correct treatment of vortex dynamics determines whether this evolution has the required
hyperbolic and causal structure necessary for numerical integrations. 
Algebraic force balance fails strong hyperbolicity for generic dissipative
counterflow, i.e., when mutual friction is included. As shown in
Sec.~\ref{sec:donnelly_glaberson}, it gives rise to a runaway instability via a complex vortex speed (with an amplifying sign), which we identify as the
unbounded short-wavelength limit of the Donnelly--Glaberson instability
\cite{Glaberson:1974zz,Glampedakis:2008ew}.
Restricted axisymmetric branches remain admissible when the momentum
counterflow along the vortex line vanishes and the additional conditions
of Sec.~\ref{sec:force_equilibrium_conditions} hold, consistent with
Newtonian analyses of the axial-flow instability
\cite{Sidery:2007ge}.
For the local bending problem in our line-energy model, tension
and inertia remove this short-wavelength pathology, as shown in
Eqs.~\eqref{eq:dg_tension_criterion} and
\eqref{eq:dg_causal_asymptote}. A finite unstable band can remain,
as in Newtonian treatments with vortex tension
\cite{Sidery:2007ge,vanEysden:2018kbc}.
\endgroup
In spirit, this parallels a similar problem with massless electrons in relativistic two-fluid plasmas \cite{Most:2026uga},
where retaining electron number, momentum and energy allows electron pressure to remove the defective modes otherwise leading to a weakly hyperbolic formulation.
With line inertia,
the electric part of canonical vorticity becomes an independent
dynamical field, so line acceleration enters together with tension,
lift and drag. The resulting system is remarkably close to force-free electrodynamics, but different due to the different force balance entering the drift speed. We supplement the shear-free line energy with a positive
array-compression term. The density dependence of the line energy mixes
fluid and array compression, placing joint bounds on the material
stiffnesses, counterflow and vortex slip. {Extending earlier
analyses of relativistic material waves
\cite{Samuelsson:2009up,Haber:2015exa,Gavassino:2022zzz,Hoult:2024cyx},
we here derive nonlinearly causal, strongly hyperbolic domains for the coupled
matter--vortex system at finite counterflow and vortex slip, including
characteristic coincidences.}

We formulate the system in flux-conservative 14-moment form so that we can apply previous insights gained in simulating relativistic multi-fluids in this context \cite{Most:2026uga,Most:2021rhr}. Owing to the close mathematical analogy of vortex dynamics to force-free electrodynamics, we decompose the relativistic vorticity tensor into electric and magnetic parts. We then introduce a new vorticity constrained-transport scheme, where a shared edge electric field advances the
canonical momentum and its face curl, preserving the circulation
geometry and the vorticity-divergence constraint to roundoff error.
Similar to force-free electrodynamics and resistive GRMHD \cite{Alic:2012df,Ripperda:2019lsi}, the effective electric field update contains a stiff current that we treat using implicit time integration.

We then validate the numerical implementation in different ways.
First, we present a full calculation of linear wave propagation, as well as the first exact solution of the relativistic superfluid Riemann problem. We then compare our numerical implementation against them in sound-wave and nonlinear Riemann tests, recovering the two acoustic
families, including the shocks and counterflow produced by finite
thermodynamic jumps. We also check that our simulations are stable for coincident sound speeds between {the neutron and charged constituents}.
Next, we validate the vortex transport sector directly by deriving the full dispersion relation of helical Kelvin waves for vortex transport with and without line tension. In the latter case, the restoring motion is supplied by the induced neutron flow. We find good agreement between the numerical evolution and the analytically calculated mode.

We also assess different limits of circulation transport
in a shear-flow setting, thereby confirming the correct behavior of the implemented Magnus, Iordanskii and mutual friction forces.

Finally, we perform fully global simulations of general-relativistic stellar two-fluid equilibria \cite{Prix:2004wq}. To this end, we solve the full Cowling problem and compare the resulting fundamental oscillation modes of the normal and superfluid components with those in the simulated two-fluid star.

We then proceed with our main demonstration of superfluid spin-down. This includes an effectively pinned crust and an unpinning mechanism for crustal and proton type-II superconductor materials. In this setting we simulate a forced pulsar spin-down and study the {release}, motion and repinning of superfluid vortex arrays inside the star for the unentrained stellar model used here. {The final configuration retains a net twist of the vortex array and differential rotation of the superfluid. Toroidal vorticity measures geometric winding, while the line energy supplies the associated bending stress.}
In short, we show that global relativistic simulations of superfluid neutron star interiors are now possible.

Future work will explore fully three-dimensional superfluid vortex arrangements and their interactions with the crust \cite{Link:2022vum}, better microphysically calibrated pinning and transport models \cite{Wlazlowski:2016yoe,Sourie:2021vsj}, and the inclusion of magnetic fields \cite{Dommes:2021sak}.

\begin{acknowledgments}
The authors gratefully acknowledge insightful discussions with Y. Levin and I. Wasserman.
YF and ERM acknowledge support from the U.S. National Science Foundation {(NSF)} under Grant No. PHY-2541792. ERM is also supported by a Research Fellowship from the Sloan Foundation and a William H. Hurt Scholarship at Caltech.
The authors also acknowledge the use of Delta at the National Center for Supercomputing Applications (NCSA) through allocation PHY210074 from the Advanced Cyberinfrastructure Coordination Ecosystem: Services \& Support (ACCESS) program, which is supported by National Science Foundation grants \#2138259, \#2138286, \#2138307, \#2137603, and \#2138296. We acknowledge support from OpenAI’s Researcher Access Program for providing access to ChatGPT models to support this work. Work in this paper used OpenAI's ChatGPT 5.6-sol, 6.1-sol and 6-astra, and Anthropic's Opus 5 and 5.5, as well as Fable 5, 5.1 and 5.2.
\end{acknowledgments}

\section*{Use of large language models}
We used large language models (LLMs), identified in the acknowledgements, to assist with algebraic derivations and symbolic checks using Mathematica and with the development,
debugging, and optimization of numerical and analysis code starting from the 19-moment multi-fluid code developed in Ref. \cite{Most:2026uga}. We also used LLMs for consistency checks between the manuscript and the implementation and for assistance in writing technical appendices that match the Mathematica derivations and code implementation. Verification of this work included explicit computer-algebra
calculations in Mathematica, comparisons with the implemented equations, and extensive numerical tests and comparisons with reference results in the literature beyond what is presented here.

\appendix
\colorlet{mainrevisionblue}{revisionblue}
\colorlet{mainblue}{blue}
\colorlet{revisionblue}{black}
\colorlet{blue}{black}
\renewcommand{\revadd}[1]{\textcolor{mainrevisionblue}{#1}}
\input{appendix_causality_hyperbolicity_tension}
\input{appendix_model_b_tension_eigenvectors}

\input{appendix_tension_constitutive_derivatives}
\input{appendix_tension_electric_adm}
\input{appendix_imex_tableau_tension}
\input{appendix_riemann_analytic_tension}
\input{appendix_kelvin_native_tension}

\input{appendix_cowling_modes_tension}

\clearpage
\colorlet{revisionblue}{mainrevisionblue}
\colorlet{blue}{mainblue}
\bibliography{GavassinoBiblio,section4_references_supplement,appendix_references_tension_snapshot,proofreading_references,introduction_references,tension_references}
\label{lastpage}
\end{document}

%% file: section_vortex_tension_model_b.tex
\subsection{Vortex tension}
\label{sec:model_b_vortex_tension}

\begin{figure*}[t]
 \centering
 \input{Plots/vortex_tension_geometry/vortex_tension_geometry.tikz}
 \caption{{Vortex geometry and the reference cell used in the vortex tension model. (a)~In a local right-handed Eulerian frame, the line tangent
 is parallel to the vorticity $B$, and its transverse drift is $v_d=E\times B/B^2$. Here $E$ and $B$ represent the electric and magnetic components of the relativistic vorticity tensor $\Omega_{\mu\nu}$.
 The displayed orientation is $B\parallel+\hat z$,
 $E\parallel+\hat y$ and $v_d\parallel+\hat x$. Dashed arrows show independent neutron and carrier velocities. The shaded plane is
 normal to the local line direction. (b)~The array-rest-frame reference
 cell defines the core scale $\xi_n$ and fixed outer cutoff $b_0$ of the vortex array. The core is enlarged for visibility. Neighboring crossings illustrate
 line density, and the circulation arrow follows the positive line
 orientation. The energy per existing line length is $F/\mathcal N$,
 while $\mathcal T$ is the marginal cost of adding line density.
 \textcolor{black}{Here $F$ is the line-energy density and $\mathcal N_0$ is
 the reference proper line density, with $\pi b_0^2=1/\mathcal N_0$.}}}
 \label{fig:model_b_line_geometry}
\end{figure*}

Section~\ref{sec:moment_vortex_closure} describes the motion of
vortex lines through balance of Magnus, Iordanskii and mutual-friction
forces. A vortex array also stores vortex line energy and carries tension
\cite{Andersson:2019fdj}, while distortion of the
relative positions of neighboring lines produces lattice shear
elasticity \cite{1969JETP...29..945T}. This shear response
supports Tkachenko waves \cite{1966JETP...23.1049T}.
{For generic dissipative counterflow, the algebraic force closure
fails strong hyperbolicity in three dimensions. Axisymmetric propagation
admits the restricted geometric and spectral branches of
Sec.~\ref{sec:force_equilibrium_conditions}.}
\textcolor{black}{Let $q_\parallel=q\cdot\ell$ be the carrier-frame momentum
counterflow along the line and
$\mathcal D=\mathcal R\nu_L-\mathcal R'\mathcal Z n_p$ the drag
combination.}
The vortex characteristics in Eq.~\eqref{eq:app_vortex_speeds} become complex for propagation with a
component along the vortex line when $\mathcal Dq_\parallel\ne0$, and
even tangent propagation can have the incomplete eigenbasis of
Eq.~\eqref{eq:app_model_b_tangent_vortex_jordan}.

We therefore extend the NC model beyond that
force-equilibrium approximation by retaining the energy and momentum
of the array. A bent neutron vortex has greater length and therefore
costs energy. We use the shear-free sector of
Ref.~\cite{Andersson:2019fdj}, retaining tension and its
momentum response while omitting lattice shear and Tkachenko modes\footnote{
The incompressible, unentrained triangular-lattice reference has
\cite[Eq.~(16)]{Haskell:2010jh}
\begin{align*}
 c_T^2&=\frac{\kappa_{\rm vel}\Omega}{8\pi},\qquad
 \kappa_{\rm vel}=\textcolor{black}{\frac{\kappa_v}{m_n}},\\
 \frac{c_T}{c}&\simeq7.4\times10^{-12}
               \left(\frac{P_{\rm rot}}{10\,\mathrm{ms}}\right)^{-1/2},
 \qquad P_{\rm rot}=\frac{2\pi}{\Omega}.
\end{align*}
Here $\Omega$ is the rotation rate and $\kappa_{\rm vel}$ is the
nonrelativistic velocity circulation, distinct from the momentum
circulation $\kappa_v$. We can see that on relativistic timescales considered in a dynamical evolution code, this is a reasonable approximation.}

\input{section_rw_constitutive_tension}

\input{section_bcs_core_tension}

\subsubsection{Forces acting on a vortex array}

{As outlined in Sec.~\ref{sec:moment_vortex_closure}, the force}
exerted on the vorticity sector is
\begin{align}
 f_{\Omega\nu}
 &=-L^\rho\Omega_{\rho\nu}+b\perp^\rho{}_{\nu}d_\rho,\notag\\
 d_\rho&=\mathcal R\nu_Lu_\rho+\mathcal R'N_{s\rho}.
 \label{eq:tension_draft_exchange}
\end{align}
Assigning
external forcing to the carrier and writing
$T_{\rm d}^{\mu\nu}=T_0^{\mu\nu}+C_\Omega^{\mu\nu}$, the Carter
balances give
\begin{align}
 f_{s\nu}&=N_s^\mu\Omega_{\mu\nu},\notag\\
 f_{c\nu}&=\mathcal ZN_c^\mu\Omega_{\mu\nu}
                          -b\perp^\rho{}_{\nu}d_\rho,\notag\\
 \nabla_\mu T_{\rm d}^\mu{}_{\nu}
 &=f_{s\nu}+f_{c\nu}+f_{{\rm ext},\nu}
   +\frac{\partial F}{\partial\nu_n}\nabla_\nu\nu_n
   +\frac{\partial F}{\partial n_p}\nabla_\nu n_p.
 \label{eq:tension_main_matter_exchange}
\end{align}
The last two terms are a result of the line energy changing when the density changes. They must accompany the
line-induced momentum shifts even when the drag vanishes.

To derive the line equation, isolate the stress due to the
two-form's proper-area response,
$\bar T_\Omega^{\mu\nu}\equiv T_\Omega^{\mu\nu}-C_\Omega^{\mu\nu}
=a_\Omega\Omega^{\mu\rho}\Omega^\nu{}_{\rho}-Fg^{\mu\nu}$.
Using $\nabla_{[\alpha}\Omega_{\mu\nu]}=0$,
\begin{equation}
 \Omega^{\mu\rho}\nabla_\mu\Omega_{\nu\rho}
     =\tfrac14\nabla_\nu(\Omega_{\alpha\beta}\Omega^{\alpha\beta}).
 \label{eq:tension_main_bianchi_contraction}
\end{equation}
Since $\Omega_{\alpha\beta}\Omega^{\alpha\beta}
=2\kappa_v^2\mathcal N^2$, multiplication by $a_\Omega$ turns
its right-hand side into
$(\partial F/\partial\mathcal N)\nabla_\nu\mathcal N$.
This cancels the vortex-density part of $\nabla_\nu F$. With the
response current defined in
Eq.~\eqref{eq:tension_vortex_response_current}, the remaining terms give
\begin{align}
 \nabla_\mu\bar T_\Omega^\mu{}_{\nu}
 &=-\mathcal J_\Omega^\rho\Omega_{\rho\nu}
   -\frac{\partial F}{\partial\nu_n}\nabla_\nu\nu_n
   -\frac{\partial F}{\partial n_p}\nabla_\nu n_p.
 \label{eq:tension_main_stress_divergence}
\end{align}
Total energy--momentum conservation and
\eqref{eq:tension_main_matter_exchange} require the same divergence
with $-\mathcal J_\Omega^\rho\Omega_{\rho\nu}$ replaced by
$f_{\Omega\nu}$, because $f_s+f_c=-f_\Omega$.
The reversible density terms therefore cancel between the two sectors,
giving $f_{\Omega\nu}=-\mathcal J_\Omega^\rho\Omega_{\rho\nu}$, akin to a generalized Lorentz force term.
Substitution of \eqref{eq:tension_draft_exchange} yields
\begin{equation}
 (N_s^\rho+\mathcal ZN_c^\rho-\mathcal J_\Omega^\rho)
       \Omega_{\rho\nu}=b\perp^\rho{}_{\nu}d_\rho.
 \label{eq:tension_draft_generalized_force}
\end{equation}
As such, including vortex tension and line energy alters the force balance, and neglecting it recovers the previous expression, Eq.
\eqref{eq:moments_vortex_force}.

\subsubsection{Vortex drift with tension}
\label{sec:tension_electric_and_drift}
We are now going to interpret vortex drift dynamics using the same electromagnetic decomposition in
Sec.~\ref{sec:moment_electromagnetic_duality} employed for force balance. At force equilibrium, the material velocities and
local force coefficients determine $E_i$ algebraically through
Eq.~\eqref{eq:moments_explicit_electric_field}. With line energy,
$\mathcal J_\Omega^\nu=\nabla_\mu(a_\Omega\Omega^{\mu\nu})$ contains
a time derivative of $a_\Omega E^i$. The line equation
\eqref{eq:tension_draft_generalized_force} therefore evolves the
electric field independently of the material velocities. It carries
two physical electric degrees of freedom after imposing $E_iB^i=0$.
The stress $T_\Omega$ is then reconstructed from $E,B$ and the
material state. {The evolution of both fields is analogous to
the electromagnetic sector of force-free electrodynamics
\cite{Alic:2012df}, with the material stress and reciprocal vortex
forces retained in the superfluid system.}
{Appendix~\ref{app:tension_electric_adm} derives the coupled
material and electric time derivatives and the vortex-energy identity.}

{To exhibit the evolution, use the Eulerian $3+1$ decomposition and write}
\begin{align*}
 \mathfrak E^i&=\sqrt\gamma\,a_\Omega E^i,\qquad
 \mathfrak B^i=\sqrt\gamma\,B^i,\\
 q_a&=\nabla_i(a_\Omega E^i)
        =\frac{\partial_i\mathfrak E^i}{\sqrt\gamma}.
\end{align*}
Let
$\mathcal J_{\Omega\perp}^i$ denote the spatial projection of
$\mathcal J_\Omega^\mu$. The spatial force equation is
$f_\Omega=-q_aE+\mathcal J_{\Omega\perp}\times B$.
It determines the current transverse to $B$. With the primitive
fluxes
\begin{align*}
 \mathcal F_E^{ji}
 &=\beta^i\mathfrak E^j-\beta^j\mathfrak E^i
       -\alpha\sqrt\gamma\,a_\Omega\epsilon^{ijk}B_k,\\
 \mathcal F_B^{ji}
 &=\beta^i\mathfrak B^j-\beta^j\mathfrak B^i
       +\alpha\sqrt\gamma\,\epsilon^{ijk}E_k,
\end{align*}
the spatial current includes a component parallel to $B$, written
in terms of $\mathcal I_\parallel=B_i\mathcal J_{\Omega\perp}^i$.
\footnote{The parallel term has zero transverse
force and, at $E_iB^i=0$, zero electric work. Put
$S=a_\Omega E_iB^i$ and $\mathcal D_t=\partial_t-\mathcal L_\beta$.
On $S=0$, the field equations give
$\mathcal D_tS=\alpha(\mathcal I_\parallel-
\mathcal I_\parallel^{\rm con})$, where in our current convention
\begin{align*}
 \mathcal I_\parallel^{\rm con}=a_\Omega\bigl[
 &\epsilon^{ijk}(E_i\nabla_jE_k-B_i\nabla_jB_k)\\
 &+2K_{ij}E^iB^j\bigr].
\end{align*}
Thus the analytic parallel current serves to preserve $E_i B^i=0$, see Ref. \cite{Alic:2012df} for a discussion.
In our numerical implementation, we instead follow Refs. \cite{Spitkovsky:2006np,Alic:2012df,Most:2020ami} in algebraically projecting out the violation.}
The field equations in conservative flux form are
\begin{align}
 \mathcal J_{\Omega\perp}^i
 &=\frac{[B\times(q_aE+f_\Omega)]^i}{B^2}
           +\mathcal I_\parallel\frac{B^i}{B^2},\notag\\
 \partial_t\mathfrak E^i+\partial_j\mathcal F_E^{ji}
 &=\sqrt\gamma\bigl(\alpha\mathcal J_{\Omega\perp}^i
                                      +\beta^i q_a\bigr),\notag\\
 \partial_t\mathfrak B^i+\partial_j\mathcal F_B^{ji}&=0.
 \label{eq:tension_main_weighted_electric_evolution}
\end{align}
The first term combines the electric and material-force contributions
into a transverse drift current, satisfying
$\mathcal J_{\Omega\perp}\times B=q_aE+f_\Omega$.
Its form remains regular at $q_a=0$. This system is almost identical to force-free electrodynamics \cite{Paschalidis:2013gma}, with the main difference being the inclusion of the material force current.
This current is distinct from
the vortex-line velocity, which is fixed geometrically by $E,B$
in Eq.~\eqref{eq:tension_primitive_adm_line_velocity}.
\textcolor{black}{With $Q_\Omega=b=\sqrt{B^2-E^2}=\kappa_v\mathcal N$ for the
invariant vorticity magnitude, solving the kernel condition
\eqref{eq:moments_line_transport} in the transverse gauge
$v_L\cdot B=0$ gives}
\begin{align}
 v_L^i&=\frac{\epsilon^{ijk}E_jB_k}{B^2},\qquad
 v_L^2=\frac{E^2}{B^2}<1,\notag\\
 W_L&=\frac{\sqrt{B^2}}{Q_\Omega},\qquad
 \ell^i=\frac{B^i}{\sqrt{B^2}}.
 \label{eq:tension_primitive_adm_line_velocity}
\end{align}
The coordinate transport velocity is
$\mathcal V^i=\alpha v_L^i-\beta^i$.
With $v_L^\mu=(0,v_L^i)$ and $\ell^\mu=(0,\ell^i)$,
$u_L^\mu=W_L({n^\mu}+v_L^\mu)$ satisfies the normalization
conditions \eqref{eq:moments_ordered_worldsheet} and
$\Omega_{\mu\nu}u_L^\nu=0$.

The force in Eq.~\eqref{eq:tension_draft_exchange} can now be
evaluated explicitly from the primitive fields. For both the force
and stress below, write $u_x^\mu=W_x(n^\mu+v_x^\mu)$ for $x=n,c$,
with $u_c^\mu=u^\mu$, $n_\mu v_x^\mu=0$,
$v_x^2=\gamma_{ij}v_x^iv_x^j$ and $W_x=(1-v_x^2)^{-1/2}$.
The currents are $N_s^\mu=\nu_n u_n^\mu$ and
$N_c^\mu=n_p u_c^\mu$. \textcolor{black}{The force coefficients are}
\begin{align*}
 \nu_L&=\nu_nW_nW_c(1-\boldsymbol v_n\cdot\boldsymbol v_c)
                                      +\mathcal Z n_p,\\
 d_n&=\mathcal R\nu_LW_c+\mathcal R'\nu_nW_n,\\
 \boldsymbol d&=\mathcal R\nu_LW_c\boldsymbol v_c
                         +\mathcal R'\nu_nW_n\boldsymbol v_n.
\end{align*}
Here $d_n=-n_\mu d^\mu$ and $\boldsymbol d$ is the spatial
projection of $d^\mu$. The Eulerian spatial force is
\begin{align}
 \boldsymbol f_\Omega={}&\nu_nW_n
       (\boldsymbol E+\boldsymbol v_n\times\boldsymbol B)\notag\\
 &+\mathcal Z n_pW_c
       (\boldsymbol E+\boldsymbol v_c\times\boldsymbol B)\notag\\
 &+Q_\Omega\left[\boldsymbol d
       -\frac{\boldsymbol B\cdot\boldsymbol d}{B^2}\boldsymbol B\right]
       \notag\\
 &-\frac{d_n-(\boldsymbol E\times\boldsymbol B)
                         \cdot\boldsymbol d/B^2}{Q_\Omega}
                           (\boldsymbol E\times\boldsymbol B).
 \label{eq:tension_primitive_force}
\end{align}
The coordinate
time component is $f_{\Omega0}=-\mathcal V^if_{\Omega i}$, with
$f_{\Omega i}=\gamma_{ij}f_\Omega^j$ and
$\mathcal V^i=\alpha(\boldsymbol E\times\boldsymbol B)^i/B^2-\beta^i$.

We can then recover the stress tensor and vortex line density from the electric and magnetic fields. Use the electric--magnetic decomposition
\eqref{eq:moments_vorticity_fields} and the simple-field identities
\eqref{eq:moments_line_transport}. The evolved fields obey
$E_iB^i=0$ and determine the proper vortex density and response through
\begin{align}
 Q_\Omega^2&=B^2-E^2>0,\qquad
 \mathcal N=Q_\Omega/\kappa_v,\notag\\
 a_\Omega&=\frac1{\kappa_vQ_\Omega}
                       \frac{\partial F}{\partial\mathcal N}.
 \label{eq:tension_primitive_line_eos}
\end{align}

For the $3+1$ evolution we decompose the line stress with respect to
the Eulerian observer $n^\mu$, the same observer used to define $E,B$.
We reserve $e,p,q^\mu,\pi^{\mu\nu}$
for fluid-frame moments. With $\gamma^{\mu\nu}=g^{\mu\nu}+n^\mu n^\nu$,
use the Eulerian decomposition
\begin{equation}
 T_\Omega^{\mu\nu}
 =\mathcal E_\Omega n^\mu n^\nu
       +2n^{(\mu}\mathcal Q_\Omega^{\nu)}+\mathcal P_\Omega^{\mu\nu}.
 \label{eq:tension_vortex_moment_stress}
\end{equation}
The projections are
\begin{align*}
 \mathcal E_\Omega&=n_\mu n_\nu T_\Omega^{\mu\nu},\\
 \mathcal Q_\Omega^\mu&=-\gamma^\mu{}_{\rho}T_\Omega^{\rho\sigma}n_\sigma,\\
 \mathcal P_\Omega^{\mu\nu}
 &=\gamma^\mu{}_{\rho}\gamma^\nu{}_{\sigma}T_\Omega^{\rho\sigma}.
\end{align*}
Thus $\mathcal E_\Omega$ is the Eulerian energy density,
$\mathcal Q_\Omega^\mu$ its spatial energy flux and
$\mathcal P_\Omega^{\mu\nu}$ the full spatial stress, including its
isotropic and anisotropic parts. Using the fields of
Eq.~\eqref{eq:moments_vorticity_fields}, they are
\begin{align}
 \mathcal E_\Omega&=F+a_\Omega E^2+
                 \sum_{x=n,c}w_x^F(W_x^2-1),\notag\\
 \mathcal Q_\Omega^\mu&=a_\Omega(E\times B)^\mu+
                 \sum_{x=n,c}w_x^FW_x^2v_x^\mu,\notag\\
 \mathcal P_\Omega^{\mu\nu}
 &=\left(a_\Omega B^2-F+\sum_{x=n,c}w_x^F\right)\gamma^{\mu\nu}\notag\\
 &\quad-a_\Omega(E^\mu E^\nu+B^\mu B^\nu)\notag\\
 &\quad+\sum_{x=n,c}w_x^FW_x^2v_x^\mu v_x^\nu.
 \label{eq:rw_primitive_stress}
\end{align}
Both flux and stress are spatial:
$n_\mu\mathcal Q_\Omega^\mu=0$ and
$n_\mu\mathcal P_\Omega^{\mu\nu}=0$.
The mean spatial stress is $\gamma_{\mu\nu}\mathcal P_\Omega^{\mu\nu}/3$;
$\mathcal P_\Omega^{\mu\nu}$ is not trace-free.
These quantities are reconstructed algebraically from the primitive
state. The vector $\mathcal Q_\Omega^\mu$ is distinct from the scalar
frictional heat $\mathcal Q_\Omega$ in Eq.~\eqref{eq:adm_entropy_work}.

%% file: Plots/vortex_tension_geometry/vortex_tension_geometry.tikz
\begingroup
\input{Plots/sketch_palette.tex}%
\begin{tikzpicture}[
 x=1cm,y=1cm,color=ink,
 arr/.style={-{Latex[length=2mm,width=1.3mm]},line width=.8pt},
 mat/.style={arr,densely dashed,line width=.8pt},
 lab/.style={font=\fontsize{9}{10.5}\selectfont,inner sep=1.5pt},
 note/.style={font=\fontsize{8}{9.5}\selectfont,inner sep=1.5pt},
 head/.style={font=\fontsize{10}{12}\selectfont\bfseries,inner sep=2pt}
]
 \path[use as bounding box] (0,-.6) rectangle (16.2,4.45);
 \node[head] at (3.8,4.2) {(a) Local line geometry};
 \node[head] at (12.0,4.2) {(b) Reference cell in the array rest frame};
 \draw[reference!55,line width=.4pt] (7.7,-.3)--(7.7,3.9);
 \filldraw[fill=fluidblue!7,draw=reference!55,line width=.45pt]
   (1.86,2.25)--(5.46,2.25)--(4.14,1.15)--(.54,1.15)--cycle;
 \draw[fluidblue,line width=1.6pt]
   (3.60,0.0)..controls (3.05,.45) and (3.00,1.10)..(3.00,1.70)
   ..controls (3.00,2.50) and (3.10,3.50)..(3.80,3.85);
 \fill (3.00,1.70) circle (.045);
 \node[lab,anchor=west] at (3.95,3.55) {vortex line};
 \draw[arr,fluidblue] (3.00,1.70)--(3.00,3.55);
 \node[lab,text=fluidblue,anchor=east] at (2.91,3.25) {$\boldsymbol B\parallel\boldsymbol\ell$};
 \draw[arr,fieldgreen] (3.00,1.70)--(5.15,1.70);
 \node[lab,text=fieldgreen,anchor=west] at (5.21,1.70) {$\boldsymbol v_d=\boldsymbol v_{L\perp}$};
 \draw[arr,dragorange] (3.00,1.70)--(1.86,.94);
 \node[lab,text=dragorange,anchor=east] at (1.82,.89) {$\boldsymbol E$};
 \draw[mat,fluidblue] (3.00,1.70)--(4.35,2.72);
 \node[lab,text=fluidblue,anchor=west] at (4.39,2.74) {$\boldsymbol v_n$};
 \draw[mat,liftpurple,dash dot] (3.00,1.70)--(1.50,2.72);
 \node[lab,text=liftpurple,anchor=east] at (1.45,2.74) {$\boldsymbol v_c$};
 \node[note,align=center] at (5.00,.50)
   {$\boldsymbol E=-\boldsymbol v_d\times\boldsymbol B$\\
    $\mathcal N=\sqrt{B^2-E^2}/\kappa_v$};
 \node[note] at (3.4,-.33) {Eulerian observer $o^\mu$; matter flows are independent};
 \begin{scope}[shift={(11.55,1.85)}]
  \draw[reference,line width=.7pt] (0,0) circle (1.45);
  \filldraw[fill=fluidblue!18,draw=fluidblue,line width=.8pt] (0,0) circle (.19);
  \node[lab,text=fluidblue] at (0,0) {$\odot$};
  \draw[arr,reference,line width=.55pt] (0,0)--(1.40,-.38);
  \node[lab,anchor=west] at (1.49,-.40) {$b_0$};
  \draw[reference,line width=.5pt] (.13,.13)--(.45,.65)--(.92,.65);
  \node[lab,anchor=west] at (.96,.65) {$\xi_n$};
  \draw[arr,fluidblue] (.86,.42) arc[start angle=26,end angle=332,radius=.96];
  \node[note,anchor=east] at (-1.55,.20) {$\boldsymbol B',\boldsymbol\ell'$};
  \node[note,anchor=east] at (-1.55,-.15) {out of the page};
  \foreach \x/\y in {2.2/0,-1.1/1.9,1.1/1.9,-2.0/-1.15,2.0/-1.15}
    {\fill[fluidblue!70] (\x,\y) circle (.055);}
 \end{scope}
 \node[note] at (14.20,3.15) {$\pi b_0^2=1/\mathcal N_0$};
 \node[lab] at (11.8,-.05) {$\displaystyle\oint\xi_i\,dx^i=\kappa_v$};
 \node[note] at (11.8,-.57) {$F/\mathcal N$: energy per length; $\mathcal T=\partial F/\partial\mathcal N$};
\end{tikzpicture}
\endgroup

%% file: section_rw_constitutive_tension.tex
To evolve the vortices independently, we specify the energy stored in
line length and array compression. We use the shear-free
vortex-fibration framework of
Refs.~\cite{Carter:1995mj,Andersson:2019fdj}.
The proper material densities retain the notation of the NC model,
$\nu_n=\sqrt{-N_s^\mu N_{s\mu}}$ and
$n_p=\sqrt{-N_c^\mu N_{c\mu}}$. With material current invariants
$I_{xy}=-g_{\mu\nu}N_x^\mu N_y^\nu$, collected into
$\boldsymbol I$, take
\begin{align}
 \Lambda_{\rm tot}&=\Lambda_0(\boldsymbol I)-F(\nu_n,n_p,\mathcal N),\notag\\
 \mathcal N&=\frac{\sqrt{\Omega_{\mu\nu}\Omega^{\mu\nu}/2}}{\kappa_v},\notag\\
 F&=A_\Omega(\nu_n,n_p)\mathcal N
       \left[L_\Omega(\nu_n,n_p)+\frac{\alpha_{\rm v}\mathcal N}{2\mathcal N_0}\right].
 \label{eq:tension_draft_master}
\end{align}
Here $\Lambda_{\rm tot}$ is the full master function,
$\Lambda_0$ is the material NC master function of
Sec.~\ref{sec:moment_carter_closure}, and $F$ is the additional
line-energy density. The invariants $\boldsymbol I$ also include
the entropy-current products when the thermal completion is retained.
The two-form $\Omega_{\mu\nu}$ is the neutron momentum vorticity,
$\kappa_v$ is the momentum circulation per line, and $\mathcal N$
is the proper number of lines per transverse area.
The positive coefficient $A_\Omega$ sets the line-energy scale,
while $L_\Omega>0$ is the dimensionless logarithmic factor;
their material-state dependence is fixed by
Eq.~\eqref{eq:regulated_line_matching} below.
The constant $\mathcal N_0>0$ is a reference proper areal line density,
and $\alpha_{\rm v}>0$ is a dimensionless compression
coefficient.
The linear term retains the isolated-line energy at that reference
spacing. The quadratic term supplies the array's positive
compression response and is an additional constitutive choice.

\paragraph{Line energy.}
The line energy is the excess energy per unit proper length of a
vortex, including the circulating neutron flow outside its core.
For a dilute array this gives $F\simeq\mathcal N\varepsilon_{\rm line}$.
The momentum response specifies the energy needed to change that
circulating flow: a spatial current perturbation changes the
conjugate neutron momentum, with the inverse response setting the
phase stiffness.
The leading neutron contribution in
Ref.~\cite[Eq.~(150)]{Rau:2020wnv} motivates the coefficient of
the logarithm. To make the matching explicit, consider a local coflow
reference state of the bare NC EOS, holding the carrier current
fixed. The spatial neutron momentum response is
\begin{equation}
 \delta\xi_i^{(0)}=\frac{H_0}{\nu_n}\delta N_{si},\qquad
 H_0=\mu_n-\mathcal A n_p>0.
 \label{eq:tension_matching_phase_response}
\end{equation}
{The single-fluid Carter--Langlois current--momentum relation
reduces in our bare, unentrained reference to
$N_s^\mu=\Phi^2\xi_{(0)}^\mu$ with $\Phi^2=\nu_n/\mu_n$.
Its spatial linearization at rest is
$\delta N_s^i=(\nu_n/\mu_n)\delta\xi_{(0)}^i$
\cite{Carter:1995mj}.}
Equation~\eqref{eq:tension_matching_phase_response} is the
fixed-carrier, entrained counterpart: $H_0$ replaces {$\mu_n$}, and
{$\nu_n/H_0$} is the corresponding neutron phase stiffness. Treating
the background as constant across the cell outside the core, the
quadratic kinetic energy at fixed mean particle numbers and the
circulation around a straight line give
\begin{align}
 \delta\varepsilon_{\rm kin}
   &=\frac{\nu_n}{2H_0}\delta\xi_i^{(0)}\delta\xi_{(0)}^i,
 &|\delta\xi_{\hat\varphi}^{(0)}|&=\frac{\kappa_v}{2\pi r},\notag\\
 \varepsilon_{\rm line}
   &=\int_{\xi_n}^{b_0}2\pi r\,\delta\varepsilon_{\rm kin}\,dr
     =\frac{\kappa_v^2}{4\pi}\frac{\nu_n}{H_0}
                                     \ln\frac{b_0}{\xi_n}.
 \label{eq:tension_matching_line_integral}
\end{align}
The matching used in \eqref{eq:tension_draft_master} is therefore
\begin{align}
 A_\Omega&=\frac{\kappa_v^2}{4\pi}\frac{\nu_n}{H_0},\qquad
 L_\Omega=\ln\frac{b_0}{\xi_n(\nu_n,n_p)},\notag\\
 b_0&=(\pi\mathcal N_0)^{-1/2}.
 \label{eq:regulated_line_matching}
\end{align}
Both $\mu_n$ and $\mathcal A$ are evaluated from the fluid master
function \eqref{eq:main_model_b_master} before adding $F$.
The core scale $\xi_n$ is the density-dependent BCS coherence length
specified in Eq.~\eqref{eq:tension_bcs_core_law} below.
The outer cutoff $b_0$ is fixed by the reference vortex
density $\mathcal N_0$, rather than the instantaneous spacing. This
freezes the logarithm's vortex-density dependence, as in the
restricted linear approximation of Carter--Langlois, while keeping
its dependence on the fluid state. Explicit density derivatives of this matching are
collected in Appendix~\ref{app:tension_constitutive_derivatives}.

All partial derivatives of $F$ hold its other arguments fixed.
At fixed fluid state, the line tension and transverse inertia per line are
\begin{align}
 \mathcal T=m_{\rm line}=\frac{\partial F}{\partial\mathcal N}
   &=A_\Omega\left(L_\Omega+
                      \frac{\alpha_{\rm v}\mathcal N}{\mathcal N_0}\right),\notag\\
 \frac{\partial^2F}{\partial\mathcal N^2}
   &=\frac{\alpha_{\rm v}A_\Omega}{\mathcal N_0},\qquad
 a_\Omega=\frac{\mathcal T}{\kappa_v^2\mathcal N}.
 \label{eq:rw_line_derivatives}
\end{align}
The proper-area dependence supplies the electric time derivative
and its inertia. At vanishing vortex density both $F/\mathcal N$ and
$\mathcal T$ approach $A_\Omega L_\Omega$. At finite density they can differ due to the
compression term. This proper-area closure ties
the transverse mass to the tension, giving bending characteristics
at $c$ along a line [Eq.~\eqref{eq:regulated_bending_speeds}].
{The equality between mass and tension is a constitutive choice
of this model. Matching the static line energy does not independently
calibrate the microscopic vortex inertia, whose evaluation requires a
dynamical calculation~\cite{ThoulessAnglin:2007VortexMass}.}

The response to the vortex two-form itself is most naturally written
as a conjugate antisymmetric tensor and its divergence:
\begin{align}
 \left.\delta F\right|_{\nu_n,n_p}
   &=\tfrac12\lambda_\Omega^{\mu\nu}\delta\Omega_{\mu\nu},\qquad
 \lambda_\Omega^{\mu\nu}=a_\Omega\Omega^{\mu\nu},\notag\\
 \mathcal J_\Omega^\nu&=\nabla_\mu\lambda_\Omega^{\mu\nu}.
 \label{eq:tension_vortex_response_current}
\end{align}
Here $\mathcal J_\Omega$ is a derived response current, containing
derivatives of the field and of its stiffness $a_\Omega$. It supplies
the tension and acceleration terms in the line equation below.

\paragraph{Vortex contribution to the stress.}
The material-density response of $F$ shifts the neutron and charged
momenta by {$(\partial F/\partial\nu_n)u_{n\nu}$} and
{$(\partial F/\partial n_p)u_\nu$}, respectively. Define
\begin{align}
 w_n^F&=\nu_n\frac{\partial F}{\partial\nu_n},\qquad
 w_c^F=n_p\frac{\partial F}{\partial n_p},\notag\\
 h_n^{\mu\nu}&=g^{\mu\nu}+u_n^\mu u_n^\nu,\qquad
 h_c^{\mu\nu}=g^{\mu\nu}+u^\mu u^\nu,\notag\\
 C_\Omega^{\mu\nu}&=w_n^Fh_n^{\mu\nu}+w_c^Fh_c^{\mu\nu}.
 \label{eq:tension_density_response_stress}
\end{align}
Thus $C_\Omega^{\mu\nu}$ is the material stress induced by the
vortex energy's shifts of the two fluid chemical potentials.
For a simple vortex two-form, the worldsheet normal projector in
Eq.~\eqref{eq:moments_lift_current} obeys
$\Omega^{\mu\rho}\Omega^\nu{}_{\rho}
 =\kappa_v^2\mathcal N^2\perp^{\mu\nu}$.
Metric variation of $-F$ then gives the vortex stress in the form
\begin{align}
 T_\Omega^{\mu\nu}
 &=F(u_L^\mu u_L^\nu-\ell^\mu\ell^\nu)\notag\\
 &\quad+(\mathcal N\mathcal T-F)\perp^{\mu\nu}
                    +C_\Omega^{\mu\nu}.
 \label{eq:tension_draft_stress}
\end{align}
The three terms describe the line energy and longitudinal stress,
the transverse response of the array, and the change in fluid
chemical potentials and pressure. We assign the last contribution
to $T_\Omega$ throughout. It remains present at coflow, and must be
included once when the primitive state and total stress are recovered.

\paragraph{Transverse compression pressure.}
The geometric form \eqref{eq:tension_draft_stress} separates the
directed line stress from the pressure normal to the worldsheet.
The coefficient of $\perp^{\mu\nu}$ in
\eqref{eq:tension_draft_stress} is the array pressure at fixed fluid
state. Equivalently, compressing a transverse area $\mathcal S$ at
fixed line number $N_{\rm lines}=\mathcal N\mathcal S$ gives
\begin{align}
 p_{\Omega,\rm reg}
 &=-\left.\frac{\partial[\mathcal S F(\nu_n,n_p,N_{\rm lines}/\mathcal S)]}
                 {\partial\mathcal S}\right|_{\nu_n,n_p,N_{\rm lines}}\notag\\
 &=\mathcal N \mathcal{T} -F
   =\frac{\alpha_{\rm v}A_\Omega\mathcal N^2}{2\mathcal N_0}.
 \label{eq:regulated_pressure_cost}
\end{align}
The isolated-line term is linear in $\mathcal N$ and therefore
contributes no such transverse pressure, but the quadratic term does.
The density-response stress $C_\Omega$ supplies an additional
contribution.

%% file: section_bcs_core_tension.tex
\paragraph{{Coherence length}}
We set the neutron core scale by the BCS coherence length
\cite{Alpar:1984zz}. Writing the effective Fermi response
as $v_{Fn}=\hbar k_{Fn}/\mu_*$\textcolor{black}{, where $v_{Fn}$ is the neutron
Fermi speed and $k_{Fn}$ its Fermi wave number,} gives
\begin{equation}
 \xi_n(\nu_n,n_p)=\frac{\hbar v_{Fn}}{\pi\Delta_{\rm pair}}
          =\frac{\hbar^2 k_{Fn}}{\pi\mu_*\Delta_{\rm pair}},
 \qquad k_{Fn}=(3\pi^2\nu_n)^{1/3}.
 \label{eq:tension_bcs_core_law}
\end{equation}
Here {$\mu_*(\nu_n,n_p)>0$} and {$\Delta_{\rm pair}(\nu_n,n_p)>0$} are the
prescribed microscopic effective Fermi response and neutron pairing
gap. With $b_0=(\pi\mathcal N_0)^{-1/2}$, substitution into the
matched line energy gives
\begin{equation}
 L_\Omega(\nu_n,n_p)=
 \ln\frac{\pi b_0\mu_*(\nu_n,n_p)\Delta_{\rm pair}(\nu_n,n_p)}
              {\hbar^2(3\pi^2\nu_n)^{1/3}}.
 \label{eq:tension_bcs_logarithm}
\end{equation}
As a result, we then get,
\begin{align}
 F={}&\frac{\kappa_v^2\nu_n\mathcal N}{4\pi H_0}
 \ln\frac{\pi b_0\mu_*\Delta_{\rm pair}}
              {\hbar^2(3\pi^2\nu_n)^{1/3}}\notag\\
 &+\frac{\kappa_v^2\nu_n\alpha_{\rm v}\mathcal N^2}
                         {8\pi H_0\mathcal N_0}.
 \label{eq:tension_bcs_line_energy}
\end{align}
The density responses of $k_{Fn}$, $\mu_*$ and $\Delta_{\rm pair}$
are retained when differentiating $F$. The microscopic functions
and their responses are independent constitutive inputs; the bulk
polytropic pressure alone does not determine them. Their explicit
contribution to the primitive EOS coefficients is given in
Appendix~\ref{app:tension_constitutive_derivatives} and
Sec.~\ref{sec:tension_evaluated_bounds}.

%% file: section_characteristics_tension.tex
\begingroup
\subsection{Characteristic structure}
\label{sec:primitive_hyperbolicity_summary}

Characteristic speeds are defined by the principal part of the
linearized equations \cite{Reula:2004xd}. To compute them, freeze a
primitive background state \textcolor{black}{$\boldsymbol W$, consisting of
independent material and vorticity primitives,} and retain every term containing a derivative of its
perturbation $\delta\boldsymbol W$. After imposing the physical
principal constraints, write the resulting system as
$\mathsf A^\mu\nabla_\mu\delta\boldsymbol W\simeq0$.
A high-frequency perturbation with phase covector $k_\mu$ has a
nonzero physical amplitude $r$ satisfying
\begin{equation}
 \mathsf A^\mu k_\mu r=0,\qquad
 k_\mu=k(s_\mu+\sigma n_\mu),\qquad r\ne0,
 \label{eq:characteristic_speed_definition}
\end{equation}
where $k>0$ is the spatial wave number, $n^\mu$ is the unit Eulerian
observer and $s_\mu n^\mu=0$, $s_\mu s^\mu=1$.
With $\mathsf A_n=-n_\mu\mathsf A^\mu$ and
$\mathsf A_s=s_\mu\mathsf A^\mu$, the characteristic speeds are therefore the roots
\begin{equation}
 \det(\mathsf A_s-\sigma\mathsf A_n)=0.
 \label{eq:characteristic_speed_pencil}
\end{equation}
The determinant is taken on the physical constraint subspace.
On a regular time branch, these roots are the eigenvalues of
$\mathsf C(s)=\mathsf A_n^{-1}\mathsf A_s$.
Their Eulerian and coordinate normal speeds obey
\begin{equation}
 \sigma=-\frac{n^\mu k_\mu}{k},\qquad
 \lambda_s=\alpha\sigma-\beta^is_i.
 \label{eq:characteristic_speed_frames}
\end{equation}

The vortex closure determines which characteristic families must
be tested. We first consider algebraic force balance for meridional
propagation in axisymmetry, then independent line motion with vortex
tension in three dimensions.

\input{section_force_equilibrium_characteristics}

\subsubsection{Vortex tension and line energy}
\label{sec:line_energy_characteristics}

On the causal, strongly hyperbolic domain of
Sec.~\ref{sec:vortex_tension_hyperbolicity}, line energy changes the
compression and vortex sectors as follows, giving eleven physical
fields in the warm NC model and ten in cold matter.
\begin{itemize}
 \item {\emph{Mixed compression waves.} The four acoustic waves of
 Sec.~\ref{sec:force_equilibrium_characteristics} couple to two array-compression
 waves through the density derivatives of $F$. Their six speeds are
 the roots of $P_6$ in Eq.~\eqref{eq:regulated_characteristic_factorization} below.
 The material-independent line-energy limit recovers the fluid quartic
 and a separate line-compression pair
 (Sec.~\ref{app:separable_characteristics}).}
 \item {\emph{Bending waves.} The vortex-transport pair
 \eqref{eq:main_model_b_general_vortex_speed} is replaced by}
\begin{equation}
 \sigma_{A\pm}=v_d\cdot s\pm\sqrt{1-|v_d|^2}\,b_s,
 \label{eq:regulated_bending_speeds}
\end{equation}
{where $v_d=E\times B/B^2$ and $b_s=B\cdot s/\sqrt{B^2}$.
These waves bend the array without changing its line density at
principal order and propagate at light speed along a resting array.
Their eigenvectors remain independent at $b_s=0$.}
\end{itemize}

For the characteristic calculation, use the local vortex rest frame
$E'=0$, $B'=\kappa_v\mathcal N\widehat b$, with the boost in
Eq.~\eqref{eq:regc_drift_frame}. \textcolor{black}{Here
$\widehat b=B'/|B'|$, $\sigma'=\omega'/|k'|$ and $s'=k'/|k'|$,
where $(\omega',k')$ is the boosted wave covector. In this frame
$v_c$ denotes the boosted carrier three-velocity.}
Put $q=\widehat b\cdot s'$,
$p^2=1-q^2$,
$c_\Omega^2=\mathcal NF_{\mathcal N\mathcal N}/F_{\mathcal N}$,
$c_\ell^2=q^2+p^2c_\Omega^2$ and
$\eta_\Omega=p^2/(\kappa_v^2a_\Omega)$.
\textcolor{black}{The quadratic material response $\mathcal Q_{xy}$, linear
material--line coupling $d_x$ and constant line response
$\mathcal S_\Omega$ have the primitive coefficients in
Eqs.~\eqref{eq:primitive_dispersion_coefficients}--\eqref{eq:primitive_dispersion_responses}.}
The eleven physical speeds obey
\begin{align}
 0&=(\sigma'-v_c\cdot s')^3[(\sigma')^2-q^2]
                         P_6(\sigma'{,}s'),\notag\\
 P_6&=[(\sigma')^2-c_\ell^2-\eta_\Omega\mathcal S_\Omega]
       (\mathcal Q_{nn}\mathcal Q_{cc}-\mathcal Q_{nc}^2)\notag\\
 &\quad+\eta_\Omega(\mathcal Q_{cc}d_n^2
       -2\mathcal Q_{nc}d_nd_c+\mathcal Q_{nn}d_c^2).
 \label{eq:regulated_characteristic_factorization}
\end{align}
\textcolor{black}{These responses include thermal entrainment and the full
density response of $F$ for a general EOS.}

Due to the constraints, the physical perturbations satisfy
\begin{equation}
 B^i\delta E_i+E_i\delta B^i=0,\qquad
 s_i\delta B^i=0,\qquad s\times\delta\xi=0.
 \label{eq:rw_principal_constraints}
\end{equation}
The canonical-curl condition supplies two constraints, with
$\delta\xi$ calculated from the full primitive EOS.

Boost the normalized wave covector $(\sigma,s)$ with
Eq.~\eqref{eq:regc_drift_frame}, writing $m=|k'|$ and
$\sigma'=\omega'/m$, $s'=k'/m$. With $\sigma_c$ denoting the carrier speed
\eqref{eq:main_model_b_material_speeds}, the Eulerian physical
polynomial is, up to a nonzero normalization,
\begin{align}
 p_s(\sigma)&\doteq(\sigma-\sigma_c)^3
       (\sigma-\sigma_{A+})(\sigma-\sigma_{A-})
                       \mathcal P_6(\sigma{,}s),\notag\\
 \mathcal P_6(\sigma{,}s)&=m^6P_6(\omega'/m{,}k'/m).
 \label{eq:rw_physical_pencil}
\end{align}
The last expression is a polynomial: the factors of $m$ cancel by
homogeneity. Its six real roots are the Eulerian compression speeds.
The corresponding physical evolution symbol $\mathsf C(s)$ is
constructed independently in Eq.~\eqref{eq:rw_projected_verification}.
At general counterflow the sextic need not factor into separate
fluid and vortex quadratics.
{The primitive right and left characteristic eigenvectors
for the NC model with line tension are computed explicitly in
Appendix~\ref{app:tension_eigenvectors}.}

\paragraph{Closed speeds at coflow.}
On the positive-inertia branch $B_n>0$, $\Delta>0$, when both
fluids are at rest with the array, the compression speeds
are $\sigma'_{j\pm}=\pm\sqrt{x_j}$. All EOS derivatives in the
following expressions hold $\varsigma=s/N$ fixed. Write
$\varepsilon_0(n,N,\varsigma)
=-\Lambda_0(n,N,I=nN,\varsigma)$ and
\begin{align}
 k_{nn}&=\partial_n^2\varepsilon_0+F_{nn},\quad
 k_{cc}=\partial_N^2\varepsilon_0+F_{NN},\notag\\
 k_{nc}&=\partial_n\partial_N\varepsilon_0+F_{nN},\quad
 f_n=F_{n\mathcal N},\quad f_c=F_{N\mathcal N},\notag\\
 B_n&=\frac{\mu_n+F_n-\mathcal A_*N}{n},\quad
 B_c=\frac{\mu_h+F_N-\mathcal A_*n}{N},\notag\\
 \Delta&=B_nB_c-\mathcal A_*^2>0,\quad
 \mathcal A_*=-\Lambda_{{\rm tot},I},
 \label{eq:primitive_coflow_responses}
\end{align}
where $\mu_n=\partial_n\varepsilon_0$ and
$\mu_h=\partial_N\varepsilon_0$.
Equivalently, $k_{nn}=\mu_nc_n^2/n+F_{nn}$ and
$k_{cc}=\mu_hc_c^2/N+F_{NN}$, with
$c_n^2=n\partial_n\mu_n/\mu_n$ and
$c_c^2=N\partial_N\mu_h/\mu_h$.
{These are the constituent adiabatic responses before line energy
is added. The mixed response $k_{nc}$ is retained.} For the
additive NC EOS, $\mathcal A_*=\mathcal A$ and
$\partial_n\partial_N\varepsilon_0=0$. Define
\begin{align}
 t_f&=\frac{B_ck_{nn}+B_nk_{cc}-2\mathcal A_*k_{nc}}{\Delta},\notag\\
 d_f&=\frac{k_{nn}k_{cc}-k_{nc}^2}{\Delta},\notag\\
 g_1&=\frac{\eta_\Omega}{\Delta}
       (B_cf_n^2+B_nf_c^2-2\mathcal A_*f_nf_c),\notag\\
 g_0&=\frac{\eta_\Omega}{\Delta}
       (k_{cc}f_n^2+k_{nn}f_c^2-2k_{nc}f_nf_c),\notag\\
 I_1&=t_f+c_\ell^2,\quad
 I_2=d_f+t_fc_\ell^2-g_1,\quad
 I_3=d_fc_\ell^2-g_0.
 \label{eq:primitive_coflow_cubic_coefficients}
\end{align}
The squared speeds solve $x^3-I_1x^2+I_2x-I_3=0$. Thus, with
$R=(I_1^2-3I_2)/9$ and
$Q=(2I_1^3-9I_1I_2+27I_3)/54$, their explicit values are
\begin{align}
 \theta&=\tfrac13\arccos(Q/R^{3/2}),\notag\\
 x_j&=\frac{I_1}{3}+2\sqrt R
                  \cos\!\left(\theta-\frac{2\pi j}{3}\right),\notag\\
 \sigma'_{j\pm}&=\pm\sqrt{x_j},\qquad j=0,1,2.
 \label{eq:primitive_coflow_closed_speeds}
\end{align}
On the causal, strongly hyperbolic domain,
$1\geq x_0\geq x_1\geq x_2>0$. These are the fast, intermediate
and slow compression pairs. {At $R=0$, use $x_0=x_1=x_2=I_1/3$.
The trigonometric expression already includes a double root.} Their eigenvectors remain complete by the
positive-energy proof below. {Since $x_0\geq c_\ell^2\geq q^2$,
$\pm\sqrt{x_0}$ bound the full coflow spectrum, including the unchanged
carrier branch.}

\input{section_donnelly_glaberson}

\subsection{Nonlinear causality and strong hyperbolicity}
\label{sec:hyperbolicity}

The characteristic speeds determine whether disturbances propagate
causally, but they do not by themselves establish a well-posed
initial-value problem. Strong hyperbolicity also requires a complete
eigenbasis whose conditioning remains bounded as the propagation
direction varies. We impose both requirements on the physical symbol
$\mathsf C(\boldsymbol W,s)$ at each admitted state.
The background may have finite counterflow and vortex slip.

For $N_{\rm phys}$ physical fields, strong hyperbolicity requires
\cite{Reula:2004xd}
\begin{align}
 \mathsf T^{-1}\mathsf C(\boldsymbol W,s)\mathsf T
 &=\operatorname{diag}(\sigma_1,\ldots,\sigma_{N_{\rm phys}}),\notag\\
 \sup_{s_is^i=1}\|\mathsf T\|\,\|\mathsf T^{-1}\|&<\infty.
 \label{eq:strong_hyperbolicity_definition}
\end{align}
{Here every $\sigma_a$ is real, and the columns of $\mathsf T$
are unit-norm right eigenvectors. The bound is uniform over propagation
directions and over a neighborhood of each admitted state.} It therefore controls the approach to
coincident speeds as well as the eigenspace at a coincidence.
Quasilinear energy estimates additionally require a sufficiently
regular symmetrizer. The positive-energy constructions below provide
one on their stated domains.

Following Ref.~\cite{Cordeiro:2023ljz}, nonlinear causality
requires real characteristic frequencies and nontimelike
characteristic covectors. In the normalization of
Eq.~\eqref{eq:characteristic_speed_definition}, these conditions are
\begin{align}
 &\sigma_a(\boldsymbol W,s)\in\mathbb R,\qquad |\sigma_a|\leq1,\label{eq:nonlinear_causality_definition}\\
 &g^{\mu\nu}k_\mu k_\nu=k^2(1-\sigma_a^2)\geq0 ,
 \quad a=1,\ldots,N_{\rm phys}.\notag
\end{align}
They constrain the signal cone to lie within the spacetime light
cone for a well-posed evolution. Their dependence on the full state
$\boldsymbol W$ is essential. Conditions obtained by setting
counterflow or slip to zero establish causality on that restricted
background.

We first give the joint conditions for algebraic force balance in
axisymmetry, where the Eulerian normals are meridional. We then
consider independent line motion with tension for every
three-dimensional propagation direction. 

\input{section_force_equilibrium_conditions}

\subsubsection{NC model with vortex tension}
\label{sec:vortex_tension_hyperbolicity}
\label{sec:tension_evaluated_bounds}

\input{section_regulated_conditions_tension}

\input{section_specialized_conditions_tension}

\input{section_tension_state_conditions}

\endgroup

%% file: section_force_equilibrium_characteristics.tex
\subsubsection{Force-equilibrium model for axisymmetric simulations}
\label{sec:force_equilibrium_characteristics}

We begin with the algebraic vortex-force model of
Sec.~\ref{sec:moment_vortex_closure} for the two-dimensional
axisymmetric calculations. It uses the material EOS without line
energy, $F=0$, and determines $E=-v_L\times B$ from force balance {\eqref{eq:moments_relativistic_line_velocity}}.
Axisymmetry restricts the Eulerian propagation normals to
the meridional plane, $\hat s\cdot e_\phi=0$\textcolor{black}{, where $\hat s$
is the unit Eulerian propagation normal and $e_\phi$ the azimuthal
unit vector. We write $z_\phi=z\cdot e_\phi$}.

The force-equilibrium system supports sound, material and vortex waves. 
The carrier Lorentz factor is $W=\sqrt{1+|z|^2}$.
\textcolor{black}{The phase projections and the carrier-frame counterflow are}
\begin{align}
 e&=\varepsilon_\xi,\quad \Xi=\sqrt{e^2+|\boldsymbol\xi|^2},\quad
 \mu=W\Xi-z\cdot\boldsymbol\xi,\notag\\
 q&=\boldsymbol\xi+\frac{z(z\cdot\boldsymbol\xi)}{W+1}-\Xi z,
 \quad Q=|q|^2=\mu^2-e^2.
 \label{eq:main_phase_kinematics}
\end{align}
For the NC model, use $n=\nu_n$, $N=n_p$, $g=\Gamma_r$,
$H=\mu_n-\mathcal AN$, $f^2=n/H$ and $\nu=ng+\mathcal ZN$.
Then $\mu=Hg+\mathcal AN$ and $Q=H^2(g^2-1)$.
{On the causal, strongly hyperbolic branches of}
Sec.~\ref{sec:force_equilibrium_conditions}, the warm NC model has nine
characteristic fields:
\begin{itemize}
 \item \emph{Sound waves.} Four compressive waves couple the neutron and
 charged currents. \textcolor{black}{Write their acoustic quartic as
 $\sum_{j=0}^4a_j\sigma^j=0$, with the EOS-response coefficients of
 Eq.~\eqref{eq:app_general_eos_acoustic_coefficients}.
 We use $b_f=a_3/a_4$, the depressed-quartic coefficients
 $P_f,Q_f,R_f$ and the largest real resolvent root $\zeta_f$ defined
 in Eqs.~\eqref{eq:app_quartic_depression}--\eqref{eq:app_finite_quartic_speeds}.}
 Their speeds are
 \begin{align}
  &\sigma_{\epsilon\tau}=-\frac{b_f}4+\frac{\epsilon\sqrt{\zeta_f}}2\notag\\
  &+\frac\tau2\sqrt{-2P_f-\zeta_f-2\epsilon Q_f/\sqrt{\zeta_f}},
  \quad\epsilon,\tau=\pm1.
  \label{eq:main_finite_flow_sound_speeds}
 \end{align}
 For $Q_f=0$,
 \begin{equation}
  \sigma_{\epsilon\tau}=-\frac{b_f}{4}
  +\epsilon\sqrt{\frac{-P_f+\tau\sqrt{P_f^2-4R_f}}{2}}.
  \label{eq:main_finite_flow_biquadratic_speeds}
 \end{equation}
 For independent constituent EOSs, including the unentrained separable
 NC model, these four speeds simplify to
 \begin{align}
  A_x&=c_x^2+(1-c_x^2)(U_x^0)^2,\notag\\
  \sigma_{x\pm}
  &=\frac{(1-c_x^2)U_x^0U_{xs}}{A_x}\notag\\
  &\quad\pm\frac{c_x}{A_x}
  \sqrt{A_x-(1-c_x^2)U_{xs}^2},\quad x=n,p.
  \label{eq:main_model_b_euler_sound_speeds}
 \end{align}
 Here $U_{xs}=\boldsymbol U_x\cdot\hat s$,
 $c_x^2=(\partial P_x/\partial e_x)_{s_x/n_x}$ is the adiabatic
 rest-frame sound speed, and the NC primitive velocities are
 \begin{align}
  U_p^0&=W,\qquad U_{ps}=z\cdot\hat s,\notag\\
  U_n^0&=gW+\frac{q\cdot z}{H},\notag\\
  U_{ns}&=g\,z\cdot\hat s+\frac{q\cdot\hat s}{H}
  +\frac{(q\cdot z)(z\cdot\hat s)}{H(W+1)}.
  \label{eq:main_model_b_sound_velocities}
 \end{align}
 \textcolor{black}{For the thermal EOS, define the charged adiabatic chemical
 response $\widehat\mu_p=\mu_p+4r/(3N)$, where
 $r=C_{\rm th}s^{4/3}$, and the stiffnesses
 $k_n=d\mu_n/dn$ and
 $k_h=(\partial\widehat\mu_p/\partial N)_{s/N}$.}
 For the specified thermal EOS, $c_n^2=nk_n/H$ and
 $c_p^2=Nk_h/\widehat\mu_p$ on this independent-fluid branch.
 At coflow, the same speed formula uses the common $U^\mu$ and the
 two coupled sound speeds $c_{1,2}$ of
 Eq.~\eqref{eq:app_general_eos_coflow_sounds}.
 \item \emph{Material waves.} Two transverse shears and entropy per
 charged particle move with the carrier at
 \begin{equation}
  \sigma_{c,1}=\sigma_{c,2}=\sigma_{c,3}
  =\frac{z\cdot\hat s}{W}.
  \label{eq:main_model_b_material_speeds}
 \end{equation}
 The entropy mode is absent in the cold NC model, which has eight modes.
 \item \emph{Vortex waves.}
\textcolor{black}{For mobile lines, let $\hat k$ be the unit carrier-frame
normal and resolve $q_\parallel=q\cdot\ell$ and
$q_\perp=q-q_\parallel\ell$. The lift response $A_L$, drag response
$C_L$, transverse line Lorentz factor $\Gamma_L$ and drift
coefficients $h_v,d_v$ are defined in
Eq.~\eqref{eq:main_mobile_sh_exact_line_coefficients}.}
The carrier-frame speeds are
\begin{align}
 \lambda_{v\pm}
 &=v_0\cdot\hat k+
 \frac{f^2q_\parallel(\ell\cdot\hat k)
          (A_L\pm iC_L\Gamma_L^2)}
      {A_L^2+C_L^2\Gamma_L^4},\notag\\
 v_0&=h_vq_\perp-d_v\ell\times q,\qquad
 q_\parallel=q\cdot\ell.
 \label{eq:main_model_b_general_vortex_pair}
\end{align}
The direction-dependent pair can be complex or defective, as derived
in Eqs.~\eqref{eq:app_vortex_physical_quadratic}--\eqref{eq:app_model_b_tangent_vortex_jordan}.
On the pinned, reactive and admitted perpendicular-drag branches,
the Eulerian vortex speed instead has the repeated form
 \begin{equation}
  \sigma_{v,1}=\sigma_{v,2}
  =\frac{[z+v+z(z\cdot v)/(W+1)]\cdot\hat s}{W+z\cdot v}.
  \label{eq:main_model_b_general_vortex_speed}
 \end{equation}
Here $v=0$ for pinning and $v=f^2q/\nu$ for zero-work reactive transport.
 Perpendicular dissipative motion instead uses
 $v=h_vq-d_v\ell\times q$, with the force-law coefficients of
 Eq.~\eqref{eq:main_mobile_sh_exact_line_coefficients} and $ng$ in $A_L$. The geometric data include the Eulerian canonical curl
$B=\boldsymbol\nabla\times\boldsymbol\xi$. Its carrier-frame direction
$\ell$ follows from Eq.~\eqref{eq:app_model_b_curl_frame_map}.
{For general dissipative line geometries, differentiating the
force law instead gives the pair in Eq.~\eqref{eq:app_vortex_speeds}.}
\end{itemize}

{Equation~\eqref{eq:main_model_b_general_vortex_pair} has
nonreal vortex speeds when
$C_Lq_\parallel(\ell\cdot\hat k)\ne0$.
Thus dissipative axial counterflow obstructs strong hyperbolicity
for nontangent propagation directions.}
In Sec. \ref{sec:donnelly_glaberson} we show that this is associated with the onset of the Donnelly--Glaberson instability, which is not bounded in the absence of vortex tension \cite{Glaberson:1974zz,Sidery:2007ge,Glampedakis:2011px,vanEysden:2018kbc}.

The warm CP model has the four sound roots above
with its own EOS responses, two carrier shears and two vortex modes.
The vortex speeds in the carrier frame are zero for pinning and
$q\cdot\hat k/\mu$ for the specified free-phase transport branch.
Thus the warm CP model has eight fields.

\paragraph{Axisymmetric admissibility.}
The expressions above are used on the corresponding complete
force-equilibrium branches. In particular, for the NC model with mobile vortex lines, and $Q>0$ and $\mathcal D>0$, the meridional reduction requires
\begin{align}
 \mathcal D&=\mathcal R\nu-\mathcal R'\mathcal ZN,\quad
 a=e_\phi+\frac{z_\phi z}{W+1},\notag\\
 q\cdot\ell&=a\cdot\ell=0,\qquad
 v\cdot(\ell\times a)\ne0.
 \label{eq:force_equilibrium_axisymmetric_geometry}
\end{align}
These geometric conditions accompany the force-admission, material
and vortex--sound tests in
{Secs.~\ref{app:model_b_finite_vortices} and
\ref{app:model_b_vortex_gap_tests}}. Pinning and reactive transport
use their own branches in Sec.~\ref{sec:force_equilibrium_conditions}.

%% file: section_donnelly_glaberson.tex
\subsubsection{Donnelly--Glaberson instability and loss of hyperbolicity}
\label{sec:donnelly_glaberson}

Counterflow along vortex lines can amplify vortex bending through the
Donnelly--Glaberson mechanism destabilizing Kelvin waves via mutual friction
\cite{Glaberson:1974zz,Glampedakis:2008ew,Sidery:2007ge}.
For carrier-frame perturbations $e^{i(k\hat k\cdot x-\omega t)}$,
the vortex pair \eqref{eq:app_vortex_speeds} has
$\max\operatorname{Im}\omega\sim k\gamma_\infty$, with
\begin{align}
 \gamma_\infty
 &=\frac{f^2|C_Lq_\parallel(\ell\cdot\hat k)|\Gamma_L^2}
             {A_L^2+C_L^2\Gamma_L^4},\quad f^2=n/H,
 \label{eq:dg_principal_growth}\\
 \gamma_\infty>0&\quad\Longleftrightarrow\quad
 \mathcal Dq_\parallel(\ell\cdot\hat k)\ne0.
 \label{eq:dg_instability_criterion}
\end{align}
{The condition $\mathcal D=0$ removes the dissipative combination
driving this principal growth, while $q_\parallel=0$ removes parallel
counterflow. Reactive vortex forces can remain at $\mathcal D=0$.}
If not suppressed, its complex principal pair fails both strong hyperbolicity
\eqref{eq:strong_hyperbolicity_definition} and nonlinear causality's
real-frequency requirement \eqref{eq:nonlinear_causality_definition}.
Without vortex tension, full three-dimensional simulations will likely become ill-posed.

{Vortex tension can supply an upper cutoff in unstable
wavenumber~\cite{Sidery:2007ge,vanEysden:2018kbc}.}
For the local prescribed-current bending problem of
Appendix~\ref{app:kelvin_tension}, take $E_0=0$,
$B_0=\kappa_v\mathcal N\ell$, $u_{L0}=u$ and currents along the
sheet. Set $\mathcal R'=0$, $\mathcal R>0$ and define
$L_\parallel=\ell\cdot L$, with $\nu_L=-u\cdot L>|L_\parallel|$.
\textcolor{black}{Let $\eta(z,t)$ be the transverse line displacement, with
$z$ measured along $\ell$ and subscripts denoting partial derivatives.}
The covariant force \eqref{eq:tension_draft_exchange} and
proper-area stress give
\begin{align}
 0={}&\mathcal T(\eta_{tt}-\eta_{zz})
       +\kappa_v\nu_L(\mathcal R I-J)\eta_t\notag\\
     &-\kappa_vL_\parallel J\eta_z,
 \label{eq:dg_displacement_equation}
\end{align}
where $J\eta=\ell\times\eta$ and
$\mathcal T=F_{\mathcal N}>0$ retains its full constitutive value.
For $J\eta_\epsilon=i\epsilon\eta_\epsilon$,
$\eta_\epsilon\propto e^{i(kz-\omega t)}$ and $\epsilon=\pm1$,
\begin{align}
 0={}&\mathcal T(\omega^2-k^2)
       +\kappa_v\nu_L(\epsilon+i\mathcal R)\omega\notag\\
     &-\epsilon\kappa_vL_\parallel k.
 \label{eq:dg_dispersion}
\end{align}
At zero line stress this recovers the aligned roots of
Eq.~\eqref{eq:app_vortex_speeds}, with
$L_\parallel=f^2q_\parallel$, $A_L=\nu_L$,
$C_L=\mathcal R\nu_L$ and $\Gamma_L=1$.
Bending exposes the axial current to transverse lift. Drag supplies
the phase lag through which one polarization extracts counterflow
energy. Its effective stiffness is
$\mathcal K_\epsilon=\mathcal T k^2+\epsilon\kappa_vL_\parallel k$:
the tension cost grows as $k^2$, while the current drive grows as $k$.
For $k>0$, growth occurs precisely when $\mathcal K_\epsilon<0$,
giving the unstable band
\begin{equation}
 \epsilon L_\parallel<0,\qquad
 0<k<k_+\equiv\frac{\kappa_v|L_\parallel|}{\mathcal T}
             <\frac{\kappa_v\nu_L}{\mathcal T}.
 \label{eq:dg_tension_criterion}
\end{equation}
Thus $k=k_+$ is neutral, and $k>k_+$ is damped. 
The allowed wavelengths must also satisfy
$d_{\rm iv}\ll2\pi/k\ll L_{\rm bg}$, where
$d_{\rm iv}\sim\mathcal N^{-1/2}$ and $L_{\rm bg}$ is the background
variation length. Coupled rotating-column modes additionally require
induced material flow and radial boundaries, which can change the band
\cite{vanEysden:2018kbc}.
On the coupled branch,
write $N_s^\mu=\nu_su_s^\mu$ and $g=-u_s\cdot u$.
\textcolor{black}{The upper bound $G$ on the material relative Lorentz factor
is defined in Eq.~\eqref{eq:tension_state_motion_caps}.}
The sufficient bounds \eqref{eq:tension_state_response_bounds} and
\eqref{eq:tension_state_motion_bounds} give $g<G$ and
\begin{equation}
 k_+=\frac{\kappa_v\nu_s\sqrt{g^2-1}}{\mathcal T}
       <\frac{\kappa_v\nu_s\sqrt{G^2-1}}{\mathcal T}.
 \label{eq:dg_motion_band_bound}
\end{equation}
For the bending roots, ${\omega}/{k}\to\pm1$,
\begin{equation}
 \operatorname{Im}\omega\to
       -\frac{\kappa_v\nu_L\mathcal R}{2\mathcal T}<0
 \quad(k\to\infty).
 \label{eq:dg_causal_asymptote}
\end{equation}
Thus tension with line inertia removes the ill-posed short-wavelength
limit.
{Causality and strong hyperbolicity do not eliminate every finite
Donnelly--Glaberson unstable band. The admitted relative motion does,
however, bound its upper wavenumber through
Eq.~\eqref{eq:dg_motion_band_bound}.}

%% file: section_force_equilibrium_conditions.tex
\subsubsection{Axisymmetric force-equilibrium systems}
\label{sec:force_equilibrium_conditions}
\begingroup

The force-equilibrium model uses the material EOS with $F=0$ and
fixes the vortex velocity algebraically. The resulting principal
symbol contains derivatives of that velocity with respect to matter
and vorticity. These derivatives determine whether the vortex modes
are real and independent of the acoustic modes.
Appendix~\ref{app:force_equilibrium_model_b} derives the conditions
below.
{Appendix~\ref{app:force_equilibrium_full_symbol} includes the
full direction-dependent vortex pair in a necessary and sufficient
primitive classification.}

\paragraph{Admitted states.}
For the NC model, {let
$\mu_h=\widehat\mu_p=\mu_p+4r/(3N)$ be the adiabatic
charged-fluid EOS response, with $r=C_{\rm th}s^{4/3}$.
Define}
$H=\mu_n-\mathcal A N$, $I_h=\widehat\mu_p-\mathcal A n$,
$k_n=d\mu_n/dn$ and
{$k_h=(\partial\mu_h/\partial N)_{s/N}
=k_p+4r/(9N^2)$, where $k_p=d\mu_p/dN$}.
The material densities are positive and both currents are future
directed. Warm states have $s,T_c>0$. The positive-inertia branch
has $H>0$ and $HI_h-\mathcal A^2nN>0$.
Pressure variables also require $J_S\ne0$ in
Eq.~\eqref{eq:main_model_b_admission_compact}.
Future phase momentum, nonzero canonical curl and nonnegative drag
heating impose
\begingroup
\begin{gather}
 \mu_n+\mathcal A N[g-1+\sqrt{g^2-1}]>0,\notag\\
 B^2>0,\notag\\
 \mathcal R\geq0,\qquad
 \mathcal R(ng+\mathcal ZN)-\mathcal R'\mathcal ZN\geq0.
 \label{eq:force_summary_physical_branch}
\end{gather}
\endgroup
Here $B^i$ is defined in Eq.~\eqref{eq:moments_vorticity_fields}.
\textcolor{black}{We use $f^2=n/H$ throughout the force-equilibrium analysis and
$a=e_\phi+z_\phi z/(W+1)$ for the meridional geometry.}

\paragraph{Material waves.}
Positive constituent stiffness requires $k_n>0$ and $k_h>0$.
{The upper bounds on the material cone are}
\begin{gather}
 H-nk_n\geq0,\qquad I_h-Nk_h\geq0,\notag\\
 (H-nk_n)(I_h-Nk_h)\geq\mathcal A^2nN.
 \label{eq:force_summary_material_causality}
\end{gather}
For $\mathcal A\ne0$, the coupled acoustic modes must also remain
independent of the carrier contact. This requires
\begingroup
\begin{align}
 &H^2(HI_h-\mathcal A^2nN)(1+z_\phi^2)\notag\\
 &\quad\times\{k_h[nk_ng^2-H(g^2-1)]
                      -\mathcal A^2n(g-1)^2\}\notag\\
 &\quad+I_h[H(H-nk_n)k_h+\mathcal A^2n^2k_n]
                                   (a\cdot q)^2>0.
 \label{eq:force_summary_material_hyperbolicity}
\end{align}
\endgroup
At $\mathcal A=0$ the constituent Euler systems are independent,
so the additional contact inequality is absent. The material
conditions retain semisimple coincident sounds. The vortex conditions
depend on whether the array is pinned, reactive or dissipative.

\paragraph{Pinned vortices.}
\textcolor{black}{Let $f_c$ denote the holding force per unit line length.}
\revadd{For static pinning, the required load must satisfy}
\begingroup
\begin{equation}
 \frac{\kappa_v n}{H}
 \sqrt{[1+(\mathcal R')^2][q^2-(q\cdot\ell)^2]}
 \leq f_c,\qquad f_c\geq0.
 \label{eq:force_summary_pin_load}
\end{equation}
\endgroup
\revadd{Strong hyperbolicity also requires} the geometry $a\times q=0$.
This condition preserves independent
vortex and carrier eigenvectors at their shared speed, as shown in
Eq.~\eqref{eq:main_model_b_pinned_primitive}.

\paragraph{Reactive transport.}
Zero-work transport has
$\mathcal D=\mathcal R(ng+\mathcal ZN)-\mathcal R'\mathcal ZN=0$.
For mobile vortices with $\mathcal Z<0$, selecting the future force
solution requires
\begingroup
\begin{align}
 N\mathcal Z&>-ng,\notag\\
 N[1+(z\cdot\widehat B)^2]\mathcal Z
 &>\Big\{[ng+n(z_n\cdot\widehat B)(z\cdot\widehat B)]^2\notag\\
 &\quad-n^2[1+(z_n\cdot\widehat B)^2]
                  [1+(z\cdot\widehat B)^2]\Big\}^{1/2}\notag\\
 &\quad-ng-n(z_n\cdot\widehat B)(z\cdot\widehat B),
 \label{eq:force_summary_future_lift}
\end{align}
\endgroup
where $\widehat B=B/|B|$ and $z_n^i$ is the neutron spatial
four-velocity. This is the admission condition in
Eqs.~\eqref{eq:app_model_b_signed_lift_admission}--\eqref{eq:app_model_b_eulerian_lift_margin}.
For $\mathcal A\ne0$ and $\mathcal Z<0$, the meridional vortex cone
additionally satisfies
\begingroup
\begin{equation}
 (1+z_\phi^2)(ng+\mathcal ZN)
 +\frac{n}{H}z_\phi(a\cdot q)
 {\geq}\frac{n}{H}|a\times q|.
 \label{eq:force_summary_reactive_causality}
\end{equation}
\endgroup
For $\mathcal Z=0$ and $\mathcal A\ne0$, a strict separation of the
vortex and acoustic speeds is
\begingroup
\begin{align}
 &(HI_h-\mathcal A^2nN)[H^2+(a\cdot q+Hgz_\phi)^2]\notag\\
 &\quad\times\{k_n[Nk_hg^2-I_h(g^2-1)]
                         -\mathcal A^2N(g-1)^2\}\notag\\
 &\quad+H[I_h(I_h-Nk_h)k_n+\mathcal A^2N^2k_h]\notag\\
 &\hspace{13mm}\times[g(a\cdot q)+H(g^2-1)z_\phi]^2>0.
 \label{eq:force_summary_reactive_hyperbolicity}
\end{align}
\endgroup
For $\mathcal A=0$ and $\mathcal Z\ne0$, the corresponding
independent-fluid gap conditions reduce to
\begingroup
\begin{align}
 &nk_n\Big\{|H(ng+\mathcal ZN)a+nz_\phi q|^2\notag\\
 &\hspace{16mm}+(\mathcal Z^2N^2-n^2)|a\times q|^2\Big\}
 >H\mathcal Z^2N^2|a\times q|^2,\notag\\
 &Nk_h|H(ng+\mathcal ZN)a+nz_\phi q|^2
 >\widehat\mu_p n^2|a\times q|^2.
 \label{eq:force_summary_independent_reactive}
\end{align}
\endgroup
The branch $\mathcal A=\mathcal Z=0$ has protected sound/vortex
crossings and needs no gap condition.

\paragraph{Dissipative transport.}
At $q^2>0$ and $\mathcal D>0$, a complete warm vortex pair in
axisymmetry requires
\begingroup
\begin{equation}
 q\cdot\ell=0,\qquad a\cdot\ell=0,\qquad
 v\cdot(\ell\times a)\ne0.
 \label{eq:force_summary_dissipative_geometry}
\end{equation}
\endgroup
The algebraic velocity is $v=h_vq-d_v\ell\times q$.
At zero entrainment, its strict acoustic gaps are
\begingroup
\begin{align}
 &nk_nH^2\{|a+z_\phi v|^2-[v\cdot(\ell\times a)]^2\}\notag\\
 &\quad>(H-nk_n)
 \{Hg\,v\cdot(\ell\times a)-q\cdot(\ell\times a)
                                      -z_\phi d_vq^2\}^2,\notag\\
 &Nk_h|a+z_\phi v|^2
            >\widehat\mu_p[v\cdot(\ell\times a)]^2,
 \label{eq:force_summary_independent_drag}
\end{align}
\endgroup
For nonzero entrainment the gap or compatibility conditions below
must also be imposed. These results are specific to meridional
propagation. In three dimensions, Eq.~\eqref{eq:app_model_b_three_dimensional_mobile_gate}
requires $\mathcal D^2Q=0$, with $Q=|q|^2$.
Dissipative counterflow with $Q>0$ and $\mathcal D>0$ therefore fails
strong hyperbolicity in three dimensions. Coflow is obtained from
the regular direct limit. 

\paragraph{Vortex--sound coincidences.}
On the reactive and perpendicular-drag branches, five carrier/vortex
fields remain independent when their speeds coincide. The cold
reactive system has four such fields. A sound meeting a vortex
requires an additional compatibility condition.
Let $K,S$ be the primitive projections of
Eq.~\eqref{eq:force_full_primitive_gap_inputs}, and set
$t_n=gS-K/H$. With $J,\Delta_M,R_n$ from
Eq.~\eqref{eq:main_model_b_material_scalars}, define
\begin{equation}
 \mathcal L_v=nH\Delta_M(1-S^2)-JR_nt_n^2.
 \label{eq:force_summary_gap_pivot}
\end{equation}
The scalar quadratic $\mathcal G_v$ has the compact primitive
coefficients in Eq.~\eqref{eq:force_full_primitive_gap_inputs}.
A strict gap
in every meridional direction is equivalent to
\begin{equation}
 \mathcal L_v>0,\qquad\mathcal G_v(1)>0.
 \label{eq:force_summary_strict_gap}
\end{equation}
Both inequalities are affine in either $k_n$ or $k_h$ separately.
The normalization $\mathcal G_v=(J/n)G$ preserves the roots and
cofactor cancellations of the existing gap polynomial $G$.
The proof and the three-dimensional projections are given in
Appendix~\ref{app:force_equilibrium_full_symbol}.
A reactive crossing can instead be retained if every acoustic-source
cofactor vanishes at the crossing, with its derivative also vanishing
at a double root. Finite drag permits at most one simple compatible
crossing. Appendix~\ref{app:force_equilibrium_coincidences} gives the
roots and the required equalities. The independent-fluid bounds
above exclude the unprotected crossings at zero entrainment.

\paragraph{CP model.}
The warm phonon EOS changes the acoustic and contact responses.
\textcolor{black}{We use the phonon pressure $p$, the cold phase-stiffness
coefficient $Y_0$ and its thermal correction coefficient $y$ defined
in Eqs.~\eqref{eq:app_model_a_general_pressure}--\eqref{eq:app_model_a_finite_energy_coefficients}.}
Its physical domain is
\begingroup\color{black}
\begin{gather}
 \Gamma>1,\quad\kappa_p>0,\quad e>1,\quad T>0,\quad\mu\geq e,\notag\\
 e^3>[\Gamma-1+(2-\Gamma)e]\mu^2,\qquad Y_0-yp>0.
 \label{eq:force_summary_model_a_admission}
\end{gather}
\endgroup
Appendix~\ref{app:model_a_force_equilibrium} expresses the time,
carrier-contact and phase-contact conditions through
$\mathcal T_A,\mathcal D_A,\mathcal C_A,\mathcal P_A,\mathcal R_A$.
For pinning, $a\times q=0$ and
\begin{equation}
 \mathcal T_A>0,\qquad\mathcal D_A>0.
 \label{eq:force_summary_model_a_pinned}
\end{equation}
For free phase transport, define
\begin{equation}
 X_c=\frac{|a\times q|^2}{1+z_\phi^2},\qquad
 X_\phi=\frac{|a\times q|^2}{\mu^2|a+z_\phi q/\mu|^2}.
 \label{eq:force_summary_model_a_contact_inputs}
\end{equation}
With $S_\Gamma=3(\Gamma-1)+2e(2-\Gamma)$, strong hyperbolicity requires
\begin{gather}
 \mathcal T_A>0,\quad\mathcal D_A>0,\notag\\
 \mathcal P_A(X_\phi)>0,\qquad\mathcal R_A(X_\phi)>0,\notag\\
 \mathcal C_A(X_c)>0\quad\text{unless }S_\Gamma=0.
 \label{eq:force_summary_model_a_free}
\end{gather}
The exceptional sheet
\begin{equation}
 S_\Gamma=0\quad\Longleftrightarrow\quad
 \Gamma=\frac{4e-3}{2e-3},\qquad e>\frac32.
 \label{eq:force_summary_model_a_protected}
\end{equation}
admits an independent carrier polarization at a sound/contact
crossing. It removes the carrier-contact exclusion, while the
phase-contact bounds still exclude a sound/vortex collision.
Repeated acoustic speeds and the carrier/vortex coincidence retain
complete eigenspaces under these conditions.

Nonlinear causality is determined by the two null-response
polynomials $N_T,N_D$ of Eq.~\eqref{eq:app_model_a_null_scalars}.
Let $\mathcal E_i$ contain both endpoints of
\begin{equation}
 I=\left[\frac{-z_\phi(a\cdot q)-|a\times q|}{1+z_\phi^2},
          \frac{-z_\phi(a\cdot q)+|a\times q|}{1+z_\phi^2}\right]
 \label{eq:force_summary_model_a_null_interval}
\end{equation}
and the real interior stationary points of $N_i$.
On either strongly hyperbolic warm branch, the strict causal condition is
\begin{equation}
 N_i(t)>0\quad(i=T,D{,}\ t\in\mathcal E_i).
 \label{eq:force_summary_model_a_finite_causality}
\end{equation}
The stationary points follow from the cubic formulas in
Eqs.~\eqref{eq:app_scalar_cubic_parameters}--\eqref{eq:app_scalar_cubic_values}.
For $\Gamma=2$ at coflow, the joint conditions reduce to the thermal
bound
\begingroup\color{black}
\begin{equation}
 e>1,\qquad T>0,\qquad
 27\kappa_p T^4\sqrt e(8e+3)<256(e-1)^{7/2}.
 \label{eq:force_summary_model_a_warm_causality}
\end{equation}
\endgroup
The cold free-phase system instead has the domain
\begingroup\color{black}
\begin{equation}
 e>1,\qquad\kappa_p>0,\qquad\Gamma>1,\qquad
 (\Gamma-2)e<\Gamma-1.
 \label{eq:force_summary_model_a_cold_causality}
\end{equation}
\endgroup
which also makes the time coefficient positive. Its vortex speeds
lie strictly between the sounds. Pinning after the entropy carrier
vanishes requires a separately specified medium.
\endgroup

%% file: section_regulated_conditions_tension.tex
\begingroup
\paragraph{Physical domain.}
Independent line motion adds electric inertia and an array-density
response to the material system. The phase momentum and vortex fields
must satisfy
\begingroup
\begin{gather}
 n>0,\quad N>0,\quad s>0,\quad T_c>0,\notag\\
 A_\Omega>0,\quad L_\Omega>0,\quad
 \alpha_{\rm v}>0,\quad\mathcal N_0>0,\quad\xi_n>0,\notag\\
 \Xi>0,\quad\Xi^2-\gamma^{ij}\xi_i\xi_j>0,\notag\\
 B^2>E^2,\qquad E_iB^i=0.
 \label{eq:tension_direct_admission}
\end{gather}
\endgroup
The line derivatives are evaluated from $\Lambda_{\rm mat}-F$.
They enter both the material inertia and the coupling between matter
and array compression.

\paragraph{Coflow.}
When both fluids and the array share a rest frame, the compression
speeds are the three pairs in
Eq.~\eqref{eq:primitive_coflow_closed_speeds}.
\textcolor{black}{Use the material compression matrix
$C_m=\bigl(\begin{smallmatrix}k_{nn}&k_{nc}\\k_{nc}&k_{cc}\end{smallmatrix}\bigr)$,
species inertia
$K_0=\bigl(\begin{smallmatrix}B_n&\mathcal A_*\\\mathcal A_*&B_c\end{smallmatrix}\bigr)$
and coupling vector $f=(f_n,f_c)^T$ from
Eq.~\eqref{eq:primitive_coflow_responses}.}
Eliminating the line
displacement gives the material response matrices
\begin{equation}
 S=C_m-\frac{ff^T}{c},\qquad
 U=K_0-C_m-\frac{ff^T}{d},
 \label{eq:tension_main_coflow_responses}
\end{equation}
The line responses are
$c=F_{\mathcal N\mathcal N}>0$ and
$d=F_{\mathcal N}/\mathcal N-c>0$.
On positive inertia, the necessary and sufficient conditions in every
propagation direction are
\begin{equation}
 S>0,\qquad U\succeq0.
 \label{eq:tension_main_coflow_conditions}
\end{equation}
Positive $S$ gives nonzero real compression speeds and a complete
basis. Positive semidefinite $U$ bounds those speeds by light speed
and retains complete luminal coincidences. The bending and carrier
blocks remain complete. Appendix~\ref{sec:tension_exact_scalar_proof}
derives these results by showing that perpendicular compression
imposes the strongest lower and upper bounds. Their scalar form is
Eq.~\eqref{eq:tension_exact_coflow_bounds}.

\paragraph{Finite counterflow and vortex slip.}
Relative motion introduces contact resonances that are absent from
the coflow calculation. For the adopted EOS, the following primitive
conditions classify the open domain on the strict causal-response
branch. Section~\ref{sec:tension_state_bounds} then gives sufficient
bounds that can be evaluated during recovery. A general smooth EOS
can be treated with the coefficient conditions of
Appendix~\ref{sec:regc_exact}, which establish root location and
uniform eigenbasis conditioning separately.
\endgroup

%% file: section_specialized_conditions_tension.tex
\begingroup
The polytropes \eqref{eq:main_model_b_master}, additive heat law
\eqref{eq:main_model_b_thermal} and BCS core prescription
\eqref{eq:tension_bcs_core_law} determine the finite-slip conditions
for the line energy \eqref{eq:tension_bcs_line_energy}. With
$n=\nu_n$, $N=n_p$ and $s=s_c$, the primitive inputs are
\begingroup
\begin{gather}
 H_0=m_n+\frac{\Gamma_nK_n}{\Gamma_n-1}n^{\Gamma_n-1}
                                  -\mathcal AN,\notag\\
 \mathcal N=\frac{\sqrt{B^2-E^2}}{\kappa_v},\qquad
 r_v=\frac{\alpha_{\rm v}\mathcal N}{\mathcal N_0},\notag\\
 L_\Omega=\ln\frac{\pi(\pi\mathcal N_0)^{-1/2}
                       \mu_*\Delta_{\rm pair}}
                   {\hbar^2(3\pi^2n)^{1/3}}>0.
 \label{eq:tension_evaluated_stellar_inputs}
\end{gather}
\endgroup
All density derivatives of the effective Fermi response and pairing
gap are retained in the coefficients
$B_n,B_c,k_{nn},k_{nc},k_{cc},f_n,f_c,c,d,t$ of
Eq.~\eqref{eq:tension_evaluated_response_pack}.

\input{section_entrained_conditions_tension}

\paragraph{Independent carrier.}
At $\mathcal A=0$, the carrier Euler system decouples if
$\mu_*$ and $\Delta_{\rm pair}$ depend only on $n$ throughout the
EOS. Then $r=f_c=k_{nc}=0$, and the carrier velocity imposes no
relative-motion restriction. Composition dependence couples the
carrier even when entrainment vanishes.
The two upper causal bounds remain
Eq.~\eqref{eq:tension_entrained_open_causality}, and the lower
compression bound is
\begingroup
\begin{equation}
 c k_{nn}>f_n^2.
 \label{eq:tension_evaluated_decoupled_stiffness}
\end{equation}
\endgroup
To state the remaining slip conditions, define
\begingroup
\begin{align}
 a&=\frac{k_{nn}}{B_n},\qquad
 b=\frac{r_v}{L_\Omega+r_v},\notag\\
 \epsilon_{n\Omega}&=\frac{f_n^2}{B_nt}.
 \label{eq:tension_evaluated_sound_parameters}
\end{align}
\endgroup
and the neutron--array transverse proper speed
\begingroup
\begin{equation}
 \mathfrak u^2=
 \frac{|W_nE+z_n\times B|^2-(E\cdot z_n)^2}{B^2-E^2}.
 \label{eq:tension_evaluated_transverse_slip}
\end{equation}
\endgroup
where $W_n=\sqrt{1+|z_n|^2}$. The acoustic intervals must overlap,
\begingroup
\begin{align}
 \mathfrak u^2(1-a)[a(1-b)+\epsilon_{n\Omega}]
 &<\left[a+\sqrt{ab-\epsilon_{n\Omega}}\right]^2 .
 \label{eq:tension_evaluated_slip_overlap}
\end{align}
\endgroup
At $\epsilon_{n\Omega}=0$ this completes the slip conditions.
For $\epsilon_{n\Omega}>0$, the central maximum of the coupled
quartic must also satisfy
\begingroup
\begin{align}
 \epsilon_{n\Omega}&<\epsilon_*(a,b,\mathfrak v),\qquad
 \mathfrak v=\frac{\mathfrak u}{\sqrt{1+\mathfrak u^2}},\notag\\
 \epsilon_*&=(b-\sigma_*^2)
 \left[a-\left(\frac{\sigma_*-\mathfrak v}
                         {1-\mathfrak v\sigma_*}\right)^2\right].
 \label{eq:tension_evaluated_slip_quartic}
\end{align}
\endgroup
The maximizing phase $\sigma_*(a,b,\mathfrak v)$ is given by
Eq.~\eqref{eq:tension_epsilon_critical_phase}, with coefficients in
Eqs.~\eqref{eq:tension_epsilon_stationary_inputs}--\eqref{eq:tension_epsilon_depressed_inputs}.
Thus no characteristic-root or angular search is needed.
At zero slip, its continuous value is $\sigma_*=0$ and
$\epsilon_*=ab$, recovering
Eq.~\eqref{eq:tension_evaluated_decoupled_stiffness}.
\endgroup

%% file: section_entrained_conditions_tension.tex
\label{sec:tension_entrained_arbitrary_slip}
\paragraph{Nonzero entrainment.}
For fixed $\mathcal A\ne0$, the neutron, carrier and array velocities
are independent. We use the full density responses in
Eq.~\eqref{eq:tension_evaluated_response_pack} and define
\begingroup
\begin{gather}
 a_n=B_n-k_{nn},\qquad a_c=B_c-k_{cc},\qquad
 r=\mathcal A-k_{nc},\notag\\
 d=\frac{\kappa_v^2nL_\Omega}{4\pi H_0\mathcal N},\qquad
 t=\frac{\kappa_v^2n(L_\Omega+r_v)}{4\pi H_0\mathcal N},\notag\\
 g=W_nW_c-z_n\cdot z_c,\qquad
 \Delta=B_nB_c-\mathcal A^2,\notag\\
 h=1+\frac{|W_cE+z_c\times B|^2-(E\cdot z_c)^2}{B^2-E^2}.
 \label{eq:tension_entrained_primitive_inputs}
\end{gather}
\endgroup
where $W_x=\sqrt{1+|z_x|^2}$. The relative Lorentz factor $g$
measures material counterflow, while $h$ measures carrier motion
across the array. The compression conditions can be written using
\begingroup
\begin{align}
 X={}&a_n(B_nk_{cc}-\mathcal A^2)+B_nr^2,\notag\\
 \mathfrak D={}&\Delta[k_{cc}(B_n-a_ng^2)-(\mathcal A-gr)^2]\notag\\
 &+B_cX(g^2-1),\notag\\
 Y={}&d\mathfrak D
 +\Delta(B_cf_n^2-2g\mathcal Af_nf_c+B_nf_c^2)\notag\\
 &-(B_nB_c-g^2\mathcal A^2)
                (a_cf_n^2-2rf_nf_c+a_nf_c^2).
 \label{eq:tension_entrained_main_response_functions}
\end{align}
\endgroup
Appendix~\ref{app:tension_entrained_primitive_list} derives these
expressions and the auxiliary scalars $\beta,\delta,\textcolor{black}{\chi_{\rm ang}}$ and
$p_6,q_6,r_6,s_6,t_6,\mathfrak H_3,\ldots,\mathfrak H_6$.
\textcolor{black}{The nonnegative angular coupling $\chi_{\rm ang}$ is
normalized in Eq.~\eqref{eq:tension_entrained_explicit_responses}
and is distinct from the chemical affinity $\chi$.}

A strict causal response requires
\begingroup
\begin{align}
 d a_n&>f_n^2,\notag\\
 (d a_n-f_n^2)a_c
 &>d r^2-2r f_nf_c+a_nf_c^2.
 \label{eq:tension_entrained_open_causality}
\end{align}
\endgroup
These inequalities bound the material--line coupling at the metric
cone. Strong hyperbolicity additionally requires the material
contact condition
\begingroup
\begin{equation}
 k_{cc}>0,\qquad
 k_{cc}(B_n-a_ng^2)>(\mathcal A-gr)^2.
 \label{eq:tension_entrained_open_hyperbolicity}
\end{equation}
\endgroup
and the carrier--array slip bound
\begingroup
\begin{equation}
 h<\frac{t\mathfrak D}{Y}.
 \label{eq:tension_entrained_line_slip}
\end{equation}
\endgroup
The preceding signs imply $\mathfrak D,Y>0$, so the displayed
quotient has a positive denominator. If $\textcolor{black}{\chi_{\rm ang}}=0$, these conditions
suffice. For $\textcolor{black}{\chi_{\rm ang}}>0$, the angular dependence imposes one of the
three alternatives
\begingroup
\begin{align}
 &\mathfrak H_6<0,\quad
 \left\{\begin{aligned}
 &(p_6>0\ \text{and}\ \mathfrak H_3\geq0),\ \text{or}\\
 &(\mathfrak H_4\geq0\ \text{and}\ (p_6\geq0\ \text{or}\ \mathfrak H_3<0)),\ \text{or}\\
 &(\mathfrak H_5>0\ \text{and}\ \mathfrak H_4\leq0){,}
 \end{aligned}\right.
 \label{eq:tension_entrained_sign_case_six}\\
 &\mathfrak H_6=\mathfrak H_5=0,\qquad \mathfrak H_4>0,\notag\\
 &\hspace{17mm}(p_6\geq0\ \text{or}\ \mathfrak H_3\leq0){,}
 \label{eq:tension_entrained_sign_case_four}\\
 &p_6>0,\quad q_6=s_6=0,\quad
 3r_6=p_6^2,\quad27t_6=p_6^3.
 \label{eq:tension_entrained_sign_case_two}
\end{align}
\endgroup
The sign expressions follow from the quadratic matrix identity
\eqref{eq:tension_entrained_angular_identity}.

%% file: section_tension_state_conditions.tex
\subsubsection{Sufficient bounds for numerical states}
\label{sec:tension_state_bounds}

Primitive recovery requires bounds that can be evaluated repeatedly
at trial states. For constant entrainment, additive carrier heat and
the adopted density-dependent line energy, the construction below
bounds material counterflow and transverse array slip without
performing the angular sign analysis. It gives a sufficient domain
within the hyperbolic region.

\paragraph{Constitutive response.}
The lower compression and upper light-cone margins are reduced by
mixed material--line derivatives. Define
\begin{align}
 c&=F_{\mathcal N\mathcal N},\qquad
 d=\frac{F_{\mathcal N}}{\mathcal N}-c,\qquad t=c+d,\notag\\
 a_n&=B_n-k_{nn},\quad a_c=B_c-k_{cc},\quad
 r=\mathcal A-k_{nc},\notag\\
 M_0&=k_{nn}k_{cc}-k_{nc}^2,\qquad
 R_0=a_na_c-r^2,\notag\\
 J_-&=k_{cc}f_n^2-2k_{nc}f_nf_c+k_{nn}f_c^2,\notag\\
 J_+&=a_cf_n^2-2rf_nf_c+a_nf_c^2,\notag\\
 \epsilon_\Omega^-&=\frac{J_-}{cM_0},\qquad
 \epsilon_\Omega^+=\frac{J_+}{dR_0}.
 \label{eq:tension_state_response_inputs}
\end{align}
Positive denominators and mixing fractions below unity require
\begin{gather}
 c>0,\quad d>0,\quad k_{nn}>0,\quad M_0>0,\notag\\
 a_n>0,\quad R_0>0,\qquad
 0\leq\epsilon_\Omega^-<1,\quad
 0\leq\epsilon_\Omega^+<1.
 \label{eq:tension_state_response_bounds}
\end{gather}
These are the Schur-complement conditions for positive density
stiffness and a strict causal response at coflow. They include the
coupling through $f_n,f_c$, which a test of the isolated line
stiffness would omit.

\paragraph{Coupled relative motion.}
When entrainment or composition-dependent line energy couples the
carrier, choose a fixed $0<\theta<1$ and define
\begin{align}
 H_{\max}&=\frac{t}{d+c\epsilon_\Omega^-}>1,\qquad
 H=1+\theta(H_{\max}-1),\notag\\
 \lambda_{\rm b}&=\frac{H}{t-Hd},\qquad
 k_{\rm b}=k_{cc}-\lambda_{\rm b}f_c^2,\notag\\
 a_{\rm b}&=a_n+\lambda_{\rm b}f_n^2,\qquad
 r_{\rm b}=r+\lambda_{\rm b}f_nf_c,\notag\\
 P_{\rm b}&=k_{\rm b}a_{\rm b}+r_{\rm b}^2,\qquad
 D_{\rm b}=k_{\rm b}(B_nP_{\rm b}-\mathcal A^2a_{\rm b}),\notag\\
 G&=\frac{\mathcal Ar_{\rm b}+\sqrt{D_{\rm b}}}{P_{\rm b}}>1.
 \label{eq:tension_state_motion_caps}
\end{align}
The constitutive inequalities make all denominators and radicands
positive. If $\mathcal Ar_{\rm b}<0$, use
$G=(B_nk_{\rm b}-\mathcal A^2)/
(\sqrt{D_{\rm b}}-\mathcal Ar_{\rm b})$ to avoid cancellation.
For $W_x=\sqrt{1+|z_x|^2}$, the sufficient motion bounds are
\begin{gather}
 1\leq g=W_nW-z_n\cdot z<G,\notag\\
 |WE+z\times B|^2-(E\cdot z)^2
       <(H-1)(B^2-E^2).
 \label{eq:tension_state_motion_bounds}
\end{gather}
The choice of $\theta$ distributes the available response margin
between material counterflow and carrier--array slip.
For example, $\theta=1/2$ divides that margin equally.
The phase momentum must remain future timelike.
At $\mathcal A<0$, this also requires
Eq.~\eqref{eq:tension_state_negative_entrainment}.
Appendix~\ref{app:tension_state_bounds} proves that the motion
bounds preserve positive compression energy in every direction and
gives their equivalent canonical-momentum ellipsoids.

\paragraph{Decoupled neutron--array system.}
If $\mathcal A=0$ and the microscopic inputs depend only on $n$,
the carrier Euler system is independent. Use the constitutive bounds
\eqref{eq:tension_state_decoupled_mixing} and define
\begin{align}
 \mu_{\rm mix}&=\sqrt{\epsilon_{n\Omega}/(ab)},\notag\\
 \mathfrak u_{\rm b}^{\,2}
 &=\frac{(1-\mu_{\rm mix})(\sqrt a+\sqrt b)^2}
 {[1-a(1-\mu_{\rm mix})][1-b(1-\mu_{\rm mix})]}.
 \label{eq:tension_state_decoupled_cap}
\end{align}
with $a,b,\epsilon_{n\Omega}$ from
Eq.~\eqref{eq:tension_evaluated_sound_parameters}. The sufficient
neutron--array slip bound is
\begin{equation}
 |\Xi E+\xi\times B|^2-(E\cdot\xi)^2
       <h_n^2\mathfrak u_{\rm b}^{\,2}(B^2-E^2).
 \label{eq:tension_state_decoupled_slip}
\end{equation}
where $h_n=nB_n$. The independent carrier velocity remains
unrestricted by this bound. Composition-dependent line energy
instead requires the coupled construction.

%% file: section_evolution_tension.tex
\begingroup
\subsection{Flux-conservative evolution equations}
\label{sec:equilibrium_3plus1}
\label{sec:common_valencia}
\label{sec:common_chart}

We write the matter and vortex equations in flux-divergence form
using the $3+1$ metric of Eq.~\eqref{eq:adm_metric_unified}.
The Eulerian observer $n^\mu$ measures the fields $E,B$ and the
vortex moments defined in Sec.~\ref{sec:tension_electric_and_drift}.
\textcolor{black}{The evolved material variables are total baryon density
$D$, carrier number density $D_R$, total momentum density
$S_{{\rm tot},i}$ and total energy minus $D$, denoted
$\tau_{\rm tot}$. Their constitutive expressions are given in
Eq.~\eqref{eq:tension_main_total_stress}.}
The evolved state is
\begin{equation}
 \widetilde{\boldsymbol U}
 =\bigl(\widetilde D,\widetilde D_R,
       \widetilde S_{{\rm tot},i},\widetilde\tau_{\rm tot},
       \xi_i,\mathfrak E^i,\mathfrak B^i\bigr)^{\!T},
 \label{eq:tension_main_complete_state}
\end{equation}
Tildes denote multiplication by $\sqrt\gamma$, and the fraktur
fields are already densitized. The fluid primitives and their
Eulerian velocities are
\begin{align}
 P^a&=(\nu_n,n_p,s,z_n^i,z^i),\qquad E_i,\ B^i,\notag\\
 W_n&=\sqrt{1+z_n^2},\qquad W=\sqrt{1+z^2},\notag\\
 v_n^i&=z_n^i/W_n,\qquad v_c^i=z^i/W.
 \label{eq:summary_density_primitives}
\end{align}
\textcolor{black}{Here $s=s_c$ is the carrier entropy density.}
Entropy moves with the carrier. The number and entropy currents
therefore take the form
\begin{align}
 (\mathcal D_n,\mathcal J_n^i)&=(\nu_nW_n,\nu_nz_n^i),\notag\\
 (\mathcal D_c,\mathcal J_c^i)&=(n_pW,n_pz^i),\notag\\
 (\mathcal D_S,\mathcal J_S^i)&=(sW,sz^i),\qquad
 I_{xy}=\mathcal D_x\mathcal D_y
                      -\gamma_{ij}\mathcal J_x^i\mathcal J_y^j.
 \label{eq:rw_primitive_currents}
\end{align}
We vary the three currents independently before imposing entropy
locking. For $x,y\in\{n,c,S\}$, the bare master function gives
\begin{align}
 K^0_{xx}&=-2\frac{\partial\Lambda_0}{\partial I_{xx}},\qquad
 K^0_{xy}=-\frac{\partial\Lambda_0}{\partial I_{xy}}\quad(x\ne y),\notag\\
 H_x^0&=\sum_yK^0_{xy}\mathcal D_y,\qquad
 p^0_{xi}=\sum_yK^0_{xy}\mathcal J_{yi},\notag\\
 \Psi_0&=\Lambda_0+\sum_x
                   (\mathcal D_xH_x^0-\mathcal J_x^ip^0_{xi}),\notag\\
 E_0&=\sum_x\mathcal D_xH_x^0-\Psi_0,\qquad
 S_{0i}=\sum_x\mathcal D_xp^0_{xi},\notag\\
 P_{0i}{}^j&=\Psi_0\delta_i{}^j+
                         \sum_x\mathcal J_x^jp^0_{xi}.
 \label{eq:rw_bare_material_stress}
\end{align}
The zero-temperature reduction omits the added entropy current.

Equation~\eqref{eq:rw_primitive_stress} supplies the line stress.
Its Eulerian momentum and spatial stress are
\begin{equation}
 m_i=\gamma_{i\mu}\mathcal Q_\Omega^\mu,\qquad
 P_{\Omega ij}=\gamma_{i\mu}\gamma_{j\nu}\mathcal P_\Omega^{\mu\nu}.
 \label{eq:tension_stress_projection_definitions}
\end{equation}
Adding these contributions to the material stress gives
\begin{align}
 D&=\mathcal D_n+\mathcal D_c,\qquad D_R=\mathcal D_c,\notag\\
 J_\perp^i&=\mathcal J_n^i+\mathcal J_c^i,\notag\\
 E_{\rm tot}&=E_0+{\mathcal{E}_\Omega},\qquad
 \tau_{\rm tot}=E_{\rm tot}-D,\notag\\
 S_{{\rm tot},i}&=S_{0i}+m_i,\qquad
 P_{{\rm tot},i}{}^j=P_{0i}{}^j+ {P_{\Omega i}{}^j}.
 \label{eq:tension_main_total_stress}
\end{align}
The total tensor $T_{\rm tot}^{\mu\nu}=T_0^{\mu\nu}+T_\Omega^{\mu\nu}$
contains the density-response stress $C_\Omega^{\mu\nu}$ once.
For the line energy \eqref{eq:tension_draft_master}, the neutron
density derivative is
\begin{align}
 \frac{\partial F}{\partial\nu_n}
 &=\frac{\kappa_v^2\mathcal N}{4\pi H_0}
 \left[\left(1-\frac{\Gamma_nK_n\nu_n^{\Gamma_n-1}}{H_0}\right)
       \left(L_\Omega+\frac{\alpha_{\rm v}\mathcal N}{2\mathcal N_0}\right)
       -\frac13\right.\notag\\
 &\left.\hspace{8mm}
       +\frac{\nu_n}{\mu_*}\frac{\partial\mu_*}{\partial\nu_n}
       +\frac{\nu_n}{\Delta_{\rm pair}}
                      \frac{\partial\Delta_{\rm pair}}{\partial\nu_n}\right].
 \label{eq:tension_explicit_neutron_line_response}
\end{align}
This derivative holds $n_p,\mathcal N$ and the reference cutoffs
fixed. The first factor comes from the phase stiffness. The
remaining terms retain the density dependence of the Fermi wave
number, effective response and pairing gap in the BCS core law
\eqref{eq:tension_bcs_core_law}. For a different bare neutron EOS,
replace $\Gamma_nK_n\nu_n^{\Gamma_n-1}$ by
$\nu_n\,\partial\mu_n/\partial\nu_n$ at constant entrainment.
The resulting neutron momentum is
\begin{align}
 \xi_i&=p^0_{ni}+\frac{\partial F}{\partial\nu_n}z_{ni},\qquad
 \Xi=H_n^0+\frac{\partial F}{\partial\nu_n}W_n,\notag\\
 \xi_0&=-\alpha\Xi+\beta^i\xi_i,\qquad
 \Phi=-\xi_0.
 \label{eq:mu_adm_unified}
\end{align}

Number conservation and the total momentum and energy balances read
\begin{align}
 \partial_t\widetilde D
 &+\partial_j[\sqrt\gamma(\alpha J_\perp^j-\beta^jD)]=0,
 \label{eq:valencia_charge_unified}\\
 \partial_t\widetilde D_R
 &+\partial_j[\sqrt\gamma(\alpha\mathcal J_c^j-\beta^jD_R)]=0,
 \label{eq:valencia_carrier_unified}\\
 \partial_t\widetilde S_{{\rm tot},i}
 &+\partial_j[\sqrt\gamma(\alpha P_{{\rm tot},i}{}^j
                              -\beta^jS_{{\rm tot},i})]\notag\\
 &=\sqrt\gamma\bigl[\tfrac\alpha2P_{\rm tot}^{jk}\partial_i\gamma_{jk}
       +S_{{\rm tot},j}\partial_i\beta^j\notag\\
 &\hspace{19mm}-E_{\rm tot}\partial_i\alpha
       +\alpha f_{{\rm ext},i}\bigr],
 \label{eq:valencia_momentum_unified}\\
 \partial_t\widetilde\tau_{\rm tot}
 &+\partial_j[\sqrt\gamma(\alpha(S_{\rm tot}^j-J_\perp^j)
                                     -\beta^j\tau_{\rm tot})]\notag\\
 &=\sqrt\gamma\bigl[\alpha P_{\rm tot}^{ij}K_{ij}
       -S_{\rm tot}^i\partial_i\alpha+\alpha\mathcal Q_{\rm ext}\bigr].
 \label{eq:valencia_energy_unified}
\end{align}
Here $f_{{\rm ext},i}=\gamma_{i\nu}f_{\rm ext}^\nu$ and
$\mathcal Q_{\rm ext}=-n_\nu f_{\rm ext}^\nu$ specify external
momentum and energy input. Internal lift and drag exchange momentum
between matter and vortices and cancel from these total balances.

For the vortex fields, we evolve the weighted electric field and
canonical vorticity of
Eq.~\eqref{eq:tension_main_weighted_electric_evolution}:
\begin{align}
 \mathfrak E^i&=\sqrt\gamma\,a_\Omega E^i,\qquad
 \mathfrak B^i=\sqrt\gamma\,B^i,\notag\\
 q_a&=\nabla_i(a_\Omega E^i)
                  =\frac{\partial_i\mathfrak E^i}{\sqrt\gamma}.
 \label{eqn:vortex_gauss}
\end{align}
The constitutive weight is evaluated at the proper area density,
$\mathcal N=\sqrt{B^2-E^2}/\kappa_v$, with
$a_\Omega=(\partial F/\partial\mathcal N)/(\kappa_v^2\mathcal N)$.
Thus $\mathfrak E^i$ depends on both the vortex and material state.
The spatial response current is
\begin{equation}
 \mathcal J_{\Omega\perp}^i
 =\frac{[B\times(q_aE+f_\Omega)]^i}{B^2}
                 +\mathcal I_\parallel\frac{B^i}{B^2},
 \label{eq:tension_evolution_drift_current}
\end{equation}
The local force $f_{\Omega\nu}$ is given by
Eq.~\eqref{eq:tension_draft_exchange}. The parallel component
$\mathcal I_\parallel$ enforces the algebraic constraint in the
implicit update of Sec.~\ref{sec:implicit}.
With the fluxes
\begin{align}
 \mathcal F_E^{ji}
 &=\beta^i\mathfrak E^j-\beta^j\mathfrak E^i
       -\alpha\sqrt\gamma\,a_\Omega\epsilon^{ijk}B_k,\notag\\
 \mathcal F_B^{ji}
 &=\beta^i\mathfrak B^j-\beta^j\mathfrak B^i
       +\alpha\sqrt\gamma\,\epsilon^{ijk}E_k,
 \label{eq:tension_evolution_field_fluxes}
\end{align}
the field equations become
\begin{align}
 \partial_t\mathfrak E^i+\partial_j\mathcal F_E^{ji}
 &=\sqrt\gamma\bigl(\alpha\mathcal J_{\Omega\perp}^i
                                      +\beta^i q_a\bigr),
 \label{eq:tension_conservative_electric_evolution}\\
 \partial_t\mathfrak B^i+\partial_j\mathcal F_B^{ji}&=0.
 \label{eq:numerical_B_continuum}
\end{align}
The weighted-current convention assigns Eulerian charge $-q_a$,
which fixes the positive electric source above. Substituting the
response current retains a derivative source through $q_a$.
The electric Gauss relation defines this charge, and constrained
transport preserves $\partial_i\mathfrak B^i=0$.

We recover the line velocity from $E,B$ using
Eq.~\eqref{eq:tension_primitive_adm_line_velocity}.
Its coordinate transport velocity is
\begin{equation}
 \mathcal V^i=\alpha v_L^i-\beta^i=u_L^i/u_L^t.
 \label{eq:numerical_vline}
\end{equation}
On $E_iB^i=0$, the induction flux can also be written as
$\mathcal F_B^{ji}=\mathcal V^j\mathfrak B^i-
\mathcal V^i\mathfrak B^j$. Define the canonical constraint by
\begin{equation}
 \Omega_{ij}=[ijk]\mathfrak B^k,\qquad
 C_{ij}=\partial_i\xi_j-\partial_j\xi_i-\Omega_{ij}.
 \label{eq:numerical_B_Phi}
\end{equation}
The canonical momentum equation then has the flux-divergence form
\begin{align}
 \partial_t\xi_i+\partial_j(\delta_i{}^j\Phi)
 &=-\alpha E_i-\beta^j\Omega_{ij}\notag\\
 &\quad+(\alpha v_n^j-\beta^j)C_{ij}.
 \label{eq:numerical_xi_continuum}
\end{align}
When $C_{ij}=0$, one edge electric field advances the canonical
momentum and its face curl. The scalar-potential gradient cancels
from the induction equation.

For the thermal model, the carrier temperature is
$T_c=-u^\mu(\mu^S_{0\mu}+\delta\mu^S_\mu)$.
We also evolve the entropy balance
\begin{align}
 \partial_t\widetilde\Sigma
 &+\partial_j[\widetilde\Sigma(\alpha v_c^j-\beta^j)]
       =\alpha\sqrt\gamma\,\Gamma_s,
 \label{eq:model_b_thermal_entropy_adm}\\
 T_c\Gamma_s&=\mathcal Q_\Omega-u_\nu f_{\rm ext}^\nu,
 \qquad \widetilde\Sigma=\sqrt\gamma sW.
 \label{eq:adm_entropy_work}
\end{align}
Here $\mathcal Q_\Omega=b\,u^\nu\perp^\rho{}_{\nu}d_\rho$
is the frictional heat. This balance distinguishes internal heating
from the prescribed external work.

%% file: section_tension_imex.tex
\begingroup
\subsection{Implicit electric update}
\label{sec:tension_imex}
\label{sec:implicit}

The local electric force can act on a timescale shorter than the
Courant timestep, as in resistive GRMHD
\cite{Palenzuela:2008sf,Ripperda:2019lsi}.
We treat this force implicitly with the four-stage, third-order
IMEX-SSP3(4,3,3) scheme of
Ref. \cite{2010arXiv1009.2757P}, commonly used in this context \cite{Palenzuela:2008sf,Alic:2012df,Ripperda:2019lsi,Most:2020ami}.
Fluxes and nonstiff sources remain explicit. The explicit operator
$\mathcal H$ includes induction, the canonical momentum equation
and both terms proportional to $q_a$. It also includes the
$E\times B$ drift, following force-free evolution
\cite{Alic:2012df,Most:2020ami}.
The implicit operator acts on the weighted electric field:
\begin{equation}
 \mathcal I_{\mathfrak E}^i
 =\alpha\sqrt\gamma\,\mathcal S^i(P,E,B),\qquad
 \boldsymbol{\mathcal S}=\frac{B\times f_\Omega}{B^2}.
 \label{eq:imex_source_partition}
\end{equation}
All other components of $\mathcal I$ vanish.
For the coefficients $a_{kl}$ and $\widetilde a_{kl}$ in
Appendix~\ref{app:imex_tableau}, each stage satisfies
\begin{align}
 U^{*,k}&=U^n+\Delta t\sum_{l<k}
       \left[a_{kl}\mathcal H(U^{(l)})
             +\widetilde a_{kl}\mathcal I(U^{(l)})\right],\notag\\
 U^{(k)}&=U^{*,k}+\Delta t\,\widetilde a_{kk}\mathcal I(U^{(k)}).
 \label{eq:imex_stage_predictor}
\end{align}
The second equation is a local backward-Euler solve. The metric,
$B$ and the conserved variables without an implicit source stay
fixed during that solve.

We first reduce the electric equation at fixed material primitives.
The coupled material recovery is given in
Sec.~\ref{sec:tension_imex_solver}.
For $B_*=\sqrt{B^2}>0$ and
$\Pi_B=\boldsymbol 1-BB/B^2$, define
\begin{align}
 \mathsf K x&=\frac{B\times x}{B_*},\qquad
 \mathsf K^2=-\Pi_B,\notag\\
 \Delta t_{\rm I}&=\alpha\widetilde a_{kk}\Delta t,\notag\\
 \widehat e_*^i&=\frac{\overline{\mathfrak E}^{*,k,i}}{\sqrt\gamma}.
 \label{eq:imex_geometry}
\end{align}
The hat denotes the undensitized transverse predictor, obtained from
\begin{equation}
 \overline{\mathfrak E}^{*,k,i}
 =\mathfrak E^{*,k,i}
   -\frac{\mathfrak E^{*,k,j}B_j}{B^2}B^i.
 \label{eq:tension_electric_algebraic_projection}
\end{equation}
This projection imposes $E_iB^i=0$ and supplies the parallel
constraint correction. The force $B\times f_\Omega$ is transverse.

To evaluate the force at large drag, \textcolor{black}{we use a bounded rescaling
following Eq.~\eqref{eq:moments_bounded_drag_inputs}, with a stage
weight $\epsilon_{\rm I}$ that bounds both drag coefficients.}
For $\mathcal R\geq0$,
\begingroup\color{black}
\begin{align}
 m_R&=\max(1,\mathcal R,|\mathcal R'|),\notag\\
 \delta_R&=m_R^{-1}+\mathcal R/m_R+|\mathcal R'|/m_R,\notag\\
 (\epsilon_{\rm I},\eta_R,\eta_{R'})
 &=\frac{(m_R^{-1},\,\mathcal R/m_R,\,\mathcal R'/m_R)}{\delta_R}.
 \label{eq:imex_bounded_drag_weights}
\end{align}
\endgroup
These ratios evaluate
$\textcolor{black}{\epsilon_{\rm I}=(1+\mathcal R+|\mathcal R'|)^{-1}}$,
$\textcolor{black}{\eta_R=\epsilon_{\rm I}\mathcal R}$ and $\textcolor{black}{\eta_{R'}=\epsilon_{\rm I}\mathcal R'}$
without forming the unbounded sum. At $\mathcal R'=0$,
$\textcolor{black}{\eta_R=1-\epsilon_{\rm I}}$.
For $\textcolor{black}{Q_\Omega=\sqrt{B^2-E^2}>0}$, the scaled force coefficients are
\begingroup\color{black}
\begin{align}
 \nu_L&=\nu_n(W_nW-z_n\cdot z)+\mathcal Z n_p,\notag\\
 \widetilde L_n&=\epsilon_{\rm I}(\nu_nW_n+\mathcal Z n_pW),\notag\\
 \widetilde L_\perp
 &=\epsilon_{\rm I}\Pi_B(\nu_n z_n+\mathcal Z n_p z),\notag\\
 \widetilde d_n&=\eta_R\nu_LW+\eta_{R'}\nu_nW_n,\notag\\
 \widetilde d&=\eta_R\nu_L z+\eta_{R'}\nu_n z_n,\qquad
 \widetilde g=\mathsf K\widetilde d/B_*.
 \label{eq:imex_force_coefficients}
\end{align}
\endgroup
Here $z^i=Wv_c^i$ and $z_n^i=W_nv_n^i$.
Multiplying the electric stage equation by $\textcolor{black}{\epsilon_{\rm I}}$ gives
\begingroup\color{black}
\begin{equation}
 \epsilon_{\rm I}\mathfrak E
 =\epsilon_{\rm I}\overline{\mathfrak E}^{*,k}
    +\Delta t_{\rm I}\sqrt\gamma\,\widetilde{\mathcal S},
 \qquad E\cdot B=0.
 \label{eq:imex_weighted_stage}
\end{equation}
\endgroup
The scaled source $\textcolor{black}{\widetilde{\mathcal S}=\epsilon_{\rm I}\mathcal S}$ is
\begingroup\color{black}
\begin{equation}
 \widetilde{\mathcal S}=\widetilde L_\perp+Q_\Omega\widetilde g
 +\frac{\widetilde L_n}{B_*}\mathsf KE
 -\frac{\widetilde d_n-E\cdot\widetilde g}{Q_\Omega}E.
 \label{eq:imex_force_reduction}
\end{equation}
\endgroup
Equivalently, in terms of the evolved field,
\begingroup\color{black}
\begin{align}
 \epsilon_{\rm I}\mathfrak E
 &+\frac{\Delta t_{\rm I}}{a_\Omega}
 \left[\frac{\widetilde L_n}{B^2}\mathfrak E\times B
 +\frac{\widetilde d_n-E\cdot\widetilde g}{Q_\Omega}\mathfrak E\right]
 \notag\\
 &=\epsilon_{\rm I}\overline{\mathfrak E}^{*,k}
   +\Delta t_{\rm I}\sqrt\gamma(\widetilde L_\perp+Q_\Omega\widetilde g).
 \label{eq:imex_expanded_weighted_stage}
\end{align}
\endgroup
The physical and evolved electric fields remain related by
$\mathfrak E=\sqrt\gamma a_\Omega E$, with
\begingroup\color{black}
\begin{equation}
 a_\Omega=a_0+\frac{a_1}{Q_\Omega},\qquad
 a_0=\frac{\alpha_{\rm v}A_\Omega}{\kappa_v^2\mathcal N_0},
 \quad a_1=\frac{A_\Omega L_\Omega}{\kappa_v}.
 \label{eq:imex_line_weight}
\end{equation}
\endgroup
The scaling applies to the electric equation. The line energy,
its material derivatives and the conserved stress retain their
physical normalization.

For fixed material primitives and a trial
$\textcolor{black}{\widetilde c=\epsilon_{\rm I} c}$, write
$\textcolor{black}{\widetilde a_j=\epsilon_{\rm I} a_j}$ and define
\begingroup\color{black}
\begin{align}
 \widetilde c&=\widetilde a_0+
 \frac{\widetilde a_1+\Delta t_{\rm I}
       (\widetilde d_n-E\cdot\widetilde g)}{Q_\Omega},\notag\\
 \widetilde\omega&=\Delta t_{\rm I}\widetilde L_n/B_*,
 \qquad
 \widetilde r_0=\epsilon_{\rm I}\widehat e_*+
                         \Delta t_{\rm I}\widetilde L_\perp,\notag\\
 \widetilde r&=\widetilde r_0+\Delta t_{\rm I}Q_\Omega\widetilde g.
 \label{eq:imex_inverse_coefficients}
\end{align}
\endgroup
The electric equation becomes
$(\widetilde c\Pi_B-\widetilde\omega\mathsf K)E=\widetilde r$.
We evaluate its inverse with normalized coefficients:
\begingroup\color{black}
\begin{align}
 \rho&=\operatorname{hypot}(\widetilde c,\widetilde\omega),
 \qquad c_\rho=\widetilde c/\rho,\quad
 \omega_\rho=\widetilde\omega/\rho,\notag\\
 t_\rho&=\Delta t_{\rm I}/\rho,\qquad
 r_{0\rho}=\widetilde r_0/\rho,\qquad
 r_\rho=r_{0\rho}+t_\rho Q_\Omega\widetilde g,\notag\\
 E&=c_\rho r_\rho+\omega_\rho\mathsf Kr_\rho.
 \label{eq:imex_electric_inverse}
\end{align}
\endgroup
The function $\operatorname{hypot}$ is the stable norm introduced
in Eq.~\eqref{eq:moments_relativistic_line_scalars}.
This form avoids squaring large inverse coefficients.
The scalar coefficient and field norm must satisfy
\begingroup\color{black}
\begin{align}
 0&=(\widetilde c-\widetilde a_0)Q_\Omega-\widetilde a_1
 -\Delta t_{\rm I}(\widetilde d_n-E\cdot\widetilde g),\notag\\
 0&=Q_\Omega^2+E^2-B_*^2.
 \label{eq:imex_two_scalar_constraints}
\end{align}
\endgroup
For stage-prescribed drag coefficients, the first constraint gives
\begin{align}
 N_\rho&=\frac{\widetilde a_1}{\rho}
 +t_\rho\bigl[\widetilde d_n
       -c_\rho\widetilde g\cdot r_{0\rho}\notag\\
 &\hspace{27mm}
       -\omega_\rho\widetilde g\cdot\mathsf Kr_{0\rho}\bigr],\notag\\
 D_\rho&=\frac{\widetilde c-\widetilde a_0}{\rho}
                  +c_\rho|t_\rho\widetilde g|^2.
 \label{eq:imex_scalar_rational_parts}
\end{align}
and therefore
\begingroup\color{black}
\begin{equation}
 Q_\Omega=\frac{N_\rho}{D_\rho},\qquad
 0=Q_\Omega^2+|r_\rho|^2-B_*^2.
 \label{eq:imex_scalar_reduction}
\end{equation}
\endgroup
The electric inverse at fixed matter thus requires a nonlinear
solve for $\widetilde c$.
For finite drag, future-causal $d^\mu$, $a_1>0$, $\textcolor{black}{Q_\Omega>0}$ and
$\Delta t_{\rm I}\geq0$, the regular branch obeys
$\widetilde c>\widetilde a_0$ and $D_\rho>0$.
If this parametrization becomes singular, we retain $\textcolor{black}{Q_\Omega}$ as an
additional unknown and solve both constraints in
Eq.~\eqref{eq:imex_two_scalar_constraints}.

\paragraph{Derivatives of the electric map.}
The Jacobian includes the density and core responses.
At fixed stage geometry, differentiating the reduced map gives
\begingroup\color{black}
\begin{align}
 \delta Q_\Omega&=\frac{\delta N_\rho-Q_\Omega\,\delta D_\rho}{D_\rho},\notag\\
 \delta E&=\delta c_\rho\,r_\rho+c_\rho\,\delta r_\rho
 +\delta\omega_\rho\,\mathsf Kr_\rho
 +\omega_\rho\mathsf K\delta r_\rho.
 \label{eq:imex_reduced_differentials}
\end{align}
\endgroup
For fixed matter and stage-prescribed drag, $\textcolor{black}{\epsilon_{\rm I}}$ is constant.
The electric block is
\begingroup\color{black}
\begin{align}
 \widetilde J_E&=\widetilde c\Pi_B
       -\widetilde{\bar\omega}\mathsf K+\widetilde k_E EE^T,\notag\\
 \widetilde{\bar\omega}
 &=\widetilde\omega-\frac{\Delta t_{\rm I}}{Q_\Omega}
       \widehat B\cdot(E\times\widetilde g),\notag\\
 \widetilde k_E&=(\widetilde c-\widetilde a_0)/Q_\Omega^2,\qquad
 \widehat B=B/B_*,\notag\\
 \rho_J&=\operatorname{hypot}(\widetilde c,\widetilde{\bar\omega}),\notag\\
 \frac{\det{}_{B^\perp}\widetilde J_E}{\rho_J^2}
 &=1+\frac{\widetilde c}{\rho_J}
             \frac{\widetilde k_E}{\rho_J}E^2.
 \label{eq:imex_electric_jacobian}
\end{align}
\endgroup
Its normalized determinant is positive on the regular branch.
The inverse is
\begin{align}
 T_J&=\frac{\widetilde c}{\rho_J}\Pi_B+
                   \frac{\widetilde{\bar\omega}}{\rho_J}\mathsf K,\qquad
 k_J=\widetilde k_E/\rho_J,\notag\\
 \widetilde J_E^{-1}
 &=\frac1{\rho_J}\left[T_J-
 \frac{k_J(T_JE)(E^TT_J)}
      {1+k_J(\widetilde c/\rho_J)E^2}\right].
 \label{eq:imex_electric_jacobian_inverse}
\end{align}

\subsection{Primitive recovery and implicit solution}
\label{sec:tension_imex_solver}
\label{sec:primitive_recovery}
\label{sec:numerical_recovery}

Primitive recovery inverts the map from the fluid and electric
fields to the evolved state. This map generally requires a
nonlinear solve \cite{Noble:2005gf,Kastaun:2020uxr}.
The line energy couples the electric field to the canonical
momentum and total stress. Following
Ref.~\cite[Sec.~III.C.2 and App.~H]{Most:2026uga}, we use
{the five-variable formulation below for the NC model
with constant nuclear entrainment and the additive thermal EOS.
Solving this system recovers all matter and electric primitives
jointly. Its $\Delta t_{\rm I}=0$ limit gives
conservative-to-primitive (C2P) recovery, while
$\Delta t_{\rm I}>0$ gives the implicit stage.}

\paragraph{Unknowns and trial state.}
After dividing the conserved densities by $\sqrt\gamma$, set
$D_n=D-D_R$, $D_c=D_R$ and $\textcolor{black}{E_{\rm tot}=\tau_{\rm tot}+D}$.
The canonical field $\xi_i$ and collocated $B^i$ are fixed.
For constant nuclear entrainment and the additive thermal EOS,
the five unknowns are
\begin{align}
 X_5&=(z_i,\chi_n,\widetilde c),\notag\\
 \chi_n&=\mu_n-\mathcal A n_p+F_{,\nu_n}>0.
 \label{eq:imex_five_unknowns}
\end{align}
The three components $z_i$ and scalar $\chi_n$ determine the
material currents. The coefficient $\widetilde c$ determines the
electric response.

\paragraph{Algebraic recovery.}
Each trial $X_5$ determines the remaining primitives in three steps.

\emph{(i) Kinematics.}
Recover the carrier density and neutron velocity from
\begin{align}
 W&=\sqrt{1+z^2},\qquad n_p=D_c/W,\notag\\
 z_{ni}&=\frac{\xi_i-\mathcal A n_p z_i}{\chi_n},\notag\\
 W_n&=\sqrt{1+z_n^2},\qquad \nu_n=D_n/W_n.
 \label{eq:imex_kinematic_elimination}
\end{align}

\emph{(ii) Electric field.}
Evaluate $a_0,a_1$ and the bounded force coefficients at these
densities and velocities. Equations~\eqref{eq:imex_scalar_reduction}
and \eqref{eq:imex_electric_inverse} give $\textcolor{black}{Q_\Omega}$ and $E$.
Then evaluate $F,F_{,\nu_n},F_{,n_p}$ at
$\textcolor{black}{(\nu_n,n_p,Q_\Omega/\kappa_v)}$, retaining the core derivatives in
Appendix~\ref{app:tension_constitutive_derivatives}.

\emph{(iii) Thermodynamics.}
With $H_0=\mu_n-\mathcal A n_p$,
$H_{c0}=\mu_p-\mathcal A\nu_n$ and
$\Psi_{{\rm B},0}$ from Eq.~\eqref{eq:main_model_b_pressure},
compute the cold and line energies:
\begin{align}
 E_{{\rm B},0}&=\nu_nH_0W_n^2+n_pH_{c0}W^2
       +2\mathcal A\nu_nn_pW_nW-\Psi_{{\rm B},0},\notag\\
 \mathcal E_\Omega&=F+a_\Omega E^2+
       \nu_nF_{,\nu_n}z_n^2+n_pF_{,n_p}z^2.
 \label{eq:imex_energy_components}
\end{align}
For $e_{\rm th}=C_{\rm th}s^{4/3}$, the energy equation gives
\begingroup\color{black}
\begin{equation}
 e_{\rm th}=\frac{3(E_{\rm tot}-E_{{\rm B},0}-\mathcal E_\Omega)}{4W^2-1},
 \qquad s=(e_{\rm th}/C_{\rm th})^{3/4}.
 \label{eq:imex_thermal_elimination}
\end{equation}
\endgroup
The entropy-primary alternative uses the advected entropy, as
described below.

\paragraph{Coupled residuals.}
The five equations for $X_5$ are
\begingroup\color{black}
\begin{align}
 R_\chi&=\chi_n-H_0-F_{,\nu_n}=0,\notag\\
 R_Q&=Q_\Omega^2+E^2-B_*^2=0,\notag\\
 R_{S_i}&=D_n\xi_i
       +D_c\left[(H_{c0}+F_{,n_p})z_i
                         +\mathcal A\nu_nz_{ni}\right]\notag\\
     &\quad+\tfrac43e_{\rm th}Wz_i
       +a_\Omega(E\times B)_i-S_{{\rm tot},i}=0.
 \label{eq:imex_five_residuals}
\end{align}
\endgroup
The neutron momentum contribution uses $R_\chi=0$ to express it
through $\xi_i$. Acceptance therefore also requires evaluating
the original stress at the converged state.
The force uses $N_s^\mu=\nu_nu_n^\mu$.
In energy-primary recovery, heating follows from total energy
without an additional internal energy source.

{The recovered state must also reproduce the
original material and electric relations. Subtract its line stress
to obtain the material targets:}
\begin{align}
 E_0&=\tau_{\rm tot}+D-\mathcal{E}_\Omega(P,E,B),\notag\\
 S_{0i}&=S_{{\rm tot},i}-m_i(P,E,B).
 \label{eq:tension_method_stress_subtraction}
\end{align}
{At convergence, the recovered primitives satisfy}
\begingroup\color{black}
\begin{align}
 \nu_nW_n&=D-D_R,\notag\\
 E_0(P)+\mathcal E_\Omega(P,E,B)&=E_{\rm tot},\notag\\
 S_{0i}(P)+m_i(P,E,B)&=S_{{\rm tot},i},\notag\\
 p^0_{ni}(P)+F_{,\nu_n}z_{ni}&=\xi_i,\notag\\
 \epsilon_{\rm I}\sqrt\gamma\,a_\Omega(P,E,B)E^i
 &=\epsilon_{\rm I}\overline{\mathfrak E}^{*,k,i}
   +\Delta t_{\rm I}\sqrt\gamma\,
                         \widetilde{\mathcal S}^{\,i}.
 \label{eq:tension_method_material_inverse}
\end{align}
\endgroup
{Evaluate $E_0(P),S_{0i}(P)$ and $p^0_{ni}(P)$
from the bare master function. These relations check the recovered
state without introducing additional nonlinear unknowns.}

{We solve the five equations~\eqref{eq:imex_five_residuals}
with a damped quasi-Newton iteration. Each trial recovers all
remaining primitives $(\nu_n,n_p,s,z_n^i,E_i)$
algebraically through Eqs.~\eqref{eq:imex_kinematic_elimination},
\eqref{eq:imex_scalar_reduction}, \eqref{eq:imex_electric_inverse}
and \eqref{eq:imex_thermal_elimination}, or the entropy alternative
below. If this solve fails, a fallback alternates a bracketed
scalar electric solve at fixed matter with material updates,
then retries the coupled Newton solve.}

\paragraph{Solution and acceptance.}
\begin{enumerate}
 \item \textit{Physical trial states.}
 Require positive species densities, $\chi_n>0$, $a_\Omega>0$
 and $\textcolor{black}{0<Q_\Omega\leq B_*}$. Check the EOS and characteristic conditions
 of Sec.~\ref{sec:vortex_tension_hyperbolicity} at every trial.
 If a trial fails, shorten the correction without changing the
 input conserved state. Failure of a sufficient characteristic
 condition does not by itself establish a nonphysical mode.

 \item \textit{Source-free/Primitive recovery.}
 {Set $\Delta t_{\rm I}=0$ and solve the five
 residual equations without evaluating drag inputs. Initialize
 from the previous valid primitives. The electric map reduces to
 $\textcolor{black}{Q_\Omega=a_1/(c-a_0)}$ and $E=\widehat e_*/c$, and the fifth coordinate
 is the unscaled $c$. This source-free limit also applies to
 primitive recovery after the final Runge--Kutta combination
 in Eq.~\eqref{eq:imex_final_combination}.}

 \item \textit{Implicit stage.}
 For $\Delta t_{\rm I}>0$, initialize from the C2P solution.
 Evaluate the bounded inputs and set $\widetilde c$ using
 Eq.~\eqref{eq:imex_inverse_coefficients}.
 Hold positive residual scales fixed during the solve, for example
 $s_\chi=\chi_{n,0}$, $s_Q=B_*^2$ and $\textcolor{black}{s_S=E_{\rm tot}>0}$, and define
\begin{equation}
  \mathcal F(X)=\left(\frac{R_\chi}{s_\chi},
       \frac{R_Q}{s_Q},\frac{R_{S_1}}{s_S},
       \frac{R_{S_2}}{s_S},\frac{R_{S_3}}{s_S}\right).
  \label{eq:imex_normalized_residuals}
 \end{equation}
 With $X=X_5$, the quasi-Newton correction satisfies
\begin{equation}
  \widehat J^{(m)}\delta X^{(m)}=-\mathcal F(X^{(m)}),
  \qquad J_{ab}=\frac{\partial\mathcal F_a}{\partial X_b}.
  \label{eq:imex_nonlinear_step}
 \end{equation}
{The initial Jacobian is computed using finite differences, and subsequent iterations can use a good Broyden update.} Repeat the algebraic recovery at every
 residual evaluation.
 In scaled coordinates, backtrack over
 $X_{\rm t}=X+\eta\delta X$, $\eta=1,1/2,\ldots$.
 Accept a trial if it is physical and decreases both
 $\|\mathcal F\|_\infty$ and
 $\varphi=\tfrac12\|\mathcal F\|_2^2$.
 If backtracking fails, rebuild the Jacobian and retry.

 \item \textit{Continuation after a failed implicit solve.}
 Multiply $\Delta t_{\rm I}$ by $\lambda\in[0,1]$ and continue
 from the C2P solution:
\begin{equation}
  \mathcal F(X,\lambda)=0,\qquad
  J\frac{dX}{d\lambda}=-\partial_\lambda\mathcal F.
  \label{eq:imex_continuation}
 \end{equation}
 Correct each prediction with the original five equations.
 A failed correction restores the last accepted state and reduces
 the step in $\lambda$. Accept a converged solution at $\lambda=1$.

 \item \textit{Thermal selection and floors.}
 Choose the thermal equation at the source-free seed and hold
 that choice fixed during the stage. The threshold $\eta_1$
 selects the energy-primary branch. Outside flagged shocks,
 evaluate the thermal indicator
\begingroup\color{black}
\begin{equation}
 f_s=\frac{sT_c(4W^2-1)/3}{|\tau_{\rm tot}|+D},
 \label{eq:imex_thermal_resolvability}
\end{equation}
\endgroup
 If $f_s>\eta_1$, or at a flagged shock, use
 Eq.~\eqref{eq:imex_thermal_elimination}. Otherwise set
\begin{equation}
 s=\frac{\widehat\Sigma}{W},\qquad
 e_{\rm th}=C_{\rm th}\,s^{4/3},
 \label{eq:imex_entropy_elimination}
\end{equation}
 Here $\widehat\Sigma$ is the undensitized stage entropy predictor.
 Retain the five residuals and record the energy mismatch
 separately. An evolved-energy deficit is not interpreted as
 negative thermal energy. 

 Near line depletion, explicit evaluations and ordinary recovery
 use a prescribed dimensionless cutoff $q_{\rm fl}$ in the weight:
\begingroup\color{black}
\begin{align}
  Q_{\Omega,\rm eff}&=\max(Q_\Omega,q_{\rm fl}B_*),\notag\\
  a_\Omega^{\rm eff}&=a_0+a_1/Q_{\Omega,\rm eff}.
  \label{eq:imex_weight_floor}
 \end{align}
\endgroup
 Evaluate the line energy and material derivatives at the true
 $\textcolor{black}{\mathcal N=Q_\Omega/\kappa_v}$. Clamp fields beyond the prescribed
 range to $\textcolor{black}{Q_\Omega=q_{\rm fl}B_*}$ and flag the correction.
 The analytic five-variable inverse applies to the unmodified
 branch.  The implicit reduction uses the
 unfloored weight and requires $\textcolor{black}{Q_\Omega\geq q_{\min}B_*}$, where
 $q_{\min}$ sets the minimum admitted ratio $\textcolor{black}{Q_\Omega/B_*}$.

 \item \textit{Final checks and source evaluation.}
 {Recompute the EOS and evaluate the line stress in
 Eq.~\eqref{eq:rw_primitive_stress}, and check the characteristic
 conditions of Sec.~\ref{sec:vortex_tension_hyperbolicity}.
 {Require the number, canonical momentum, total
 momentum and electric equations in
 Eq.~\eqref{eq:tension_method_material_inverse} to hold for the
 prescribed inputs.} Enforce energy through
 Eq.~\eqref{eq:imex_thermal_elimination}, or record its mismatch
 when using Eq.~\eqref{eq:imex_entropy_elimination}.
 Check the coupled Jacobian of Eq.~\eqref{eq:imex_nonlinear_step}
 for nonsingularity and require the normalized residuals of
 Eq.~\eqref{eq:imex_normalized_residuals} to meet the prescribed
 tolerance.}
 For $\widetilde a_{kk}>0$, store the implicit source as
\begingroup\color{black}
\begin{align}
 \mathcal I_{\mathfrak E}^{(k)}
 &=\frac{\mathfrak E^{(k)}-\overline{\mathfrak E}^{*,k}}
              {\widetilde a_{kk}\Delta t},\notag\\
 \epsilon_{\rm I}\mathcal I_{\mathfrak E}^{(k)}
 &=\alpha\sqrt\gamma\,\widetilde{\mathcal S}^{(k)}.
 \label{eq:imex_source_replay}
\end{align}
\endgroup
 This avoids dividing the scaled force by a small $\textcolor{black}{\epsilon_{\rm I}}$.
 At $\textcolor{black}{\epsilon_{\rm I}=0}$, evaluate the limiting constraint.
 A zero implicit diagonal uses ordinary recovery.
 Record projection and floor corrections separately, then repeat
 recovery and the source check. A failed source-free recovery,
 continuation or final check rejects the stage.
\end{enumerate}

\endgroup

%% file: section_numerics_tension.tex
\subsection{Matter and line transport}
\label{sec:numerical_state}

At each face, we reconstruct left and right primitive states with
five-point interpolation and WENO-Z weights \cite{2008JCoPh.227.3191B}.
The states share the stored normal vorticity. The fluid primitives
and $E_i$ determine $u_L(E,B)$ and the total stress
\eqref{eq:tension_main_total_stress}.
Cell fluxes advance number, total momentum and total energy using
Sec.~\ref{sec:equilibrium_3plus1}. The conservative electric balance
\eqref{eq:tension_conservative_electric_evolution} advances
$\mathfrak E^i$, and constrained transport advances circulation.

Before evaluating a flux, we check the physical state, recovery
Jacobian and characteristic conditions of
Sec.~\ref{sec:vortex_tension_hyperbolicity}.
For the adopted line energy, we use the fourteen sufficient
inequalities in Eq.~\eqref{eq:regc_scalar_certificate}, evaluating
the pivots with Eq.~\eqref{eq:regc_three_scalar_pivots}.

For admitted face states, the Harten--Lax--van Leer flux is
\cite{Amiram:2006zjz}
\textcolor{black}{written using the reconstructed conserved states $U_L,U_R$
and their physical fluxes $F_L,F_R$ as}
\begin{equation}
 \widehat F=\frac{c^-F_R+c^+F_L-c^+c^-(U_R-U_L)}{c^++c^-},
 \qquad c^\pm\geq0.
 \label{eq:numerical_hll}
\end{equation}
The extreme eigenvalues of the coupled primitive matrix determine
the signal speeds:
\begin{equation}
 \sigma_{\min}=\min\operatorname{spec}\mathsf C(s),\qquad
 \sigma_{\max}=\max\operatorname{spec}\mathsf C(s).
 \label{eq:numerical_hll_full_spectrum_endpoints}
\end{equation}
The spectrum must be real and have a complete, uniformly bounded
eigenbasis, including at repeated roots. Fluid and vortex waves
mix, so their ordering depends on the EOS and local state.
A decoupled vortex speed does not supply a separate bound.
For a face normal to $x^d$, set
$s_i=\delta_i^d/\sqrt{\gamma^{dd}}$. The coordinate speeds are
\begin{equation}
 \lambda_{\min,\max}
 =\alpha\sqrt{\gamma^{dd}}\,\sigma_{\min,\max}-\beta^d.
 \label{eq:numerical_hll_coordinate_endpoints}
\end{equation}
The bounds enclosing both face spectra are
\begin{align}
 c^+&=\max(0,\lambda_{\max,L},\lambda_{\max,R}),\notag\\
 c^-&=\max(0,-\lambda_{\min,L},-\lambda_{\min,R}).
 \label{eq:numerical_hll_exact_signal_bounds}
\end{align}
These bounds enter the cell fluxes, edge transport and Courant
timestep. Algebraic drag changes the source timescale without
changing the principal matrix. The HLL flux alone does not
guarantee preservation of the allowed state domain.
For a state already known to be causal and strongly hyperbolic,
the light-cone endpoints provide a simpler bound:
\begin{equation}
 \lambda_-^{\rm cone}=-\beta^d-\alpha\sqrt{\gamma^{dd}},\qquad
 \lambda_+^{\rm cone}=-\beta^d+\alpha\sqrt{\gamma^{dd}}.
 \label{eq:regulated_hll_cone_bounds}
\end{equation}

\subsection{Constrained transport of circulation}
\label{sec:numerical_ct}

We evolve $\xi_i$ and derive $\mathfrak B^i$ from its discrete
curl, as in vector-potential constrained transport
\cite{Etienne:2010ui,Etienne:2011re}.
A shared electric field at each edge preserves both the canonical
momentum--vorticity relation and the divergence constraint to
roundoff \cite{Evans:1988qd}.
The evolved electric field determines transverse line motion.
It is not reset to instantaneous force equilibrium when line tension and energy are included.

\begin{figure*}[tbp]
 \centering
 \includegraphics[width=0.98\textwidth]{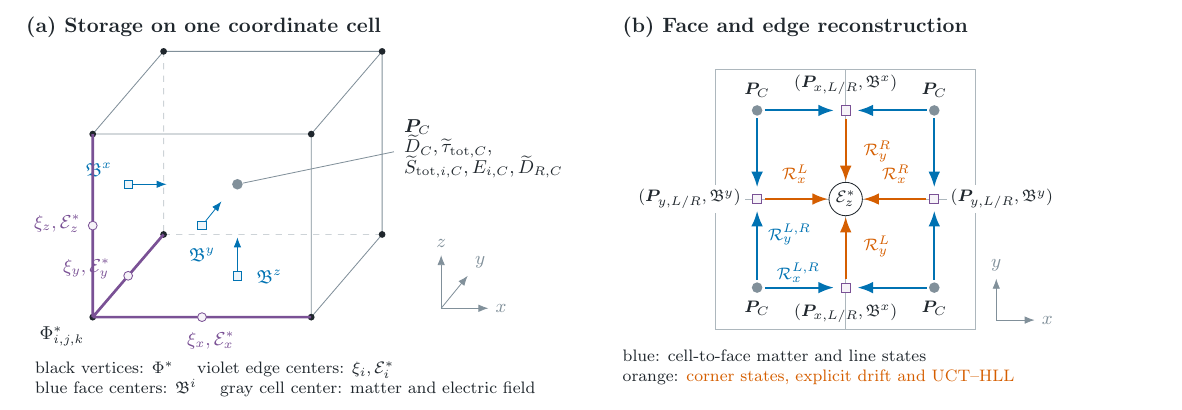}
 \caption{Staggering and reconstruction for canonical constrained
 transport. (a) Number densities, total energy and momentum, and
 the evolved weighted electric field are cell centered.
 Canonical momentum is edge centered, vorticity face centered,
 and the shared scalar potential vertex centered.
 (b) Reconstruction gives four states adjoining a $z$ edge.
 Their line velocities follow from $(E,B)$, and the UCT--HLL
 average gives one electric field for all adjoining faces.}
 \label{fig:canonical-ct-schematic}
\end{figure*}

Figure~\ref{fig:canonical-ct-schematic}(a) shows the staggered cell.
Vertices carry integer indices $(i,j,k)$, and the cell center is
$C=(i+\tfrac12,j+\tfrac12,k+\tfrac12)$.
The recovered state, total energy and momentum, number densities
and weighted electric field are stored at $C$.
The recovered electric covector $E_i$ is collocated there.
Canonical momentum and numerical electric fields lie on edges,
vorticity on faces and the shared scalar potential on vertices:
\begin{align}
 \xi_x,\mathcal E_x^*&:(i+\tfrac12,j,k),\notag\\
 \xi_y,\mathcal E_y^*&:(i,j+\tfrac12,k),\notag\\
 \xi_z,\mathcal E_z^*&:(i,j,k+\tfrac12),\notag\\
 \mathfrak B^x&:(i,j+\tfrac12,k+\tfrac12),\notag\\
 \mathfrak B^y&:(i+\tfrac12,j,k+\tfrac12),\notag\\
 \mathfrak B^z&:(i+\tfrac12,j+\tfrac12,k),\qquad
 \Phi^*:(i,j,k).
 \label{eq:numerical_ct_locations}
\end{align}
On a uniform coordinate grid, the curl is
\begin{align}
 \mathfrak B^x_{i,j+\frac12,k+\frac12}
 &=\frac{\xi_{z;i,j+1,k+\frac12}-\xi_{z;i,j,k+\frac12}}{\Delta y}\notag\\
 &\quad-\frac{\xi_{y;i,j+\frac12,k+1}-\xi_{y;i,j+\frac12,k}}{\Delta z},\notag\\
 \mathfrak B^y_{i+\frac12,j,k+\frac12}
 &=\frac{\xi_{x;i+\frac12,j,k+1}-\xi_{x;i+\frac12,j,k}}{\Delta z}\notag\\
 &\quad-\frac{\xi_{z;i+1,j,k+\frac12}-\xi_{z;i,j,k+\frac12}}{\Delta x},\notag\\
 \mathfrak B^z_{i+\frac12,j+\frac12,k}
 &=\frac{\xi_{y;i+1,j+\frac12,k}-\xi_{y;i,j+\frac12,k}}{\Delta x}\notag\\
 &\quad-\frac{\xi_{x;i+\frac12,j+1,k}-\xi_{x;i+\frac12,j,k}}{\Delta y}.
 \label{eq:numerical_ct_index_curl}
\end{align}
The separator in each subscript distinguishes the component from
its grid location. The cell-centered divergence is
\begin{align}
 (\nabla_h\!\cdot\mathfrak B)_C
 &=\frac{\mathfrak B^x_{i+1,j+\frac12,k+\frac12}
           -\mathfrak B^x_{i,j+\frac12,k+\frac12}}{\Delta x}\notag\\
 &\quad+\frac{\mathfrak B^y_{i+\frac12,j+1,k+\frac12}
           -\mathfrak B^y_{i+\frac12,j,k+\frac12}}{\Delta y}\notag\\
 &\quad+\frac{\mathfrak B^z_{i+\frac12,j+\frac12,k+1}
           -\mathfrak B^z_{i+\frac12,j+\frac12,k}}{\Delta z}=0.
 \label{eq:numerical_ct_index_divergence}
\end{align}
Substitution of the curl cancels mixed differences in pairs.
A suppressed coordinate removes its directional difference.

The continuum canonical equation is
\begin{equation}
 \partial_t\xi_d=-\mathcal E_d-\partial_d\Phi,\qquad
 \mathcal E_d=-[dab]{\mathcal V^a}\mathfrak B^b,\qquad \Phi=-\xi_0.
 \label{eq:numerical_ct_continuum}
\end{equation}
The coordinate line velocity is given by
Eq.~\eqref{eq:numerical_vline}. For example, the $z$-edge update is
\begin{align}
 \partial_t\xi_{z;i,j,k+\frac12}
 &=-\mathcal E^*_{z;i,j,k+\frac12}\notag\\
 &\quad-\frac{\Phi^*_{i,j,k+1}-\Phi^*_{i,j,k}}{\Delta z},
 \label{eq:numerical_full_force_edge_update}
\end{align}
Cyclic permutations give the other components.
Every face meeting an edge uses one $\mathcal E_d^*$, and every
edge meeting a vertex uses one $\Phi^*$.
The endpoint differences cancel around each face, giving
\begin{equation}
 \nabla_h\times\nabla_h\Phi^*=0,\qquad
 \nabla_h\!\cdot(\nabla_h\times\xi)=0.
 \label{eq:numerical_mimetic_identities}
\end{equation}
These identities also hold for oriented edge integrals.

The cell input is
\begin{equation}
 \boldsymbol P_C=(\boldsymbol P_{{\rm mat},C},E_{i,C}),
 \label{eq:numerical_ct_primitive_inputs}
\end{equation}
with $B_C$ collocated from the bounding faces. For example,
\begin{equation}
 B_C^x=\frac{\mathfrak B^x_{i,j+1/2,k+1/2}
                  +\mathfrak B^x_{i+1,j+1/2,k+1/2}}
                  {2\sqrt{\gamma_C}},
 \label{eq:numerical_ct_vorticity_cell_collocation}
\end{equation}
with cyclic permutations. At a face, the normal vorticity remains
the stored value, and the transverse components are reconstructed:
\begin{equation}
 \mathfrak B_{q,s}^{j}=\begin{cases}
 \mathfrak B^q,&j=q,\\
 \mathcal R_q^s\mathfrak B_C^j,&j\ne q.
 \end{cases}
 \label{eq:numerical_ct_face_primitives}
\end{equation}
Collocate the metric and force inputs with these states.
Project the reconstructed $E_i$ into $B^\perp$ and check the vortex
conditions. The {Eulerian} drift $v=E\times B/B^2$
{and $W_L=(1-|v|^2)^{-1/2}$ give}
\begin{equation}
 {u_L=W_L(n+v)},\qquad
 \mathcal V^i=\alpha v^i-\beta^i,
 \label{eq:numerical_full_force_input}
\end{equation}
This reconstructs the line velocity without an instantaneous
force-law solve.

Reconstruction transverse to a $z$ edge gives four valid states,
$s_xs_y\in\{L,R\}^2$, sharing the two side limits of the stored
face vorticity. In each state, evaluate
\begin{equation}
 \mathcal E_z^{s_xs_y}
 =-\mathcal V_{s_xs_y}^{x}\mathfrak B^{y,s_x}
   +\mathcal V_{s_xs_y}^{y}\mathfrak B^{x,s_y}.
 \label{eq:numerical_ct_transverse_reconstruction}
\end{equation}
Take the maximum surrounding face bounds for each edge direction
and define
\begin{equation}
 \omega_q^L=\frac{c_q^+}{c_q^++c_q^-},\quad
 \omega_q^R=\frac{c_q^-}{c_q^++c_q^-},\quad
 \eta_q=\frac{c_q^+c_q^-}{c_q^++c_q^-}.
 \label{eq:numerical_ct_hll_weights}
\end{equation}
The UCT--HLL electric field is
\cite{Londrillo:2003qi,DelZanna:2007pk,Mignone:2020qec}
\begin{align}
 \mathcal E_z^*
 &=\sum_{s_x,s_y}\omega_x^{s_x}\omega_y^{s_y}
                          \mathcal E_z^{s_xs_y}\notag\\
 &\quad+\eta_x(\mathfrak B^{y,R}-\mathfrak B^{y,L})
       -\eta_y(\mathfrak B^{x,R}-\mathfrak B^{x,L}).
 \label{eq:numerical_ct_uct_edge}
\end{align}
The signs follow from
$\mathcal E_z=-F^x(\mathfrak B^y)=F^y(\mathfrak B^x)$.
Cyclic permutations give the other edges. In a suppressed direction,
the side states coincide and the jump vanishes, recovering the
one-dimensional HLL flux. Jump terms supply numerical dissipation
and do not enter physical vortex heating.

We evaluate the cell scalar potential after primitive recovery,
using the collocated canonical momentum.
In an $(x,y)$ calculation with $\partial_z=0$, set
$C_2=(i+\tfrac12,j+\tfrac12)$.
The in-plane components average two parallel edges, and the
out-of-plane component averages four:
\begin{align}
 \xi_{x,C_2}&=\tfrac12\bigl(
    \xi_{x;i+\frac12,j}+\xi_{x;i+\frac12,j+1}\bigr),\notag\\
 \xi_{y,C_2}&=\tfrac12\bigl(
    \xi_{y;i,j+\frac12}+\xi_{y;i+1,j+\frac12}\bigr),\notag\\
 \xi_{z,C_2}&=\tfrac14\bigl(
    \xi_{z;i,j}+\xi_{z;i+1,j}\notag\\
 &\hspace{22mm}+\xi_{z;i,j+1}+\xi_{z;i+1,j+1}\bigr).
 \label{eq:numerical_ct_cell_collocation_2d}
\end{align}
In three dimensions, each component is centered along its edge
and nodal in the transverse coordinates. Each average uses four
parallel edges:
\begin{align}
 \xi_{x,C}&=\tfrac14\bigl(
    \xi_{x;i+\frac12,j,k}+\xi_{x;i+\frac12,j+1,k}\notag\\
 &\hspace{17mm}+\xi_{x;i+\frac12,j,k+1}
    +\xi_{x;i+\frac12,j+1,k+1}\bigr),\notag\\
 \xi_{y,C}&=\tfrac14\bigl(
    \xi_{y;i,j+\frac12,k}+\xi_{y;i+1,j+\frac12,k}\notag\\
 &\hspace{17mm}+\xi_{y;i,j+\frac12,k+1}
    +\xi_{y;i+1,j+\frac12,k+1}\bigr),\notag\\
 \xi_{z,C}&=\tfrac14\bigl(
    \xi_{z;i,j,k+\frac12}+\xi_{z;i+1,j,k+\frac12}\notag\\
 &\hspace{17mm}+\xi_{z;i,j+1,k+\frac12}
    +\xi_{z;i+1,j+1,k+\frac12}\bigr).
 \label{eq:numerical_ct_cell_collocation}
\end{align}
The recovered $\varepsilon_{\xi,C}$ and cell metric determine
\begin{align}
 \Xi_C&=\sqrt{\varepsilon_{\xi,C}^2
                 +\gamma_C^{ij}\xi_{i,C}\xi_{j,C}},\notag\\
 \Phi_C&=-\xi_{0,C}=\alpha_C\Xi_C-\beta_C^i\xi_{i,C},\qquad
 \xi_C^0=\Xi_C/\alpha_C.
 \label{eq:numerical_ct_scalar_potential}
\end{align}
Apply boundary data consistently to the canonical field and
potential. Reconstruct $\Phi_C$ successively in $x$, $y$ and $z$
to obtain two, four and eight side states at each vertex:
\begin{equation}
 \Phi_{s_xs_ys_z}
 =\mathcal R_z^{s_z}\mathcal R_y^{s_y}\mathcal R_x^{s_x}\Phi_C.
 \label{eq:numerical_ct_potential_reconstruction}
\end{equation}
Reconstruct each $\xi_q$ along its edge direction to the vertex.
The resulting jumps enter the shared nodal potential
\begin{equation}
 \Phi^*=\sum_{s\in\{L,R\}^3}
          \left(\prod_{q=x,y,z}\omega_q^{s_q}\right)\Phi_s
        -\sum_{q=x,y,z}\eta_q(\xi_{q,R}-\xi_{q,L}).
 \label{eq:numerical_ct_uct_potential}
\end{equation}
The nodal bounds enclose the spectra of all eight adjoining cells.
The input is the recovered scalar $\Phi_C$ from
Eq.~\eqref{eq:numerical_ct_scalar_potential}, so this reconstruction
requires no fluid solve at a vertex.
The upwind value $\Phi^*$ includes the jump terms and enters every
endpoint difference in Eq.~\eqref{eq:numerical_full_force_edge_update}.
Its gradient therefore cancels from the curl for any reconstruction
weights.

\subsection{Conservative electric evolution}
\label{sec:numerical_electric}

We advance $\mathfrak E^i_C$ using
Eq.~\eqref{eq:tension_conservative_electric_evolution}.
Evaluate the electric flux $\mathcal F_E^{di}$ from
Eq.~\eqref{eq:tension_evolution_field_fluxes} at the admitted
matter face states. The HLL combination uses the jump in
$\mathfrak E^i$ and the bounds of Eq.~\eqref{eq:numerical_hll}.
The residual is
\begin{align}
 (L_{\mathfrak E})^i_C
 &=-\sum_d\frac{\widehat{\mathcal F}_E^{di}|_{d,+}
                    -\widehat{\mathcal F}_E^{di}|_{d,-}}{\Delta x^d}\notag\\
 &\quad+\sqrt\gamma\bigl(\alpha\mathcal J_{\Omega\perp}^i
                                      +\beta^iq_a\bigr)_C.
 \label{eq:tension_electric_numerical_residual}
\end{align}
Use the current in Eq.~\eqref{eq:tension_evolution_drift_current}
with $\mathcal I_\parallel=0$ in the provisional update.
Similar to resistive GRMHD codes \cite{Ripperda:2019lsi}, we compute the charge density from the Gauss constraint, instead of evolving it independently.
Evaluate the Gauss derivative source with the point derivative
\begin{align}
 (\partial_d^hX)_C&=\frac{\{X\}_{d,+}-\{X\}_{d,-}}{\Delta x^d},\notag\\
 \{X\}_{d,\pm}&=\tfrac12(X_{L,d,\pm}+X_{R,d,\pm}),
 \label{eq:tension_electric_point_difference}
\end{align}
so that $q_{a,C}=\sum_d(\partial_d^h\mathfrak E^d)_C/
\sqrt{\gamma_C}$. When wider point derivatives are enabled,
use them throughout the update. 

The provisional stage is followed by the projection
\eqref{eq:tension_electric_algebraic_projection} and joint recovery.

%% file: appendix_causality_hyperbolicity_tension.tex
\section{{Characteristic structure, causality and hyperbolicity}}
\label{app:causality_hyperbolicity}
\label{app:condition_reductions}

This appendix derives the joint causality and hyperbolicity conditions
summarized in Sec.~\ref{sec:hyperbolicity}. We freeze a smooth primitive
state $\boldsymbol W$ and impose the physical principal constraints
before forming the square evolution system. With the Eulerian observer
$n^\mu$ and unit spatial normal $s_\mu$ of
Eq.~\eqref{eq:characteristic_speed_definition}, define
\begin{align}
 \mathsf A_n&=-n_\mu\mathsf A^\mu,\qquad
 \mathsf A_s=s_\mu\mathsf A^\mu,\notag\\
 \mathsf C(\boldsymbol W,s)&=\mathsf A_n^{-1}\mathsf A_s,\qquad
 p(\sigma)=\det(\sigma I-\mathsf C).
 \label{eq:appendix_characteristic_problem}
\end{align}
The time matrix must be invertible on this physical subspace.
Its dimension is $N_{\rm phys}$, and the roots of $p$ are the
Eulerian characteristic speeds. The allowed normals $\mathcal S$
form the unit sphere in three dimensions or the meridional unit
circle in axisymmetry. An evolution away from the constraint surface
can have additional modes and requires a separate test
\cite{Abalos:2018uwg}.

Strong hyperbolicity requires real speeds and a uniformly bounded
basis of physical eigenvectors \cite{Reula:2004xd,
Abalos:2018uwg}. More precisely, each admitted state
$\boldsymbol W_0$ must have a neighborhood $\mathcal U_0$ on which
\begin{align}
 \mathsf C\mathsf T
 &=\mathsf T\operatorname{diag}(\sigma_1,\ldots,\sigma_{N_{\rm phys}}),\notag\\
 \sup_{\boldsymbol W\in\mathcal U_0,\ s\in\mathcal S}
       \|\mathsf T\|\,\|\mathsf T^{-1}\|&\leq K_0<\infty.
 \label{eq:appendix_uniform_hyperbolicity}
\end{align}
Here the columns of $\mathsf T$ have unit norm and $\|\cdot\|$ is
an operator norm. At a root of algebraic multiplicity $m_\sigma$,
completeness requires
$\dim\ker(\mathsf C-\sigma I)=m_\sigma$.
The supremum also excludes eigenvectors becoming parallel as a
coincidence is approached. Pointwise reality and completeness at
the coincident state are therefore insufficient. Quasilinear energy
estimates additionally require a sufficiently regular symmetrizer,
as discussed in Sec.~\ref{sec:hyperbolicity}.

Following Ref.~\cite{Cordeiro:2023ljz}, nonlinear causality
requires real characteristic frequencies and nontimelike
characteristic covectors. For every physical mode, every
$s\in\mathcal S$ and every state in the admitted domain, this gives
\begin{align}
 \sigma_a&\in\mathbb R,\qquad |\sigma_a|\leq1,\notag\\
 g^{\mu\nu}k_\mu k_\nu&=k^2(1-\sigma_a^2)\geq0,\qquad
 k_\mu=k(s_\mu+\sigma_a n_\mu).
 \label{eq:appendix_nonlinear_causality}
\end{align}
These conditions place the signal cone inside the light cone for
a well-posed evolution. They are evaluated on finite counterflow
and vortex slip, without linearizing the background about coflow.
A causal repeated root must still satisfy
Eq.~\eqref{eq:appendix_uniform_hyperbolicity}.

Section~\ref{app:force_equilibrium_model_b} treats the NC model with
algebraic force balance. The acoustic construction in
Sec.~\ref{app:finite_flow_sounds} supplies the material roots.
Sections~\ref{app:primitive_characteristics} and
\ref{app:force_equilibrium_coincidences} determine the vortex
modes and compatibility at sound--vortex crossings.
Section~\ref{app:force_equilibrium_full_symbol} combines these
requirements into the primitive classification for the allowed
normals. Section~\ref{app:force_equilibrium_model_a} then gives
the changed material and contact conditions for the CP model.

Section~\ref{app:model_b_tension_changes} adds independent line
motion and compression. Sections~\ref{app:regulated_characteristics}
and \ref{app:rw_derivation} construct the response matrices and
coflow bounds. Sections~\ref{app:tension_primitive_bounds} and
\ref{app:tension_state_bounds} give the adopted EOS's primitive
conditions and sufficient motion bounds.
Section~\ref{sec:tension_conservative_certificate} treats
coincident characteristics. Finally, Sec.~\ref{sec:regc_exact}
tests real causal roots and uniform eigenvector conditioning for
a general EOS without assuming positive canonical energy.

\begingroup
\let\mergedsubsection\subsection
\let\mergedsubsubsection\subsubsection
\let\mergedparagraph\paragraph
\def\section{\par
 \ifdim\pagegoal<\maxdimen
  \ifdim\dimexpr\pagegoal-\pagetotal\relax<6\baselineskip
   \newpage
  \fi
 \fi
 \mergedsubsubsection}
\let\subsection\mergedparagraph
\let\subsubsection\mergedparagraph

\mergedsubsection{{Force-equilibrium NC model}}
\label{app:force_equilibrium_model_b}
\begingroup
With constant entrainment, additive locked heat and $F=0$, the line
velocity is fixed by force balance. Its derivatives contribute to
the principal equations. We derive the material conditions, evaluate
the vortex response and determine when a sound/vortex coincidence
retains a bounded eigenbasis.
The main-text inequalities are collected in
Sec.~\ref{sec:force_equilibrium_conditions}.

Axisymmetry restricts Eulerian normals to
$\hat s\cdot e_\phi=0$, while azimuthal currents and momenta are
retained. Results stated for three dimensions use the full normal
sphere. Removing entrainment decouples the material Euler systems,
but the vortex response still requires the force-law conditions.
\endgroup
\input{appendix_force_equilibrium_bounds_tension}
\input{appendix_shared_acoustics_tension}
\input{appendix_characteristics_tension}
\input{appendix_finite_flow_characteristics_tension}
\input{appendix_acoustic_coincidence_tension}
\input{appendix_force_equilibrium_coincidences_tension}
\input{appendix_force_equilibrium_full_symbol_tension}
\input{appendix_force_balance_electric_tension}
\input{appendix_vortex_electric_tension}
\mergedsubsection{{Force-equilibrium CP model}}
\label{app:force_equilibrium_model_a}
\input{appendix_model_a_force_equilibrium_tension}

\mergedsubsection{{NC model with vortex tension}}
\label{app:model_b_tension_changes}
\begingroup
With independent $E,B$, the line energy changes the matter momenta,
provides electric inertia and couples material and array compression.
The derivatives are those of $\Lambda_{\rm mat}-F$.
We first construct the response and derive the coflow lower and
upper bounds. A positive-energy estimate then gives sufficient
finite-motion conditions. The adopted BCS EOS permits the primitive
reduction and recovery bounds that follow.
The final coefficient analysis treats a general regular frozen
symbol without assuming positive canonical energy.
All tension results use every three-dimensional normal.
\endgroup
\input{appendix_regulated_characteristics_tension}

\input{appendix_rw_tension}
\input{appendix_condition_reductions_tension}

\input{appendix_tension_state_bounds}
\input{appendix_separable_characteristics_tension}
\input{appendix_tension_model}
\input{appendix_general_coefficient_tests_tension}
\endgroup

%% file: appendix_force_equilibrium_bounds_tension.tex
\section{{Set-up and physical branch}}
\label{app:force_equilibrium_condition_reductions}

\begingroup
The force-equilibrium analysis uses the material EOS
\eqref{eq:main_model_b_master} with entropy locked to the charged
current. In current variables,
\begin{align}
 -\Lambda_{\rm mat}
 &=\rho_n(n)+\rho_c(N,\varsigma)
   +\mathcal A(I-nN),\notag\\
 I&=-j_n\cdot j_c,\qquad \mathcal A=\mathrm{constant}.
 \label{eq:regc_material_class}
\end{align}
Here $j_n=nu_n$, $j_c=Nu_c$, $\varsigma=s/N$, and $\rho_c$ includes
the additive thermal energy $r=C_{\rm th}s^{4/3}\geq0$. The matter
responses are $\mu_n=\partial\rho_n/\partial n$,
$k_n=\partial\mu_n/\partial n$,
$\mu_h=(\partial\rho_c/\partial N)_{\varsigma}=\widehat\mu_p$
and $k_h=(\partial\mu_h/\partial N)_{\varsigma}$.
For the NC model, $k_x=\Gamma_xK_xn_x^{\Gamma_x-2}$,
$\widehat\mu_p=\mu_p+4r/(3N)$ and $k_h=k_p+4r/(9N^2)$,
with positive EOS masses and polytropic coefficients and $\Gamma_x>1$.
We use $n=\nu_n$, $N=n_p$, $g=\Gamma_r$ and the phase kinematics of
Eq.~\eqref{eq:main_phase_kinematics}, so that $H=\mu_n-\mathcal AN$,
$\mu=Hg+\mathcal AN$ and $Q=H^2(g^2-1)$.
\revadd{The material response is $f^2=n/H$, as in
Sec.~\ref{sec:force_equilibrium_conditions}. The line energy $F$ vanishes
in this approximation.}

\paragraph{Admitted states.}
The physical branch has $n,N>0$, $g\geq1$, finite responses and a
future-directed carrier. Hot states also have $s>0$. It requires
\begin{align}
 &\mu_n+\mathcal AN[g-1+\sqrt{g^2-1}]>0,\notag\\
 &J_S=H[k_nk_p-\mathcal A^2(g-1)^2]
 +\mathcal A^2k_nn(g^2-1).
 \label{eq:main_model_b_admission_compact}
\end{align}
The inequality selects a future-timelike phase momentum. The
pressure coordinates are regular only where $J_S\ne0$.
We further require $H>0$ and positive combined inertia,
$J=HI_h-\mathcal A^2nN>0$ with $I_h=\widehat\mu_p-\mathcal An$.
Indeed, the cold current coefficients are $\mathcal B_n=H/n$
and $\mathcal B_p=(\mu_p-\mathcal An)/N$. The warm inertia
determinant is $\mathcal B_n(\mathcal B_p+4r/(3N^2))-\mathcal A^2
=J/(nN)$.
Positive affinity is not required, so $\chi=0$ is allowed.
Compressibility, characteristic and common-inverse tests
remain separate.

\paragraph{Force branch.}
\revadd{Write
$a_L=\widetilde{\boldsymbol{\mathcal U}}_v/\widetilde{\mathcal U}_v^0$
for the transverse line velocity of
Eq.~\eqref{eq:moments_relativistic_line_velocity}.}
At nonzero curl we require the regular future force branch
\begin{align}
 B^2&>0,\quad \widetilde{\mathcal U}_v^0>0,\notag\\
 (\widetilde{\mathcal U}_v^0)^2
       -|\widetilde{\boldsymbol{\mathcal U}}_v|^2&>0,\qquad
 E=-a_L\times B.
 \label{eq:force_drift_primitive_gate}
\end{align}
Full force projection enforces the electric equality and gives
$E\cdot B=0$ and $B^2>E^2$. These kinematic conditions admit a
timelike line velocity. The derivative of that velocity must still
give real characteristics and a uniformly complete eigenbasis.
\endgroup

%% file: appendix_shared_acoustics_tension.tex
\section{{Acoustic construction}}

\label{app:finite_flow_sounds}
\begingroup
We reduce the sound sector of either force-equilibrium EOS, or of a
regular general two-current EOS, to a symmetric two-potential pencil
and give its four speeds in closed form.

\revadd{Let $\Lambda_{\rm red}$ be the material master function with
the entropy current locked to the charged current.}
At fixed entropy per charged particle $\varsigma=s_c/N$, use
$p^x_\mu=\partial\Lambda_{\rm red}/\partial j_x^\mu$ with
$S_c^\mu=\varsigma j_p^\mu$, $j_n^\mu=n_n^\mu$ and $j_p^\mu=n_p^\mu$.
For our thermal completion,
$p^n_\mu=\xi_\mu$ and $p^p_\mu=\mu^p_\mu+\varsigma T_cu_\mu$.
On a regular current--momentum branch, let $\mathcal H_{xy}^{\mu\nu}
=(\partial j_x^\mu/\partial p^y_\nu)_{\varsigma}$, $x,y=n,p$,
be the inverse response, including density cross-responses and
state-dependent entrainment.
Its projections along the Eulerian normal $n_\mu$ and the spatial
face covector $s_\mu$ are
\begin{align}
 \mathsf T_{xy}&=-n_\mu\mathcal H_{xy}^{\mu\nu}n_\nu,\notag\\
 \mathsf L_{xy}&=(n_\mu s_\nu+s_\mu n_\nu)
                 \mathcal H_{xy}^{\mu\nu},\notag\\
 \mathsf E_{xy}&=s_\mu\mathcal H_{xy}^{\mu\nu}s_\nu.
 \label{eq:app_general_eos_projected_response}
\end{align}
The potential reduction describes the compressive modes. For
$k_\mu=s_\mu+\sigma n_\mu$, perturbations
$\delta p^x_\mu=k_\mu\phi_x$ reduce the two continuity equations to
$\det\mathsf Q=0$ with
$\mathsf Q=\mathsf E+\sigma\mathsf L-\sigma^2\mathsf T$.
With $\operatorname{adj}$ denoting the adjugate, the
quartic coefficients are
\begin{align}
 a_4&=\det\mathsf T,\qquad a_0=\det\mathsf E,\notag\\
 a_3&=-\operatorname{tr}[(\operatorname{adj}\mathsf T)\mathsf L],\notag\\
 a_2&=\det\mathsf L-
       \operatorname{tr}[(\operatorname{adj}\mathsf T)\mathsf E],\notag\\
 a_1&=\operatorname{tr}[(\operatorname{adj}\mathsf E)\mathsf L].
 \label{eq:app_general_eos_acoustic_coefficients}
\end{align}
For the NC EOS, the response is congruent to
Eq.~\eqref{eq:app_model_b_potential_pencil}.

At coflow, $\mathsf L=0$ in the common rest frame, giving two
coupled sound speeds:
\begin{equation}
 c_{1,2}^2=\frac12\left\{
 \operatorname{tr}(\mathsf T^{-1}\mathsf E)
 \pm\sqrt{[\operatorname{tr}(\mathsf T^{-1}\mathsf E)]^2
 -4\frac{\det\mathsf E}{\det\mathsf T}}\right\}.
 \label{eq:app_general_eos_coflow_sounds}
\end{equation}
Their four boosted speeds follow
Eq.~\eqref{eq:main_model_b_euler_sound_speeds} with the common velocity.

\paragraph{Closed-form speeds.}
For either EOS, normalize with $b_f=a_3/a_4$, $c_2=a_2/a_4$,
$d=a_1/a_4$ and $f=a_0/a_4$, and define
\begin{align}
 P_f&=c_2-3b_f^2/8,\quad
 Q_f=d-b_fc_2/2+b_f^3/8,\notag\\
 R_f&=f-b_fd/4+b_f^2c_2/16-3b_f^4/256.
 \label{eq:app_quartic_depression}
\end{align}
For $Q_f\ne0$, let $\revadd{\zeta_f}$ be the largest real cubic root for coefficients
$(1,2P_f,P_f^2-4R_f,-Q_f^2)$, evaluated with
Eqs.~\eqref{eq:app_scalar_cubic_parameters}--\eqref{eq:app_scalar_cubic_values}.
The four sounds are
\begin{align}
 \lambda_{\sigma\tau}&=-\frac{b_f}4+\frac{\sigma\sqrt{\revadd{\zeta_f}}}2\notag\\
 &\quad+\frac\tau2\sqrt{-2P_f-\revadd{\zeta_f}-2\sigma Q_f/\sqrt{\revadd{\zeta_f}}},
 \quad\sigma,\tau=\pm1.
 \label{eq:app_finite_quartic_speeds}
\end{align}
At $Q_f=0$, use Eq.~\eqref{eq:main_finite_flow_biquadratic_speeds}.
Repeated real roots are retained.

\paragraph{Real and causal roots.}
The following tests locate the acoustic roots without testing their
eigenvectors. A regular acoustic time branch requires $a_4\ne0$.
Four real sounds, counting multiplicity, and their strict upper
light-cone bound are then equivalent to
\begin{align}
 P_f\leq0,\quad P_f^2-4R_f\geq0,\quad\mathfrak D_f&\geq0,
 \label{eq:app_primitive_sound_reality}\\
 \operatorname{sgn}(a_4)\sum_{i=j}^4\binom ij a_i>0,
 \qquad j&=0,1,2,3,
 \label{eq:app_finite_causal_tests}
\end{align}
respectively, with $\mathfrak D_f=16R_f(P_f^2-4R_f)^2
-4P_f(P_f^2-36R_f)Q_f^2-27Q_f^4$.
Reversing the normal changes each speed's sign and supplies the
lower bound. The material and vortex analyses below impose the
additional conditions for a complete coupled basis.
For meridional normals, substitute $s=(1-t^2,2t)/(1+t^2)$ and
test the omitted normal $s=(-1,0)$ separately. Clear positive
denominators. A weak sign allows the zero polynomial. Otherwise,
require a positive leading coefficient and no real roots of the
odd-multiplicity square-free part. A strict sign requires a positive
leading coefficient and no roots of the full square-free part.
Use the Sturm sequence $R_0=f$, $R_1=f'$,
$R_{i+1}=-\operatorname{rem}(R_{i-1},R_i)$ and equal variation
counts at $\pm\infty$. On the carrier-frame interval
$[-\sqrt Q,\sqrt Q]$, a weak sign instead requires a positive
first nonzero inward coefficient at the left endpoint and no odd
interior roots, or the zero polynomial. A strict sign requires
positive endpoints and no interior roots. At $Q=0$, test one value.

\paragraph{Cubic roots.}
For cubic coefficients $(A,B,C,D)$ with $A\ne0$, define
\begin{align}
 h_c&=-B/(3A),\quad
 p_c=(3AC-B^2)/(3A^2),\notag\\
 q_c&=(2B^3-9ABC+27A^2D)/(27A^3),\notag\\
 \delta_c&=(q_c/2)^2+(p_c/3)^3.
 \label{eq:app_scalar_cubic_parameters}
\end{align}
For $\delta_c>0$, the only real root is
$h_c+\sqrt[3]{-q_c/2+\sqrt{\delta_c}}+
\sqrt[3]{-q_c/2-\sqrt{\delta_c}}$, using real cube roots.
For $\delta_c=0$, use $h_c+2u,h_c-u$ with $u=\sqrt[3]{-q_c/2}$.
For $\delta_c<0$, set
$\theta_c=\frac13\arccos[(3q_c/2p_c)\sqrt{-3/p_c}]$.
the three roots are
\begin{equation}
 h_c+2\sqrt{-p_c/3}\cos(\theta_c-2\pi m/3),\quad m=0,1,2.
 \label{eq:app_scalar_cubic_values}
\end{equation}
At $A=0$, use $(-C\pm\sqrt{C^2-4BD})/(2B)$ for $B\ne0$
and nonnegative radicand, or $-D/C$ if only $C\ne0$.
\endgroup

%% file: appendix_characteristics_tension.tex
\begingroup
\allowdisplaybreaks[1]
\section{{Vortex eigenvectors and loss of hyperbolicity}}
\label{app:primitive_characteristics}
\label{app:characteristic_material}

The algebraic line velocity can remain subluminal even when its
vorticity perturbations lack a complete real basis.
We calculate that response in the carrier frame, using the phase
kinematics \eqref{eq:main_phase_kinematics} and
$\revadd{f^2=n/H}$, $\nu=ng+\mathcal ZN$,
$\mathcal D=\mathcal R\nu-\mathcal R'\mathcal ZN$.
Resolve $q$ along the carrier-frame curl direction $\ell$ as
$q_\parallel=q\cdot\ell$, $q_\perp=q-q_\parallel\ell$ and
$Q_\perp=|q_\perp|^2$. The velocity coefficients are
\begingroup
\begingroup\colorlet{revisionblue}{mainrevisionblue}\color{revisionblue}
\begin{align}
 A_L&=\frac{\nu+\mathcal R'(\mathcal R'ng+\mathcal R\nu)}
 {1+(\mathcal R')^2},\notag\\
 C_L&=\frac{\mathcal D}{1+(\mathcal R')^2},\qquad
 \Delta_L=A_L^2-C_L^2-f^4Q_\perp,\notag\\
 \Gamma_L^2&=\frac{-\Delta_L+
 \sqrt{\Delta_L^2+4A_L^2C_L^2}}{2C_L^2},\notag\\
 h_v&=\frac{f^2A_L}{A_L^2+C_L^2\Gamma_L^2},\qquad
 d_v=\frac{f^2C_L\Gamma_L}{A_L^2+C_L^2\Gamma_L^2}.
 \label{eq:main_mobile_sh_exact_line_coefficients}
\end{align}
\endgroup
\endgroup
The unit line representative
$u_L^{(u)}=\Gamma_L(u+h_vq_\perp-d_v\ell\times q)$
is perpendicular to the carrier-frame curl. It spans the worldsheet
of Eq.~\eqref{eq:moments_relativistic_line_velocity} and gives the
corresponding force projector and electric field.
For $C_L=0$, use $\revadd{h_v=f^2/\nu}$, $d_v=0$ and
$\revadd{\Gamma_L=(1-f^4Q_\perp/\nu^2)^{-1/2}}$.
The work is $b_{\rm prop}C_L(\Gamma_L^2-1)$, where
$b_{\rm prop}=|\mathcal B_{(u)}|/\Gamma_L$.
The Eulerian fields determine the carrier-frame curl by
\begingroup
\begin{equation}
 \mathcal B_{(u)}=WB-z\times E-\frac{z(z\cdot B)}{W+1},
 \qquad\ell=\frac{\mathcal B_{(u)}}{|\mathcal B_{(u)}|}.
 \label{eq:app_model_b_curl_frame_map}
\end{equation}
\endgroup

\subsection{Real speeds and independent vortex modes}
\begingroup
For this local calculation let $B=\mathcal B_{(u)}$,
$\ell=B/|B|$ and hold the matter state fixed. The force-law electric
response and the equation determining its transverse Lorentz factor are
\begingroup\colorlet{revisionblue}{mainrevisionblue}\color{revisionblue}
\begin{align}
 E(B)&=-h_vq\times B+d_v|B|q_\perp,\notag\\
 (\Gamma_L^2-1)(A_L^2+C_L^2\Gamma_L^2)
   &=f^4Q_\perp\Gamma_L^2,\notag\\
 \frac{\partial\Gamma_L}{\partial Q_\perp}
   &=\frac{f^4\Gamma_L^3}{2(A_L^2+C_L^2\Gamma_L^4)}.
 \label{eq:app_vortex_force_response_derivative}
\end{align}
\endgroup
Thus the derivative of $E$ includes the dependence of $h_v,d_v$
on the curl direction through $Q_\perp$. Freezing those coefficients
would give a different principal matrix.
Faraday evolution gives the physical vortex symbol
$\mathsf P_v=[\hat k\times](\partial E/\partial B)$ restricted
to $\hat k\cdot\delta B=0$. With
$v_0=h_vq_\perp-d_v\ell\times q$ and
$\mathfrak D_L=A_L^2+C_L^2\Gamma_L^4$, its determinant is
\begingroup\colorlet{revisionblue}{mainrevisionblue}\color{revisionblue}
\begin{align}
 \det(\lambda I_2-\mathsf P_v)
 &=\left[\lambda-v_0\cdot\hat k
       -\frac{f^2A_Lq_\parallel(\ell\cdot\hat k)}{\mathfrak D_L}\right]^2
       \notag\\
 &\quad+\left[\frac{f^2C_L\Gamma_L^2q_\parallel
                         (\ell\cdot\hat k)}{\mathfrak D_L}\right]^2.
 \label{eq:app_vortex_physical_quadratic}
\end{align}
\endgroup
The two carrier-frame vortex speeds are therefore
\endgroup
\begingroup
\begingroup\colorlet{revisionblue}{mainrevisionblue}\color{revisionblue}
\begin{align}
 \lambda_{v\pm}
 &= (h_vq_\perp-d_v\ell\times q)\cdot\hat k\notag\\
 &\quad+
 \frac{f^2q_\parallel(\ell\cdot\hat k)
       (A_L\pm iC_L\Gamma_L^2)}
      {A_L^2+C_L^2\Gamma_L^4}.
 \label{eq:app_vortex_speeds}
\end{align}
\endgroup
\endgroup
For $\mathcal Dq_\parallel\ne0$, the frequencies are complex when
$\ell\cdot\hat k\ne0$ because $C_L$ is proportional to
$\mathcal D$. Tangent propagation can instead give real defective
roots. In particular, choose $\ell=e_3$,
$q=q_te_1+q_\parallel e_3$ and $\hat k=e_1$.
On divergence-free amplitudes $(\delta B_2,\delta B_3)$, the symbol is
\begingroup
\begin{equation}
 P_v^{\rm tan}=h_vq_t I_2+
 \begin{pmatrix}0&0\\-d_vq_\parallel&0\end{pmatrix}.
 \label{eq:app_model_b_tangent_vortex_jordan}
\end{equation}
\endgroup
The nilpotent part leaves one eigenvector at the repeated speed.
A complete real vortex pair therefore requires
$\mathcal Dq_\parallel=0$.
For $q_\parallel=0$, the pair is scalar transport at
$v=h_vq-d_v\ell\times q$, with $|v|^2=1-\Gamma_L^{-2}<1$.
For $\mathcal D=0$, both speeds are $\revadd{f^2q\cdot\hat k/\nu}$ even
when the curl is oblique to the counterflow.

The tangent Jordan block gives a real repeated speed but only one
independent vortex mode. Real subluminal speeds therefore impose
weaker conditions than a complete, uniformly bounded eigenbasis.
For $\mathcal D>0$, their geometric condition is
\begingroup
\begin{equation}
 \begin{cases}
 q\cdot\ell=0,&\text{3D},\\
 (q\cdot\ell)^2|\ell\times b_0|^2=0,&\text{axisym.},
 \end{cases}
 \label{eq:app_model_b_weak_vortex_causality}
\end{equation}
\endgroup
Here $v_0=h_vq_\perp-d_v\ell\times q$,
$a=e_\phi+z_\phi z/(W+1)$ and $b_0=a+z_\phi v_0$.
The meridional propagation plane satisfies $b_0\cdot k=0$.
The additional meridional case $q_\parallel\ne0$ and
$\ell\parallel b_0$ has a repeated real speed with one independent
vortex eigenvector. Its Jordan block permits growth unbounded in
wave number. The real-speed condition
\eqref{eq:app_model_b_weak_vortex_causality} therefore does not replace
the strong-hyperbolicity conditions of
Sec.~\ref{app:model_b_finite_vortices}.

For the unentrained stellar EOS with
$(\Gamma_n,\Gamma_p)=(2,4)$, the neutron rest-frame sound is causal
for $e>m_n$. The charged sound additionally requires
\begingroup
\begin{equation}
 24K_pN^4<9m_pN+8r.
 \label{eq:main_nonlinear_causality}
\end{equation}
\endgroup

\subsection{Regularity at zero curl}
\label{app:zero_curl_regularity}
The zero-curl limit must also admit a differentiable constitutive
map. For regular matter-dependent coefficients and
$B=\mathcal B_{(u)}$, the carrier-frame electric response is
\begingroup
\begin{equation}
 E(B)=-h_vq\times B+d_v|B|q_\perp.
 \label{eq:app_model_b_zero_curl_response}
\end{equation}
\endgroup
The coefficients depend on $Q_\perp$ and are even under curl
reversal. Thus $E(B)+E(-B)=2d_v|B|q_\perp$.
At finite slip and $\mathcal D>0$, approaching $B=0$ along opposite
directions perpendicular to $q$ gives incompatible derivatives.
The obstruction remains for divergence-free meridional
perturbations. Defining $E(0)=0$ makes the map continuous but does
not give a principal symbol there.

At coflow, $|E|=O(|q||B|)$ gives a zero joint derivative, although
there is no open $C^1$ neighborhood of dissipative states.
The identically reactive full-lift response
$\revadd{E=-f^2q\times B/\nu}$ and ideal pinning $E=0$ extend smoothly to
zero curl without a line direction. Coefficients with explicit
curl dependence require their own limit analysis.
\endgroup

%% file: appendix_finite_flow_characteristics_tension.tex
\section{{Material response and counterflow}}

\label{app:model_b_finite_acoustic}

\begingroup
We now derive the NC conditions at finite counterflow. The EOS
responses entering the acoustic and vortex tests are
\begingroup\colorlet{revisionblue}{mainrevisionblue}\color{revisionblue}
\begin{align}
 \partial_n(f^2)&=(H-nk_n)/H^2,\qquad \partial_N(f^2)=\mathcal A n/H^2,\notag\\
 \beta&=\mu_p-\mathcal A n-\mathcal A^2Nf^2+4r/(3N),\notag\\
 \partial_n\beta&=-\mathcal A-\mathcal A^2N\partial_n(f^2),\notag\\
 \partial_N\beta&=k_p-\mathcal A^2f^2-\mathcal A^2N\partial_N(f^2)+4r/(9N^2).
 \label{eq:app_model_b_finite_scalar_coefficients}
\end{align}
\endgroup
\endgroup
\input{appendix_model_b_material_tension}

\subsection{{Entrained pinned and mobile vortex modes}}
\label{app:model_b_finite_vortices}

\begingroup
A complete material basis must also satisfy vortex and force-admission
tests, for signed $\mathcal Z,\mathcal R'$.
Define the lift density and reactive transport velocity by
\begingroup\colorlet{revisionblue}{mainrevisionblue}\color{revisionblue}
\begin{align}
 \nu&=ng+\mathcal ZN,
 &\mathcal D&=\mathcal R\nu-\mathcal R'\mathcal ZN,\notag\\
 v&=\eta q,\quad\eta=f^2/\nu,
 &Z_\xi&=\mathcal Z-\mathcal Af^2.
 \label{eq:app_entrained_force_density}
\end{align}
\endgroup
Thus $\revadd{\nu=f^2\mu+Z_\xi N}$.
Equations~\eqref{eq:main_mobile_sh_exact_line_coefficients} and
\eqref{eq:app_vortex_speeds} give the line solution and vortex pair,
with $ng$ as the neutron-current term in $A_L$.

\paragraph{Force admission.}
The force prescription requires $\mathcal R\geq0$, nonnegative
vortex heating $\mathcal D\geq0$, $\nu>0$ and a future-timelike
projected lift in the Eulerian frame. With the Eulerian curl
direction $\widehat B=B/|B|$, define
\begingroup\colorlet{revisionblue}{mainrevisionblue}\color{revisionblue}
\begin{align}
 u_B&=z\cdot\widehat B,\qquad
 j_B=f^2(\boldsymbol\xi-\mathcal ANz)\cdot\widehat B,\notag\\
 a_B&=n^2+j_B^2,\quad b_B=ng+j_Bu_B,\quad c_B=1+u_B^2.
 \label{eq:app_model_b_signed_lift_admission}
\end{align}
\endgroup
The projected constituent currents are future timelike, so
$a_B,b_B,c_B>0$ and $b_B^2-a_Bc_B\geq0$.
The projected-lift margin is
\begin{equation}
 (L^0)^2-|L_\perp|^2
 =a_B+2Nb_B\mathcal Z+N^2c_B\mathcal Z^2.
 \label{eq:app_model_b_eulerian_lift_margin}
\end{equation}
Its future component selects the upper root,
\begin{equation}
 \mathcal Z>\max\left\{-\frac{ng}{N},\,
 \frac{\sqrt{b_B^2-a_Bc_B}-b_B}{Nc_B}\right\},
 \label{eq:main_model_b_force_admission}
\end{equation}
which includes $\nu>0$. The carrier-rest bound
$\revadd{\nu>f^2\sqrt{Q_\perp}}$, with $Q_\perp=Q-(q\cdot\ell)^2$ for the
local curl direction $\ell$, cannot replace this test in a general
Eulerian frame.

\paragraph{Pinned.}
\revadd{For a unit carrier-frame wave normal $\hat k$, write
$q_k=q\cdot\hat k$ and
$\delta\xi_T=\delta\boldsymbol\xi-\hat k(\hat k\cdot\delta\boldsymbol\xi)$
for the longitudinal counterflow and transverse momentum perturbation.}
Beyond the material tests, pinned completeness is equivalent to
\begin{equation}
 Q=0\quad(\text{3D}),\qquad H_q=0\quad(\text{axisym.}).
 \label{eq:main_model_b_pinned_primitive}
\end{equation}
At the shared zero root, the transverse forcing $\revadd{f^2q_k\delta\xi_T}$
requires $q_k=0$. The remaining lift is bounded without a contact
denominator. The pinned vortex speeds are causal.
Static pinning also requires the holding-force bound
\begin{equation}
 \kappa_vf^2\sqrt{[1+(\mathcal R')^2]Q_\perp}\leq f_c,
 \qquad f_c\geq0,
 \label{eq:main_primitive_pinned_load}
\end{equation}
\revadd{Here $f^2=n/H$.} The reported ideal static branch takes $\mathcal R'=0$.

\paragraph{Mobile.}
In three dimensions, coupled-system completeness requires
\begin{equation}
 \mathcal D^2Q=0,\qquad Q=|q|^2.
 \label{eq:app_model_b_three_dimensional_mobile_gate}
\end{equation}
This is stronger than the curl-pair requirement
$\mathcal Dq_\parallel=0$. Otherwise the pair is complex or
defective, or perpendicular slip removes a vortex mode at normals
with $k\cdot v=0$ and $q_k\ne0$.
Coflow is regular. At $Q>0$, three-dimensional transport must be
reactive.

Laboratory axisymmetry permits warm dissipative counterflow
with $Q>0$, $\mathcal D>0$ only when
$q\cdot\ell=a\cdot\ell=0$ and $S_0=v\cdot(\ell\times a)\ne0$.
Here $v=h_vq-d_v\ell\times q$ is the line velocity of
Eq.~\eqref{eq:main_model_b_general_vortex_speed}.
With $b=a+z_\phi v$, define
\begin{align}
 K&=\frac{q\cdot(\ell\times a)+z_\phi d_vQ}{|b|},
 &S&=\frac{S_0}{|b|}.
 \label{eq:app_entrained_dissipative_projections}
\end{align}
The reachable normals have $q_k=Kt$, $v_k=St$ with
$-1\leq t\leq1$. The defective family admitted by
Eq.~\eqref{eq:app_model_b_weak_vortex_causality} cannot replace these
strong-hyperbolicity conditions.
\begingroup
At a shared line/carrier speed, the carrier-frame normal is parallel
to $\ell$, since $a,v$ are independent vectors in $\ell^\perp$.
In that frame choose $\ell=e_3$ and $q=\sqrt Qe_1$.
At zero frequency the independent principal equations reduce to
\begingroup\colorlet{revisionblue}{mainrevisionblue}\color{revisionblue}
\begin{align}
 \delta j_c^3&=N\delta z^3=0,\notag\\
 \delta j_n^3&=f^2(\delta\xi_3-\mathcal AN\delta z^3)=0,\notag\\
 \delta\Psi_{\rm mat}&=0,\notag\\
 \delta\Xi-h_v\sqrt Q\,\delta\xi_1
                 +d_v\sqrt Q\,\delta\xi_2&=0.
 \label{eq:app_model_b_dissipative_contact_minor}
\end{align}
\endgroup
The remaining stress equations follow from the two number fluxes.
In density variables $(n,N,r,z,\xi)$ these four equations are
independent: $\partial_r\Psi_{\rm mat}=1/3$,
$\partial_r\Xi=0$ and
$(\partial_n\Xi)_{N,z,\xi}=k_n/g>0$ at the chosen background.
Their rank is four, giving five independent warm contacts without
dividing by the pressure-chart Jacobian. Regular time evolution and
the material sound gap give a bounded local completion.
\endgroup
The case $S=0$ is defective.
For reactive transport, use $S=\eta K$ with $K=\sqrt Q$ in
three dimensions or $K=\sqrt{H_q}/|b|$ in axisymmetry.
At $K=0$, take the regular direct limit.
\endgroup

\subsection{{Finite primitive vortex--sound cancellation tests}}
\label{app:model_b_vortex_gap_tests}

\begingroup
Coincident vortex and sound speeds require a bounded completion
of their coupled eigenvectors. Set $x=t^2$ and define, with the
responses of Eq.~\eqref{eq:app_model_b_finite_scalar_coefficients},
\begingroup\colorlet{revisionblue}{mainrevisionblue}\color{revisionblue}
\begin{align}
 c_{n0}&=\partial_n\beta+\mathcal A\mu \partial_n(f^2),\notag\\
 c_{n1}&=-\mathcal A SK \partial_n(f^2),\notag\\
 c_{N0}&=\partial_N\beta+\mathcal A\mu \partial_N(f^2),\notag\\
 c_{N1}&=-\mathcal A SK \partial_N(f^2)-\beta S^2/N,\notag\\
 a_{n0}&=Hk_n+\mathcal A^2NQ\partial_n(f^2)/\beta,\notag\\
 a_{n1}&=-\mathcal A^2NK^2\partial_n(f^2)/\beta,\notag\\
 a_{N0}&=\mathcal AH(g-1)+\mathcal A^2NQ\partial_N(f^2)/\beta,\notag\\
 a_{N1}&=-\mathcal A SK-\mathcal A^2NK^2\partial_N(f^2)/\beta,\notag\\
 \delta&=HgS-K,\quad j=c_{n0}\partial_N(f^2)-c_{N0}\partial_n(f^2),\notag\\
 L_0&=c_{n0}a_{N0}-c_{N0}a_{n0},\quad G_0=-f^2L_0,\notag\\
 L_1&=c_{n0}a_{N1}+c_{n1}a_{N0}
 -c_{N0}a_{n1}-c_{N1}a_{n0},\notag\\
 G_1&=\delta^2j+f^2(S^2L_0-L_1),\notag\\
 G_2&=\delta^2\beta S^2\partial_n(f^2)/N+f^2S^2L_1.
 \label{eq:app_entrained_vortex_gap_coefficients}
\end{align}
\endgroup
The gap $G(x)=G_0+G_1x+G_2x^2$ is quadratic, since
$c_{n1}a_{N1}-c_{N1}a_{n1}=0$, and $G_0=n\Delta_M/J\ne0$ on the
complete material domain.
Write $c_y(x)=c_{y0}+c_{y1}x$ and $a_y(x)=a_{y0}+a_{y1}x$ for
$y=n,N$, and $\delta\xi_k=\hat k\cdot\delta\boldsymbol\xi$ in the
fixed local carrier frame. Eliminating the transverse material
perturbations from continuity, longitudinal stress and phase
response at $\lambda=St$ gives
\begingroup\colorlet{revisionblue}{mainrevisionblue}\color{revisionblue}
\begin{equation}
 \begin{pmatrix}
 \delta t\partial_n(f^2)&\delta t\partial_N(f^2)&f^2(S^2x-1)\\
 c_n(x)&c_N(x)&0\\
 a_n(x)&a_N(x)&-\delta t
 \end{pmatrix}
 \begin{pmatrix}\delta n\\\delta N\\\delta\xi_k\end{pmatrix}
 =\begin{pmatrix}tU\\V\\W_f\end{pmatrix}.
 \label{eq:app_entrained_forced_acoustic_system}
\end{equation}
\endgroup
The right-hand side is the forcing from a normalized vortex
perturbation.
The matrix determinant is $G(x)$, with $G(0)\ne0$. No division
by $t$ is needed at the zero-projection normal. Its adjugate gives
$(\delta n,\delta N,\delta\xi_k)
=(C_n,C_N,tC_\psi)/G$, with
\begingroup\colorlet{revisionblue}{mainrevisionblue}\color{revisionblue}
\begin{align}
 C_n(x)&=-\delta Uxc_N(x)+\delta^2\partial_N(f^2)Vx\notag\\
 &\quad+f^2(S^2x-1)[a_N(x)V-c_N(x)W_f],\notag\\
 C_N(x)&=\delta Uxc_n(x)-\delta^2\partial_n(f^2)Vx\notag\\
 &\quad+f^2(S^2x-1)[c_n(x)W_f-a_n(x)V],\notag\\
 C_\psi(x)&=\delta\{[\partial_n(f^2)c_N(x)-\partial_N(f^2)c_n(x)]W_f\notag\\
 &\qquad+[\partial_N(f^2)a_n(x)-\partial_n(f^2)a_N(x)]V\}\notag\\
 &\quad+(L_0+L_1x)U.
 \label{eq:app_entrained_vortex_cofactors}
\end{align}
\endgroup
A bounded lift requires these to cancel each gap zero to its
multiplicity. Reactive transport has one source:
\begingroup\colorlet{revisionblue}{mainrevisionblue}\color{revisionblue}
\begin{equation}
 (U,V,W_f)=\left(-\eta Sf^2,\ \eta Z_\xi,
 N\left[\frac{\mathcal AZ_\xi}{\beta}
 -\eta\left(\frac{Z_\xi}{f^2}+\mathcal A\right)\right]\right).
 \label{eq:app_entrained_reactive_source}
\end{equation}
\endgroup
The gap roots in $[0,1]$ are $-G_0/G_1$ for $G_2=0$, and
otherwise
\begin{equation}
 x_\pm=\frac{-G_1\pm\sqrt{G_1^2-4G_0G_2}}{2G_2}.
 \label{eq:app_entrained_gap_roots}
\end{equation}
There are no roots if $G_1=G_2=0$.
Reactive transport requires vanishing cofactors at every root
$x_*$ and, for each $C=d_0+d_1x+d_2x^2$, a vanishing slope at a
double root:
\begin{align}
 C_n(x_*)=C_N(x_*)=C_\psi(x_*)&=0,
 \label{eq:force_preserving_coincidences}\\
 d_1+2xd_2&=0.
 \label{eq:app_entrained_gap_cancellations}
\end{align}
A strict vortex/sound gap is instead equivalent to
\begin{equation}
 G_0(G_0+G_1+G_2)>0,
 \label{eq:force_preserving_gap_endpoints}
\end{equation}
with $4G_0G_2>G_1^2$ also required when
$G_0G_1<0$ and $G_0(G_1+2G_2)>0$.
These signs cover either sign of $G_0$.
At $Q=0$ use the regular coflow limit rather than divide by $K$.
Full force projection transports each accepted or defective physical
coincidence into the electric variables with the same multiplicity.

\paragraph{Zero Iordanskii weight.}
For $\mathcal Z=0$ and $\mathcal A\ne0$, let $X_*=Q$ in
three dimensions or $X_*=H_q/|a+z_\phi q/(Hg)|^2$ in
laboratory axisymmetry.
Here $K=1$ and $S=1/(Hg)$ factorize the gap as
$G(x)=[1-x/(H^2g^2)](n/J)
[\Delta_M-x(\Delta_M+HR_p/N)/(H^2g^2)]$.
The first factor is positive on $0\leq x\leq X_*\leq Q$, and the
charged-density cofactor $C_N$ does not vanish there.
Every gap zero therefore makes the lift unbounded, and the
reactive criterion reduces to one endpoint bound:
\begin{equation}
 \Delta_M[(H^2g^2-X_*)\Delta_M-HR_pX_*/N]>0.
 \label{eq:main_model_b_zero_lift_linear_gap}
\end{equation}
On the causal domain, $\Delta_M>0$ removes the outer sign factor.
In three dimensions,
$\Delta_M-H(g^2-1)R_p/N=(J/N)G_v$ with
$G_v=k_n[Nk_hg^2-B(g^2-1)]-\mathcal A^2N(g-1)^2$,
so the material and vortex bounds become $G_\parallel,G_v>0$.
With $\alpha_n=nk_n/H$, $\alpha_h=Nk_h/B$,
$c_*=\min(\alpha_n,\alpha_h)$ and
$b_*=\mathcal A^2nN/[HB\max(\alpha_n,\alpha_h)]$,
together these bounds give
$c_*g^2-(g^2-1)-b_*(g-1)^2>0$, or the counterflow ceiling
$g_*=[b_*+\sqrt{1-c_*+c_*b_*}]/(1-c_*+b_*)$.
The causal EOS signs ensure $0<c_*<1$, $b_*>0$ and $g_*>1$.
For $\mathcal A<0$, the independent future-phase ceiling is
$g<g_{\rm ph}=1+\mu_n^2/(-2\mathcal AHN)$.
$g_{\rm ph}=\infty$ for $\mathcal A\geq0$.
Thus the standard nonzero-entrainment three-dimensional reactive
branch requires $1\leq g<\min(g_*,g_{\rm ph})$ and both causal EOS signs.
\endgroup
{{At $\mathcal A=\mathcal Z=0$, the
protected reactive branch needs no counterflow gap. Other force
branches still require their vortex geometry tests. At finite
$\mathcal Z$, the general gap tests above also apply when
$\mathcal A=0$.}}

\begingroup
\paragraph{Warm finite drag.}
For warm finite drag, compatibility of the two independent sources
of the perpendicular dissipative branch reduces to three affine
equalities. Set $\rho=K/S$, $y=S^2x$ and
\begingroup\colorlet{revisionblue}{mainrevisionblue}\color{revisionblue}
\begin{align}
 \Theta&=1+\frac{\mathcal Af^2(\mu-\rho)}\beta,\qquad
 \Pi=4Hg-\Theta(3\mu+\rho),\notag\\
 Z(y)&=(Hg-\rho)\Pi+3\mathcal AN\Theta(Hg-\rho y)\notag\\
 &\quad+\frac{4\mathcal A^2f^2N}{\beta}(Q-\rho^2y).
 \label{eq:app_model_b_drag_affine_primitives}
\end{align}
\endgroup
A compatible nonzero crossing requires exactly
\begingroup\colorlet{revisionblue}{mainrevisionblue}\color{revisionblue}
\begin{align}
 E_\psi(y)&=(y-1)\Pi-4y(Hg-\rho)=0,\notag\\
 E_n(y)&=\frac{H-nk_n}{H^2}Z(y)
             -3\mathcal AN\Theta+4nk_n=0,\notag\\
 E_N(y)&=\frac{\mathcal An}{H^2}Z(y)
 +3N\Theta\left(k_p+\frac{4r}{9N^2}-\mathcal A^2f^2\right)\notag\\
 &\quad-3\beta\Theta y+4\mathcal Af^2[H(g-1)-\rho y]=0.
 \label{eq:app_model_b_drag_affine_equalities}
\end{align}
\endgroup
Each is affine in $y$.
On a complete material branch, finite drag permits at most one
simple root $x_*\in(0,1]$, with $y=S^2x_*$.
\revadd{Two distinct crossings and double crossings are excluded, since}
\begingroup\colorlet{revisionblue}{mainrevisionblue}\color{revisionblue}
\begin{align*}
 4G={}&-(y-1)c_NE_n+(y-1)c_nE_N\notag\\
 &+(Hg-\rho)[c_N\partial_n(f^2)-c_n\partial_N(f^2)]E_\psi
\end{align*}
\endgroup
\revadd{and $G_0\ne0$.}
A semisimple acoustic double also cannot meet the vortex pair:
its rank-one density block cannot contain both drag sources.
Force admission and $S\ne0$ remain required.

\paragraph{Arbitrary-boost reactive transport.}
For zero-work reactive transport, set $\mathcal D=0$,
$\nu>0$ and $\revadd{\eta=f^2/\nu}$.
Then $A_L=\nu$ and $d_\mu=\mathcal R'L_\mu$ give current advection
for any local curl direction $\ell$, even with two nonzero drag
channels.
\revadd{Let $\mathsf P_9(\sigma,s)$ denote the primitive principal
pencil of the warm nine-field force-equilibrium system.}
The warm nine-field determinant is
\begin{equation}
 \det\mathsf P_9(\sigma,s)
 =-\frac{N^4\beta}{gH}\,
 \omega^3(\omega-\eta j)^2
 \mathcal P_{\rm lab}(\sigma,s).
 \label{eq:app_model_b_general_reactive_full_factor}
\end{equation}
The three $\omega=0$ modes are carrier contacts.
The double $\omega=\eta j$ mode transports curl with the laboratory
velocity $v_{\rm adv,lab}=L^i/L^0$, where
$1-|v_{\rm adv,lab}|^2=(1-\eta^2Q)/(W+\eta u)^2$ and $u=q\cdot z$.
The future branch requires $W+\eta u>0$.
At $\omega=j=0$, the transverse pencil has five independent
contacts. Elsewhere, the lift is regular unless
$\mathcal P_{\rm lab}$ vanishes at the line speed. There it is a
positive multiple of $\beta G(x)$, with $K=1$, $S=\eta$ and
$0\leq x\leq H_q/|b|^2$ for $b=a+z_\phi\eta q$.
On a complete material branch, the gap and cofactor tests above are
then necessary and sufficient, given force admission and a regular
future time factor.
{Strictly subluminal reactive curl propagation imposes the bounds}
\begingroup\colorlet{revisionblue}{mainrevisionblue}\color{revisionblue}
\begin{equation}
 \begin{cases}
 \nu>f^2\sqrt Q,&\text{3D},\\
 (1+z_\phi^2)\nu+f^2z_\phi(a\cdot q)>f^2\sqrt{H_q},
 &\text{axisym.}
 \end{cases}
 \label{eq:app_model_b_signed_reactive_causality}
\end{equation}
\endgroup
{For the closed cone replace the upper
cone signs by nonstrict inequalities, retaining the strict future
force-admission conditions. The complete primitive list is in
Appendix~\ref{app:force_equilibrium_full_symbol}.}
Both are linear in $\mathcal Z$.
For the full sphere, the bound is
$\mathcal ZN>-n/[g+\sqrt{g^2-1}]$.
For $\mathcal Z\geq0$, force and reactive cone bounds are automatic,
since $\revadd{\nu^2-f^4Q=n^2+2ng\mathcal ZN+\mathcal Z^2N^2>0}$.

\paragraph{Zero entrainment.}
At $\mathcal A=0$ the constituent sounds decouple and follow
Eq.~\eqref{eq:main_model_b_euler_sound_speeds}, with
$c_n^2=nk_n/H$ and $c_p^2=Nk_h/\widehat\mu_p$.
In a fixed frame, material strong hyperbolicity requires positive
time coefficients $HU^2-nk_n(U^2-1)$ and
$\widehat\mu_pW^2-Nk_h(W^2-1)$. The causal bounds $H-nk_n>0$ and
$\mu_p-Nk_p+8r/(9N)>0$ make both positive in every frame.
For zero-work $\mathcal Z=0$ transport, the transverse lift is
bounded, since $\eta=1/(Hg)$ gives
$1-\eta^2q_k^2\geq g^{-2}>0$. No charged-sound gap is needed.

For $\mathcal Z\ne0$, the density sources
$\revadd{-f^2\mathcal ZN/\nu}$ and $\revadd{f^2\mathcal Z/\nu}$ are nonzero and strict
gaps are required. With
$S=\eta K$, $\delta=(Hg\eta-1)K$ and $x=t^2\in[0,1]$,
the two acoustic factors at the line speed are affine:
\begingroup\colorlet{revisionblue}{mainrevisionblue}\color{revisionblue}
\begin{align}
 c_N(x)&=k_h-\widehat\mu_pS^2x/N,\notag\\
 P_n(x)&=\delta^2\frac{H-nk_n}{H^2}x
             +f^2 Hk_n(S^2x-1).
 \label{eq:direct_reactive_affine_poles}
\end{align}
\endgroup
Using $Hg\eta-1=-\mathcal ZN/\nu$, the determinant and density
cofactors factor for every $x$ as $G=-c_NP_n$,
$C_n=-N\eta\mathcal Z\,c_N$ and $C_N=-\eta\mathcal Z\,P_n$.
Either acoustic pole therefore destroys a bounded density lift.
Simultaneous poles do not restore completeness.
On the positive causal constituent branch,
\begin{equation}
 P_n(x)=-nk_n+
 \frac{K^2x[n^3k_n+\mathcal Z^2N^2(H-nk_n)]}{H^2\nu^2}.
 \label{eq:direct_reactive_neutron_affine}
\end{equation}
Thus $P_n$ starts negative and increases, while $c_N$ starts
positive and decreases, and only the two endpoint signs are
required. Clearing their positive geometric denominators gives
Eq.~\eqref{eq:force_summary_independent_reactive}.
The first sign also implies the meridional curl-causality margin
$|H\nu a+nz_\phi q|^2>n^2|a\times q|^2$.
At $\mathcal Z=0$ both density residues vanish, and
these two strict gaps must be dropped.

With finite drag and zero entrainment, every nonzero line--sound
coincidence is excluded.
With $\delta=HgS-K$ and the projections of
Eq.~\eqref{eq:app_entrained_dissipative_projections}, the required
strict bounds are linear in the stiffnesses:
\begin{align}
 nk_n[H^2(1-S^2)+\delta^2]&>H\delta^2,\notag\\
 Nk_p-\mu_pS^2+\frac{4r}{9N}(1-3S^2)&>0.
 \label{eq:app_model_b_zero_entrainment_drag_linear_gap}
\end{align}
They remain strict even at coincident material sounds.
\endgroup

\subsection{{Cold NC boundary}}
\label{app:model_b_cold_characteristics}

\begingroup
Remove entropy before taking $r=0$.
The cold system has four sounds, two carrier shears and two vortices.
Its material conditions specialize directly, including acoustic
coincidences and pressure-chart folds.
For reactive transport, $q\cdot E=0$ identically and the warm lift
reduces regularly to the eight-field cold lift, so the reactive
gap and cancellation tests apply at $r=0$.
Pinned transport retains $Q=0$ or $H_q=0$, with full contact rank
four even at $J_S=0$.
Positive slip work generates thermal energy.
A closed cold system therefore uses pinning or identically zero work.
\endgroup

%% file: appendix_model_b_material_tension.tex
\subsubsection{{Material conditions}}
\label{app:model_b_definite_material}

\begingroup
We first give the conditions for the material subsystem: its four
sounds and contacts. The charged inertia is
$\revadd{I_h=\widehat\mu_p-\mathcal An}$.
Set $U=gW+q\cdot z/H$, $a=e_\phi+z_\phi z/(W+1)$ and
$H_q=|a\times q|^2$, and define
\begingroup\colorlet{revisionblue}{mainrevisionblue}\color{revisionblue}
\begin{align}
 X&=H-nk_n,\quad Y=I_h-Nk_h,\notag\\
 J&=HI_h-\mathcal A^2nN>0,\quad C_\ell=XY-\mathcal A^2nN,\notag\\
 \Delta_M&=Jk_nk_h+\mathcal A^2[2J(g-1)-C_\ell(g^2-1)],\notag\\
 R_n&=HXk_h+\mathcal A^2n^2k_n,\quad
 R_p=I_hYk_n+\mathcal A^2N^2k_h,\notag\\
 \mathcal C_M&=nH^2\Delta_M(1+z_\phi^2)-I_hR_nH_q,\notag\\
 \mathcal T_n&=Hn\Delta_M+U^2JR_n,\notag\\
 \mathcal T&=nN\Delta_M+HnW^2R_p+I_hNU^2R_n\notag\\
 &\quad-2\mathcal A^2nNUW(J-gC_\ell)+JC_\ell U^2W^2.
 \label{eq:main_model_b_material_scalars}
\end{align}
\endgroup
We require $n,N,H,k_n,k_h,J>0$, $g\geq1$ and future-directed
currents and phase. The pressure-chart sign $J_S$ is unrestricted.
For $\mathcal A\ne0$, warm and cold material strong hyperbolicity
on the meridional circle holds exactly when
$\Delta_M\mathcal C_M>0$, $\Delta_M\mathcal T_n>0$ and
$\Delta_M\mathcal T>0$.
These signs retain semisimple coincident sounds and acausal
components with $\Delta_M<0$, and are affine in either stiffness
separately. For the full normal sphere, replace $\mathcal C_M$ by
\begin{equation}
 G_\parallel=k_h[nk_ng^2-H(g^2-1)]-\mathcal A^2n(g-1)^2.
 \label{eq:app_model_b_material_exact_gates}
\end{equation}
Full-system strong hyperbolicity also requires the vortex conditions
of Sec.~\ref{app:model_b_finite_vortices} and the vortex--sound tests
of Sec.~\ref{app:model_b_vortex_gap_tests}, also at $\mathcal A=0$.

\paragraph{Acoustic reduction.}
\revadd{In a local orthonormal frame, write
$\eta=\operatorname{diag}(-1,1,1,1)$ for the Minkowski metric.}
At fixed entropy per charged particle, the current-to-momentum
Hessian is $\mathsf K\otimes\eta+\mathsf U\mathsf C\mathsf U^T$.
Here the columns of $\mathsf U$ contain the covariant constituent
velocities, $\mathsf K$ is symmetric with $K_{nn}=H/n$,
$K_{np}=\mathcal A$, $\revadd{K_{pp}=I_h/N}$ and $\det\mathsf K=J/(nN)>0$,
and $\mathsf C$ is $\mathsf K$ with $\revadd{(H,I_h)}$ replaced by $(X,Y)$.
Its rank-two inverse gives the symmetric response $\mathsf M$, with
$M_{nn}=JR_n/(n^2\Delta_M)$, $M_{pp}=JR_p/(N^2\Delta_M)$,
$M_{np}=\mathcal AJ(J-gC_\ell)/(nN\Delta_M)$ and
$\det\mathsf M=C_\ell J^2/(n^2N^2\Delta_M)$.
For a unit wave normal $s$, define
\begin{align}
 \omega&=W\sigma-z\cdot s,\quad
 j=q\cdot s+(q\cdot z)\left[\frac{z\cdot s}{W+1}-\sigma\right],\notag\\
 \omega_n&=g\omega-j/H,\qquad
 \mathsf O=\operatorname{diag}(\omega_n,\omega).
 \label{eq:app_model_b_general_lab_primitives}
\end{align}
The four sounds are determined by the symmetric acoustic quadratic
\begin{align}
 \mathsf Q(\sigma,s)&=(1-\sigma^2)\mathsf K-\mathsf O\mathsf M\mathsf O,\notag\\
 \mathcal P_{\rm lab}(\sigma,s)
 &=\frac{n^2N\Delta_M}{HJ}\det\mathsf Q(\sigma,s).
 \label{eq:app_model_b_potential_pencil}
\end{align}
Equation~\eqref{eq:app_finite_quartic_speeds} gives the roots of
$\mathcal P_{\rm lab}$. For $\Delta_M\ne0$ the momentum potentials
lift boundedly to the four acoustic modes, independently of
pressure coordinates.

At the carrier contact $\sigma_c=z\cdot s/W$, contact separation on
the connected meridional circle is exactly
$\Delta_M\mathcal C_M>0$. For the full sphere it is
$\Delta_MG_\parallel>0$. It is necessary at finite entrainment:
the shear forcing at a sound/contact collision obstructs a uniform
eigenbasis. Acoustic time evolution is controlled by
$\mathsf T=\mathsf K+\operatorname{diag}(U,W)\mathsf M
\operatorname{diag}(U,W)$, with
$T_{nn}=\mathcal T_n/(n^2\Delta_M)$ and
$\det\mathsf T=J\mathcal T/(n^2N^2\Delta_M)$.
Over a connected circle or sphere of normals, a separated contact
has two real sounds on each side, and a nonpositive time eigenvalue
violates that count or makes time characteristic.
Thus $\mathsf T>0$ gives the other two signs.
Conversely, the three signs make $\mathsf Q(\sigma_c)$ and
$\mathsf T$ positive definite. These symmetrize the acoustic
quadratic uniformly on the compact normal set and, with the bounded
lift, give a complete seven-field warm or six-field cold
material basis, including semisimple double sounds.

\paragraph{Nonlinear causality.}
On the complete material branch, all sounds lie in $(-1,1)$
exactly when $\mathsf Q(\pm1)<0$.
Since $\mathsf Q(\pm1)=-\mathsf O\mathsf M\mathsf O$ and both
current frequencies are nonzero, this requires $\mathsf M>0$.
In primitive variables, $\Delta_M>0$, $X>0$, $Y>0$ and $C_\ell>0$.
The three EOS signs are equivalent to $\Sigma_c=X+Y>0$ and
$C_\ell>0$, and imply $J=C_\ell+nk_nY+Nk_hX+nNk_nk_h>0$.
In the EOS responses, set $d_n=\mu_n-nk_n$,
$d_h=\widehat\mu_p-Nk_h$ and
\begin{align}
 \Sigma_c&=d_n+d_h-\mathcal A(n+N),\notag\\
 C_\ell&=d_nd_h-\mathcal A(nd_n+Nd_h).
 \label{eq:main_model_b_causal_eos_scalars}
\end{align}
For $\mathcal A\ne0$, the material system is {strictly subluminal} and strongly
hyperbolic for all laboratory meridional directions exactly when
\begin{equation}
 {\Sigma_c>0,\qquad C_\ell>0,\qquad\mathcal C_M>0.}
 \label{eq:main_model_b_causal_material_combined}
\end{equation}
The first two bounds are linear in entrainment and exclude
light-speed sounds, since $\mathcal P_{\rm lab}(\pm1,s)=
\beta C_\ell\omega_\pm^2(Hg\omega_\pm-j_\pm)^2/(H^2N)$.
The last bound separates sounds from contacts. Since $R_n,R_p>0$,
it implies $\Delta_M>0$, $\mathcal T_n>0$ and, through
$R_nR_p-\mathcal A^2(J-gC_\ell)^2=C_\ell\Delta_M$, $\mathcal T>0$.
For all three-dimensional normals replace $\mathcal C_M>0$ by
$G_\parallel>0$, using
$\mathcal C_M|_{z_\phi=0,H_q=Q}=H^2JG_\parallel$.
{The closed cone also admits $C_\ell=0$.
Keep $X,Y\geq0$ and the contact exclusion, as stated in
Eq.~\eqref{eq:force_full_primitive_material_cone}.
The acoustic contact and time forms remain positive while the null
response becomes semidefinite, retaining complete luminal modes.}
Pinning has $H_q=0$, reducing contact separation to $\Delta_M>0$.
At three-dimensional coflow, $\Delta_M=Jk_nk_h$ already ensures it.
At $\mathcal A=0$, only the two EOS bounds remain in the material
sector: the independent Euler blocks retain their complete bases at
sound/contact coincidence.

\paragraph{Pressure coordinates.}
The pressure-field Jacobian at fixed $(z,\xi_i)$ is
$-T_c\Xi J_S/(3s\varepsilon_\xi H U)$, with carrier temperature
$T_c$. Either nonzero sign is a regular coordinate change.
Physicality does not require $J_S>0$. Use density fields at a fold.
For example, let $n=N=H=\mu_p=1$, $\mathcal A=1/5$, $k_n=1/2$,
$k_p=71/100$, $q=\sqrt{g^2-1}e_\phi$, $z=0$ and $r=0$. The
acoustic quartic $P_4$ in the speed $\lambda$ and the chart sign are
\begin{align}
 5000P_4&=(1368+1520g^2)\lambda^4\notag\\
 &\quad-(2414+304g+1500g^2)\lambda^2\notag\\
 &\quad+1046+304g-g^2,\notag\\
 J_S&=(59+16g-4g^2)/200.
 \label{eq:app_model_b_cold_chart_fold_family}
\end{align}
At $g=2+5\sqrt3/2$, the pressure chart loses one rank, but the
density system retains four simple subluminal sounds and four
independent contacts for meridional normals.
\endgroup

%% file: appendix_acoustic_coincidence_tension.tex
\begingroup
\subsection{Coflow and coincident material sounds}
\label{app:single_fluid_resonance}
At coflow the currents share a velocity, entropy moves with the
charged particles, and the material quartic becomes a quadratic in
the squared sound speed. We use $k_n=d\mu_n/dn>0$ and
$k_h=(\partial\widehat\mu_p/\partial N)_{s/N}>0$. For additive heat,
$\mu_h=\mu_p+4r/(3N)$ and $k_h=k_p+4r/(9N^2)$.
Put $H=\mu_n-\mathcal AN$, $I_h=\mu_h-\mathcal An$,
$h_n=H/n$, $h_h=I_h/N$ and $\delta_H=h_nh_h-\mathcal A^2$.
The longitudinal particle and Euler principal equations are
$\partial_t\delta n+\partial_x\delta j=0$ and
$\mathsf H\partial_t\delta j+\mathsf K\partial_x\delta n=0$, with
$\mathsf H=\left(\begin{smallmatrix}h_n&\mathcal A\\
\mathcal A&h_h\end{smallmatrix}\right)$ and
$\mathsf K=\operatorname{diag}(k_n,k_h)$.
The squared sounds are
\begin{equation}
 c_\pm^2=\frac{k_nh_h+k_hh_n
 \pm\sqrt{(k_nh_h-k_hh_n)^2+4\mathcal A^2k_nk_h}}
 {2\delta_H}.
 \label{eq:app_entrained_coflow_speeds}
\end{equation}
With $X=H-nk_n$ and $Y=I_h-Nk_h$, the closed-cone conditions on the
positive-density branch are $X\geq0$, $Y\geq0$ and
$XY\geq\mathcal A^2nN$, that is, $\mathsf H-\mathsf K\geq0$. They
include luminal sound modes and, with positive stiffnesses, imply positive
material inertia. The positive energy
$\operatorname{diag}(\mathsf K,\mathsf H)$ retains complete repeated
sounds.

For identical cold EOSs at equal density $n_\star$\revadd{, write
$\mu_\star=\mu(n_\star)$ and $k_\star=(d\mu/dn)_{n_\star}$}.
\revadd{Common and relative}
motion have squared speeds
\begin{equation}
 c_+^2=\frac{n_\star k_\star}{\mu_\star},\qquad
 c_-^2=\frac{n_\star k_\star}{\mu_\star-2\mathcal An_\star}.
 \label{eq:singlefluid_identical_speeds}
\end{equation}
At $\mathcal A=0$ they coincide at $c_{\rm ac}$ with independent
polarizations.
Common motion defines $e_{\rm sf}(n_B)=2e(n_B/2)$. Relative motion
and composition remain independent. Thus, equal speeds do not remove
a physical mode.
\endgroup

%% file: appendix_force_equilibrium_coincidences_tension.tex
\begingroup
\allowdisplaybreaks[1]
\section{{Compatibility at vortex--sound crossings}}
\label{app:force_equilibrium_coincidences}

A vortex speed can cross an acoustic speed without losing an
eigenvector if the forcing of the acoustic kernel vanishes.
Strong hyperbolicity further requires that the forced solution remain
bounded as the normal approaches the crossing.
We evaluate these conditions for the force-equilibrium branches of
Sec.~\ref{sec:force_equilibrium_conditions}.

For $Q>0$, the drag projections $K,S$ are defined in
Eq.~\eqref{eq:app_entrained_dissipative_projections}.
For reactive transport, use
$K=|a\times q|/|a+z_\phi v|$, $S=f^2K/\nu$,
$v=f^2q/\nu$ and $\nu=ng+\mathcal ZN$.
The coefficients in
Eq.~\eqref{eq:app_entrained_vortex_gap_coefficients} give
$G(x)=G_0+G_1x+G_2x^2$ on $0\leq x\leq1$.
The causal material branch has $G_0=n\Delta_M/J>0$.
Equation~\eqref{eq:force_summary_strict_gap} excludes all crossings.
To retain a crossing, evaluate each real root in $(0,1]$,
\begin{equation}
 x_\pm=\frac{-G_1\pm\sqrt{G_1^2-4G_0G_2}}{2G_2}.
 \label{eq:force_summary_gap_roots}
\end{equation}
For $G_2=0$ and $G_1\ne0$, the root is $x_*=-G_0/G_1$.
A nonzero constant $G$ has no crossing.

\paragraph{Reactive transport.}
Write the acoustic-source cofactors of
Eqs.~\eqref{eq:app_entrained_vortex_cofactors} and
\eqref{eq:app_entrained_reactive_source} as
$C_\alpha(x)=\sum_{j=0}^2C_{\alpha j}x^j$, with
$\alpha=n,N,\psi$.
Every retained root must satisfy
\begin{align}
 C_{\alpha0}+C_{\alpha1}x_*+C_{\alpha2}x_*^2&=0,\notag\\
 C_{\alpha1}+2C_{\alpha2}x_*&=0\quad\text{for a double root}.
 \label{eq:force_summary_reactive_coincidences}
\end{align}
The first equality removes the resonant forcing.
At a double root the derivative equality also removes the
singularity in its approach, as required by
Eqs.~\eqref{eq:force_preserving_coincidences} and
\eqref{eq:app_entrained_gap_cancellations}.
The protected $\mathcal A=\mathcal Z=0$ branch needs no gap test.

\paragraph{Finite drag.}
Drag permits at most one simple compatible crossing and excludes
a vortex meeting a double sound. At its root $x_*$, put
$y=S^2x_*$, $\rho=K/S$ and
$\beta=I_h-\mathcal A^2Nf^2$. Define
\begin{align}
 \Theta&=1+\frac{\mathcal Af^2(\mu-\rho)}\beta,\notag\\
 \Pi&=4Hg-\Theta(3\mu+\rho),\notag\\
 Z(y)&=(Hg-\rho)\Pi+3\mathcal AN\Theta(Hg-\rho y)\notag\\
 &\quad+\frac{4\mathcal A^2f^2N}{\beta}(Q-\rho^2y).
 \label{eq:force_summary_drag_coincidence_inputs}
\end{align}
The three affine compatibility relations are
\begin{align}
 (y-1)\Pi-4y(Hg-\rho)&=0,\notag\\
 \frac{H-nk_n}{H^2}Z(y)-3\mathcal AN\Theta+4nk_n&=0,\notag\\
 \frac{\mathcal An}{H^2}Z(y)+3N\Theta(k_h-\mathcal A^2f^2)
 &\notag\\[-2pt]
 \quad-3\beta\Theta y+4\mathcal Af^2[H(g-1)-\rho y]&=0.
 \label{eq:force_summary_drag_coincidences}
\end{align}
Two distinct crossings or a double gap root are inadmissible.
At zero entrainment,
Eqs.~\eqref{eq:force_summary_independent_reactive} and
\eqref{eq:force_summary_independent_drag} already exclude the
unprotected crossings.

The other coincidences preserve independent modes under the stated
material and geometric conditions. These include semisimple
coincident sounds, independent-fluid sound/contact crossings at
$\mathcal A=0$, and the pinned carrier/vortex coincidence.
For reactive transport, five carrier/vortex modes remain independent
at zero carrier-frame projection of $q$, as shown by
Eq.~\eqref{eq:app_model_b_general_reactive_full_factor}.
{The perpendicular-drag contact reduction}
\eqref{eq:app_model_b_dissipative_contact_minor} gives five modes on
that branch as well. Removing entropy leaves four on the cold
reactive branch.
\endgroup

%% file: appendix_force_equilibrium_full_symbol_tension.tex
\begingroup
\section{Primitive classification with the direction-dependent vortex pair}
\label{app:force_equilibrium_full_symbol}

We complete the force-equilibrium classification using scalar
primitive conditions. The EOS is that of the additive NC model with constant
$\mathcal A$, positive polytropic stiffnesses and a regular force
solution. Force coefficients may depend smoothly on matter but have
no explicit curl dependence. Work in density variables at a
pressure-chart fold. The time matrix must be regular, and $B^2>0$.
The conditions below are necessary and sufficient for a causal
spectrum and a uniformly complete real basis of the regular frozen
system on the selected normal sphere or meridional circle.
The material, force and coincidence conditions are imposed jointly.
They do not establish preservation of the admitted state set or
well-posedness of a pinning switch.

\paragraph{Eulerian vortex pair.}
With $\revadd{f^2=n/H}$, define the primitive coefficients
\begingroup\colorlet{revisionblue}{mainrevisionblue}\color{revisionblue}
\begin{align}
 a_v&=\frac{f^2A_Lq_\parallel}{A_L^2+C_L^2\Gamma_L^4},\qquad
 b_v=\frac{f^2C_L\Gamma_L^2q_\parallel}{A_L^2+C_L^2\Gamma_L^4},\notag\\
 L_z&=\ell\cdot z,\qquad T_v=W+v_0\cdot z,\notag\\
 U_s&=\left[z+v_0+\frac{z(v_0\cdot z)}{W+1}\right]\cdot s,\notag\\
 L_s&=\left[\ell+\frac{z(\ell\cdot z)}{W+1}\right]\cdot s,\notag\\
 A_v&=T_v+a_vL_z,\qquad B_v=U_s+a_vL_s,\notag\\
 D_v&=A_v^2+b_v^2L_z^2>0.
 \label{eq:force_full_eulerian_vortex_inputs}
\end{align}
\endgroup
Here $v_0,q,\ell$ are carrier-frame vectors.
Boosting the wave covector gives
$\omega'=W\sigma-z\cdot s$ and
$k'=s+z(z\cdot s)/(W+1)-\sigma z$.
The Eulerian vortex polynomial, speeds and discriminant are
\begin{align}
 P_v(\sigma,s)&=(A_v\sigma-B_v)^2
                       +b_v^2(L_s-L_z\sigma)^2,\notag\\
 \sigma_{v\pm}&=\frac{A_vB_v+b_v^2L_zL_s
               \pm i b_v(T_vL_s-U_sL_z)}{D_v},\notag\\
 \operatorname{disc}_{\sigma}P_v
   &=-4b_v^2(T_vL_s-U_sL_z)^2.
 \label{eq:force_full_eulerian_vortex_pair}
\end{align}
Thus, for one real normal, vortex reality and cone containment are
exactly
\begin{align}
 b_v(T_vL_s-U_sL_z)&=0,\notag\\
 |A_vB_v+b_v^2L_zL_s|&\leq D_v.
 \label{eq:force_full_vortex_cone}
\end{align}
These two signs do not test completeness. The tangent Jordan block
\eqref{eq:app_model_b_tangent_vortex_jordan} can obey both.
The all-direction primitive reductions follow below.

\paragraph{Material conditions.}
Use $H,I_h,J,X,Y,C_\ell,\Delta_M,R_n,R_p,\mathcal C_M$ of
Eq.~\eqref{eq:main_model_b_material_scalars} and $n,N,H,J>0$.
The closed metric cone is equivalent to
\begin{equation}
 X=H-nk_n\geq0,\quad Y=I_h-Nk_h\geq0,\quad
 XY\geq\mathcal A^2nN.
 \label{eq:force_full_primitive_material_cone}
\end{equation}
For $\mathcal A\ne0$, additionally require
$\mathcal C_M>0$ in axisymmetry, or $G_\parallel>0$ for the full
sphere. For $\mathcal A=0$, this contact exclusion is absent.
Each inequality is affine in either stiffness separately.
The nonstrict cone signs retain complete luminal acoustic modes.
Indeed, the causal response is positive semidefinite while the
contact form and time form remain positive definite. The acoustic
energy therefore remains positive at the null boundary.
The joint cone and contact conditions imply $\Delta_M>0$, so no
density-response denominator vanishes in these reductions.

\paragraph{Pinned, reactive and dissipative branches.}
At $Q=0$, take the direct coflow limit, which needs no additional
vortex gap. The selected force branch must satisfy the following
geometric and cone conditions:
\begin{itemize}
 \item Ideal pinning requires $Q=0$ for the full sphere or
 $|a\times q|^2=0$ in axisymmetry. Its vortex speed equals the
 carrier speed, and the material conditions complete the basis.
 The holding-load inequality remains a separate force-admission test.
 \item Reactive transport has $\mathcal D=0$, $\nu>0$ and
 $\revadd{\eta=f^2/\nu}$. Its necessary and sufficient vortex cone bound is
 \begingroup\colorlet{revisionblue}{mainrevisionblue}\color{revisionblue}
\begin{equation}
 \begin{cases}
 \nu\geq f^2\sqrt Q,&\text{3D},\\
 (1+z_\phi^2)\nu+f^2z_\phi(a\cdot q)
                 \geq f^2|a\times q|,&\text{axisym.}
 \end{cases}
 \label{eq:force_full_primitive_reactive_cone}
 \end{equation}
\endgroup
 These are linear lower bounds on $\mathcal Z$ because
 $\nu=ng+\mathcal ZN$. They are automatic for $\mathcal Z\geq0$.
 The source cancellation tests below still apply at a luminal
 vortex/sound coincidence.
 \item Dissipative counterflow has $\mathcal D>0$.
 No $Q>0$ state has a complete basis for every three-dimensional
 normal. The necessary and sufficient meridional geometry is
 \begin{equation}
 q\cdot\ell=0,\qquad a\cdot\ell=0,\qquad
 v_0\cdot(\ell\times a)\ne0.
 \label{eq:force_full_primitive_drag_geometry}
 \end{equation}
 Its vortex cone is already strict because $|v_0|<1$.
 A closed cold system excludes this heating branch.
\end{itemize}
Here $a=e_\phi+z_\phi z/(W+1)$.
The force-admission and heating signs in
Secs.~\ref{app:force_equilibrium_condition_reductions} and
\ref{app:model_b_finite_vortices} remain required.

\paragraph{A simpler vortex/sound gap.}
For reactive transport use $K=\sqrt Q$, $S=\eta K$ in three
 dimensions, and
$K=|a\times q|/|a+z_\phi\eta q|$, $S=\eta K$ in axisymmetry.
For the admitted dissipative geometry use
Eq.~\eqref{eq:app_entrained_dissipative_projections}.
Set $t_n=gS-K/H$ and define
\begin{align}
 B_g&=\frac{I_hR_n}{n}t_n^2+\frac{HR_p}{N}S^2
       -2\mathcal A^2(J-gC_\ell)t_nS,\notag\\
 C_g&=\frac{C_\ell J}{nN}t_n^2S^2,\notag\\
 \mathcal G_v(x)&=\Delta_M(1-S^2x)^2
                  -B_gx(1-S^2x)+C_gx^2,\notag\\
 \mathcal L_v&=nH\Delta_M(1-S^2)-JR_nt_n^2.
 \label{eq:force_full_primitive_gap_inputs}
\end{align}
The quadratic $\mathcal G_v$ equals $(J/n)G$ from
Eq.~\eqref{eq:app_entrained_vortex_gap_coefficients}.
A strict gap for every allowed normal is equivalent to just
\begin{equation}
 \mathcal L_v>0,\qquad\mathcal G_v(1)>0.
 \label{eq:force_full_primitive_gap_endpoints}
\end{equation}
Both are affine in either $k_n$ or $k_h$ separately. No angular
search or quartic root is needed.
To see the equivalence, evaluate the acoustic form at
$\omega=St$, $j=Kt$ and write $x=t^2$.
It is affine in $x$ and starts positive at $x=0$.
The causal response makes its derivative negative semidefinite.
Positivity for all $x\in[0,1]$ is therefore equivalent to
positivity at $x=1$. Its first principal pivot has the sign of
$\mathcal L_v$, and its determinant has the sign of
$\mathcal G_v(1)$. This also proves equivalence with the quadratic minimum test
following Eq.~\eqref{eq:force_preserving_gap_endpoints}.

\paragraph{Compatible coincidences and the complete criterion.}
A strict gap is not necessary at a protected crossing.
If Eq.~\eqref{eq:force_full_primitive_gap_endpoints} fails, compute
the at most two roots of $\mathcal G_v$ in $(0,1]$.
They are the roots \eqref{eq:app_entrained_gap_roots}, since the
normalization is positive. For reactive transport, all three
primitive source cofactors must vanish at every root, with their
first derivatives vanishing at a double root
[Eq.~\eqref{eq:force_summary_reactive_coincidences}].
For finite drag, retain at most one simple root and impose the
three affine equalities \eqref{eq:force_summary_drag_coincidences}.
These cancellations remove the full pole, including its approach,
and are necessary and sufficient for a bounded coupled completion.
They test uniform hyperbolicity, not just the kernel at one normal
\cite{Abalos:2018uwg}.

Thus the material conditions, the selected force geometry and cone
bound, and either the two strict gap signs or all required source
cancellations form a necessary and sufficient primitive list.
At $\mathcal A=\mathcal Z=0$, reactive transport has protected
crossings and needs no gap condition.
At $\mathcal Z=0$, $\mathcal A\ne0$, the gap reduces to the single
bound \eqref{eq:force_summary_reactive_hyperbolicity} in axisymmetry.
At zero entrainment, use the simpler separate constituent bounds
\eqref{eq:force_summary_independent_reactive} or
\eqref{eq:force_summary_independent_drag} as appropriate.
Their dependence on each stiffness is linear.

\paragraph{Why the direction-dependent cases are excluded.}
The vortex discriminant forces $\mathcal Dq_\parallel=0$ for all
three-dimensional normals. A meridional family with
$\mathcal Dq_\parallel\ne0$ can have real repeated speeds, but its
tangent response is defective.
Even when $q_\parallel=0$, a complete vortex block does not suffice
for the coupled system. At a shared carrier/vortex speed, choose a
carrier-frame normal with $v_0\cdot\hat k=0$ and
$q_s=q\cdot\hat k\ne0$. \revadd{Let $\delta J_n,\delta J_c$ be
the spatial number-current perturbations, $\delta\Xi$ the phase
energy perturbation in this carrier frame, and $\delta\Psi_m$ the
material pressure perturbation.} Number and stress conservation give
\begingroup\colorlet{revisionblue}{mainrevisionblue}\color{revisionblue}
\begin{align}
 \delta J_c\cdot\hat k&=\delta J_n\cdot\hat k=0,\notag\\
 f^2q_s\delta\Xi&=0,\notag\\
 \delta\Psi_m\hat k+f^2q_s\delta\xi&=0.
 \label{eq:force_full_perpendicular_contact_obstruction}
\end{align}
\endgroup
Both transverse canonical polarizations are removed.
Such normals exist for every dissipative $Q>0$ state in three
 dimensions, giving the necessary condition $\mathcal D^2Q=0$.
In axisymmetry, Eq.~\eqref{eq:force_full_primitive_drag_geometry}
ensures that the shared normal is parallel to $\ell$ and that the
ratio of the carrier and vortex approaches remains bounded.
Equation~\eqref{eq:app_model_b_dissipative_contact_minor} then gives
five independent warm contacts.
A dissipative coflow point is regular as a frozen symbol, but it
does not supply an open three-dimensional domain of dissipative
counterflow. Local evolution requires the conditions throughout a
smooth preserved branch and the symmetrizer regularity stated in
Sec.~\ref{sec:hyperbolicity}.
\endgroup

%% file: appendix_force_balance_electric_tension.tex
\begingroup
\section{{Electric variables in force equilibrium}}
\label{app:tension_force_equilibrium_transport}
\label{app:tension_legacy_domain}
\label{sec:force_drift_electric}
We now ask whether evolving the electric field can change the
characteristic structure of the force-equilibrium closure, without
line tension. The primitive force-law velocity is
$a_L(\mathcal V,B)=\widetilde{\boldsymbol{\mathcal U}}_v/
\widetilde{\mathcal U}_v^0$, with $a_L\cdot B=0$. Here $\mathcal V$
denotes the matter primitives and the numerator is
Eq.~\eqref{eq:moments_relativistic_line_velocity}. The future-timelike force solution
has $|a_L|<1$. Exact force balance and the line kernel require
\begin{equation}
 E=-a_L(\mathcal V,B)\times B.
 \label{eq:force_drift_matching_exact}
\end{equation}
Consequently, $E\cdot B=0$ and $B^2>E^2$ on the regular
nonzero-curl branch.
\endgroup

%% file: appendix_vortex_electric_tension.tex
\begingroup
\subsection{Force-preserving electric evolution}
\label{app:vortex_electric_evolution}
\label{app:force_preserving_evolution}
\label{app:force_preserving_symbol}
Differentiating Eq.~\eqref{eq:force_drift_matching_exact} with all
state feedback in $\partial_ta_L$, and using induction on
$\nabla\cdot B=0$, gives
\begin{align}
 \partial_tE_i+\partial_j(a_L^jE_i)
 &=E_i\partial_ja_L^j-E_j\partial_i a_L^j\notag\\
 &\quad-\epsilon_{ijk}(\partial_ta_L^j)B^k.
 \label{eq:force_electric_flux_balance}
\end{align}
For a primitive variation, the paired force roots of
Eq.~\eqref{eq:moments_relativistic_line_velocity} obey
\begin{align}
 \delta\alpha_v&=
  \frac{\alpha_v\delta C_v+2\lambda_v\delta p_v}{2H_v},\notag\\
 \delta\lambda_v&=
  \frac{2\alpha_v\delta p_v-\lambda_v\delta C_v}{2H_v},\notag\\
 \delta a_L&=
  \frac{\delta\widetilde{\boldsymbol{\mathcal U}}_v
              -a_L\delta\widetilde{\mathcal U}_v^0}
       {\widetilde{\mathcal U}_v^0},
 \label{eq:force_electric_primitive_derivatives}
\end{align}
where $H_v=\alpha_v^2+\lambda_v^2>0$ on the regular branch.
For an externally prescribed smooth timelike
$a_L(t,x)$ the divergence-completed system is symmetric hyperbolic
and causal in three dimensions. For a state-dependent force velocity
its source contains principal derivatives, which cannot be dropped
or, for a general nonlinear force law, absorbed into a local flux.
With $V^i=\alpha a_L^i-\beta^i$ and
$\Omega_{ij}=[ijk]\sqrt\gamma B^k$, the force residual
$C_i=E_i+\Omega_{ij}a_L^j$ is then transported,
$(\partial_t+\mathcal L_V)C_i=0$, and the augmented principal matrix
is similar to
$[\alpha A_{\rm force}(\hat n)-(\beta\cdot\hat n)I]
\oplus(V\cdot\hat n)I_C$.
Thus, the electric extension preserves exact force balance and the
symmetrizer supplied by the original primitive conditions, but a
complex or defective physical mode also survives it.
In particular, at carrier rest with $q=q_\parallel\ell$, $E_0=0$ and
$B_0=B\ell$, differentiating the force solution gives
$\delta E_\perp=-b_d\delta B_\perp-a_d\ell\times\delta B_\perp$ with
$a_d=h_vq_\parallel$ and $b_d=d_vq_\parallel$, and Faraday evolution
has the speeds $a_d\pm ib_d$.
Both perturbations obey $E\cdot B=0$ and $\nabla\cdot B=0$, so no
algebraic projection removes this pair. The allowed states therefore
satisfy the same three-dimensional or restricted meridional
conditions as the original force closure.

\subsection{Independent drift-current model}
\label{app:independent_force_drift_test}
\label{app:force_drift_spectrum}
\label{app:force_drift_coincidences}
\label{app:force_drift_invariance}
An independent electric predictor with only the $E\cdot B$
projection defines a different model. With $\rho_\Omega=\revadd{\nabla_i}E^i$ and
$v_E=E\times B/B^2$, it evolves $\partial_tB=-\nabla\times E$ and
$\partial_tE=\nabla\times B-a_L\nabla\cdot E$,
followed by the reset $E^+=E_*-(E_*\cdot B_*)B_*/B_*^2$ at fixed
$B^+=B_*$ (compare the stiff driver and algebraic reset of
Ref.~\cite[Sec.~2.3 and Appendix~A]{Alic:2012df}).
On the physical CT subspace $\hat n\cdot\delta B=0$, the projected
characteristic polynomial is $(\sigma^2-1)P_D(\sigma)$, with
$P_D(\sigma)=(\sigma-a_n)(\sigma-v_n^E)-c_D(b\cdot\hat n)^2$,
$b=B/|B|$, $a_n=a_L\cdot\hat n$, $v_n^E=v_E\cdot\hat n$ and
$c_D=1-a_L\cdot v_E$.
For all three-dimensional normals, the roots are real exactly when
$c_D\geq0$.
On $B^2>0$, $E\cdot B=0$ and the strictly timelike force branch,
each of the following frozen-field properties is equivalent to
$a_L=v_E$, the matching in
Eq.~\eqref{eq:force_drift_matching_exact}: (i) all physical speeds
lie in the closed metric cone, and (ii) the physical principal symbol is
strongly hyperbolic for the full normal sphere.
The matching locus has empty interior in the state space constrained
only by $E\cdot B=0$, and the predictor leaves it even from
force-matched smooth data. For a stationary coflow shear
$u^\mu=\gamma(1,0,V(x),0)$ with constant proper densities and
chemical potentials and $V=\omega_0x+d_0x^2$, $\omega_0>0$,
$d_0\ne0$, it gives
$\partial_t(E+a_L\times B)=-2\mu_nd_0e_2\ne0$ at the origin,
perpendicular to $B$ and thus untouched by the projection. Thus, the weaker projection supplies no preserved open
three-dimensional causal, strongly hyperbolic domain, and a
counterflow bound cannot replace the missing matching condition.
Resetting the entire field to $E=-a_L\times B$ imposes matching but
returns the instantaneous force closure with its complex pair.
\endgroup

%% file: appendix_model_a_force_equilibrium_tension.tex
\begingroup
\allowdisplaybreaks[1]
\section{{CP model: contact conditions and the causal cone}}
\label{app:model_a_force_equilibrium}

For the CP model, phase and entropy potentials replace the two material
potentials in the acoustic reduction. We analyze two prescribed
transport choices: pinning to the entropy carrier and free phase
transport at $q\cdot\hat k/\mu$. Pinning uses the geometry in
Eq.~\eqref{eq:main_model_b_pinned_primitive}. The calculation below
tests the characteristics of these choices. It does not derive a
microscopic pinning or drag prescription for the CP model.
We first require regular time evolution and independent contact
modes, then bound the acoustic response on the metric cone.

\subsection{Phonon pressure and its derivatives}
\label{app:model_a_finite_energy}
For $\Gamma>1$ and $\revadd{\kappa_p>0}$, the warm phonon branch has
$e>1$, $T>0$, $\mu\geq e$ and $\Delta>0$.
Its pressure and counterflow dependence are
\begingroup
\begin{align}
 \chi_\Gamma&=\Gamma-1+(2-\Gamma)e,\quad
 \Delta=e^3-\chi_\Gamma\mu^2,\ \ Q=\mu^2-e^2,\notag\\
 B&=e^3(8e-9)+[(3\Gamma-2)e-3(\Gamma-1)]\mu^2,\notag\\
 p&=\frac{27T^4\mu^4e^{3/2}\sqrt{(\Gamma-1)(e-1)}}{256\Delta^2}.
 \label{eq:app_model_a_general_pressure}
\end{align}
\endgroup
Differentiation gives the coefficients
\begingroup
\begingroup\colorlet{revisionblue}{mainrevisionblue}\color{revisionblue}
\begin{align}
 \alpha_\Gamma&=\frac{\Gamma-1}{\Gamma\kappa_p^{1/\Gamma}},\qquad
 R=\frac{4e^3}{\mu\Delta},\notag\\
 Y_0&=\frac{\Gamma\alpha_\Gamma^{\Gamma/(\Gamma-1)}}{e(\Gamma-1)}
                      (e-1)^{1/(\Gamma-1)},\notag\\
 Y_{0e}&=Y_0\left[\frac1{(\Gamma-1)(e-1)}-\frac1e\right],\quad
 y=\frac{B}{2e^2(e-1)\Delta},\notag\\
 \Delta_e&=3e^2-(2-\Gamma)\mu^2,\quad
 R_e=R(3/e-\Delta_e/\Delta),\notag\\
 R_m&=R(-1/\mu+2\chi_\Gamma\mu/\Delta),\quad \ell=R-\mu y,\notag\\
 y_e&=y\left[\frac{e^2(32e-27)+(3\Gamma-2)\mu^2}{B}
 -\frac2e\right]\notag\\
 &\quad-y\left[\frac1{e-1}+\frac{\Delta_e}{\Delta}\right],\notag\\
 y_m&=\frac{2y\mu[(3\Gamma-2)e-3(\Gamma-1)]}{B}
 +\frac{2y\chi_\Gamma\mu}{\Delta}.
 \label{eq:app_model_a_finite_energy_coefficients}
\end{align}
\endgroup
\endgroup
The subscripts $e$ and $m$ denote responses to $e$ and $\mu$.
Define $Y=Y_0-yp>0$, $c=4+\mu R$, $D=R^2/4-R_m$ and $L=c-QD$.
The physical EOS has $B,c,L>0$.

\subsection{Time coefficients and contact resonances}
\label{app:model_a_full_system}
Let $A_e=Y_{0e}+p(ey^2-y_e)$ and define the acoustic functions
\begingroup
\begin{align}
 d&=\mu\omega-j,\quad a_\omega=R_ed/e-D\omega,\notag\\
 b_\omega&=j+QR\omega/4,\notag\\
 A(\omega,j,m)&=Y(m-\omega^2)-A_ed^2/e\notag\\
 &\quad+pD\omega^2-2pR_ed\omega/e+pQa_\omega^2/L,\notag\\
 B(\omega,j,m)&=\omega(dy-R\omega/4)+a_\omega b_\omega/L,\notag\\
 C(\omega,j,m)&=m/c-3\omega^2/4+jR\omega/(4c)\notag\\
 &\quad+(R\omega/4+jD/c)b_\omega/L.
 \label{eq:app_model_a_acoustic_scalar_functions}
\end{align}
\endgroup
Here $\omega,j$ are the projections of
Eq.~\eqref{eq:app_model_b_general_lab_primitives}, evaluated with
the CP phase variables. For $m=1-\sigma^2+\omega^2$, the
acoustic quartic is $P_4(\sigma)=4ecL[AC-pB^2]$.
Its roots follow from Sec.~\ref{app:finite_flow_sounds}.

At $(\omega,j,m)=(W,-q\cdot z,W^2-1)$, define
\begingroup
\begin{equation}
 \mathcal T_A=-A,\qquad \mathcal D_A=AC-pB^2.
 \label{eq:app_model_a_time_scalars}
\end{equation}
\endgroup
Both signs must be positive for the acoustic time form to be positive.
At the carrier contact the remaining scalar is
\begingroup
\begin{equation}
 \mathcal C_A(x)=
 \frac{(L+Dx)(Y-A_ex/e)+pR_e^2x(Q-x)/e^2}{cL}.
 \label{eq:app_model_a_carrier_contact_scalar}
\end{equation}
\endgroup
At the phase contact the corresponding functions are
\begingroup
\begin{align}
 A_\phi(x)&=Y(1-x)+pDcx/L,\notag\\
 B_\phi(x)&=-x(Rc/4+\mu D)/L,\notag\\
 C_\phi(x)&=1/c+x[-3/4+\mu R/(4c)\notag\\
 &\qquad+(R/4+\mu D/c)(\mu+QR/4)/L],\notag\\
 \mathcal P_A(x)&=A_\phi(x),\qquad
 \mathcal R_A(x)=A_\phi(x)C_\phi(x)-pB_\phi(x)^2.
 \label{eq:app_model_a_phase_contact_scalars}
\end{align}
\endgroup
The squared projections range to $X_c=Q$ and $X_\phi=Q/\mu^2$
for the full normal sphere. In axisymmetry their endpoints are
\begingroup
\begin{equation}
 X_c=\frac{H_q}{1+z_\phi^2},\qquad
 X_\phi=\frac{H_q}{\mu^2|a+z_\phi q/\mu|^2}.
 \label{eq:app_model_a_contact_intervals}
\end{equation}
\endgroup
where $a=e_\phi+z_\phi z/(W+1)$ and $H_q=|a\times q|^2$.

The pinned eight-field system has a complete characteristic basis
if and only if
\begingroup
\begin{equation}
 Q=0\ (\mathrm{3D})\ \text{or}\ H_q=0\ (\mathrm{axisym.}),
 \qquad\mathcal T_A>0,\quad\mathcal D_A>0.
 \label{eq:app_model_a_exact_pinned}
\end{equation}
\endgroup
At the shared carrier/vortex speed, transverse forcing
$Yq_k\delta\xi_T/p$ requires $q_k=0$.
For free phase transport, set $S_\Gamma=3(\Gamma-1)+2e(2-\Gamma)$.
The eight-field completeness conditions are
\begingroup
\begin{align}
 &\mathcal T_A>0,\quad\mathcal D_A>0,\quad
 \mathcal P_A(X_\phi)>0,\quad\mathcal R_A(X_\phi)>0,\notag\\
 &\mathcal C_A(X_c)>0\quad\text{unless }S_\Gamma=0.
 \label{eq:app_model_a_exact_free}
\end{align}
\endgroup
To obtain them, use the symmetric acoustic form
$\mathsf Q=\left(\begin{smallmatrix}A&pB\\pB&pC\end{smallmatrix}\right)$.
At each contact it is affine in the squared projection and starts
from $\operatorname{diag}(Y,p/c)>0$.
For $S_\Gamma\ne0$, the derivative
$y_m=4e\mu S_\Gamma/\Delta^2$ couples a sound to the carrier contact.
When $S_\Gamma=0$, the polarization
$(\delta e,\delta T/T,\delta u_T,\delta\xi)=(0,Rq_T/4,1,0)$
provides a smooth independent contact mode.
It removes the carrier-contact exclusion on that sheet.
The phase-vortex gap remains necessary.

\subsection{Causal acoustic response}
The metric-cone conditions follow by setting the wave covector null.
Define
\begingroup
\begin{align}
 N_T(t)&=-A(1,t,1)-Y_0C(1,t,1),\notag\\
 N_D(t)&=A(1,t,1)C(1,t,1)-pB(1,t,1)^2.
 \label{eq:app_model_a_null_scalars}
\end{align}
\endgroup
On either complete warm branch, strict causality is equivalent to
$N_T,N_D>0$. In three dimensions the angular parameter ranges over
$I=[-\sqrt Q,\sqrt Q]$. Its meridional interval is
\begingroup
\begin{equation}
 I=\left[\frac{-z_\phi(a\cdot q)-\sqrt{H_q}}{1+z_\phi^2},
          \frac{-z_\phi(a\cdot q)+\sqrt{H_q}}{1+z_\phi^2}\right]
 \label{eq:app_model_a_null_interval}
\end{equation}
\endgroup
The polynomials have degrees at most two and four.
Their minima occur at interval endpoints or interior stationary
points. If $\mathcal E_i$ contains those points, causality reduces to
\begingroup
\begin{equation}
 N_i(t)>0\qquad(i=T,D{,}\ t\in\mathcal E_i),
 \label{eq:main_model_a_causal_gates}
\end{equation}
\endgroup
The cubic formulas in
Eqs.~\eqref{eq:app_scalar_cubic_parameters}--\eqref{eq:app_scalar_cubic_values}
give the stationary points. The signs $N_T,N_D$ are the trace and
determinant conditions for positive $-\mathsf Q$ at null wave
covectors. With a positive contact form, this causal response also
implies the positive time coefficients.

\subsection{Coflow and the thermal-pressure bound}
\label{app:model_a_characteristics}
For $\Gamma=2$, differentiate the phonon pressure before setting
$q=0$. Write
$\revadd{\theta=27\kappa_p T^4\sqrt e/[256(e-1)^{7/2}]}$ and
$L=2e+3\theta(4e-3)$. The squared-speed sum and product are
\begingroup
\begin{align}
 \Sigma&=\frac{(e-1)[2(7e-3)+3\theta(7+24e-16e^2)]}
 {3(2e-1)L},\notag\\
 \Pi&=\frac{2(e-1)^2[1-(8e+3)\theta]}{3(2e-1)L}.
 \label{eq:app_model_a_sound_coefficients}
\end{align}
\endgroup
The four acoustic speeds are
$\lambda=\pm[(\Sigma\pm\sqrt{\Sigma^2-4\Pi})/2]^{1/2}$.
The larger and smaller magnitudes give first and second sound.
The four zero-speed modes consist of two carrier shears and two
transverse canonical momenta. Physical admission, strong
hyperbolicity and causality jointly reduce to
\begingroup
\begin{equation}
 {e>1,\quad T>0,\quad
 (8e+3)\theta<1.}
 \label{eq:app_model_a_coflow_domain}
\end{equation}
\endgroup
The thermal bound is equivalent to $f^2>0$.
Within it, $\Pi>0$, $\Sigma>0$, $\Sigma^2-4\Pi>0$,
$1-\Sigma+\Pi>0$ and $\Sigma<2$.
Finite counterflow still requires the contact and null-response
conditions above.

\subsection{Cold free-phase limit}
\label{app:model_a_cold_characteristics}
At $T=0$, the entropy carrier disappears, leaving four phase-fluid
fields. In the Eulerian frame, use $\mu=\Xi$,
$q=\boldsymbol\xi$, $t=q\cdot\hat k$, $Q=\mu^2-e^2$,
$c_s^2=(\Gamma-1)(e-1)/e$ and $D_c=\mu^2-c_s^2Q$.
The acoustic speeds are
\begingroup
\begin{equation}
 \lambda_\pm=\frac{(1-c_s^2)\mu t
 \pm c_se\sqrt{D_c-(1-c_s^2)t^2}}{D_c}.
 \label{eq:app_model_a_cold_speeds}
\end{equation}
\endgroup
The two remaining modes advect canonical vorticity at $t/\mu$.
On $e>1$, $\revadd{\kappa_p>0}$ and $\Gamma>1$, the first line below is
equivalent to strong hyperbolicity. The second line includes the
additional strict causal bound.
\begingroup\colorlet{revisionblue}{mainrevisionblue}\color{revisionblue}
\begin{align}
 &D_c>0\quad\Longleftrightarrow\quad
 e\mu^2>(\Gamma-1)(e-1)Q,
 \label{eq:app_model_a_cold_strong_domain}\\
 &e>1,\quad\kappa_p>0,\quad\Gamma>1,\quad
 (\Gamma-2)e<\Gamma-1.
 \label{eq:app_model_a_cold_domain}
\end{align}
\endgroup
The time determinant is $Y_0D_c/(ec_s^2\mu)$.
For $D_c<0$, transverse sound speeds are imaginary.
For $D_c>0$, the radicand lies between $D_c$ and $e^2$, while
$e^2$ bounds the vortex-lift denominators from below.
The causal bound ensures $D_c>0$ and places the vortex speed
strictly between the sounds. Pinning in the absence of an entropy
carrier requires a separately specified medium.
\endgroup

%% file: appendix_regulated_characteristics_tension.tex
\begingroup
\allowdisplaybreaks[1]
\newcommand{\regcI}{\mathbb I}
\section{{Line inertia, coflow and positive wave energy}}
\label{app:regulated_characteristics}

The density-dependent line energy changes the material momenta and
adds inertia for independent array motion. We obtain these responses
from the current Hessian, evaluate the coflow compression problem,
and then construct a sufficient positive-energy domain at finite
relative motion. In each case the mixed derivatives of $F$ enter
the material--array coupling.

\subsection{Current and line responses}
\label{sec:regc_branches}
Use $n=\nu_n$, $N=n_p$, $\varsigma=s_c/N$,
$j_n^a=nu_n^a$ and $j_c^a=Nu_c^a$.
\revadd{The circulation quantum is $\kappa_v$, and the line-compression
coefficient is $\alpha_{\rm v}$, as in Eq.~\eqref{eq:tension_draft_master}.}
For the material EOS \eqref{eq:regc_material_class}, adding the
line energy gives $\Lambda=\Lambda_{\rm mat}-F$ and the responses
\begin{align}
 h_n&=\mu_n+F_n-\mathcal A N,&
 h_c&=\mu_h+F_N-\mathcal A n,\notag\\
 B_n&=h_n/n,& B_c&=h_c/N,\notag\\
 k_n^*&=k_n+F_{nn},& k_c^*&=k_c+F_{NN},\notag\\
 a_n&=B_n-k_n^*,&a_c&=B_c-k_c^*,\notag\\
 r&=\mathcal A-F_{nN}.&&
 \label{eq:regc_scalar_responses}
\end{align}
Here $\mu_h=\widehat\mu_p$ and $k_c=k_h$.
All fluid-density derivatives hold $\mathcal N$, $\varsigma$,
$\revadd{\alpha_{\rm v}}$ and $\mathcal N_0$ fixed and include the density dependence
of the line coefficient. Appendix~\ref{app:tension_constitutive_derivatives}
evaluates the required first, mixed and second derivatives.
The EOS and remaining material derivatives are specified in
Sec.~\ref{app:force_equilibrium_condition_reductions}.

The proper array density is
$\revadd{\mathcal N=\sqrt{B^2-E^2}/\kappa_v>0}$, with $E\cdot B=0$ and
$B^2>E^2$. \revadd{In the vortex rest frame, decompose each material
four-velocity as $u_x'=(\gamma_x,U_x)$, with
$\gamma_x=\sqrt{1+|U_x|^2}$.} In the vortex rest frame introduced below, write the
line-corrected neutron momentum as
$\Theta_n=h_n\gamma_n+\mathcal A N\gamma_c$ and
$P_n=h_nU_n+\mathcal A NU_c$.
Future timelike phase momentum and positive electric inertia require
\begingroup\colorlet{revisionblue}{mainrevisionblue}\color{revisionblue}
\begin{align}
 \Theta_n&>0,\qquad \Theta_n^2-|P_n|^2>0,
 \label{eq:regc_phase_branch}\\
 a_\Omega&=\frac{F_{\mathcal N}}{\kappa_v^2\mathcal N}>0.
 \label{eq:regc_electric_coefficient}
\end{align}
\endgroup
If the EOS also requires a future timelike carrier momentum, impose
the first two inequalities with the species interchanged.
Positive $h_n,h_c$ and $\mathcal A\geq0$ imply these momentum
conditions. Negative entrainment requires their direct evaluation.

For a general smooth master function
$\Lambda(n,N,I,\varsigma,\mathcal N)$, differentiating the invariants
$q_r\in\{n,N,I\}$ gives the current Hessian
\begin{align}
 \Lambda_{,\revadd{j_x^a}\revadd{j_y^b}}
 &=\sum_{r,t}\Lambda_{,q_rq_t}
      q_{r,\revadd{j_x^a}}q_{t,\revadd{j_y^b}}
 +\sum_r\Lambda_{,q_r}q_{r,\revadd{j_x^a}\revadd{j_y^b}},\notag\\
 n_{,j_n^a}&=-u_{na},\qquad
 n_{,j_n^aj_n^b}=-(g_{ab}+u_{na}u_{nb})/n,\notag\\
 I_{,j_n^a}&=-j_{ca},\qquad
 I_{,j_n^aj_c^b}=-g_{ab}.
 \label{eq:regc_current_hessian}
\end{align}
Species exchange gives the carrier-density derivatives.
The remaining mixed invariant derivatives vanish.
The full master function supplies the derivatives for thermal
counterflow, including the entropy force.

\subsection{Vortex rest frame and regular time evolution}
\label{sec:regc_frame}
Magnetic dominance permits a local Lorentz boost to $E'=0$.
The drift velocity and transformation of a four-vector $Q^\mu$ are
\begin{align}
 v_d&=\frac{E\times B}{B^2},\qquad
 \gamma_d=(1-v_d^2)^{-1/2},\notag\\
 Q'^0&=\gamma_d(Q^0-v_d\cdot\boldsymbol Q),\notag\\
 \boldsymbol Q'&=\boldsymbol Q+
 \left[\frac{\gamma_d^2}{\gamma_d+1}
            v_d\cdot\boldsymbol Q-\gamma_dQ^0\right]v_d.
 \label{eq:regc_drift_frame}
\end{align}
The transformation is regular at $v_d=0$ and gives
$\revadd{B'=\kappa_v\mathcal N\widehat b}$. Apply it to both material currents
and write $u_x=(\gamma_x,U_x)$ and $v_x=U_x/\gamma_x$.
The warm primitive system has fourteen fields: eight current
components, entropy per carrier, two transverse electric components
and three magnetic components. The cold system omits entropy.

In this frame the primitive time matrix is invertible if and only if
$\det K\ne0$ and $a_\Omega\ne0$, where $K$ is defined below.
A regular transformation to conserved variables preserves the
characteristics. Replacing entropy by total energy also requires a
nonzero temperature response.

\input{appendix_coflow_hyperbolicity_tension}

\subsection{Response matrices at finite relative motion}
\label{sec:regc_domain}
For $z=(j_n'^0,j_c'^0,B'_\parallel)$ and
$w=(\boldsymbol j_n',\boldsymbol j_c')$, the additive EOS gives
kinetic block $K$, mixed block $C$ and density block
$M=-\Lambda_{zz}$ in the form
\begingroup\colorlet{revisionblue}{mainrevisionblue}\color{revisionblue}
\begin{align}
 K&=K_0\otimes\regcI_3+VRV^T,\qquad C=D_0V^T,\notag\\
 R&=\begin{pmatrix}a_n&r\\r&a_c\end{pmatrix},\qquad
 V=\begin{pmatrix}U_n&0\\0&U_c\end{pmatrix},\notag\\
 D_0&=\begin{pmatrix}
 -\gamma_na_n&-\gamma_nr\\
 -\gamma_cr&-\gamma_ca_c\\
 F_{n\mathcal N}/\kappa_v&F_{N\mathcal N}/\kappa_v
 \end{pmatrix},
 \label{eq:regc_lowrank_blocks}\\
 M_{11}&=k_n^*\gamma_n^2-B_n(\gamma_n^2-1),\notag\\
 M_{22}&=k_c^*\gamma_c^2-B_c(\gamma_c^2-1),\notag\\
 M_{12}&=\mathcal A-r\gamma_n\gamma_c,\quad
 M_{33}=F_{\mathcal N\mathcal N}/\kappa_v^2,\notag\\
 M_{13}&=\gamma_nF_{n\mathcal N}/\kappa_v,\quad
 M_{23}=\gamma_cF_{N\mathcal N}/\kappa_v,
 \label{eq:regc_explicit_blocks}
\end{align}
\endgroup
where
$K_0=\left(\begin{smallmatrix}B_n&\mathcal A\\\mathcal A&B_c\end{smallmatrix}\right)$.
The dependence on the two spatial currents is a rank-two correction.
This allows the energy inequalities to be reduced to small scalar
systems without assuming aligned velocities.

Normalize the canonical energy with positive isotropic weights.
For the additive thermal EOS, one choice is
\begin{align}
 c_n^2&=nk_n/\mu_n,&g_n&=\mu_nc_n/n,\notag\\
 c_c^2&=Nk_c/\mu_h,&g_c&=\mu_hc_c/N,\notag\\
 G_z&=\operatorname{diag}(g_n,g_c,a_\Omega),\notag\\
 G_w&=\operatorname{diag}(g_n^{-1}\regcI_3,g_c^{-1}\regcI_3),
 \label{eq:regc_weights}
\end{align}
Let $v_*=\max\{|v_n|,|v_c|\}$.
The weights set the normalization of a sufficient energy estimate.
They are not the characteristic speeds of the coupled system.
For pure canonical evolution of neutron constraint errors, $|v_c|$
can replace $v_*$.
A sufficient positive and causal wave energy is obtained from
\begin{align}
 \mathcal U&=\begin{pmatrix}G_z-M&C\\C^T&K-G_w^{-1}\end{pmatrix}>0,\notag\\
 \mathcal P&=\begin{pmatrix}
 M-v_*G_z&\sqrt{v_*}C\\
 \sqrt{v_*}C^T&G_w^{-1}-v_*K
 \end{pmatrix}>0.
 \label{eq:regc_domain_lmi}
\end{align}
At fixed weights, these matrix inequalities are affine in $K,C,M$.
For a general EOS those blocks follow from the full Hessian.
For the additive EOS, set $X_n=|U_n|^2$, $X_c=|U_c|^2$,
$X_{nc}=U_n\cdot U_c$, $R_+=R$ and $R_-=-v_*R$, and define
\begin{align}
 B_+&=K_0-\operatorname{diag}(g_n,g_c),\notag\\
 B_-&=\operatorname{diag}(g_n,g_c)-v_*K_0,\notag\\
 \Delta_\pm&=\det B_\pm,\qquad
 b_{\pm,ij}=(B_\pm)_{ij},\notag\\
 W_\pm&=\frac1{\Delta_\pm}
 \begin{pmatrix}
 b_{\pm,22}X_n&-b_{\pm,12}X_{nc}\\
 -b_{\pm,12}X_{nc}&b_{\pm,11}X_c
 \end{pmatrix},
 \label{eq:regc_base_species}\\
 \tau_\pm&=2+\operatorname{tr}(R_\pm W_\pm),\notag\\
 \delta_\pm&=1+\operatorname{tr}(R_\pm W_\pm)
                    +\det R_\pm\det W_\pm,\notag\\
 T_\pm&=\frac{W_\pm+\det W_\pm\operatorname{adj}R_\pm}
             {\delta_\pm},\notag\\
 S_+&=G_z-M-D_0T_+D_0^T,\notag\\
 S_-&=M-v_*G_z-v_*D_0T_-D_0^T.
 \label{eq:regc_small_schur}
\end{align}
A symmetric three-by-three matrix $S=(s_{ij})$ is positive definite
when its successive Schur pivots satisfy
\begin{align}
 s_{11}&>0,\qquad s_{22}>\frac{s_{12}^2}{s_{11}},\notag\\
 s_{33}&>
 \frac{s_{22}s_{13}^2-2s_{12}s_{13}s_{23}+s_{11}s_{23}^2}
 {s_{11}s_{22}-s_{12}^2}.
 \label{eq:regc_three_scalar_pivots}
\end{align}
The scalar form of the sufficient energy conditions is therefore
\begin{equation}
 {\begin{gathered}
 a_\Omega>0,\qquad b_{\pm,11}>0,\quad\Delta_\pm>0,\\
 \tau_\pm>0,\quad\delta_\pm>0,\\
 S_+\ \hbox{and}\ S_-\ \hbox{satisfy
 Eq.~\eqref{eq:regc_three_scalar_pivots}}.
 \end{gathered}}
 \label{eq:regc_scalar_certificate}
\end{equation}
On the physical branch with $a_\Omega>0$, the fourteen comparisons
are equivalent to Eq.~\eqref{eq:regc_domain_lmi}.
Their displayed order ensures positive denominators.
The reduction remains regular for vanishing or parallel currents.
These conditions describe the chosen sufficient energy domain.
Failure of a comparison excludes a state from this domain, but
does not establish loss of hyperbolicity. The coefficient test of
Sec.~\ref{sec:regc_exact} can admit additional states.

\subsection{Positive energy and the characteristic basis}
\label{sec:regc_proof}
A normal perpendicular to both $U_n$ and $U_c$ exists in three
dimensions. Along that normal,
Eq.~\eqref{eq:regc_domain_lmi} requires $B_\pm>0$.
Normalize by $B_\pm\otimes\regcI_3$.
The remaining rank-two forms are
$I_2+W_\pm^{1/2}R_\pm W_\pm^{1/2}$, whose trace and determinant
are $\tau_\pm$ and $\delta_\pm$.
Their positivity gives the corresponding scalar signs.
No square root is inverted, so the Gram matrices $W_\pm$ may lose
rank without making the reduction singular.

The identity
$W(I_2+RW)^{-1}=(W+\det W\operatorname{adj}R)/\det(I_2+RW)$
gives $T_\pm$ and the Schur complements $S_\pm$.
These conditions are equivalent to $K>0$ and
$v_*G_a<H_a<G_a$, with $G_a=\operatorname{diag}(G_z,G_w)$ and
$H_a=\bigl(\begin{smallmatrix}M+CK^{-1}C^T&-CK^{-1}\\-K^{-1}C^T&K^{-1}\end{smallmatrix}\bigr)$.
The matrix $H_a$ is the canonical energy of the two number densities,
the parallel magnetic component and six fluid momenta.
The four transverse line coordinates have unit energy after
reciprocal scaling of $B,D$, because
$\Lambda_{E_AE_B}=-\Lambda_{B_AB_B}=a_\Omega\delta_{AB}$ at $E'=0$.
The symmetric canonical differential matrices then give real causal
wave speeds and an energy-orthogonal basis at repeated speeds.

The advected fields require an additional estimate.
Rotate the propagation normal to $(p,0,q)$, where
$q=\widehat b\cdot\widehat s$\revadd{ and $\widehat s$ is the unit
normal in the vortex rest frame}.
The two bending fields decouple with speeds $\pm q$.
\revadd{Here $H_6$ is the compression energy in normalized canonical
coordinates and $J_6$ its symmetric differential matrix.}
The six compression waves have $A_6=J_6H_6$, with
$J_6=J_6^T$, $J_6^2=I_6$ and $v_*I_6<H_6\leq I_6$.
A uniform lower energy bound $h_*$ therefore gives
$|\lambda_6|\geq h_*>v_*$.
Carrier vorticity and entropy advect at $v_c\cdot\widehat s$.
Neutron canonical-curl errors advect at $v_n\cdot\widehat s$ for
the Euler extension and have zero speed for the pure canonical
extension. Magnetic divergence also has zero speed.

These fields force the compression equations, whose resolvent is
bounded by $(h_*-v_*)^{-1}$. Their eigenvectors can therefore be
completed uniformly without a vanishing spectral denominator.
The independent bending eigenspace remains complete at zero and
luminal coincidences. The warm extension has fourteen fields, of
which eleven are physical. Adding a cleaning field changes the
principal system and requires a separate characteristic analysis.
\endgroup

\input{appendix_primitive_speeds_tension}

%% file: appendix_coflow_hyperbolicity_tension.tex
\begingroup
\allowdisplaybreaks[1]
\subsection{Compression at coflow}
\label{sec:tension_exact_scalar_proof}

The coflow conditions of
Eqs.~\eqref{eq:tension_main_coflow_responses}--\eqref{eq:tension_main_coflow_conditions}
follow from a symmetric displacement problem. For the adopted line
energy, the two responses
$\mathfrak c_\Omega=F_{\mathcal N\mathcal N}$ and
$\mathfrak d_\Omega=F_{\mathcal N}/\mathcal N-F_{\mathcal N\mathcal N}$
are
\begingroup\colorlet{revisionblue}{mainrevisionblue}\color{revisionblue}
\begin{equation}
 \mathfrak c_\Omega=\frac{A_\Omega\alpha_{\rm v}}{\mathcal N_0}>0,\qquad
 \mathfrak d_\Omega=\frac{A_\Omega L_\Omega}{\mathcal N}>0.
 \label{eq:tension_exact_line_responses}
\end{equation}
\endgroup
where $\revadd{r_v=\alpha_{\rm v}\mathcal N/\mathcal N_0}$. At fixed
$\varsigma=s/N$, the mixed and material density responses are
\begin{align}
 f_x&=(L_\Omega+r_v)\partial_x A_\Omega
                         +A_\Omega\partial_xL_\Omega,\notag\\
 k_{xy}&=\partial_x\partial_y\varepsilon_0
   +\mathcal N\big[(L_\Omega+r_v/2)\partial_x\partial_y A_\Omega
                         +A_\Omega\partial_x\partial_yL_\Omega\notag\\
 &\hspace{10mm}+\partial_xA_\Omega\partial_yL_\Omega
                         +\partial_yA_\Omega\partial_xL_\Omega\big],\notag\\
 S_{xy}&=k_{xy}-\frac{f_xf_y}{\mathfrak c_\Omega},\notag\\
 U_{xy}&=K_{xy}-k_{xy}-\frac{f_xf_y}{\mathfrak d_\Omega},
 \label{eq:tension_exact_screened_responses}
\end{align}
Here $K_{nn}=B_n$, $K_{NN}=B_c$ and $K_{nN}=\mathcal A_*$ are the
inertia coefficients in Eq.~\eqref{eq:primitive_coflow_responses}.
The constituent stiffnesses obey
$\partial_n^2\varepsilon_0=\mu_nc_n^2/n$ and
$\partial_N^2\varepsilon_0=\mu_hc_c^2/N$.

At $u_n=u_c=u_L$, define
$C_m=\left(\begin{smallmatrix}k_{nn}&k_{nc}\\k_{nc}&k_{cc}\end{smallmatrix}\right)$,
$K_0=\left(\begin{smallmatrix}B_n&\mathcal A_*\\\mathcal A_*&B_c\end{smallmatrix}\right)$
and $f=(f_n,f_c)^T$. Normalize the line displacement and choose a
normal $(p,0,q)$ with $p^2+q^2=1$. The squared compression speeds
are the generalized eigenvalues of
\begingroup\colorlet{revisionblue}{mainrevisionblue}\color{revisionblue}
\begin{align}
 M&=\operatorname{diag}(K_0,a_\Omega),\notag\\
 V_s&=\begin{pmatrix}
 C_m&-pf/\kappa_v\\
 -pf^T/\kappa_v&a_\Omega q^2+p^2F_{\mathcal N\mathcal N}/\kappa_v^2
 \end{pmatrix}.
 \label{eq:tension_exact_coflow_pencil}
\end{align}
\endgroup
For $B_n>0$ and $B_nB_c-\mathcal A_*^2>0$, the inertia $M$ is
positive. The symmetric pencil then gives real, nonzero compression
speeds with a complete energy-orthogonal basis if and only if
$V_s>0$. A zero squared compression speed produces a length-two
Jordan chain. A negative squared speed gives a nonreal pair.
Causality imposes the independent upper condition
$M-V_s\succeq0$.

Both inequalities can be reduced to perpendicular propagation.
The matrix $M-V_s$ is its value at $p=1$ conjugated by
$\operatorname{diag}(I_2,p)$. Eliminating the perpendicular line
coordinate gives
\begin{equation}
 S=C_m-ff^T/\mathfrak c_\Omega,\qquad U=K_0-C_m-ff^T/\mathfrak d_\Omega.
 \label{eq:tension_exact_coflow_schur}
\end{equation}
For a general normal, the Schur complement of $V_s$ equals $S$ plus
$\revadd{\kappa_v^2a_\Omega q^2ff^T/
[\mathfrak c_\Omega(\kappa_v^2a_\Omega q^2+p^2\mathfrak c_\Omega)]\succeq0}$.
Perpendicular compression therefore gives the strongest lower bound
as well as the strongest upper bound. The necessary and sufficient
conditions for causal, strongly hyperbolic coflow are
\begin{align}
 S_{nn}&>0,\qquad S_{NN}>\frac{S_{nN}^2}{S_{nn}},\notag\\
 U_{nn}&\geq0,\qquad U_{NN}\geq0,\qquad
 U_{nn}U_{NN}\geq U_{nN}^2.
 \label{eq:tension_exact_coflow_bounds}
\end{align}
The strict lower signs impose $S>0$. The nonstrict upper signs impose
$U\succeq0$ and retain luminal compression modes. The bending block
is independent and semisimple at $q=0$, while positive energy bounds
the acoustic eigenbasis through repeated speeds. No further angular
or coincidence condition is needed at coflow.

When the lower and upper conditions are imposed jointly, positive
species inertia follows from
$K_0=S+U+(\mathfrak c_\Omega^{-1}+\mathfrak d_\Omega^{-1})ff^T>0$.
It need not be checked independently in that joint list.
\endgroup

%% file: appendix_primitive_speeds_tension.tex
\begingroup
\allowdisplaybreaks[1]
\subsection{Explicit primitive dispersion relation}
\label{app:primitive_dispersion}

We now calculate the scalar coefficients in
Eq.~\eqref{eq:regulated_characteristic_factorization} from
$\Lambda=\Lambda_{\rm tot}(n,N,I,\varsigma,\mathcal N)$, with
$I=-j_n\cdot j_c=nN\Gamma$ and $\Gamma=-u_n\cdot u_c$. All partial
derivatives hold the other scalar arguments fixed. Define
\begin{align}
 B_n&=-\Lambda_n/n,\quad B_c=-\Lambda_N/N,\quad
 \mathcal A_*=-\Lambda_I,\notag\\
 \Delta&=B_nB_c-\mathcal A_*^2,\notag\\
 \beta_{nn}&=B_c/\Delta,\quad
 \beta_{nc}=-\mathcal A_*/\Delta,\quad
 \beta_{cc}=B_n/\Delta,\notag\\
 R_{11}&=\Lambda_{nn}+B_n,\quad
 R_{22}=\Lambda_{NN}+B_c,\quad R_{33}=\Lambda_{II},\notag\\
 R_{12}&=\Lambda_{nN},\quad
 R_{13}=\Lambda_{nI},\quad R_{23}=\Lambda_{NI},\notag\\
 G_{11}&=-\beta_{nn},\quad G_{22}=-\beta_{cc},\quad
 G_{12}=-\Gamma\beta_{nc},\notag\\
 G_{13}&=-N\Gamma\beta_{nn}-n\beta_{nc},\notag\\
 G_{23}&=-N\beta_{nc}-n\Gamma\beta_{cc},\notag\\
 G_{33}&=-N^2\beta_{nn}-2nN\Gamma\beta_{nc}-n^2\beta_{cc},
 \label{eq:primitive_dispersion_coefficients}
\end{align}
with $R_{rs},G_{rs}$ symmetric and $r,s,t$ running over $(n,N,I)$.
For the additive class \eqref{eq:regc_material_class},
$R_{11}=a_n$, $R_{22}=a_c$, $R_{12}=r$ and
$R_{13}=R_{23}=R_{33}=0$. With
$A_{rs}=\delta_{rs}+\sum_tR_{rt}G_{ts}$, put
$T_{rs}=\sum_t(A^{-1})_{rt}R_{ts}$, which is symmetric. On the
positive-energy domain $\Delta>0$ and $\det A\ne0$. In the vortex
rest frame let $\ell_x=u_x\cdot k=-\gamma_x\sigma'+U_x\cdot s'$ for
$k_a=(-\sigma',s')$, and
$f_r^\Omega=(-F_{n\mathcal N},-F_{N\mathcal N},0)_r$. Then
\begin{align}
 Z_{1n}&=-\beta_{nn}\ell_n,\quad
 Z_{1c}=-\beta_{nc}\ell_n,\notag\\
 Z_{2n}&=-\beta_{nc}\ell_c,\quad
 Z_{2c}=-\beta_{cc}\ell_c,\notag\\
 Z_{3n}&=-N\beta_{nn}\ell_c-n\beta_{nc}\ell_n,\notag\\
 Z_{3c}&=-N\beta_{nc}\ell_c-n\beta_{cc}\ell_n,\notag\\
 b_r&=f_r^\Omega-\sum_{s,t}T_{rs}G_{st}f_t^\Omega,\notag\\
 \mathcal S_\Omega&=\sum_{r,s}f_r^\Omega G_{rs}b_s,\notag\\
 \mathcal Q_{xy}(\sigma')
  &=(1-(\sigma')^2)\beta_{xy}
                    -\sum_{r,s}Z_{rx}T_{rs}Z_{sy},\notag\\
 d_x(\sigma')&=\sum_rZ_{rx}b_r,\qquad x,y\in\{n,c\}.
 \label{eq:primitive_dispersion_responses}
\end{align}
These formulas do not divide by a relative velocity, a Gram
determinant, or $R$, and remain valid at coflow and for parallel
currents.

\paragraph{Primitive elimination.}
The current Hessian \eqref{eq:regc_current_hessian} is
$\Lambda_{j_x^aj_y^b}=B_{xy}g_{ab}+\sum_{r,s}U_{r,xa}R_{rs}U_{s,yb}$,
with $B_{nn}=B_n$, $B_{cc}=B_c$, $B_{nc}=\mathcal A_*$ and
$U_{r,xa}$ the current gradients of $(n,N,I)$. Their contractions
with the base inverse $\beta_{xy}g^{ab}$ give $G_{rs}$, and $T$
supplies the rank-three correction to that inverse. Contracting the
inverse twice with $k_a$ gives $\mathcal Q_{xy}$, once with $k_a$
and once with
$\Lambda_{j_x^a\mathcal N}=\sum_rU_{r,xa}f_r^\Omega$ gives $d_x$,
and twice with the latter gives $\mathcal S_\Omega$. Off the
carrier contacts, eliminating the two potential amplitudes and
$\delta\mathcal N$ from number conservation and the transverse line
equation gives $P_6$, which extends to contact and light-cone
coincidences and to $p=0$ by continuity. Completeness follows from
the positive-energy construction of Sec.~\ref{sec:regc_proof},
rather than root reality alone. At $u_n=u_c=u_L$, with
$x=(\sigma')^2$ and the responses of
Eq.~\eqref{eq:primitive_coflow_responses},
\begingroup
the normalization is fixed by the displacement pencil:
\begin{align}
 (\det C_m)P_6(\sigma',s')
 &=\frac{\det[xM-V_{s'}]}{a_\Omega\Delta}\notag\\
 &=x^3-I_1x^2+I_2x-I_3.
 \label{eq:primitive_coflow_polynomial_normalization}
\end{align}
Here $C_m,M,V_{s'}$ are defined in
Eq.~\eqref{eq:tension_exact_coflow_pencil} and its preceding text.
Thus the monic cubic has the coefficients of
Eq.~\eqref{eq:primitive_coflow_cubic_coefficients}.
\endgroup
The Rayleigh quotient of a pure line-compression perturbation bounds
the largest squared compression speed from below by the unmixed
line value, $x_0\geq c_\ell^2$.
\endgroup

%% file: appendix_rw_tension.tex
\begingroup
\section{{Density-dependent line stress and coflow waves}}
\label{app:rw_derivation}
We now derive the stress of the density-dependent action
\eqref{eq:tension_draft_master} and construct its characteristic
matrix from the primitive equations, retaining every material
derivative of the matched line energy.

\subsection{Effective action and stress}
For independent currents $j_x$, the invariants vary as
$\delta I_{xy}=-j_{y\nu}\delta j_x^\nu-j_{x\nu}\delta j_y^\nu
-j_x^\mu j_y^\nu\delta g_{\mu\nu}$. The resulting momentum
correction is
$\delta\mu^x_\nu=\sum_yK^\Omega_{xy}j_{y\nu}$, with the diagonal
factor of two in Eq.~\eqref{eq:rw_material_coefficient_matrix}.
Varying the metric at fixed current three-forms and vortex two-form
gives Eq.~\eqref{eq:tension_draft_stress}, with the $3+1$
projections \eqref{eq:rw_primitive_stress}. Symmetry of the
density-response coefficients makes the energy flux equal to the
momentum density. For a general set of invariants $I_A$, the density
terms of Eqs.~\eqref{eq:tension_main_stress_divergence} and
\eqref{eq:tension_main_matter_exchange} are
$-\sum_AF_A\nabla_\nu I_A$ and $+\sum_AF_A\nabla_\nu I_A$,
respectively\revadd{, where $F_A=(\partial F/\partial I_A)_{\mathcal N}$}.
The total stress therefore obeys
$\nabla_\mu T_{\rm tot}^{\mu\nu}=f_{\rm ext}^\nu$. In the energy
equation \eqref{eq:tension_main_adm_energy} these terms transfer
energy reversibly as the fluid state changes.

\subsection{Characteristic matrix}
For our line energy,
Eq.~\eqref{eq:rw_line_derivatives} gives $c_\Omega^2\in(0,1)$, and
we write $r_\Omega=(c_\Omega^2-1)/(B^2-E^2)<0$.
The pure line response holds the fluid state fixed. Hyperbolicity
of the coupled system also depends on the mixed derivatives
$F_{x\mathcal N}$ and $F_{xy}$ in
Sec.~\ref{app:regulated_characteristics}.

To construct the spectrum of Eq.~\eqref{eq:rw_physical_pencil}
independently, freeze $\alpha=1$, $\beta=0$ and $\gamma_{ij}$ at a
point, choose a unit $s_i$, and retain only the coefficients of
$s^i\revadd{\nabla_i}W$ in the balance laws
\eqref{eq:valencia_charge_unified}--\eqref{eq:valencia_energy_unified}
and \eqref{eq:numerical_xi_continuum} and in the electric and
induction equations of
Appendix~\ref{app:tension_electric_adm}. In the variable order
$W=(P,E,B)$\revadd{, define the undensitized material map
$\mathcal U=(D,D_R,E_{\rm tot},S_{{\rm tot},i},\xi_i)$ and its
Jacobians $\mathcal A=\partial_P\mathcal U$,
$\mathcal B_E=\partial_E\mathcal U$ and
$\mathcal B_B=\partial_B\mathcal U$ at fixed remaining arguments.
The electric--material response row is
$\zeta=\sum_Ab_A\partial_PI_A$, with
$b_A=(\partial\ln a_\Omega/\partial I_A)_{\mathcal N}$.
The coupled time matrix is}
\begin{equation}
 M_t=\begin{pmatrix}
 \mathcal A&\mathcal B_E&\mathcal B_B\\
 E\zeta&I-r_\Omega EE^T&0\\
 0&0&I
 \end{pmatrix}.
 \label{eq:rw_full_time_matrix}
\end{equation}
Here $E\zeta$ is a $3\times9$ outer product. The spatial coefficients
define $M(s)$ through
$M_t\partial_t\delta W+M(s)\partial_s\delta W=0$. On $E\cdot B=0$, the
derivative of the stage projection maps
$\delta E_i\mapsto\delta E_i-(B^j\delta E_j+E_j\delta B^j)B_i/B^2$.
The other primitive perturbations are unchanged at first order,
because a parallel electric variation gives $\delta\mathcal N=0$ and
$\mathcal B_E B=0$ when $E\cdot B=0$. This specifies the linearized
projection $\mathsf P$. Simplicity, magnetic divergence and the two
canonical-curl conditions leave an eleven-dimensional physical
perturbation space. For an orthonormal basis $R(s)$ of that space,
\begin{align}
 \mathsf C(s)&=R(s)^T\mathsf P M_t^{-1}M(s)R(s),\notag\\
 p_s(\sigma)&=\det[\sigma I-\mathsf C(s)].
 \label{eq:rw_projected_verification}
\end{align}
Its roots agree with the explicit scalar polynomial
\eqref{eq:rw_physical_pencil}, whose primitive coefficients are
Eq.~\eqref{eq:primitive_dispersion_coefficients}.

\subsection{Coflow compression and the unregulated logarithm}
\label{app:regulated_coflow}
\label{app:rw_coflow}
At a common rest state, a perpendicular compression of the two
fluids and the array at fixed entropy per carrier has three squared
speeds, the generalized eigenvalues of the pencil
\eqref{eq:tension_exact_coflow_pencil}.
Sec.~\ref{sec:tension_exact_scalar_proof} reduces their reality,
completeness and causality to
Eq.~\eqref{eq:tension_exact_coflow_bounds}. The admitted domain
contains states with independent vortex slip. For $F=\epsilon f$
with positive line responses, the line and mixed entries of the
pencil are $O(\epsilon)$ and the Schur corrections vanish with
$\epsilon$. With positive fluid stiffness and a strict fluid cone
margin, sufficiently small positive $\epsilon$ therefore satisfies
the bounds, and continuity admits nearby noncollinear fluid and
vortex velocities.

For comparison, consider the full logarithmic dependence on array
density in Ref.~\cite[Eq.~(150)]{Rau:2020wnv}, without the
positive compressional regulator. At fixed fluid state,
$F_{\log}=\mathcal N(A_{\log}L_{\log}+M_{\log})$ with
$L_{\log}=\ln[0.0712/(\mathcal N\xi_n^2)]$ has
$F_{\log,\mathcal N\mathcal N}=-A_{\log}/\mathcal N<0$. For positive
full inertia\revadd{, use the inertia $M$ and stiffness $V_s$ of
Eq.~\eqref{eq:tension_exact_coflow_pencil}, evaluated with $F_{\log}$
at perpendicular propagation. The index $L$ denotes the line
displacement coordinate.} The perpendicular physical compression Rayleigh
quotient of the line coordinate obeys
\begin{equation}
 \sigma_{\min}^2\leq
 \revadd{\frac{(V_s)_{LL}}{M_{LL}}}
 =-\frac{A_{\log}}{F_{\log,\mathcal N}}<0.
 \label{eq:rw_negative_squared_speed}
\end{equation}
Mixed fluid-density derivatives cannot remove this negative Rayleigh
quotient. This conclusion applies to the stated local action with
independent line motion. It is not an
instability claim for microscopic neutron matter, the complete
electromagnetic multifluid model or force-equilibrium HVBK.
\endgroup

%% file: appendix_condition_reductions_tension.tex
\begingroup
\allowdisplaybreaks[1]
\section{{Primitive conditions for the adopted line energy}}
\label{app:tension_primitive_bounds}

We now insert the polytropic material EOS and BCS line coefficients
into the characteristic conditions. The coflow result above admits
nonstrict upper signs. At finite relative motion we distinguish the
strict causal-response branch and the structurally decoupled carrier.
All density derivatives of the microscopic inputs are retained.

The thermodynamic requirement of nonnegative internal heating remains
separate from the principal-symbol conditions. The projected
velocities give
\begin{align}
 \mathcal Q_\Omega/b
 &=\mathcal R\nu_L r_c^2+\mathcal R' r_c\cdot r_s\geq0,\notag\\
 r_c^\mu&=\perp^\mu{}_\nu u^\nu,\qquad
 r_s^\mu=\perp^\mu{}_\nu N_s^\nu.
 \label{eq:rw_physical_work}
\end{align}
Finite smooth algebraic sources leave the characteristic speeds
unchanged. This does not apply to the density-gradient forces
already included in the principal line response.
\endgroup

\input{appendix_specialized_bounds_tension}
\input{appendix_entrained_primitive_tension}

%% file: appendix_specialized_bounds_tension.tex
\begingroup
\allowdisplaybreaks[1]
\subsection{BCS density responses and the independent-carrier limit}
\label{app:tension_evaluated_bounds}

Insert the polytropic EOS and BCS core law
\eqref{eq:tension_bcs_core_law} into the line energy.
The masses and polytropic coefficients are positive,
$\Gamma_x>1$, and the material entrainment is constant.
All parameters remain symbolic. Define
\begin{gather}
 C_\kappa=\frac{\kappa_v^2}{4\pi},\qquad
 H_0=m_n+\frac{\Gamma_nK_n}{\Gamma_n-1}n^{\Gamma_n-1}-\mathcal AN,
 \notag\\
 k_n=\Gamma_nK_nn^{\Gamma_n-2},\qquad
 r_v=\frac{\alpha_{\rm v}\mathcal N}{\mathcal N_0},\notag\\
 P_v=L_\Omega+\frac{r_v}{2},\qquad Q_v=L_\Omega+r_v,\notag\\
 \mu_h=m_p+\frac{\Gamma_pK_p}{\Gamma_p-1}N^{\Gamma_p-1}
                      +\frac{4C_{\rm th}s^{4/3}}{3N},\notag\\
 k_h=\Gamma_pK_pN^{\Gamma_p-2}
                      +\frac{4C_{\rm th}s^{4/3}}{9N^2}.
 \label{eq:tension_evaluated_general_inputs}
\end{gather}
The carrier derivative $k_h$ holds $s/N$ fixed.
\revadd{Write $\mathcal L^{(x)}=\partial_xL_\Omega$ and
$\mathcal L^{(xy)}=\partial_x\partial_yL_\Omega$ for $x,y\in\{n,N\}$.}
Using $\mathcal L^{(n)},\ldots,\mathcal L^{(NN)}$ from
Eq.~\eqref{eq:tension_evaluated_bcs_derivatives}, the primitive
response coefficients become
\begin{align}
 B_n={}&\frac{H_0}{n}
 +\frac{C_\kappa\mathcal N}{nH_0^2}
       [(H_0-nk_n)P_v+nH_0\mathcal L^{(n)}],\notag\\
 B_c={}&\frac{\mu_h-\mathcal An}{N}
 +\frac{C_\kappa n\mathcal N}{NH_0^2}
       [\mathcal AP_v+H_0\mathcal L^{(N)}],\notag\\
 k_{nn}={}&k_n+\frac{C_\kappa\mathcal N}{H_0^3}
 \Big[\Gamma_nk_n(2K_nn^{\Gamma_n-1}-H_0)P_v\notag\\
 &+2H_0(H_0-nk_n)\mathcal L^{(n)}
                    +nH_0^2\mathcal L^{(nn)}\Big],\notag\\
 k_{nc}={}&\frac{C_\kappa\mathcal N}{H_0^3}
 \Big[\mathcal A(H_0-2nk_n)P_v
           +H_0(H_0-nk_n)\mathcal L^{(N)}\notag\\
 &+\mathcal AnH_0\mathcal L^{(n)}
                    +nH_0^2\mathcal L^{(nN)}\Big],\notag\\
 k_{cc}={}&k_h+\frac{C_\kappa n\mathcal N}{H_0^3}
 \Big[2\mathcal A^2P_v+2\mathcal AH_0\mathcal L^{(N)}
                    +H_0^2\mathcal L^{(NN)}\Big],\notag\\
 f_n={}&\frac{C_\kappa}{H_0^2}
              [(H_0-nk_n)Q_v+nH_0\mathcal L^{(n)}],\notag\\
 f_c={}&\frac{C_\kappa n}{H_0^2}
                     [\mathcal AQ_v+H_0\mathcal L^{(N)}],\notag\\
 c={}&\frac{C_\kappa n\alpha_{\rm v}}{H_0\mathcal N_0},\qquad
 d=\frac{C_\kappa nL_\Omega}{H_0\mathcal N},\notag\\
 t={}&\frac{C_\kappa nQ_v}{H_0\mathcal N}=c+d.
 \label{eq:tension_evaluated_response_pack}
\end{align}
Even constant microscopic inputs leave the term $-1/(3n)$ in
$\mathcal L^{(n)}$.

\paragraph{Upper response and coflow stiffness.}
On the regular physical inertia orientation, the null compression
pencil is a negative congruence of the feedback response of
\begin{equation}
 R_{\rm aug}=
 \begin{pmatrix}
 B_n-k_{nn}&\mathcal A-k_{nc}&-f_n\\
 \mathcal A-k_{nc}&B_c-k_{cc}&-f_c\\
 -f_n&-f_c&d
 \end{pmatrix}.
 \label{eq:tension_evaluated_bare_null_response}
\end{equation}
The update identities in
Eqs.~\eqref{eq:primitive_dispersion_coefficients}--\eqref{eq:primitive_dispersion_responses}
preserve its inertia, including semidefinite boundaries.
Given strong hyperbolicity, $R_{\rm aug}\succeq0$ therefore imposes
the coflow upper inequalities at arbitrary slip.
The strict causal-response branch has $R_{\rm aug}>0$.
Composition derivatives must be retained in $k_{nc}$ and $f_c$ even
when $\mathcal A=0$.

At coflow, $g=h=1$, eliminating the positive line response $d$ gives
Eq.~\eqref{eq:tension_entrained_open_causality}.
Its nonstrict boundary additionally requires $da_c\geq f_c^2$.
Eliminating $c>0$ from the density-stiffness form gives
\begin{align}
 c k_{nn}&>f_n^2,\notag\\
 (c k_{nn}-f_n^2)k_{cc}
 &>c k_{nc}^2-2k_{nc}f_nf_c+k_{nn}f_c^2.
 \label{eq:tension_evaluated_stellar_hyperbolicity}
\end{align}
Together with the upper signs, these imply positive material inertia.
The finite-slip entrained conditions reduce to these coflow
conditions with $\mathfrak D=\Delta(k_{nn}k_{cc}-k_{nc}^2)$ and a
positive contact matrix.

\paragraph{Decoupling the carrier.}
For $\mathcal A=0$ and microscopic inputs depending only on $n$,
$F$ is independent of $N$ and the carrier Euler system decouples.
The neutron--array equations depend on neutron motion through
$\mathfrak u^2$ in Eq.~\eqref{eq:tension_evaluated_transverse_slip}.
Choose the transverse flow along $x$ in an array rest frame.
For $a,b,\epsilon_{n\Omega}\geq0$ defined in
Eq.~\eqref{eq:tension_evaluated_sound_parameters}, the coflow
stiffness and cone bounds become
\begin{equation}
 ab>\epsilon_{n\Omega},\qquad
 (1-a)(1-b)\geq\epsilon_{n\Omega}.
 \label{eq:tension_evaluated_dimensionless_coflow}
\end{equation}
Propagation along the transverse flow gives, up to positive
normalization, the characteristic quartic
\begin{align}
 P_4(\sigma)={}&[1+(1-a)\mathfrak u^2]\sigma^4
 -2(1-a)\mathfrak u\sqrt{1+\mathfrak u^2}\,\sigma^3\notag\\
 &-\left[a+b-\big((1-a)(1-b)-\epsilon_{n\Omega}\big)
                                      \mathfrak u^2\right]\sigma^2\notag\\
 &+2[(1-a)b+\epsilon_{n\Omega}]
                  \mathfrak u\sqrt{1+\mathfrak u^2}\,\sigma\notag\\
 &+ab-\epsilon_{n\Omega}
      -[(1-a)b+\epsilon_{n\Omega}]\mathfrak u^2 .
 \label{eq:tension_evaluated_explicit_quartic}
\end{align}
For nonzero mixing, its symmetric two-amplitude pencil must have a
positive central interval. This is equivalent to the interval overlap
\eqref{eq:tension_evaluated_slip_overlap} and a positive discriminant
of $P_4$. Four real roots along the flow can also occur in a
disconnected region that fails for another propagation normal.

The angular extension makes this distinction explicit.
Normalize the fluid spatial and line time coefficients of $Q$ to
unity. With $p^2=s_x^2+s_y^2$ and
$D_s=\operatorname{diag}(s_x,p)$, one obtains
\begin{align}
 Q(w s_x,s)
 &=D_sQ(w,e_x)D_s\notag\\
 &\quad+\operatorname{diag}
      (s_y^2+s_z^2,\ w^2s_y^2+s_z^2).
 \label{eq:tension_evaluated_all_angle_lift}
\end{align}
A positive $Q(w,e_x)$ therefore gives a positive wave energy for
every normal. Conversely, loss of the central component as the
normal rotates gives a repeated acoustic root. The nonzero mixing
column makes that root defective. The bending and carrier blocks
remain complete at their coincidences.
At zero mixing, the overlap condition suffices for the open-domain
limit. Disjoint or tangent diagonal intervals are not interior
because the coupling becomes nonzero in nearby primitive states.

\paragraph{The limiting mixing amplitude.}
On the strict causal branch, $0<a,b<1$,
$0\leq\epsilon_{n\Omega}<ab$ and
$\epsilon_{n\Omega}<(1-a)(1-b)$.
For $0<\mathfrak v=\mathfrak u/\sqrt{1+\mathfrak u^2}<1$, define
\begin{equation}
 \mathcal R(\sigma)=(b-\sigma^2)
       \left[a-\left(\frac{\sigma-\mathfrak v}
                           {1-\mathfrak v\sigma}\right)^2\right].
 \label{eq:tension_epsilon_proof_ratio}
\end{equation}
The characteristic equation $P_4=0$ is equivalent to
$\mathcal R(\sigma)=\epsilon_{n\Omega}$.
The largest admissible mixing amplitude is set by the central
maximum of this function. Its stationary coefficients are
\begin{align}
 \eta_3&=-\frac{2+(1-3a)\mathfrak v^2}
                    {\mathfrak v(1-a\mathfrak v^2)},\quad
 \eta_2=\frac{3(1-a)}{1-a\mathfrak v^2},\notag\\
 \eta_1&=\frac{a+b-(1+b)\mathfrak v^2}
                    {\mathfrak v(1-a\mathfrak v^2)},\quad
 \eta_0=-\frac{b(1-\mathfrak v^2)}{1-a\mathfrak v^2},
 \label{eq:tension_epsilon_stationary_inputs}\\
 p_*&=\eta_2-\frac38\eta_3^2,\qquad
 q_*=\eta_1-\frac12\eta_3\eta_2+\frac18\eta_3^3,\notag\\
 r_*&=\eta_0-\frac14\eta_3\eta_1
                 +\frac1{16}\eta_3^2\eta_2-\frac3{256}\eta_3^4,
 \notag\\
 I_*&=p_*^2+12r_*,\qquad
 J_*=2p_*^3-72p_*r_*+27q_*^2.
 \label{eq:tension_epsilon_depressed_inputs}
\end{align}
and differentiation gives
\begin{align}
 \frac{d\mathcal R}{d\sigma}
 &=-\frac{2\mathfrak v(1-a\mathfrak v^2)}
                      {(1-\mathfrak v\sigma)^3}
 \bigl(\sigma^4+\eta_3\sigma^3+\eta_2\sigma^2
                              +\eta_1\sigma+\eta_0\bigr).
 \label{eq:tension_epsilon_proof_stationarity}
\end{align}
When the uncoupled speeds $\pm\sqrt b$ and
$(\mathfrak v\pm\sqrt a)/(1\pm\mathfrak v\sqrt a)$ are distinct,
the stationary quartic has one zero in each of the three intervening
intervals and one above $1/\mathfrak v$.
The second zero gives the central maximum and is
\begin{align}
 m_*&=\frac23\left[
 \sqrt{I_*}\cos\!\left(\frac13\arccos
                   \frac{J_*}{2I_*^{3/2}}\right)-p_*\right],\notag\\
 \sigma_*&=-\frac{\eta_3}{4}-\frac{\sqrt{m_*}}2
       +\frac12\sqrt{-m_*-2p_*+\frac{2q_*}{\sqrt{m_*}}}.
 \label{eq:tension_epsilon_critical_phase}
\end{align}
All radicals and the inverse cosine are real on this branch, with
$m_*>0$. Thus $\epsilon_*=\mathcal R(\sigma_*)$.
The other two interior extrema are nonpositive, while the external
minimum exceeds $(1-a)(1-b)$.
These stationary values are the zeros of the discriminant of $P_4$
viewed as a polynomial in $\epsilon_{n\Omega}$.
Its positive leading coefficient makes a positive discriminant on
the causal interval equivalent to $\epsilon_{n\Omega}<\epsilon_*$.
Coincident uncoupled speeds follow by continuity.
For nonzero mixing, equality produces a defective repeated acoustic
root.
\endgroup

%% file: appendix_entrained_primitive_tension.tex
\begingroup
\allowdisplaybreaks[1]
\subsection{Finite slip at nonzero entrainment}
\label{app:tension_entrained_primitive_list}

We derive the primitive conditions of
Sec.~\ref{sec:tension_entrained_arbitrary_slip} for constant
$\mathcal A\ne0$ and the symbolic BCS EOS.
The domain has a strict causal response and positive physical
inertia. The warm constrained system has eleven fields, and the
cold system has ten. The derivation reduces the carrier-contact
problem to a two-by-two compression matrix before eliminating its
angular dependence.

\paragraph{Response coefficients and relative geometry.}
From Eq.~\eqref{eq:tension_evaluated_response_pack} and the
primitive inputs \eqref{eq:tension_entrained_primitive_inputs}, define
\begin{align}
 \mathfrak D={}&\Delta^2-\Delta(B_ca_n+B_na_c-2g\mathcal Ar)\notag\\
 &+(B_nB_c-g^2\mathcal A^2)(a_na_c-r^2),\notag\\
 X={}&\Delta a_n-B_n(a_na_c-r^2),\notag\\
 J={}&(\Delta-B_na_c+g\mathcal Ar)f_n
                +(B_nr-g\mathcal Aa_n)f_c,\notag\\
 Y={}&d\mathfrak D+
 [B_c\Delta-(B_nB_c-g^2\mathcal A^2)a_c]f_n^2\notag\\
 &-2[g\mathcal A\Delta-(B_nB_c-g^2\mathcal A^2)r]f_nf_c\notag\\
 &+[B_n\Delta-(B_nB_c-g^2\mathcal A^2)a_n]f_c^2,\notag\\
 K={}&\Delta(a_nd-f_n^2)\notag\\
 &-B_n[d(a_na_c-r^2)-a_cf_n^2+2rf_nf_c-a_nf_c^2],\notag\\
 \beta={}&\frac{B_cX}{\mathfrak D},\qquad
 \delta=\frac{Y}{t\mathfrak D},\notag\\
 \revadd{\chi_{\rm ang}}={}&\frac{B_c\Delta J^2}{t\mathfrak D^2},\qquad
 e=\frac{B_cK}{t\mathfrak D}=\beta\delta-\revadd{\chi_{\rm ang}}.
 \label{eq:tension_entrained_explicit_responses}
\end{align}
The Eulerian fields determine the relative geometry through
\begin{gather}
 Q=B^2-E^2,\qquad V_x=W_xE+z_x\times B\quad(x=n,c),\notag\\
 Z_c=|V_c|^2-(E\cdot z_c)^2,\notag\\
 Z_{nc}=V_n\cdot V_c-(E\cdot z_n)(E\cdot z_c),\notag\\
 h=1+\frac{Z_c}{Q},\qquad U^2=g^2-1,\notag\\
 U_b^2=\frac{
 [z_n\cdot(W_cB-z_c\times E)-W_n(B\cdot z_c)]^2}{Qh},\notag\\
 U_L^2=\frac{(gZ_c-Z_{nc})^2}{Q^2h},
 \label{eq:tension_entrained_primitive_geometry}
\end{gather}
where spatial products use $\gamma_{ij}$.
In the carrier rest frame, $U$ is the neutron spatial four-velocity,
$h=1+|L|^2$, $U_b^2=(U\cdot\hat b)^2$ and
$U_L^2=(U\cdot L)^2$, with $L\cdot\hat b=0$.

\paragraph{Compression at the carrier contact.}
Write
$R=\bigl(\begin{smallmatrix}a_n&r\\r&a_c\end{smallmatrix}\bigr)$,
$f=(-f_n,-f_c)^T$ and
$G=-\bigl(\begin{smallmatrix}B_c&-g\mathcal A\\-g\mathcal A&B_n\end{smallmatrix}\bigr)/\Delta$.
The strict upper response gives $a_n,a_c>0$ and
$a_na_c-r^2>0$. Let $q_m=B_n-a_ng^2$ and $w=\mathcal A-gr$.
With $k_{cc}>0$ and $k_{cc}q_m>w^2$, completing squares gives
$\Delta>0$ and the identities
\begin{align}
 &q_mX=(q_mr-a_ngw)^2+a_nB_n(k_{cc}q_m-w^2)>0,
 \label{eq:tension_entrained_positive_orientation_pivot}\\
 &\mathfrak D-B_cX(g^2-1)\notag\\
 &\quad=\Delta[k_{cc}(B_n-a_ng^2)-(\mathcal A-gr)^2]>0.
 \label{eq:tension_entrained_material_cancellation}
\end{align}
Since $g\geq1$, these signs imply $X,\mathfrak D>0$.
They also give $Y>0$, so that the carrier--array slip bound is
$t\mathfrak D>hY$ with no additional denominator condition.

The causal inequalities
\eqref{eq:tension_entrained_open_causality} state that
$R_{\rm aug}=\bigl(\begin{smallmatrix}R&f\\f^T&d\end{smallmatrix}\bigr)>0$.
Eliminating the material response gives
\begin{gather}
 N=\begin{pmatrix}T&b\\b^T&D\end{pmatrix}
   =[I+R_{\rm aug}\operatorname{diag}(G,0)]^{-1}R_{\rm aug}>0,\notag\\
 T=(I+RG)^{-1}R,\qquad b=(I+RG)^{-1}f,\notag\\
 D=d-f^TGb,
 \label{eq:tension_entrained_proof_feedback}
\end{gather}
The entries satisfy $T_{nn}=\Delta X/\mathfrak D$,
$b_n=-\Delta J/\mathfrak D$ and $D=Y/\mathfrak D$.
With $A_0=\Delta/B_c>0$, the normalized coefficients are
$\beta=T_{nn}/A_0$, $\delta=D/t$ and
$\revadd{\chi_{\rm ang}}=b_n^2/(A_0t)$. The identity
$XY-\Delta J^2=\mathfrak DK$ gives
$e=\beta\delta-\revadd{\chi_{\rm ang}}$. Thus $\beta,\delta,e>0$ and $\revadd{\chi_{\rm ang}}\geq0$.

At the carrier contact, eliminate the positive potential pivot.
For a unit carrier-frame normal $s$, put $x=(U\cdot s)^2$,
$y=s\cdot P_{\rm sp}s$ and
$P_{\rm sp}=I+LL^T-\hat b\hat b^T$.
The remaining compression matrix is
\begin{equation}
 P_c(s)=\begin{pmatrix}
 A_0-T_{nn}x&-b_n(U\cdot s)\sqrt y\\
 -b_n(U\cdot s)\sqrt y&t-Dy
 \end{pmatrix},
 \label{eq:tension_entrained_contact_proof_matrix}
\end{equation}
Its normalized determinant is $1-\beta x-\delta y+exy$.
If $P_c(s)>0$ for every normal, the compression energy is uniformly
positive. The acoustic roots are then real, complete and separated
from the carrier contact. A bounded resolvent completes the carrier
shear and entropy eigenvectors. The bending block is independent,
and the positive upper response excludes timelike characteristic
covectors.

Conversely, a normal perpendicular to the two spatial velocities
requires positive density response and fixes the inertia orientation.
On this orientation, causality requires a nonnegative upper response
and $X>0$. The necessary material contact gap and
Eq.~\eqref{eq:tension_entrained_positive_orientation_pivot} give
$q_m>0$ and $k_{cc}>0$.
For $U\ne0$ and $b_n\ne0$, a compression root meeting the carrier
contact couples to the carrier shear and makes the spectral
projector unbounded. The number of compression roots on either side
of the contact is therefore constant over the normal sphere.
The sign at $s=\hat b$ fixes $\det P_c>0$.
Zeros of $b_n$ add no open domain, because $J$ cannot vanish on a
state-open set. Zeros of the determinant or material gap are also
excluded from its interior. Hence an open physical hyperbolic domain
on this branch requires $P_c(s)>0$ for every normal.

\paragraph{Eliminating the propagation angle.}
The diagonal conditions on $P_c$ give
Eqs.~\eqref{eq:tension_entrained_open_hyperbolicity} and
\eqref{eq:tension_entrained_line_slip}.
For $\revadd{\chi_{\rm ang}}=0$ they suffice. For $\revadd{\chi_{\rm ang}}>0$, introduce a real auxiliary
variable $a$ and the quadratic matrix
$Q(a)=a^2(\delta I-eUU^T)-2\revadd{\chi_{\rm ang}} aI+\revadd{\chi_{\rm ang}}(\beta I-eP_{\rm sp})$.
For each unit normal,
\begin{align}
 s\cdot Q(a)s={}&(\delta-ex)
       \left(a-\frac{\revadd{\chi_{\rm ang}}}{\delta-ex}\right)^2\notag\\
 &+\frac{\revadd{\chi_{\rm ang}} e(1-\beta x-\delta y+exy)}{\delta-ex}.
 \label{eq:tension_entrained_angular_identity}
\end{align}
The square completion shows that the all-angle condition is
equivalent to $\det Q(a)$ having no real zero.
The coefficients in $\det Q(a)=\sum_{j=0}^6z_ja^j$ are
\begin{align}
 z_0={}&\revadd{\chi_{\rm ang}}^3\beta(\beta-e)(\beta-eh),\notag\\
 z_1={}&2\revadd{\chi_{\rm ang}}^3[-3\beta^2+2\beta e(h+1)-e^2h],\notag\\
 z_2={}&-\revadd{\chi_{\rm ang}}^2[-3\beta^2\delta+\beta^2eU^2-12\beta\revadd{\chi_{\rm ang}}\notag\\
 &+2\beta\delta e(h+1)-\beta e^2hU^2-\beta e^2U_b^2
                 +\beta e^2U_L^2\notag\\
 &+4\revadd{\chi_{\rm ang}} e(h+1)-\delta e^2h+e^3hU_b^2],\notag\\
 z_3={}&2\revadd{\chi_{\rm ang}}^2[-6\beta\delta+2\beta eU^2-4\revadd{\chi_{\rm ang}}
                  +2\delta e(h+1)\notag\\
 &-e^2hU^2-e^2U_b^2+e^2U_L^2],\notag\\
 z_4={}&-\revadd{\chi_{\rm ang}}[-3\beta\delta^2+2\beta\delta eU^2
                     -12\revadd{\chi_{\rm ang}}\delta+4\revadd{\chi_{\rm ang}} eU^2\notag\\
 &+\delta^2e(h+1)-\delta e^2hU^2-\delta e^2U_b^2
                          +\delta e^2U_L^2],\notag\\
 z_5={}&-2\revadd{\chi_{\rm ang}}\delta(3\delta-2eU^2),\notag\\
 z_6={}&\delta^2(\delta-eU^2),\qquad c_j=z_j/z_6
                     \quad(j=0,\ldots,5).
 \label{eq:tension_entrained_primitive_sextic_entries}
\end{align}
The diagonal slip bounds give $z_0,z_6>0$.
Shifting $a\mapsto a-c_5/6$ removes the quintic term and gives
\begin{align}
 p_6={}&c_4-\frac5{12}c_5^2,\notag\\
 q_6={}&c_3-\frac23c_4c_5+\frac5{27}c_5^3,\notag\\
 r_6={}&c_2-\frac12c_3c_5+\frac16c_4c_5^2
                                      -\frac5{144}c_5^4,\notag\\
 s_6={}&c_1-\frac13c_2c_5+\frac1{12}c_3c_5^2
                  -\frac1{54}c_4c_5^3+\frac1{324}c_5^5,\notag\\
 t_6={}&c_0-\frac16c_1c_5+\frac1{36}c_2c_5^2
        -\frac1{216}c_3c_5^3\notag\\
 &+\frac1{1296}c_4c_5^4-\frac5{46656}c_5^6.
 \label{eq:tension_entrained_depressed_entries}
\end{align}
The sign polynomials used to count its real zeros are
\endgroup

\begingroup
\input{appendix_entrained_sign_polynomials_tension}
\endgroup

\begingroup
\paragraph{Multiplicity and the open domain.}
The signature of the multiplicity-weighted Hermite trace form counts
the distinct real zeros of the depressed sextic.
Frobenius' rule, including vanishing leading minors
\cite[pp.~206--207, Eqs.~(13)--(17)]{doi:10.1515/crll.1895.114.187},
then gives the three alternatives in
Eqs.~\eqref{eq:tension_entrained_sign_case_six}--\eqref{eq:tension_entrained_sign_case_two}.
In the rank-two alternative the polynomial has a conjugate pair of
triple roots,
\begin{equation}
 a^6+p_6a^4+q_6a^3+r_6a^2+s_6a+t_6
                  =(a^2+p_6/3)^3,\qquad p_6>0,
 \label{eq:tension_entrained_rank_two_cube}
\end{equation}
These zeros belong to the auxiliary elimination polynomial and do
not remove physical eigenvectors.

The resulting list classifies the strict causal-response branch.
The nonstrict coflow upper conditions also admit complete luminal
EOS boundaries. For general density-dependent BCS inputs, the
present argument does not establish that all such boundaries are
noninterior.
\endgroup

%% file: appendix_entrained_sign_polynomials_tension.tex
\begin{widetext}
\allowdisplaybreaks[1]
\begin{align}
 \mathfrak H_3 &=- 16 p_{6}^{3} + 48 p_{6} r_{6} - 54 q_{6}^{2},\notag\\
 \mathfrak H_4 &=32 p_{6}^{4} r_{6} - 12 p_{6}^{3} q_{6}^{2} + 96 p_{6}^{3} t_{6} - 224 p_{6}^{2} r_{6}^{2} + 300 p_{6} s_{6}^{2} - 81 q_{6}^{4} + 324 q_{6}^{2} t_{6} + 384 r_{6}^{3}\notag\\
 &\quad- 120 p_{6}^{2} q_{6} s_{6} + 324 p_{6} q_{6}^{2} r_{6} - 288 p_{6} r_{6} t_{6} - 720 q_{6} r_{6} s_{6},\notag\\
 \mathfrak H_5 &=32 p_{6}^{4} r_{6}^{3} - 384 p_{6}^{4} t_{6}^{2} - 256 p_{6}^{2} r_{6}^{4} - 2592 p_{6} t_{6}^{3} - 54 q_{6}^{4} r_{6}^{2} + 512 r_{6}^{5} + 3750 s_{6}^{4}\notag\\
 &\quad+144 p_{6}^{5} s_{6}^{2} - 4800 p_{6} q_{6} s_{6}^{3} - 2688 p_{6} r_{6}^{3} t_{6} + 162 q_{6}^{5} s_{6} + 3456 r_{6}^{2} t_{6}^{2} - 10800 r_{6} s_{6}^{2} t_{6}\notag\\
 &\quad- 128 p_{6}^{5} r_{6} t_{6} - 1232 p_{6}^{3} r_{6} s_{6}^{2} - 648 p_{6} q_{6}^{2} t_{6}^{2} - 2268 q_{6}^{3} s_{6} t_{6} - 2688 q_{6} r_{6}^{3} s_{6}\notag\\
 &\quad+48 p_{6}^{4} q_{6}^{2} t_{6} + 24 p_{6}^{3} q_{6}^{3} s_{6} - 8 p_{6}^{3} q_{6}^{2} r_{6}^{2} + 324 p_{6} q_{6}^{4} t_{6} + 288 p_{6} q_{6}^{2} r_{6}^{3}\notag\\
 &\quad+1184 p_{6}^{3} r_{6}^{2} t_{6} + 1116 p_{6}^{2} q_{6}^{2} s_{6}^{2} + 864 p_{6}^{2} r_{6} t_{6}^{2} + 2160 p_{6}^{2} s_{6}^{2} t_{6} + 1296 q_{6}^{2} r_{6}^{2} t_{6}\notag\\
 &\quad- 1296 p_{6}^{2} q_{6}^{2} r_{6} t_{6} + 2880 p_{6} r_{6}^{2} s_{6}^{2} + 3240 q_{6}^{2} r_{6} s_{6}^{2} + 6480 q_{6} s_{6} t_{6}^{2}\notag\\
 &\quad- 112 p_{6}^{4} q_{6} r_{6} s_{6} - 144 p_{6}^{3} q_{6} s_{6} t_{6} + 864 p_{6}^{2} q_{6} r_{6}^{2} s_{6} - 972 p_{6} q_{6}^{3} r_{6} s_{6}\notag\\
 &\quad+3024 p_{6} q_{6} r_{6} s_{6} t_{6},\notag\\
 \mathfrak H_6 &=- 13824 p_{6}^{3} t_{6}^{4} - 8748 q_{6}^{4} t_{6}^{3} - 1024 r_{6}^{6} t_{6} - 13824 r_{6}^{3} t_{6}^{3} + 3125 s_{6}^{6} - 46656 t_{6}^{5}\notag\\
 &\quad- 1024 p_{6}^{6} t_{6}^{3} + 108 p_{6}^{5} s_{6}^{4} + 729 q_{6}^{6} t_{6}^{2} + 108 q_{6}^{5} s_{6}^{3} + 34992 q_{6}^{2} t_{6}^{4} + 256 r_{6}^{5} s_{6}^{2}\notag\\
 &\quad- 8640 p_{6}^{3} q_{6}^{2} t_{6}^{3} - 17280 p_{6}^{2} r_{6}^{2} t_{6}^{3} - 32400 p_{6} s_{6}^{2} t_{6}^{3} - 22500 r_{6} s_{6}^{4} t_{6}\notag\\
 &\quad- 4352 p_{6}^{3} r_{6}^{3} t_{6}^{2} - 3750 p_{6} q_{6} s_{6}^{5} - 8640 q_{6}^{2} r_{6}^{3} t_{6}^{2} - 1600 q_{6} r_{6}^{3} s_{6}^{3}\notag\\
 &\quad- 192 p_{6}^{4} s_{6}^{2} t_{6}^{2} - 900 p_{6}^{3} r_{6} s_{6}^{4} - 128 p_{6}^{2} r_{6}^{4} s_{6}^{2} - 1350 q_{6}^{3} s_{6}^{3} t_{6}\notag\\
 &\quad- 64 p_{6}^{4} r_{6}^{4} t_{6} + 16 p_{6}^{4} r_{6}^{3} s_{6}^{2} + 16 p_{6}^{3} q_{6}^{3} s_{6}^{3} - 27 q_{6}^{4} r_{6}^{2} s_{6}^{2}\notag\\
 &\quad+512 p_{6}^{5} r_{6}^{2} t_{6}^{2} + 108 p_{6}^{3} q_{6}^{4} t_{6}^{2} + 512 p_{6}^{2} r_{6}^{5} t_{6} + 108 q_{6}^{4} r_{6}^{3} t_{6}\notag\\
 &\quad+825 p_{6}^{2} q_{6}^{2} s_{6}^{4} + 1500 p_{6}^{2} s_{6}^{4} t_{6} + 9216 p_{6} r_{6}^{4} t_{6}^{2} + 2000 p_{6} r_{6}^{2} s_{6}^{4} + 2250 q_{6}^{2} r_{6} s_{6}^{4}\notag\\
 &\quad+9216 p_{6}^{4} r_{6} t_{6}^{3} + 62208 p_{6} r_{6} t_{6}^{4} + 27000 q_{6} s_{6}^{3} t_{6}^{2} + 43200 r_{6}^{2} s_{6}^{2} t_{6}^{2}\notag\\
 &\quad- 27540 p_{6} q_{6}^{2} s_{6}^{2} t_{6}^{2} - 10560 p_{6} r_{6}^{3} s_{6}^{2} t_{6} - 77760 q_{6} r_{6} s_{6} t_{6}^{3}\notag\\
 &\quad- 6480 p_{6}^{2} r_{6} s_{6}^{2} t_{6}^{2} - 4860 p_{6} q_{6}^{4} r_{6} t_{6}^{2} - 9720 q_{6}^{2} r_{6}^{2} s_{6}^{2} t_{6}\notag\\
 &\quad- 576 p_{6}^{5} r_{6} s_{6}^{2} t_{6} - 576 p_{6}^{4} q_{6}^{2} r_{6} t_{6}^{2} - 630 p_{6} q_{6}^{3} r_{6} s_{6}^{3} - 576 p_{6} q_{6}^{2} r_{6}^{4} t_{6}\notag\\
 &\quad- 72 p_{6}^{4} q_{6} r_{6} s_{6}^{3} - 4 p_{6}^{3} q_{6}^{2} r_{6}^{2} s_{6}^{2} - 120 p_{6}^{3} q_{6} s_{6}^{3} t_{6} - 486 q_{6}^{5} r_{6} s_{6} t_{6}\notag\\
 &\quad+24 p_{6}^{4} q_{6}^{2} s_{6}^{2} t_{6} + 16 p_{6}^{3} q_{6}^{2} r_{6}^{3} t_{6} + 162 p_{6} q_{6}^{4} s_{6}^{2} t_{6} + 144 p_{6} q_{6}^{2} r_{6}^{3} s_{6}^{2}\notag\\
 &\quad+768 p_{6}^{5} q_{6} s_{6} t_{6}^{2} + 4816 p_{6}^{3} r_{6}^{2} s_{6}^{2} t_{6} + 560 p_{6}^{2} q_{6} r_{6}^{2} s_{6}^{3} + 3888 p_{6} q_{6}^{2} r_{6} t_{6}^{3}\notag\\
 &\quad+5832 p_{6}^{2} q_{6}^{3} s_{6} t_{6}^{2} + 8208 p_{6}^{2} q_{6}^{2} r_{6}^{2} t_{6}^{2} + 21384 q_{6}^{3} r_{6} s_{6} t_{6}^{2} + 6912 q_{6} r_{6}^{4} s_{6} t_{6}\notag\\
 &\quad- 5760 p_{6}^{3} q_{6} r_{6} s_{6} t_{6}^{2} - 4536 p_{6}^{2} q_{6}^{2} r_{6} s_{6}^{2} t_{6} + 46656 p_{6}^{2} q_{6} s_{6} t_{6}^{3}\notag\\
 &\quad- 72 p_{6}^{3} q_{6}^{3} r_{6} s_{6} t_{6} - 2496 p_{6}^{2} q_{6} r_{6}^{3} s_{6} t_{6} - 3456 p_{6} q_{6} r_{6}^{2} s_{6} t_{6}^{2}\notag\\
 &\quad+320 p_{6}^{4} q_{6} r_{6}^{2} s_{6} t_{6} + 2808 p_{6} q_{6}^{3} r_{6}^{2} s_{6} t_{6} + 19800 p_{6} q_{6} r_{6} s_{6}^{3} t_{6}.\label{eq:tension_entrained_explicit_sign_polynomials}
\end{align}
\end{widetext}

%% file: appendix_tension_state_bounds.tex
\begingroup
\allowdisplaybreaks[1]
\section{{Sufficient motion bounds for primitive recovery}}
\label{app:tension_state_bounds}

The contact matrix provides direct sufficient bounds on material
counterflow and transverse array slip. We derive the bounds used in
Sec.~\ref{sec:tension_state_bounds} and express them in both velocity
and canonical-momentum variables.
The allocation of a shared margin follows
Ref.~\cite[App.~C, Eqs.~(C36)--(C37)]{Most:2026uga}.
Here the superfluid contact response determines the margin.
All coefficients are evaluated at the trial state using
Eq.~\eqref{eq:tension_evaluated_response_pack}.

\paragraph{Physical and constitutive conditions.}
Assume a smooth EOS with finite responses and future unit material
four-velocities. In the normal frame, write $z^i=Wv_c^i$,
$z_n^i=W_nv_n^i$ and $W_x=\sqrt{1+|z_x|^2}$.
The positive-parameter and magnetic-dominance conditions are
\begingroup\colorlet{revisionblue}{mainrevisionblue}\color{revisionblue}
\begin{gather}
 n>0,\quad N>0,\quad s>0,\quad T_c>0,\notag\\
 H_0=\mu_n-\mathcal AN>0,\quad
 \mu_*>0,\quad\Delta_{\rm pair}>0,\notag\\
 \kappa_v>0,\quad\alpha_{\rm v}>0,\quad\mathcal N_0>0,
 \quad0<\xi_n<b_0=(\pi\mathcal N_0)^{-1/2},\notag\\
 Q_\Omega=\sqrt{B^2-E^2}>0,\quad E_iB^i=0,\quad
 \mathcal N=Q_\Omega/\kappa_v>0.
 \label{eq:tension_state_physical}
\end{gather}
\endgroup
They give $A_\Omega>0$ and $L_\Omega=\ln(b_0/\xi_n)>0$.
The core length $\xi_n$ is distinct from the phase momentum $\xi_i$.
With $h_n=nB_n$, the phase branch requires
\begin{gather}
 \xi_i=h_nz_{ni}+\mathcal ANz_i,\qquad
 \Xi=h_nW_n+\mathcal ANW,\notag\\
 \Xi>0,\qquad\Xi^2-\gamma^{ij}\xi_i\xi_j>0.
 \label{eq:tension_state_phase}
\end{gather}
Cold states set $s=T_c=0$ and omit entropy and its recovery equation.

The response and mixing conditions are
Eqs.~\eqref{eq:tension_state_response_inputs}--\eqref{eq:tension_state_response_bounds}.
The fractions $\epsilon_\Omega^-$ and $\epsilon_\Omega^+$ quantify
the reduction of density stiffness and null response by
material--line coupling. Positive denominators and fractions below
unity are equivalent to the strict coflow lower and upper signs.
For the adopted line law,
$c=A_\Omega\alpha_{\rm v}/\mathcal N_0$ and
$d=A_\Omega L_\Omega/\mathcal N$ also give
$a_\Omega=(c+d)/\kappa_v^2>0$.

\paragraph{Allocating the relative-motion margin.}
Let $q=B_n-a_ng^2$ and $w=\mathcal A-rg$, and define
\begin{align}
 M(g)&=k_{cc}q-w^2,\notag\\
 J_f(g)&=k_{cc}g^2f_n^2-2gwf_nf_c+qf_c^2.
 \label{eq:tension_state_joint_responses}
\end{align}
The determinant of the normalized contact matrix is
$1-\beta x-\delta y+(\beta\delta-\revadd{\chi_{\rm ang}})xy$,
with $x\leq g^2-1$ and $y\leq h$.
A positive lower bound over these magnitudes follows from
$h<tM/(dM+J_f)$.
Choose $0<\theta<1$ and the intermediate slip limit $H$ in
Eq.~\eqref{eq:tension_state_motion_caps}.
For $h\leq H$, positivity follows from
\begin{align}
 M-\lambda_{\rm b}J_f
 &=k_{\rm b}(B_n-a_{\rm b}g^2)
                         -(\mathcal A-r_{\rm b}g)^2>0.
 \label{eq:tension_state_allocated_gap}
\end{align}
This is a quadratic inequality in $g$.
It is positive at coflow, and its upper zero is the bound $G$ in
Eq.~\eqref{eq:tension_state_motion_caps}.
The response conditions make
$t-Hd,k_{\rm b},P_{\rm b},D_{\rm b}$ positive.
The conditions $g<G$ and $h<H$ therefore give positive compression
energy for every normal and preserve the carrier and bending
bases. The construction applies at $\mathcal A=0$ if line-energy
composition dependence still couples the carrier.

For $\mathcal Ar_{\rm b}<0$, evaluate the upper root as
$G=(B_nk_{\rm b}-\mathcal A^2)/
(\sqrt{D_{\rm b}}-\mathcal Ar_{\rm b})$ to avoid cancellation.
With $\revadd{h=1+[|WE+z\times B|^2-(E\cdot z)^2]/Q_\Omega^2}$,
the resulting velocity and field conditions are
Eq.~\eqref{eq:tension_state_motion_bounds}.

\paragraph{Canonical momentum and the phase branch.}
For $D_i=\xi_i-(\mathcal AN+h_nG)z_i$, the condition $g<G$ is
equivalent to the ellipsoid
\begin{equation}
 D_i(\gamma^{ij}-v_c^iv_c^j)D_j<h_n^2(G^2-1).
 \label{eq:tension_state_xi_bound}
\end{equation}
At fixed neutron velocity, its carrier form is
\begin{equation}
 (z_i-Gz_{ni})(\gamma^{ij}-v_n^iv_n^j)(z_j-Gz_{nj})<G^2-1.
 \label{eq:tension_state_carrier_bound}
\end{equation}
These ellipsoids are alternative expressions of the counterflow
bound, so they impose no additional restriction.
For $\mathcal A\geq0$, the positive momentum sum in
Eq.~\eqref{eq:tension_state_phase} is future timelike.
For $\mathcal A<0$, also require $h_n+\mathcal AN>0$ and
\begin{equation}
 g<G_{\rm phase}
 =\frac{h_n^2+(\mathcal AN)^2}{-2h_n\mathcal AN}.
 \label{eq:tension_state_negative_entrainment}
\end{equation}
The ellipsoids then use the smaller of $G$ and $G_{\rm phase}$.

\paragraph{Independent carrier and neutron--array slip.}
If $\mathcal A=0$ and the microscopic inputs depend only on $n$,
then $f_c=k_{nc}=r=0$ and the carrier Euler system is independent.
The strict constitutive conditions become
\begin{gather}
 B_n>0,\quad B_c>0,\quad0<k_{cc}/B_c<1,\notag\\
 0<a=k_{nn}/B_n<1,\qquad0<b=c/t<1,\notag\\
 0\leq\epsilon_{n\Omega}=\frac{f_n^2}{B_nt}
     <\min\{ab,(1-a)(1-b)\}.
 \label{eq:tension_state_decoupled_mixing}
\end{gather}
In this limit,
$\epsilon_\Omega^-=\epsilon_{n\Omega}/(ab)$ and
$\epsilon_\Omega^+=\epsilon_{n\Omega}/[(1-a)(1-b)]$.
Use $\mu_{\rm mix}$ and $\mathfrak u_{\rm b}$ from
Eq.~\eqref{eq:tension_state_decoupled_cap}.
Since $\xi^\mu=h_nu_n^\mu$, the canonical-field condition
\eqref{eq:tension_state_decoupled_slip} is equivalent to
$\mathfrak u^2<\mathfrak u_{\rm b}^{\,2}$.

To establish sufficiency, reduce the diagonal acoustic speeds to
$\sqrt{a(1-\mu_{\rm mix})}$ and
$\sqrt{b(1-\mu_{\rm mix})}$.
Below their relativistic speed sum, their intervals overlap at a
phase where both diagonal factors are positive.
Their product exceeds
$ab\mu_{\rm mix}^2=\epsilon_{n\Omega}$.
Equation~\eqref{eq:tension_evaluated_all_angle_lift} extends the
positive energy to every normal.
At zero mixing, the interval-overlap limit gives the continuous
value in Eq.~\eqref{eq:tension_state_decoupled_cap}.
In the array rest frame the bound is
$|\xi_{\perp B}|^2<h_n^2\mathfrak u_{\rm b}^{\,2}$.
The carrier velocity adds no relative-motion condition.
Material-independent line energy instead has the separable
conditions of Sec.~\ref{sec:separable_hyperbolicity}.

\paragraph{Constraints, sources and recovery.}
For either constitutive case, enforce
\begin{gather}
 B^i=\epsilon^{ijk}\nabla_j\xi_k,\qquad\nabla_iB^i=0,\notag\\
 \mathfrak B^i=\sqrt\gamma B^i,\qquad
 \mathfrak E^i=\sqrt\gamma\,a_\Omega E^i.
 \label{eq:tension_state_constraint_checks}
\end{gather}
Corrections to canonical edges must preserve the discrete curl
relation. Nonnegative internal heating requires
\begin{equation}
 \mathcal Q_\Omega
 =\kappa_v\mathcal N\,u^\nu\perp^\rho{}_{\nu}d_\rho\geq0,
 \label{eq:tension_state_source_check}
\end{equation}
A closed cold evolution requires zero net heating.
Finite smooth algebraic sources impose no additional principal-speed
bound. In conserved variables, the independent recovery equations
\eqref{eq:tension_method_material_inverse} must have a nonsingular
Jacobian, and the recovered state must reproduce every supplied
conserved component. Replacing warm entropy by energy also requires
a nonzero temperature response.

Apply the physical, phase and response conditions before the motion
bounds. Then use either the coupled inequalities
\eqref{eq:tension_state_motion_bounds} or the decoupled conditions
\eqref{eq:tension_state_decoupled_mixing} and
\eqref{eq:tension_state_decoupled_slip}, followed by the constraint,
source and recovery checks. Strict inequalities require an inward
numerical margin. A correction to $E$ changes $\mathcal N$, the
EOS coefficients and the motion bounds.
Recompute the full constitutive response and repeat the inequalities
after each correction. Record changes to the conserved image and use
the accepted state for $\mathfrak E$, stress and induction flux.
\endgroup

%% file: appendix_separable_characteristics_tension.tex
\begingroup
\section{{Separable line energy: waves, energy and constraints}}
\label{app:separable_characteristics}
\label{app:tension_draft}
\revadd{We now take the line energy $F=F(\mathcal N)$ to be independent of
the fluid state.} The mixed derivatives then vanish: the material block is that
of Sec.~\ref{app:force_equilibrium_model_b}, and the line waves
separate.

\subsection{Characteristic structure}
\label{sec:separable_characteristic_summary}
The unchanged material speeds are
Eqs.~\eqref{eq:main_finite_flow_sound_speeds}--%
\eqref{eq:main_model_b_material_speeds}. For the line waves put
\begin{align}
 v&=\frac{E\times B}{B^2},\quad V^2=|v|^2,\quad
 b_s=\frac{B\cdot s}{|B|},\quad v_s=v\cdot s,\notag\\
 c&=c_\Omega^2,\qquad \delta=1-V^2,\qquad d=1-cV^2.
 \label{eq:tension_coupled_line_primitives}
\end{align}
The two bending and two compressive line speeds are
\begin{align}
 \sigma_{A\pm}&=v_s\pm\sqrt\delta\,b_s,\notag\\
 \sigma_{F\pm}&=\frac{(1-c)v_s\pm\sqrt{\delta\mathcal L_s}}d,\notag\\
 \mathcal L_s&=d[c+(1-c)b_s^2]-c(1-c)v_s^2.
 \label{eq:tension_coupled_line_speeds}
\end{align}
With monic acoustic quartic $P_4$, the physical polynomial is
\begin{equation}
 P_{\rm phys}=(\sigma-\sigma_c)^3P_4(\sigma)
       \prod_{j=A,F}(\sigma-\sigma_{j-})(\sigma-\sigma_{j+}),
 \label{eq:tension_coupled_physical_polynomial}
\end{equation}
with a squared carrier factor in cold matter. The exact outer
speeds are $\sigma_{\min}=\min(\sigma_{\rm m}^-,\sigma_{F-})$ and
$\sigma_{\max}=\max(\sigma_{\rm m}^+,\sigma_{F+})$, where
$\sigma_{\rm m}^\pm$ are the extrema of the four sounds and the
carrier speed. For the quadratic line energy, $c=1$ and
$\sigma_{F\pm}=\pm1$.

\subsection{Nonlinear causality and strong hyperbolicity}
\label{sec:separable_hyperbolicity}
We now state the primitive conditions for the separable reference.
They hold for every three-dimensional normal and are derived in
Sec.~\ref{sec:tension_conservative_certificate}.

\paragraph{Physical branch.}
Use the matter branch and EOS responses of
Sec.~\ref{app:force_equilibrium_model_b}. Write
\begin{align}
 H&=\mu_n-\mathcal AN,\qquad I_h=\widehat\mu_p-\mathcal An,\notag\\
 S_n&=nk_n=\mu_n c_{n,\rm EOS}^2,\notag\\
 S_c&=Nk_h=\widehat\mu_p c_{c,\rm EOS}^2,\qquad
 \mathcal E=\mathcal A^2nN,\notag\\
 g&=-u_n\cdot u\geq1,\qquad w^2=1-g^{-2}.
 \label{eq:tension_draft_bulk_primitives}
\end{align}
The constituent sound responses allow arbitrary separable EOSs at
constant $\mathcal A$. The phase condition is equivalently
\begin{equation}
 \mu_n^2+2\mathcal AHN(g-1)>0.
 \label{eq:tension_draft_phase_bound}
\end{equation}
It is automatic for $\mathcal A\geq0$ and linear in $g$ otherwise.

\paragraph{Vortex lines.}
\revadd{Primes on $F$ denote derivatives with respect to $\mathcal N$.}
The complete primitive line bounds are
\begingroup\colorlet{revisionblue}{mainrevisionblue}\color{revisionblue}
\begin{gather}
 \mathcal N>0,\quad F'>0,\quad
 0<\mathcal NF''\leq F',\notag\\
 E\cdot B=0,\qquad B^2>E^2.
 \label{eq:tension_draft_line_bounds}
\end{gather}
\endgroup
Strictly linear line energy has $c_\Omega^2=0$ and a physical
compression Jordan block, so nonzero tension alone is insufficient.
The divergence and canonical compatibility constraints must be
preserved, Eq.~\eqref{eq:tension_cons_bare_jordan}.

\paragraph{Coupled stellar EOS with additive heat.}
At $\mathcal A\ne0$, all material conditions reduce to one charged
stiffness range and one neutron stiffness interval:
\begin{gather}
 0<S_c<I_h,\notag\\
 Hw^2+\frac{\mathcal E}{S_c}(1-g^{-1})^2
       < S_n\leq H-\frac{\mathcal E}{I_h-S_c}.
 \label{eq:tension_draft_bulk_bounds}
\end{gather}
The lower endpoint is the carrier-contact exclusion and the upper
endpoint is the light-cone bound. A luminal sound is allowed at the
upper equality. Replace that equality by a strict inequality if
strictly subluminal matter waves are required. If $\mathcal A=0$
identically, the complete replacement is
\begin{equation}
 0<S_n\leq\mu_n,\qquad 0<S_c\leq\widehat\mu_p,
 \label{eq:tension_draft_independent_bounds}
\end{equation}
with no relative-speed restriction. More general density or entropy
entrainment uses the test of Sec.~\ref{app:tension_draft_eos}.
Isolated constituent sound signs do not replace that coupled test.

\paragraph{Coincidences and constraint modes.}
Equations~\eqref{eq:tension_draft_line_bounds}--%
\eqref{eq:tension_draft_bulk_bounds} already include repeated sounds,
the double bending root, null fast/bending coincidences and every
matter/line coincidence: the constrained matter and line principal
operators have a positive direct-sum energy. At finite entrainment,
the strict lower bound excludes the nonsemisimple carrier-contact
crossing. Zero entrainment is the protected independent-fluid
exception.

\begingroup
\endgroup

\endgroup

%% file: appendix_tension_model.tex
\begingroup
\allowdisplaybreaks[1]
\section{{Line force and flux-conservative vortex equations}}
\label{app:tension_coupling}
\label{sec:tension_flux_conservative}
We now derive the conservative line equations for the shear-free
stress \cite{Carter:1995mj,Andersson:2019fdj}
\revadd{with line energy $F=F(\mathcal N)$}. The density terms of
Eq.~\eqref{eq:tension_main_stress_divergence} are then absent, and
Eq.~\eqref{eq:tension_draft_generalized_force} gives the normal force
\begin{align}
 \frac{L^\mu\Omega_{\mu\nu}}{\mathcal N}
 &=\perp_\nu{}^\alpha\bigl[
 \mathcal T(\ell^\mu\nabla_\mu\ell_\alpha
              -u_L^\mu\nabla_\mu u_{L\alpha})
                  -\nabla_\alpha\mathcal T\bigr]\notag\\
 &\quad+\kappa_v\perp^\rho{}_{\nu}d_\rho.
 \label{eq:tension_draft_curvature}
\end{align}
The curvature and line acceleration appear together, as in
Ref.~\cite[Eq.~(127)]{Andersson:2019fdj}. Omitting lattice
shear removes neither. The line stress is invariant under the
constant chemical-gauge changes of
Sec.~\ref{sec:theory_force_specialization}
\cite[Sec.~4.3]{Gavassino:2020zfk}, so that
$L^\mu-\mathcal J_\Omega^\mu$ has the same form in both bases. This
fixes the transformation of the lift coefficient, not its
microphysical value, which is not derived from $\revadd{F}$.

For the characteristic proof we use, instead of $E_i$, the
equivalent momentum $m_i=a_\Omega(E\times B)_i$. The flux and
stress conservation are those of the shear-free string fluid
\cite[Sec.~II]{Schubring:2014ena}, whose quadratic law
$\revadd{F=\tfrac12K_\Omega\mathcal N^2}$, with
$\mathcal T=K_\Omega\mathcal N$ and $c_\Omega^2=1$, serves as a
reference. No neutron-core calibration is assumed. Choose the
transverse representative $v=E\times B/B^2$ and put
$W_\Omega=(1-v^2)^{-1/2}$ and $\mathcal L=|B|/\kappa_v$, so that
$\mathcal N=\mathcal L/W_\Omega$ and
$m_i=\mathcal N\mathcal TW_\Omega^2v_i$. The energy
$\mathcal{E}_\Omega$ and stress $P_{\Omega i}{}^j$ are those of
Eqs.~\eqref{eq:rw_primitive_stress} and
\eqref{eq:tension_stress_projection_definitions} with $w_x^F=0$.
In an orthonormal inertial frame the conservation equations are
\begin{align}
 \partial_t\mathcal U_\Omega+\partial_j\mathcal F_\Omega^j
   &=\mathcal S_\Omega,\qquad
 \mathcal U_\Omega=(B^i,m_i,\mathcal{E}_\Omega)^T,\notag\\
 \mathcal F_\Omega^j&=(v^jB^i-B^jv^i,P_{\Omega i}{}^j,m^j)^T,\notag\\
 \mathcal S_\Omega&=(0,f_{\Omega i},v^if_{\Omega i})^T.
 \label{eq:tension_conservative_evolution}
\end{align}
Here the force $f_{\Omega i}$ is algebraic and normal to the
worldsheet. Curvature and tension gradients are contained in
$\partial_jP_{\Omega i}{}^j$ and must not be added again as sources.
On smooth constrained solutions, the energy equation follows from
induction and line-momentum conservation. It introduces no
additional degree of freedom. The speed $V=|v|$ is recovered from $|m|$ at
fixed $\mathcal L$, which is regular on the admitted domain because
$d|m|/dV=\mathcal L\mathcal TW_\Omega^3(1-c_\Omega^2V^2)>0$. Then
$v^i=Vm^i/|m|$ and $E=B\times v$. The algebraic
constraint $m_iB^i=0$ gives $E\cdot B=0$ and obeys
$\partial_t(m\cdot B)|_{m\cdot B=0}=a_\Omega B^2\,\nabla\cdot B$.
Thus, it is preserved when the magnetic-divergence constraint is
preserved.

\section{{Hyperbolicity and nonlinear causality at coincidences}}
\label{sec:tension_conservative_certificate}
We now show that, at every finite state satisfying
Eq.~\eqref{eq:tension_draft_line_bounds}, the conservative vortex
system has a complete, uniformly bounded physical characteristic
basis in every three-dimensional direction, including all
coincidences, and that its physical signals lie in the closed metric
cone. The physical subspace retains the algebraic constitutive
constraints and $\revadd{\nabla_i}B^i=0$.

To do so, rotate the orthonormal spatial axes so that $B=B_0e_3$,
$E=VB_0e_1$, $0\leq V<1$. The line velocity is $w=-Ve_2$. Use
$c,\delta,d$ of Eq.~\eqref{eq:tension_coupled_line_primitives} and
normalize variations by $\boldsymbol b=\delta B/B_0$ and
$\boldsymbol M=\delta m/(a_\Omega B_0^2)$. The tangent constitutive
constraints determine $\delta E$ and $M_z=Vb_y$, so
$Y=(M_x,M_y,b_x,b_y,b_z)^T$ are five regular local coordinates. Let
$A_C(k)$ be the symbol of the linearized fluxes of
Eq.~\eqref{eq:tension_conservative_evolution} in these coordinates.
With the divergence covector $L(k)=(0,0,k_x,k_y,k_z)$ and
$p=(0,0,0,-V,0)^T$, define the completion
$\widehat A(k)=A_C(k)+pL(k)$. Then $LA_C=0$ and
$L\widehat A=-Vk_yL$. Both operators leave $\ker L$ invariant and
agree there. The rank-one term controls a divergence-error mode and
is absent from the conservative physical fluxes. Such distinctions
are essential in force-free formulations
\cite[Secs.~3--4]{Pfeiffer:2013wza}. Define $H_C$ by the
direction-independent quadratic energy
\begin{align}
 \mathcal E_C[Y]=Y^TH_CY
 &=\frac{\delta^2}{d}
       [M_y+(1+c)Vb_z]^2+c\delta^2b_z^2\notag\\
 &\quad+\delta(M_x^2+b_y^2)+\delta^2b_x^2.
 \label{eq:tension_cons_energy}
\end{align}
It is positive definite, with $\det H_C=c\delta^8/d$, and direct
multiplication gives $H_C\widehat A^i=(\widehat A^i)^TH_C$ for all
three spatial matrices. Thus, the conservative physical operator is
self-adjoint in this positive energy and has an $H_C$-orthonormal
eigenbasis on $\ker L$, even at repeated roots. If $c\geq c_0>0$ and
$\delta\geq\delta_0>0$, then
$(c_0\delta_0^2/9)|Y|^2\leq\mathcal E_C[Y]\leq9|Y|^2$. Hence the
physical diagonalizer has condition number at most
$9/(\sqrt{c_0}\delta_0)$, with no angular gap denominator.

For a unit normal $k$, the physical characteristic polynomial is
\begin{align}
 P_{\rm phys}(\sigma,k)&=\mathcal B(\sigma,k)\mathcal F(\sigma,k)/d,
 \notag\\
 \mathcal B&=(\sigma+Vk_y)^2-\delta k_z^2,\notag\\
 \mathcal F&=d\sigma^2+2(1-c)Vk_y\sigma+(1-c)V^2k_y^2\notag\\
 &\quad-\delta\{c(k_x^2+k_y^2)+k_z^2\}.
 \label{eq:tension_cons_physical_polynomial}
\end{align}
The bending speeds $-Vk_y\pm\sqrt\delta k_z$ are causal because the
vectors $-Ve_2\pm\sqrt\delta e_3$ have unit norm. The fast quadratic
obeys
\begin{align}
 \frac{\operatorname{disc}\mathcal F}{4}
 &=\delta\{cdk_x^2+c\delta k_y^2+dk_z^2\}>0,\notag\\
 \mathcal F(\pm1,k)
 &=(1-c)[(k_y\pm V)^2+\delta k_x^2]\geq0,\notag\\
 \mathcal F&=(1-c)\mathcal B+\delta c(\sigma^2-1).
 \label{eq:tension_cons_causal_endpoints}
\end{align}
Its vertex is strictly between $-1$ and $1$, because
$d-(1-c)V=(1-V)(1+cV)>0$. Both real fast roots therefore lie in
$[-1,1]$. At a bending root the last identity gives
$\mathcal F(\sigma_A)=\delta c(\sigma_A^2-1)\leq0$, so the fast pair
encloses both bending speeds. For any future timelike time covector
with $t_0>|t|$, the completed time form obeys
$H_C[t_0I-\widehat A(t)]\geq(t_0-|t|)H_C>0$, so the result holds on
every metric-spacelike ADM slicing. The compression bounds are
sharp. In the line rest frame, transverse propagation has
$\sigma_F^2=c$: negative $c$ gives complex speeds, $c>1$
superluminal speeds and $c=0$ a physical Jordan chain. For the
quadratic law, the magnetic-null boundary also has a defective
physical double root.

\paragraph{Coincidences and the bare divergence-error mode.}
The two bending roots coincide for $k_z=0$, at $\sigma_A=-Vk_y$. By the last identity in
Eq.~\eqref{eq:tension_cons_causal_endpoints}, a common fast--bending
root must be null, $\sigma=\epsilon=\pm1$, with normals
$k=(0,-\epsilon V,\pm\sqrt\delta)$. Both roots are exactly double
and, by self-adjointness, each has two independent eigenvectors in
$\ker L$. In the completed system the drift eigenvector
$Y_D=H_C^{-1}L^T$ of speed $-Vk_y$ is energy-orthogonal to the
physical vectors, so its threefold root with the bending pair at
$k_z=0$ is semisimple as well. The physical restriction
is essential. For $k=e_1$, $q=(0,0,1,0,0)^T$ and
$r=(0,0,0,1,0)^T$,
\begin{equation}
 A_Cq=Vr,\qquad A_Cr=0,\qquad Lq=1,\quad Lr=0.
 \label{eq:tension_cons_bare_jordan}
\end{equation}
For every $V\ne0$ this is a Jordan chain, despite all physical
primitive inequalities being strict. Thus, the bare conservative
five-coordinate extension is not strongly hyperbolic without the
divergence constraint. The completion transports the divergence
error with the line velocity $w$ and obeys a linearized energy
estimate with no inverse frequency or wave-gap factor.

\paragraph{Canonical compatibility and coupled characteristics.}
\label{app:tension_coupled_reduction}
The nine Euler perturbations include two transverse neutron modes
that describe errors in $\nabla\times\xi-B$, rather than independent
vortex motion. A constrained reduction and an unrestricted extension
can have different hyperbolicity domains
\cite{Abalos:2018uwg}. Let $B_9(s)$ be the
time-normalized Euler matrix in the material primitive variables
$U_E=(n,N,\varsigma,z^i,\xi_i)$, $\varsigma=s/N$, for a unit normal
$s$. Define
\begin{align}
 T_s&=\begin{pmatrix}I_6&0\\0_{3\times6}&s\end{pmatrix},\quad
 P_s=T_sT_s^T,\quad Q_s=I_9-P_s,\notag\\
 Q_sB_9&=\sigma_nQ_s,\qquad
 \sigma_n=\frac{u_n\cdot s}{u_n^0},\notag\\
 B_7&=T_s^TB_9T_s,\qquad B_9T_s=T_sB_7.
 \label{eq:tension_coupled_physical_material_symbol}
\end{align}
Canonical compatibility at principal order is $Q_s\delta U_E=0$.
The physical material matrix $B_7$ is thus the restriction of
$B_9$. It has four
sounds and three carrier modes, with six fields in cold matter. The
removed factor is $(\sigma-\sigma_n)^2$. At spatial frequency
$\rho>0$ the lower-order part of compatibility is
$\delta\xi_T=\delta B\times s/(i\rho)$, which contributes only
$O(1)$ coupling to a first-order symbol.
\revadd{Let $B_{\Omega,4}(s)$ denote the restriction of the line
symbol $A_C(s)$ to the four-dimensional divergence-free space
$\ker L(s)$, expressed in orthonormal coordinates.}
The physical principal
matrix is therefore $B_{\rm phys}(s)=B_7(s)\oplus B_{\Omega,4}(s)$.
On a causal, uniformly strongly hyperbolic material domain with
symmetrizer $H_7$, the coupled energy is
$H_7\oplus\revadd{\left.H_C\right|_{\ker L(s)}}$: when a
fluid and a line speed coincide, the independent eigenspaces combine
and their dimensions add. The constraint-adapted extension
$\widehat B_9=P_sB_9P_s+\sigma_nQ_s$ is symmetrized by
$T_sH_7T_s^T+Q_s>0$ without assuming strong hyperbolicity of the
Euler error modes. Its correction $-P_sB_9Q_s$ is proportional to
the canonical constraint.

\paragraph{General entrained equations of state.}\label{app:tension_draft_eos}
Entrainment couples the fluid modes, so the isolated constituent
sound-speed signs alone are insufficient. At a regular finite state,
$B_7$ must then satisfy the definitions of
Sec.~\ref{sec:hyperbolicity} directly. With an invertible time
matrix and primitive Jacobian, $Z_E=(x+i\epsilon)I_7-B_7(s)$ and
$R_E=\operatorname{adj}Z_E$, the exact material test is
$\det[\sigma I_7-B_7(s)]\ne0$ for $|\sigma|>1$, together with
$\det Z_E\ne0$ and
$\epsilon^2\|R_E\|_F^2\leq C_E|\det Z_E|^2$, where one constant
$C_E$ must work for every unit $s$, every real $x$ and every
$\epsilon>0$. Cold matter uses the six-dimensional matrices. Applying
the test to $B_9$ instead tests the unrestricted Euler extension, a
stronger requirement. We caution that a test at one frozen state
does not establish the symmetrizer regularity used in
variable-coefficient estimates at an arbitrary EOS crossing. For
independent fluids, $\Lambda_{\rm mat}=-e_n(n)-e_c(N,s)$, the test
reduces to positive enthalpies and adiabatic sound speeds
$0<c_x^2\leq1$ at arbitrary relative flow, as in
Eq.~\eqref{eq:tension_draft_independent_bounds}.

\paragraph{Stellar EOS.}\label{app:tension_draft_bulk}
We now reduce the material test for the stellar EOS to
Eq.~\eqref{eq:tension_draft_bulk_bounds}. Use $X,Y,J,C_\ell$ from
Eq.~\eqref{eq:main_model_b_material_scalars} and $S_n,S_c,\mathcal E,w$
from Eq.~\eqref{eq:tension_draft_bulk_primitives}, and put
$\chi_n=S_n/H$, $\chi_c=S_c/I_h$ and
$\epsilon_{\mathcal A}=\mathcal E/(HI_h)$. In primitive stiffnesses,
$C_\ell$, the physical carrier-contact form $G_n$ and the auxiliary
neutron-contact form $G_c$ are
\begin{align}
 C_\ell&=HI_h[(1-\chi_n)(1-\chi_c)-\epsilon_{\mathcal A}],\notag\\
 G_n&=\frac{g^2}{N}
       [S_c(S_n-Hw^2)-\mathcal E(1-g^{-1})^2],\notag\\
 G_c&=\frac{g^2}{n}
       [S_n(S_c-I_hw^2)-\mathcal E(1-g^{-1})^2].
 \label{eq:tension_draft_dimensionless_identity}
\end{align}
Here $G_n$ is $G_\parallel$ of
Eq.~\eqref{eq:app_model_b_material_exact_gates}. For the physical
seven-field restriction and $\mathcal A\ne0$, the causal, complete branch is exactly $X>0$,
$Y>0$, $C_\ell\geq0$, $G_n>0$. The potential matrix and energy are
those of Sec.~\ref{app:force_equilibrium_model_b}. The basis remains
regular at the null boundary $C_\ell=0$ and at repeated acoustic
roots. At nonzero entrainment, a zero carrier-contact form
either has nonzero shear forcing or an unbounded limiting
eigenvector at an endpoint, so $G_n>0$ is necessary. With
$0<S_c<I_h$, the signs $C_\ell\geq0$ and $G_n>0$ are exactly the
upper and lower bounds on $S_n$ in
Eq.~\eqref{eq:tension_draft_bulk_bounds}. The lower bound gives
$S_n>0$, while the upper bound implies $X>0$ when $\mathcal A\ne0$.
Together they also ensure $J>0$. The phase condition
\eqref{eq:tension_draft_phase_bound} follows from
$(Hg+\mathcal AN)^2-H^2(g^2-1)=\mu_n^2+2\mathcal AHN(g-1)$, and
future orientation from $Hg+\mathcal AN=\mu_n+H(g-1)>0$. We caution
that $G_c>0$ is needed only by the unrestricted nine-field
extension, which permits two canonical-constraint errors, and must
not be imposed on physical circulation data. There are warm
states with $C_\ell>0$, $G_n>0$ and $G_c<0$ at which the raw
nine-field Euler symbol has a triple root with only two
eigenvectors, while the seven-field restriction, the eleven-field
symbol and the constraint-adapted extension retain complete bases.

\paragraph{Auxiliary electric constraint completion.}
\label{app:tension_full_certificate}
For the separable line EOS, the same physical system admits a
six-field electric completion. Put $Q_\Omega^2=B^2-E^2$,
$r_\Omega=(c-1)/Q_\Omega^2$, $G_\Omega=\tfrac12\nabla Q_\Omega^2$,
$M_\Omega=I-r_\Omega EE^T$ and $\Pi_B=I-BB^T/B^2$. Then
\begin{align}
 \Pi_BM_\Omega\partial_tE
 &=\Pi_B\bigl[\nabla\times B+r_\Omega G_\Omega\times B\notag\\
 &\quad-(\nabla\cdot E+r_\Omega E\cdot G_\Omega)v
 +r_\Omega E(B\cdot\nabla\times E)\notag\\
 &\quad+B\times f_\Omega/(a_\Omega B^2)\bigr],\notag\\
 \partial_t(E\cdot B)&=-v\cdot\nabla(E\cdot B),\notag\\
 \partial_tB&=-\nabla\times E-v\nabla\cdot B
                    -\frac B{B^2}\times\nabla(E\cdot B).
 \label{eq:tension_augmented_electric}
\end{align}
The added terms vanish on physical solutions. They construct a
symmetrizer, rather than prescribe the numerical parallel current
or replace the provisional electric update. At $c=1$ the principal
part is the augmented force-free system of
Ref.~\cite{Pfeiffer:2013wza}. In the aligned frame used
above, set $e_i=\delta E_i/B_0$ and $b_i=\delta B_i/B_0$. The energy
\begin{align}
 \mathcal H_\Omega={}&d(e_x-Vb_z)^2+c\delta^2b_z^2
                   +\delta(e_y^2+b_y^2)\notag\\
 &+(e_z+Vb_x)^2+(Ve_z+b_x)^2
 \label{eq:tension_moving_symmetrizer}
\end{align}
has determinant $cd\delta^6>0$ and symmetrizes every spatial
matrix, including the linearized constraint directions.
The physical cones are
\begin{align}
 (u_L\cdot k)^2-(\ell\cdot k)^2&=0,\notag\\
 (u_L\cdot k)^2-(\ell\cdot k)^2
                 -c\perp^{\mu\nu}k_\mu k_\nu&=0.
 \label{eq:tension_covariant_cones}
\end{align}
In the sheet rest frame they give
$\sigma_A=\pm\cos\theta$ and
$\sigma_F=\pm\sqrt{\cos^2\theta+c\sin^2\theta}$,
consistent with the line bounds above.
The compressive group speed obeys
$|\nabla_k\omega|^2=(k_\parallel^2+c^2k_\perp^2)/
(k_\parallel^2+ck_\perp^2)\leq1$. The bending rays follow
the null worldsheet directions $u_L\pm\ell$.
\endgroup

%% file: appendix_general_coefficient_tests_tension.tex
\begingroup
\allowdisplaybreaks[1]
\section{{Causality and a uniform eigenbasis for a general EOS}}
\label{sec:regc_exact}

The sufficient energy inequalities can fail while the physical
principal symbol remains causal and diagonalizable. We therefore
provide coefficient conditions that test a regular frozen state
without assuming positive canonical energy. The EOS is smooth and
the primitive time matrix is invertible. The analysis applies either
to the physical restriction or to a specified unrestricted evolution.
\revadd{We retain the circulation quantum $\kappa_v$ of
Eq.~\eqref{eq:superfluid_circulation_quantization}.}

\paragraph{Principal symbol.}
In the vortex rest frame, rotate $B'$ to $\revadd{\kappa_v\mathcal N e_3}$.
Use $Z=(j_n'^0,j_c'^0,B'_1,B'_2,B'_3)$,
$V=(\boldsymbol j_n',\boldsymbol j_c',E'_1,E'_2)$ and
$\varsigma=s_c/N$. Derivatives of the master function at $E'=0$
follow from Eq.~\eqref{eq:regc_current_hessian}. Put
$\mathcal K=\Lambda_{VV}=\operatorname{diag}(K,a_\Omega I_2)$,
where $K$ is the six-spatial-current Hessian, and let
$y=(Z,\varsigma)$. \revadd{Use the Legendre variables}
$w=(Z,\pi_V,\varsigma)$, with
$\pi_V=(\boldsymbol\pi_n,\boldsymbol\pi_c,D_{E1},D_{E2})$,
\revadd{where $\pi_x$ is the spatial canonical momentum conjugate to
$j_x'$ and $D_{Ei}=\Lambda_{E_i}$. We use
$\gamma_c=u_c'^0=\sqrt{1+|\boldsymbol u_c'|^2}$ for the carrier
Lorentz factor in this frame. The partial Legendre transform}
has Hessian
\begin{align}
 \mathcal E_{\pi_V\pi_V}&=\mathcal K^{-1},\qquad
 \mathcal E_{y\pi_V}=-\Lambda_{yV}\mathcal K^{-1},\notag\\
 \mathcal E_{yy}&=-\Lambda_{yy}+\Lambda_{yV}\mathcal K^{-1}\Lambda_{Vy}.
 \label{eq:tension_explicit_energy_hessian}
\end{align}
The inversion requires $a_\Omega\det K\ne0$.
Positivity of $\mathcal E$ and invertibility of the eight-current
constitutive Hessian are not assumed. The symmetric differential
array $S(s)$ and the Euler transport array $V(s)$ are specified by
\begin{align}
 S_{D_n,\boldsymbol\pi_n}&=s^T,\qquad
 S_{D_c,\boldsymbol\pi_c}=s^T,\qquad
 S_{B,D_E}=[s\times]_{:,1:2},
 \label{eq:tension_explicit_differential_array}\\
 V_{\pi_x\pi_x}(s)&=(v_x\cdot s)I_3-sv_x^T,
                         \quad x=n,c,\notag\\
 V_{\pi_c\varsigma}(s)&=-sT_c/\gamma_c,\qquad
 V_{\varsigma\varsigma}(s)=v_c\cdot s,
 \label{eq:tension_explicit_transport_array}
\end{align}
and the transposes of the first line. Here $D_x=j_x'^0$.
The canonical neutron extension sets
$V_{\pi_n\pi_n}=0$. The time-normalized symbol of the number,
canonical, carrier, Faraday and electric equations is
$A(s)=S(s)\mathcal E+V(s)$.

The orthogonal projector
$P(s)=\operatorname{diag}(I_2,I_3-ss^T,ss^T,I_3,I_2,1)$
has rank eleven and imposes $s\cdot\delta B=0$ and
$s\times\delta\pi_n=0$. Because $AP=PAP$, the physical symbol is
invariant on its image. In the coefficient expressions below, the
physical restriction replaces $A^i$ by $A^iP$, including
$A^0P=P$, and uses $d=11$. The unrestricted warm system uses
$P=I_{14}$ and $d=14$. Deleting entropy gives $d=10$ or $13$.
{All traces and Frobenius products below are taken on
$\operatorname{im}P$. Equivalently, use an orthonormal basis $R$
and the square matrix $R^TAR$. In the ambient representation the
zeroth power is $P$, with $\mu_0=\operatorname{tr}P=d$.
Using the ambient identity would add three spurious zero modes to
the physical characteristic and conditioning tests.}

For the warm Euler extension, divide $\det(xI-A)$ by
$x(x-v_n\cdot s)^2(x-v_c\cdot s)^3(x^2-q^2)$, with
$q=\widehat b\cdot s$. The quotient is the monic compression sextic
$p_6(x,s)=\sum_{k=0}^6a_k(s)x^k$.
The physical restriction omits $x(x-v_n\cdot s)^2$.
The canonical neutron extension replaces that factor by $x^3$.

\paragraph{Polynomial signs on the normal sphere.}
The root and eigenbasis conditions require polynomial inequalities
for every unit normal. Weighted trace forms provide a finite
real-root count for these inequalities
\cite{2019arXiv190108364P}.
Adjoin positive infinitesimals ordered as
$0<\delta\ll\tau_3\ll\tau_2\ll\tau_1\ll\epsilon\ll1$
to $\mathbb R$ and take the real closure.
\revadd{For a real polynomial $f(s)$ and Lagrange multiplier $z$, define}
\begin{align}
 F_f&=f(s)+\delta(1+\tau_1s_1+\tau_2s_2+\tau_3s_3),
 \label{eq:tension_positive_perturbation}\\
 I_f&=\left\langle s\cdot s-1,\quad
       \partial_{s_i}F_f-2z s_i\ (i=1,2,3)\right\rangle,
 \label{eq:tension_critical_ideal}\\
 \mathfrak n[f]&=
   \tfrac12\big[\operatorname{sig}\mathcal T_{F_f^2}
                         -\operatorname{sig}\mathcal T_{F_f}\big].
 \label{eq:tension_finite_negative_count}
\end{align}
Here $(\mathcal T_g)_{ab}=\operatorname{tr}M_{g v_av_b}$,
$\{v_a\}$ is a standard-monomial basis of the quotient algebra of
$I_f$, and $M_g$ denotes multiplication by $g$ in that algebra.
\revadd{For a symmetric matrix, $\operatorname{sig}$ is the number
of positive eigenvalues minus the number of negative eigenvalues.}
The positive perturbation makes the critical set finite. Its weighted
signatures count real critical points with $F_f<0$, including every
negative minimum. Consequently, $\mathfrak n[f]=0$ is equivalent
to $f\geq0$ on the unit sphere.

\paragraph{Real causal roots.}
Let $\nu_k$ be the Newton sums of $p_6$, computed recursively from
$a_k$\revadd{, with $\nu_0=6$ and
$\nu_k=\sum_{\alpha=1}^6\lambda_\alpha^k$ for $k\geq1$,
where the roots $\lambda_\alpha$ are counted with multiplicity}, and define
\begin{equation}
 \Delta_j(s)=\det(\nu_{a+b})_{a,b=0}^{j-1},\qquad
 b_j(s)=\sum_{k=j}^{6}\binom{k}{j}a_k(s).
 \label{eq:tension_exact_scalar_reality}
\end{equation}
If $m$ is the largest $j\leq6$ for which $\Delta_j$ has a nonzero
coefficient, the compression roots are real and causal in every
direction if and only if
\begin{align}
 \Delta_j&\not\equiv0,\quad \mathfrak n[\Delta_j]=0,
                    &&j=2,\ldots,m,\notag\\
 \mathfrak n[b_j]&=0,&&j=0,\ldots,5.
 \label{eq:tension_exact_scalar_cone}
\end{align}
The reality conditions follow from the positive semidefiniteness of
the Newton matrix $(\nu_{i+j})_{i,j=0}^5$.
The coefficients $b_j$ of $p_6(1+t,s)$ impose the upper light-cone
bound. The lower bound follows by reversing the normal, since
$A(-s)=-A(s)$. Transport and bending roots are already causal on
the admitted timelike-current and magnetically dominated branch.
These polynomial conditions determine root location. They do not
constrain the eigenvectors.

\paragraph{A uniformly bounded characteristic basis.}
\revadd{We use $\|\cdot\|_F$ for the Frobenius norm.}
For the selected physical or unrestricted symbol, set
$\mu_k=\operatorname{tr}A^k$, $\mu_0=d$ and
$r=\max_{1\leq j\leq d}\{j:D_j\not\equiv0\}$.
With $0\leq a,b<r$ in $G$, define
\begin{align}
 L_j&=(\mu_{a+b})_{a,b=0}^{j-1},\qquad D_j=\det L_j,\notag\\
 g&=\operatorname{adj}(L_r)(\mu_r,\ldots,\mu_{2r-1})^T,\notag\\
 G_{ab}&=\operatorname{tr}[(A^a)^TA^b],\qquad
 T_r=\operatorname{tr}[\operatorname{adj}(L_r)G],\notag\\
 \rho_r&=\left\|D_rA^r-\sum_{i=0}^{r-1}g_iA^i\right\|_F^2.
 \label{eq:tension_exact_scalar_arrays}
\end{align}
In addition to Eq.~\eqref{eq:tension_exact_scalar_cone}, a uniformly
bounded real diagonalization requires and is implied by
\begin{equation}
 {\rho_r\equiv0,\qquad
                  \mathfrak n[D_r-\epsilon T_r]=0.}
 \label{eq:tension_exact_scalar_uniform}
\end{equation}
On the dense stratum $D_r>0$, the identity $\rho_r\equiv0$
excludes nontrivial Jordan blocks. Moreover, $T_r/D_r$ is the trace
of the symmetrizer
$H_*=\sum_\lambda P_\lambda^TP_\lambda/m_\lambda\succeq I_d/d$,
where $P_\lambda$ is the spectral projector and $m_\lambda$ its
multiplicity. The second condition bounds this trace uniformly.
The positive infinitesimal $\epsilon$ detects divergence of the
projectors as the normal approaches a coincidence. Pointwise
semisimplicity at the coincident speed would not exclude that
divergence.

The rank $r$ is at most eleven for the Euler extension, ten for the
canonical neutron extension and nine for the physical system.
For the physical restriction, reduce the polynomials modulo
$s\cdot s-1$, because $P$ is not homogeneous away from the sphere.

\paragraph{Scope of the coefficient conditions.}
These conditions classify the regular frozen-symbol problem,
including complete luminal modes. They are finite but can be
expensive to evaluate, since the conditioning polynomials can have
degree $110$. The numerical method therefore uses the sufficient
primitive bounds. A state that fails those bounds must be evaluated
with the coefficient conditions before hyperbolicity is excluded.
A singular selected time matrix, a nonreal physical root, an
incompatible contact resonance or an unbounded eigenbasis violates
the corresponding necessary condition.

Quasilinear energy estimates additionally require a sufficiently
regular local symmetrizer \cite{Reula:2004xd}.
The positive-energy domain constructs one. The coefficient
conditions alone establish uniform diagonalizability and do not
establish the required regularity in state space.
Finite smooth lift, drag, Magnus and reaction forces are algebraic
sources. Tension and regulator forces containing density gradients
are already part of the principal EOS response.
Instantaneous force balance, discontinuous pinning switches,
vortex-free interfaces and added cleaning fields each require a
separate principal analysis.
\endgroup

%% file: appendix_model_b_tension_eigenvectors.tex
\allowdisplaybreaks[1]
\section{Characteristic eigenvectors and hyperbolicity}
\label{app:tension_eigenvectors}

Appendix~\ref{app:causality_hyperbolicity} provides the characteristic
speeds and causal domain. Here we construct the primitive right and
left eigenvectors of the NC model with line tension in the Eulerian frame, {following Teukolsky's method for transforming constrained
eigensystems between frames
\cite{Teukolsky:2025erz,Teukolsky:2025yle} and the presentation
of Ref.~\cite{Most:2026uga}.}
At a repeated root or a coincidence between characteristic families,
the individual columns are replaced by bases of the joint eigenspace.
{Repeated speeds use the joint normalization in
Eq.~\eqref{eq:tev_joint_normalization}.}

For a characteristic covector $k_\mu$, let
$\mathbb A(k)=\mathbb A^\mu k_\mu$ be the primitive principal
symbol. A right eigenvector is a nonzero column $R_I$ satisfying
$\mathbb A(k_I)R_I=0$. It gives the primitive perturbation carried
by the mode. A left kernel row $L_I$ satisfies
$L_I\mathbb A(k_I)=0$. The modal row of the evolution problem is
$\ell_I=L_I\mathbb A_n$, with
$\mathbb A_n=-n_\mu\mathbb A^\mu$.
For simple roots we normalize $\ell_IR_J=\delta_{IJ}$ on the
physical tangent space.

The state satisfies the material and line conditions of
Sec.~\ref{sec:vortex_tension_hyperbolicity}, with finite counterflow
and vortex slip. The material master function is
Eq.~\eqref{eq:regc_material_class},
with constant $\mathcal A$, and the line energy is
$F(n,N,\mathcal N)$. Its density dependence is retained throughout.
The formulas allow smooth constituent EOSs and the microscopic
responses in Appendix~\ref{app:tension_constitutive_derivatives}.
All perturbations obey
\begin{align}
 B\cdot\delta E+E\cdot\delta B&=0,\qquad
 s\cdot\delta B=0,\notag\\
 s\times\delta\pi_n&=0,\qquad \pi_{ni}=\xi_i.
 \label{eq:tev_constraints}
\end{align}
Thus the fifteen displayed Eulerian components below describe
eleven physical fields. Deleting entropy gives ten fields in cold
matter. Finite smooth algebraic vortex forces affect the source
terms, and do not alter this principal decomposition.

\subsection{Eulerian representation}
\label{app:tev_eulerian}

Let $(s,e_1,e_2)$ be an orthonormal Eulerian face triad and write
$k_\mu=(-\sigma,s_i)$ in local normal coordinates. The coordinate
speed is $\lambda=\alpha\sigma-\beta^is_i$.
The metric, triad and density weights are fixed in the principal
symbol. Eulerian and coordinate speeds therefore use the same
primitive columns. We use the vortex rest frame obtained by the canonical boost with
\begin{align}
 v_d&=\frac{E\times B}{B^2},\qquad
 \gamma_d=(1-v_d^2)^{-1/2},\notag\\
 \omega'&=\gamma_d(\sigma-v_d\cdot s),\notag\\
 k'&=s+\left[\frac{\gamma_d^2}{\gamma_d+1}v_d\cdot s
                         -\gamma_d\sigma\right]v_d,\notag\\
 m&=|k'|,\qquad a=\omega'/m,\qquad \widehat s=k'/m.
 \label{eq:tev_covector_map}
\end{align}
Here $a$ is the vortex-frame speed, and $m>0$ for a causal
characteristic. The aberrated normal $\widehat s$ depends on the
mode. All background components are expressed in one fixed boosted
frame, without a separate boost for each mode.

\subsubsection{Independent amplitudes and frame map}

In this frame, $E'=0$, $\revadd{B'=\kappa_v\mathcal N\widehat b}$ and
$u_x'=(\gamma_x,U_x)$, \revadd{with $x=n,c$ and the circulation
quantum $\kappa_v$ of Eq.~\eqref{eq:superfluid_circulation_quantization}.}
All background coefficients are obtained from Eulerian primitives:
\begingroup\colorlet{revisionblue}{mainrevisionblue}\color{revisionblue}
\begin{align}
 W_x&=\sqrt{1+|z_x|^2},\qquad
 \mathcal N=\frac{\sqrt{B^2-E^2}}{\kappa_v},\qquad
 \widehat b=\frac B{|B|},\notag\\
 \gamma_x&=\gamma_d(W_x-v_d\cdot z_x),\notag\\
 U_x&=z_x+\left[\frac{\gamma_d^2}{\gamma_d+1}v_d\cdot z_x
                         -\gamma_dW_x\right]v_d.
 \label{eq:tev_background_primitives}
\end{align}
\endgroup
The independent amplitudes are
\begin{align}
 \widehat V=\bigl(&\delta D_n,\delta J_n^i,
 \delta D_c,\delta J_c^i,\delta E'_1,\delta E'_2,
 \delta B'^i,\delta\varsigma\bigr)^T.
 \label{eq:tev_rest_amplitudes}
\end{align}
Here $D_x=j_x'^0$, $J_x=j_x'^i$ and
$\delta E'\cdot\widehat b=0$. The proper-density variations are
\begin{equation}
 \delta n_x=\gamma_x\delta D_x-U_x\cdot\delta J_x.
 \label{eq:tev_density_variations}
\end{equation}
The two current normalizations are therefore already imposed.

For later use define the fixed boost matrices
\begin{align}
 \mathsf L_d&=I+\frac{\gamma_d^2}{\gamma_d+1}v_dv_d^T,\notag\\
 \mathsf G_d&=\gamma_dI-
                   \frac{\gamma_d^2}{\gamma_d+1}v_dv_d^T,\qquad
 \mathsf C_dx=\gamma_dv_d\times x.
 \label{eq:tev_boost_blocks}
\end{align}
A rest-frame column gives the Eulerian current and field components
through
\begin{align}
 \delta D_x^{\rm E}&=\gamma_d(\delta D_x+v_d\cdot\delta J_x),\notag\\
 \delta J_x^{\rm E}&=\mathsf L_d\delta J_x+
                                     \gamma_dv_d\delta D_x,\notag\\
 \delta E&=\mathsf G_d\delta E'-\mathsf C_d\delta B',\notag\\
 \delta B&=\mathsf G_d\delta B'+\mathsf C_d\delta E',\notag\\
 \delta z_x&=\frac{\delta J_x^{\rm E}-z_x\delta n_x}{n_x}.
 \label{eq:tev_forward_map}
\end{align}
The background $z_x=W_xv_x$ is Eulerian. The scalar
$\delta\varsigma$ is unchanged. If entropy density is used instead,
$\delta s=N\delta\varsigma+\varsigma\delta N$.

\subsubsection{Explicit Eulerian components}

Every mode below specifies a tuple
$(\delta D_n,\delta J_n,\delta D_c,\delta J_c,
\delta E',\delta B',\delta\varsigma)$.
Equations~\eqref{eq:tev_density_variations} and
\eqref{eq:tev_forward_map} give its complete public column:
\begin{widetext}
\begin{equation}
 R^{\rm E}[\widehat r]=
 \bigl(\delta n,\delta N,\delta\varsigma,
 \delta z_n^s,\delta z_n^1,\delta z_n^2,
 \delta z_c^s,\delta z_c^1,\delta z_c^2,
 \delta E^s,\delta E^1,\delta E^2,
 \delta B^s,\delta B^1,\delta B^2\bigr)^T.
 \label{eq:tev_public_column}
\end{equation}
\end{widetext}

The left rows require the inverse primitive map and the Eulerian
time symbol. For an arbitrary Eulerian differential, first form
\begin{align}
 dD_x^{\rm E}&=W_x\,dn_x+n_xv_x\cdot dz_x,\notag\\
 dJ_x^{\rm E}&=z_x\,dn_x+n_x\,dz_x,\notag\\
 dD_x&=\gamma_d(dD_x^{\rm E}-v_d\cdot dJ_x^{\rm E}),\notag\\
 dJ_x&=\mathsf L_d\,dJ_x^{\rm E}-\gamma_dv_d\,dD_x^{\rm E},\notag\\
 dE'&=\mathsf G_d\,dE+\mathsf C_d\,dB,\notag\\
 dB'&=\mathsf G_d\,dB-\mathsf C_d\,dE.
 \label{eq:tev_coframe}
\end{align}
These are inverse maps on the algebraic tangent constraint in
Eq.~\eqref{eq:tev_constraints}. The boost matrices are held fixed
when differentiating.

The primitive material responses are
\begingroup\colorlet{revisionblue}{mainrevisionblue}\color{revisionblue}
\begin{align}
 B_n&=\frac{\mu_n+\partial_nF-\mathcal AN}{n},\qquad
 B_c=\frac{\mu_h+\partial_NF-\mathcal An}{N},\notag\\
 a_n&=B_n-k_n-\partial_n^2F,\qquad
 a_c=B_c-k_c-\partial_N^2F,\notag\\
 r&=\mathcal A-\partial_n\partial_NF,\qquad
 f_n=\partial_n\partial_{\mathcal N}F,\quad
 f_c=\partial_N\partial_{\mathcal N}F,\notag\\
 a_\Omega&=\frac{\partial_{\mathcal N}F}{\kappa_v^2\mathcal N},\qquad
 d_\Omega=\partial_{\mathcal N}^2F,\qquad
 \mu_\varsigma=\left(\frac{\partial\mu_h}{\partial\varsigma}\right)_N.
 \label{eq:tev_responses}
\end{align}
\endgroup
Here $\mu_n=d\rho_n/dn$,
$\mu_h=(\partial\rho_c/\partial N)_{\varsigma}$ and
$T_c=N^{-1}\partial\rho_c/\partial\varsigma$.
All density derivatives hold the other density,
$\mathcal N$ and $\varsigma$ fixed. Equivalently,
$k_n=\mu_nc_n^2/n$ and $k_c=\mu_hc_c^2/N$ in terms of the
constituent adiabatic sound responses.
For the adopted additive thermal EOS,
$\mu_\varsigma=4T_c/3$. No microscopic density derivative of $F$
is set to zero.

For $\theta_n=-a_n\,dn-r\,dN+f_n\,d\mathcal N$ and
$\theta_c=-r\,dn-a_c\,dN+f_c\,d\mathcal N+
\mu_\varsigma\,d\varsigma$, the momentum and line-stress
one-forms in the rest frame are
\begingroup\colorlet{revisionblue}{mainrevisionblue}\color{revisionblue}
\begin{align}
 d\pi_{x0}&=-B_x\,dD_x-\mathcal A\,dD_y-\gamma_x\theta_x,\notag\\
 d\pi_x&=B_x\,dJ_x+\mathcal A\,dJ_y+U_x\theta_x,
                         \qquad y\ne x,\notag\\
 d\mathcal N&=\frac{B\cdot dB-E\cdot dE}{\kappa_v^2\mathcal N}
              =\widehat b\cdot dB'/\kappa_v,\notag\\
 dH'&=a_\Omega(I-\widehat b\widehat b^T)dB'\notag\\
 &\quad+\widehat b\left[
 \frac{d_\Omega}{\kappa_v^2}\widehat b\cdot dB'
                 +\frac{f_n\,dn+f_c\,dN}{\kappa_v}\right].
 \label{eq:tev_momentum_forms}
\end{align}
\endgroup
The Eulerian spatial momentum forms are
\begin{equation}
 d\pi_x^{\rm E}=\mathsf L_d\,d\pi_x-
                                      \gamma_dv_d\,d\pi_{x0}.
 \label{eq:tev_momentum_boost}
\end{equation}

Let $v_x'=U_x/\gamma_x$. The Eulerian time forms, with a common
factor $\gamma_d$ suppressed, are
\begin{align}
 \mathcal D_x&=dD_x+v_d\cdot dJ_x,\notag\\
 \mathcal P_c&=(1+v_d\cdot v_c')d\pi_c\notag\\
 &\quad-v_d\left[d\pi_{c0}+v_c'\cdot d\pi_c
                         +\frac{T_c}{\gamma_c}d\varsigma\right],\notag\\
 \mathcal B&=dB'+v_d\times dE',\notag\\
 \mathcal E&=a_\Omega dE'-v_d\times dH',\notag\\
 \mathcal S&=(1+v_d\cdot v_c')d\varsigma.
 \label{eq:tev_time_forms}
\end{align}
They follow by transforming both the temporal and spatial
coefficients of the primitive equations. In particular, the
$\mathcal E$ form contains the material response through $dH'$.
It cannot be replaced by a field-only boost of a left row.

\subsubsection{The explicit three-row chart}
\label{app:tev_three_row}

We reduce the spatial-current inverse to two scalar responses.
Put
\begin{align}
 \Delta&=B_nB_c-\mathcal A^2,\qquad
 \beta=\frac1\Delta
       \begin{pmatrix}B_c&-\mathcal A\\-\mathcal A&B_n\end{pmatrix},\notag\\
 W&=\begin{pmatrix}
 \beta_{nn}|U_n|^2&\beta_{nc}U_n\cdot U_c\\
 \beta_{nc}U_n\cdot U_c&\beta_{cc}|U_c|^2
 \end{pmatrix},\notag\\
 d_v&=1+a_nW_{11}+2rW_{12}+a_cW_{22}\notag\\
 &\quad+(a_na_c-r^2)(W_{11}W_{22}-W_{12}^2).
 \label{eq:tev_spatial_feedback}
\end{align}
The three entries of the symmetric feedback matrix are
\begin{align}
 T_{11}&=\frac{a_n+(a_na_c-r^2)W_{22}}{d_v},\notag\\
 T_{12}&=\frac{r-(a_na_c-r^2)W_{12}}{d_v},\notag\\
 T_{22}&=\frac{a_c+(a_na_c-r^2)W_{11}}{d_v}.
 \label{eq:tev_feedback_entries}
\end{align}
Define $W_e=W-WTW$ and $J_v=I_2-WT$. Their explicit entries are
\begin{align}
 W_e&=\frac1{d_v}\begin{pmatrix}
 W_{11}+a_c\det W&W_{12}-r\det W\\
 W_{12}-r\det W&W_{22}+a_n\det W
 \end{pmatrix},\notag\\
 J_v&=\frac1{d_v}\begin{pmatrix}
 1+rW_{12}+a_cW_{22}&-rW_{11}-a_cW_{12}\\
 -a_nW_{12}-rW_{22}&1+a_nW_{11}+rW_{12}
 \end{pmatrix}.
 \label{eq:tev_feedback_completions}
\end{align}
{This chart uses $B_n>0$, $\Delta>0$, $a_\Omega>0$ and
positive spatial-current Hessian $K_{\rm sp}$, denoted by $K$ in
Eq.~\eqref{eq:regc_lowrank_blocks}. Then
$d_v=\det K_{\rm sp}/\Delta^3>0$.
These signs hold on the coflow and sufficient positive-energy
domains of Appendix~\ref{app:regulated_characteristics}.
The general coefficient conditions of Sec.~\ref{sec:regc_exact}
can also test regular states outside this chart.}
Neither a relative velocity nor a current Gram determinant is
divided out.

For spatial covectors $g_n,g_c$, define their primitive inverse by
\begin{align}
 \zeta_n&=U_n\cdot(\beta_{nn}g_n+\beta_{nc}g_c),\notag\\
 \zeta_c&=U_c\cdot(\beta_{nc}g_n+\beta_{cc}g_c),\qquad
 w=T(\zeta_n,\zeta_c)^T,\notag\\
 \mathcal J_n[g]&=\beta_{nn}(g_n-U_nw_n)
                         +\beta_{nc}(g_c-U_cw_c),\notag\\
 \mathcal J_c[g]&=\beta_{nc}(g_n-U_nw_n)
                         +\beta_{cc}(g_c-U_cw_c).
 \label{eq:tev_spatial_inverse}
\end{align}
Every coefficient is an algebraic function of the primitives and
EOS responses in Eq.~\eqref{eq:tev_responses}.

Complete the normalized rest-frame normal by
\begin{align}
 q&=\widehat b\cdot\widehat s,\qquad p=\sqrt{1-q^2},\notag\\
 e_p&=(\widehat s-q\widehat b)/p,\qquad
 e_A=\widehat b\times e_p,\qquad t=qe_p-p\widehat b.
 \label{eq:tev_wave_triad}
\end{align}
At $p=0$ choose any fixed orthonormal pair $(e_p,e_A)$ transverse
to $\widehat b$. The formulas below contain no division by $p$.
Set
\begin{widetext}
\begingroup\colorlet{revisionblue}{mainrevisionblue}\color{revisionblue}
\begin{align}
 Z&=\operatorname{diag}(U_n\cdot\widehat s,
                                  U_c\cdot\widehat s)\,\beta,\notag\\
 D_0&=\begin{pmatrix}
 -\gamma_na_n&-\gamma_nr\\
 -\gamma_cr&-\gamma_ca_c\\
 -pf_n/\kappa_v&-pf_c/\kappa_v
 \end{pmatrix},\notag\\
 M_0&=\begin{pmatrix}
 B_n-a_n\gamma_n^2&\mathcal A-r\gamma_n\gamma_c&-p\gamma_nf_n/\kappa_v\\
 \mathcal A-r\gamma_n\gamma_c&B_c-a_c\gamma_c^2&-p\gamma_cf_c/\kappa_v\\
 -p\gamma_nf_n/\kappa_v&-p\gamma_cf_c/\kappa_v&a_\Omega q^2+p^2d_\Omega/\kappa_v^2
 \end{pmatrix},\notag\\
 M&=M_0+D_0W_eD_0^T,\qquad C=-D_0J_vZ,\notag\\
 \mathsf I&=\beta-Z^TTZ,\qquad
 d_I=I_{11}I_{22}-I_{12}^2,\notag\\
 \mathsf I^{-1}&=\frac1{d_I}
                 \begin{pmatrix}I_{22}&-I_{12}\\-I_{12}&I_{11}\end{pmatrix}.
 \label{eq:tev_density_coefficients}
\end{align}
\endgroup
\end{widetext}
Here $M$ is three by three, $C$ is three by two, and $\mathsf I$ is two by
two. The positive spatial-current inertia gives $\mathsf I>0$.

The three density amplitudes are $z=(\delta D_n,\delta D_c,b)^T$,
where $\delta B'=bt$. Let
$\mathfrak p=(\Pi_n,\Pi_c,\delta(a_\Omega E'_A))^T$\revadd{,
where $\Pi_x=\widehat s\cdot\delta\pi_x$ are the longitudinal
canonical-momentum amplitudes and $E'_A=e_A\cdot E'$.
We collect them in $\Pi=(\Pi_n,\Pi_c)^T$} and define
\begin{align}
 D_3&=\operatorname{diag}(1,1,-1),\qquad C_p=(C\ \ 0),\notag\\
 H_p&=\operatorname{diag}(\mathsf I,a_\Omega^{-1}),\notag\\
 H_6&=\begin{pmatrix}M&C_p\\C_p^T&H_p\end{pmatrix},\qquad
 J_6=\begin{pmatrix}0&D_3\\D_3&0\end{pmatrix}.
 \label{eq:tev_six_wave_blocks}
\end{align}
With the carrier amplitudes set to zero, the six principal equations are
\begin{equation}
 a\begin{pmatrix}z\\\mathfrak p\end{pmatrix}
       =J_6H_6\begin{pmatrix}z\\\mathfrak p\end{pmatrix}.
 \label{eq:tev_six_wave_equations}
\end{equation}
The first three give
$\mathfrak p=H_p^{-1}(aD_3-C_p^T)z$.
Its third component is $\delta(a_\Omega E'_A)=-a_\Omega ab$.
Substitution into the last three gives the symmetric pencil
\begin{align}
 B_{i\alpha}(a)&=a\delta_{i\alpha}-C_{i\alpha},
                         \qquad \alpha=1,2,\notag\\
 K_{ij}(a)&=M_{ij}-\mathcal Q(B_i,B_j)
                            -a_\Omega a^2\delta_{i3}\delta_{j3},\notag\\
 \mathcal Q(x,y)&=\frac{I_{22}x_1y_1-I_{12}(x_1y_2+x_2y_1)
                                      +I_{11}x_2y_2}{d_I}.
 \label{eq:tev_density_pencil}
\end{align}
Thus all nine entries are
\begin{widetext}
\begin{align}
 k_{11}&=M_{11}-\mathcal Q(B_1,B_1),&
 k_{12}&=M_{12}-\mathcal Q(B_1,B_2),\notag\\
 k_{13}&=M_{13}-\mathcal Q(B_1,B_3),&
 k_{21}&=k_{12},\notag\\
 k_{22}&=M_{22}-\mathcal Q(B_2,B_2),&
 k_{23}&=M_{23}-\mathcal Q(B_2,B_3),\notag\\
 k_{31}&=k_{13},\qquad k_{32}=k_{23},&
 k_{33}&=M_{33}-\mathcal Q(B_3,B_3)-a_\Omega a^2.
 \label{eq:tev_nine_entries}
\end{align}
\end{widetext}
The monic mixed-compression polynomial is
\begin{equation}
 \mathcal P_6^{\rm prim}(a,\widehat s)
                   =-\frac{d_I}{a_\Omega}\det K(a).
 \label{eq:tev_primitive_sextic}
\end{equation}
Indeed, block elimination gives
$\det(aI_6-J_6H_6)=-(d_I/a_\Omega)\det K$.
It has the roots and multiplicities of $P_6$ in
Eq.~\eqref{eq:regulated_characteristic_factorization}.
The density chart does not require inversion of the eight-current
constitutive map used in the potential reduction of
Sec.~\ref{app:primitive_dispersion}.

\clearpage
At a simple root, the three polarization charts are
\begin{widetext}
\begin{align}
 z^{(12)}&=(k_{12}k_{23}-k_{13}k_{22},
             k_{13}k_{12}-k_{11}k_{23},
             k_{11}k_{22}-k_{12}^2)^T,\notag\\
 z^{(23)}&=(k_{22}k_{33}-k_{23}^2,
             k_{23}k_{13}-k_{12}k_{33},
             k_{12}k_{23}-k_{22}k_{13})^T,\notag\\
 z^{(31)}&=(k_{23}k_{13}-k_{33}k_{12},
             k_{33}k_{11}-k_{13}^2,
             k_{12}k_{13}-k_{23}k_{11})^T.
 \label{eq:tev_polarization_charts}
\end{align}
\end{widetext}
They are the cross products of pairs of rows of $K$.
At rank two at least one is nonzero. A vanishing chart is replaced
by another nonzero chart. Nonzero columns can be rescaled to unit
norm before forming their duals. When all cofactors vanish, use an
independent basis of $\ker K$.

\subsection{NC model with vortex tension}
\label{app:tev_model_b}

The physical determinant and mode count are
\begin{align}
 \det\mathbb A(k)&\propto
 (a-v_c'\cdot\widehat s)^3(a^2-q^2)
                         \mathcal P_6^{\rm prim}(a,\widehat s),\notag\\
 11&=3_{\rm carrier}+2_{\rm bend}+6_{\rm comp}.
 \label{eq:tev_count}
\end{align}
The carrier factor is squared in cold matter. For an Eulerian face,
Eq.~\eqref{eq:tev_covector_map} is inserted before solving for
$\sigma$. Homogeneity gives the Eulerian compression polynomial
$m^6\mathcal P_6^{\rm prim}(\omega'/m,k'/m)$.
No factor of $a$, $p$ or $q$ is cancelled after specializing the
state. Such a cancellation can remove physical multiplicities.

\subsubsection{All right columns}
\label{app:tev_right}

We first give one reconstruction that applies to the compression
and carrier families. Let $z=(z_1,z_2,b)^T$ and let
$\eta=(\eta_1,\eta_2,\eta_\varsigma)^T$ specify the two transverse
carrier momenta and entropy. The longitudinal momenta are
\begin{equation}
 \Pi=\mathsf I^{-1}\left[
 a\begin{pmatrix}z_1\\z_2\end{pmatrix}-C^Tz-F_p\eta\right].
 \label{eq:tev_longitudinal_momenta}
\end{equation}
The matrix $F_p$, needed for carrier modes, is given below.
Define
\begingroup\colorlet{revisionblue}{mainrevisionblue}\color{revisionblue}
\begin{align}
 g_n&=\widehat s\Pi_n+U_n\left[
 \gamma_na_nz_1+\gamma_crz_2+\frac{pf_n}{\kappa_v}b\right],\notag\\
 g_c&=\widehat s\Pi_c+t\eta_1+e_A\eta_2\notag\\
 &\quad+U_c\left[
 \gamma_nrz_1+\gamma_ca_cz_2+\frac{pf_c}{\kappa_v}b
                         -\mu_\varsigma\eta_\varsigma\right].
 \label{eq:tev_reconstruction_covectors}
\end{align}
\endgroup
All nonzero amplitudes of the rest-frame column are
\begin{align}
 \delta D_n&=z_1,\qquad \delta D_c=z_2,\notag\\
 \delta J_n&=\mathcal J_n[g],\qquad
 \delta J_c=\mathcal J_c[g],\notag\\
 \delta E'&=-ab e_A,\qquad \delta B'=bt,\qquad
 \delta\varsigma=\eta_\varsigma.
 \label{eq:tev_reconstructed_column}
\end{align}
In particular,
$\delta\pi_n=\widehat s\Pi_n$ and
$\delta\pi_c=\widehat s\Pi_c+t\eta_1+e_A\eta_2$.
Equations~\eqref{eq:tev_forward_map} and
\eqref{eq:tev_public_column} print every Eulerian component.

For each of the six compression roots, set $\eta=0$ and use any
nonzero polarization of Eq.~\eqref{eq:tev_polarization_charts}.
The first two number equations eliminate $\Pi$, while induction
gives $\delta E'=-ab e_A$. The remaining three equations are
$K(a)z=0$. Thus the construction solves the primitive equations
without dividing by the wave frequency.

At each bending root, the complete nonzero entries are
\begin{equation}
 a=\epsilon q:\qquad
 \delta E'=\epsilon e_p,\qquad
 \delta B'=e_A,\qquad \epsilon=\pm1.
 \label{eq:tev_bending_columns}
\end{equation}
All material amplitudes and $\delta\mathcal N$ vanish. The two
columns remain independent at $q=0$. Their Eulerian speeds are
Eq.~\eqref{eq:regulated_bending_speeds}.

To obtain the carrier columns, first evaluate three spatial
responses at zero $z$ and zero longitudinal momenta:
\begin{align}
 g_n^{(j)}&=0,\qquad
 (g_c^{(1)},g_c^{(2)},g_c^{(\varsigma)})
                   =(t,e_A,-\mu_\varsigma U_c),\notag\\
 J_x^{(j)}&=\mathcal J_x[g^{(j)}],\qquad
 n_j=-U_n\cdot J_n^{(j)},\quad N_j=-U_c\cdot J_c^{(j)}.
 \label{eq:tev_carrier_seed_response}
\end{align}
The two-by-three matrix $F_p$ and three-by-three matrix $G_z$ have
columns
\begingroup\colorlet{revisionblue}{mainrevisionblue}\color{revisionblue}
\begin{align}
 (F_p)_j&=\begin{pmatrix}
 \widehat s\cdot J_n^{(j)}\\\widehat s\cdot J_c^{(j)}
 \end{pmatrix},\notag\\
 (G_z)_j&=\begin{pmatrix}
 -\gamma_n(a_nn_j+rN_j)\\
 -\gamma_c(rn_j+a_cN_j-\mu_\varsigma\delta_{j\varsigma})\\
 -p(f_nn_j+f_cN_j)/\kappa_v
 \end{pmatrix},\notag\\
 V_p&=\begin{pmatrix}
 0&0&0\\-v_c'\cdot t&-v_c'\cdot e_A&-T_c/\gamma_c\\0&0&0
 \end{pmatrix},\notag\\
 h(a,\eta)&=-\left[B(a)\mathsf I^{-1}F_p+G_z+D_3V_p\right]\eta.
 \label{eq:tev_carrier_forcing}
\end{align}
\endgroup
These are the inhomogeneous terms left by carrier vorticity and
entropy after eliminating the longitudinal momenta.

We choose the carrier seeds in the Eulerian face frame. Write
$v_c^{\rm E}=z_c/W_c$ and define covectors
\begin{align}
 w^{\rm E}_{A\mu}&=(-v_c^{\rm E}\cdot e_A^{\rm E},e_A^{\rm E}),
                            \qquad A=1,2,\notag\\
 w^{\rm E}_{\varsigma\mu}&=(-T_c/W_c,0),\notag\\
 w'_{j0}&=\gamma_d(w^{\rm E}_{j0}+v_d\cdot w_j^{\rm E}),\notag\\
 w'_j&=\mathsf L_d w_j^{\rm E}+\gamma_dv_dw^{\rm E}_{j0},\notag\\
 \eta_j&=(t\cdot w'_j,e_A\cdot w'_j,\delta_{j\varsigma})^T.
 \label{eq:tev_carrier_seeds}
\end{align}
The unprimed $e_A^{\rm E}$ are the two Eulerian face tangents,
whereas $e_A$ without a superscript is the rest-frame vector in
Eq.~\eqref{eq:tev_wave_triad}.
For $\sigma=\sigma_c=v_c^{\rm E}\cdot s$, evaluate
{$(a_{\rm adv},\widehat s)$} with Eq.~\eqref{eq:tev_covector_map}.
{Here $a_{\rm adv}=U_c\cdot\widehat s/\gamma_c$.} Set
\begin{equation}
 K({a_{\rm adv}})z_j=h({a_{\rm adv}},\eta_j).
 \label{eq:tev_carrier_completion}
\end{equation}
On the strict contact-gap branch this inverse is explicit. With
$\mathfrak k=\det K$,
\begin{widetext}
\begin{align}
 \mathfrak k&=k_{11}k_{22}k_{33}+2k_{12}k_{13}k_{23}
 -k_{11}k_{23}^2-k_{22}k_{13}^2-k_{33}k_{12}^2,\notag\\
 A_K&=\begin{pmatrix}
 k_{22}k_{33}-k_{23}^2&k_{13}k_{23}-k_{12}k_{33}&k_{12}k_{23}-k_{13}k_{22}\\
 k_{13}k_{23}-k_{12}k_{33}&k_{11}k_{33}-k_{13}^2&k_{12}k_{13}-k_{11}k_{23}\\
 k_{12}k_{23}-k_{13}k_{22}&k_{12}k_{13}-k_{11}k_{23}&k_{11}k_{22}-k_{12}^2
 \end{pmatrix},\notag\\
 z_j&=A_Kh({a_{\rm adv}},\eta_j)/\mathfrak k.
 \label{eq:tev_carrier_inverse}
\end{align}
\end{widetext}
Inserting $(z_j,\eta_j)$ into
Eqs.~\eqref{eq:tev_longitudinal_momenta}--\eqref{eq:tev_reconstructed_column}
gives the three carrier columns $C_j^{\rm E}$.
At a compatible zero of $\mathfrak k$, use a particular solution of
Eq.~\eqref{eq:tev_carrier_completion} and a basis of $\ker K$.

\subsubsection{All left rows}
\label{app:tev_left}

The carrier rows are the following primitive one-forms:
\begin{equation}
 \zeta_1=e_1^{\rm E}\cdot d\pi_c^{\rm E},\qquad
 \zeta_2=e_2^{\rm E}\cdot d\pi_c^{\rm E},\qquad
 \zeta_\varsigma=d\varsigma.
 \label{eq:tev_carrier_rows}
\end{equation}
The chosen seeds give $\zeta_i(C_j^{\rm E})=\delta_{ij}$.
These rows annihilate every compression and bending column.
They therefore already have the required normalization.

For a compression column $I$, let $(\Pi_{nI},\Pi_{cI},b_I)$
be its reconstructed momenta and magnetic amplitude. Define
\begin{widetext}
\begin{align}
 H_{sI}&=\widehat s_I\cdot\delta H'_I,\notag\\
 \Phi_I&=\Pi_{nI}\mathcal D_n+\Pi_{cI}\mathcal D_c\notag\\
 &\quad+\frac{m_I}{\gamma_d}
 \left[\delta D_{nI}^{\rm E}s\cdot d\pi_n^{\rm E}
       +\delta D_{cI}^{\rm E}s\cdot d\pi_c^{\rm E}\right]\notag\\
 &\quad+a_\Omega a_Ib_It_I\cdot\mathcal B
       -H_{sI}v_d\cdot dB'-b_Ie_{AI}\cdot\mathcal E.
 \label{eq:tev_compression_row}
\end{align}
\end{widetext}
All the differential forms on the right are given in
Eqs.~\eqref{eq:tev_coframe}--\eqref{eq:tev_time_forms}.
The coefficients $\delta D_{xI}^{\rm E}$ and $\delta H'_I$ are
computed from the displayed right column, with no new inversion.

For a bending column, take
\begin{equation}
 \Phi_{A\epsilon}=\frac12\left[
 e_A\cdot\mathcal B+\frac{\epsilon}{a_\Omega}e_p\cdot\mathcal E
 \right],\qquad \epsilon=\pm1.
 \label{eq:tev_bending_rows}
\end{equation}
Both expressions include the Eulerian time symbol. To see why no
frequency pole occurs in Eq.~\eqref{eq:tev_compression_row}, the
physical magnetic constraint gives
\begin{equation}
 \widehat s_I\cdot\mathcal B=-a_Iv_d\cdot dB'.
 \label{eq:tev_divergence_cancellation}
\end{equation}
It removes the apparent $1/a_I$ in the longitudinal magnetic
part of the left row. Number conservation and neutron canonical
compatibility similarly remove the apparent neutron-contact pole.

Off the carrier contact, direct elimination instead gives the
carrier equation weights
\begin{align}
 A_{cI}&=\frac{\delta J_{cI}-v_c'\delta D_{cI}}
                         {a_I-v_c'\cdot\widehat s_I},\notag\\
 B_{\varsigma I}&=\frac{\mu_\varsigma\delta N_I
                            -(T_c/\gamma_c)\delta D_{cI}}
                         {a_I-v_c'\cdot\widehat s_I}.
 \label{eq:tev_intermediate_left_weights}
\end{align}
The corresponding row uses $A_{cI}\cdot\mathcal P_c+
B_{\varsigma I}\mathcal S$. Its restriction to
$\zeta_1=\zeta_2=\zeta_\varsigma=0$ is
Eq.~\eqref{eq:tev_compression_row}.
We use the carrier columns to complete that restriction without
dividing by a wave--contact separation:
\begin{equation}
 \widehat\Phi_I=\Phi_I-\sum_{j=1,2,\varsigma}
                              \Phi_I(C_j^{\rm E})\zeta_j.
 \label{eq:tev_left_completion}
\end{equation}
The same completion can be applied to the bending rows.
For a simple wave root, the normalized row is
\begin{equation}
 \ell_I^{\rm E}=\frac{\widehat\Phi_I}
                         {\widehat\Phi_I(R_I^{\rm E})},\qquad
 \ell_{C_j}^{\rm E}=\zeta_j.
 \label{eq:tev_normalized_left}
\end{equation}
Thus every scalar normalization is an explicit primitive pairing.
No full eigenvector-matrix inverse is required.

{For a semisimple repeated wave speed, collect the independent
right columns $R_a^{\rm E}$ from every wave family at that speed
and set}
\begin{equation}
 G_{ab}=\widehat\Phi_a(R_b^{\rm E}),\qquad
 \ell_a^{\rm E}=\sum_b(G^{-1})_{ab}\widehat\Phi_b.
 \label{eq:tev_joint_normalization}
\end{equation}
Its inverse is given by
$(G^{-1})_{ab}=(-1)^{a+b}\det G_{(b|a)}/\det G$, where
$G_{(b|a)}$ deletes row $b$ and column $a$.
This finite Gram normalization replaces the individual scalar
normalizations.
{The carrier rows remain dual at a wave--contact coincidence
provided every carrier completion is compatible.}
Deleting the entropy row and column gives the cold formulas.

The complete Eulerian basis, in family order, is therefore
\begin{widetext}
\begin{align}
 \mathsf R_{\rm B}^{\rm E}
 &=\bigl(C_1^{\rm E},C_2^{\rm E},C_\varsigma^{\rm E},
 R_{A-}^{\rm E},R_{A+}^{\rm E},
 R_1^{\rm E},R_2^{\rm E},R_3^{\rm E},
 R_4^{\rm E},R_5^{\rm E},R_6^{\rm E}\bigr),\notag\\
 \mathsf L_{\rm B}^{\rm E}
 &=\bigl(\zeta_1,\zeta_2,\zeta_\varsigma,
 \ell_{A-}^{\rm E},\ell_{A+}^{\rm E},
 \ell_1^{\rm E},\ell_2^{\rm E},\ell_3^{\rm E},
 \ell_4^{\rm E},\ell_5^{\rm E},\ell_6^{\rm E}\bigr)^T.
 \label{eq:tev_complete_basis}
\end{align}
\end{widetext}
The last six members are the compression waves. The matrices have
sizes $15\times11$ and $11\times15$, because the public tuple
retains the constrained field components. Their restriction to
independent physical components is square. At a degenerate speed,
Eq.~\eqref{eq:tev_joint_normalization} supplies the dual basis of
that family, including coincidences between different families.

%% file: appendix_tension_constitutive_derivatives.tex
\allowdisplaybreaks[1]
\par
\ifdim\pagegoal<\maxdimen
 \ifdim\dimexpr\pagegoal-\pagetotal\relax<8\baselineskip
  \newpage
 \fi
\fi
\section{Density derivatives of the vortex-line energy}
\label{app:tension_constitutive_derivatives}

\begingroup
The density dependence of the line energy shifts the material
momenta and couples matter to array compression. We evaluate the
derivatives of Eq.~\eqref{eq:tension_draft_master} needed by the
primitive responses in Sec.~\ref{sec:tension_evaluated_bounds}.
For $x,y\in\{n,N\}$, $\partial_x$ holds the other density,
$\mathcal N$ and the entropy per carrier fixed. The parameters
$\alpha_{\rm v}$ and $\mathcal N_0$ are constant. With
$r_v=\alpha_{\rm v}\mathcal N/\mathcal N_0$ and
$P_v=L_\Omega+r_v/2$, differentiating
$F=\mathcal NA_\Omega P_v$ gives
\begin{align}
 \partial_xF
 &=\mathcal NA_\Omega
       (P_v\partial_x\ln A_\Omega+\partial_xL_\Omega),\notag\\
 \partial_x\partial_{\mathcal N}F
 &=A_\Omega[(L_\Omega+r_v)\partial_x\ln A_\Omega
                                  +\partial_xL_\Omega],\notag\\
 \partial_x\partial_yF
 &=\mathcal NA_\Omega\bigl[
 P_v\partial_x\ln A_\Omega\partial_y\ln A_\Omega\notag\\
 &\qquad+P_v\partial_x\partial_y\ln A_\Omega
       +\partial_x\ln A_\Omega\partial_yL_\Omega\notag\\
 &\qquad+\partial_y\ln A_\Omega\partial_xL_\Omega
       +\partial_x\partial_yL_\Omega\bigr].
 \label{eq:rw_material_derivatives}
\end{align}
For the constant-entrainment NC model of
\eqref{eq:main_model_b_master}, $\mu_n=\mu_n(n)$ and
$H_0=\mu_n-\mathcal AN>0$. Primes denote $d/dn$.
The matching in Eq.~\eqref{eq:regulated_line_matching} gives
\endgroup
\begin{align}
 \partial_n\ln A_\Omega&=1/n-\mu_n'/H_0,\qquad
 \partial_N\ln A_\Omega=\mathcal A/H_0,\notag\\
 \partial_n^2\ln A_\Omega
 &=-1/n^2-\mu_n''/H_0+\mu_n'^2/H_0^2,\notag\\
 \partial_n\partial_N\ln A_\Omega
 &=-\mathcal A\mu_n'/H_0^2,\qquad
 \partial_N^2\ln A_\Omega=\mathcal A^2/H_0^2,\notag\\
 \partial_xL_\Omega
 &=-\partial_x\ln\xi_n,\qquad
 \partial_x\partial_yL_\Omega=-\partial_x\partial_y\ln\xi_n.
 \label{eq:regulated_matching_derivatives}
\end{align}
{The BCS core law supplies the remaining
logarithmic responses. For Eq.~\eqref{eq:tension_bcs_logarithm}, write}
$\mathcal L^{(x)}=\partial_xL_\Omega$ and
$\mathcal L^{(xy)}=\partial_x\partial_yL_\Omega$. The five density
responses are
\begin{align}
 \mathcal L^{(x)}
 &=\partial_x\ln(\mu_*\Delta_{\rm pair})-\frac{\delta_{xn}}{3n},\notag\\
 \mathcal L^{(xy)}
 &=\partial_x\partial_y\ln(\mu_*\Delta_{\rm pair})
   +\frac{\delta_{xn}\delta_{yn}}{3n^2}.
 \label{eq:tension_evaluated_bcs_derivatives}
\end{align}
\begingroup
The explicit $1/n$ and $1/n^2$ terms come from
$k_{Fn}\propto n^{1/3}$. The logarithmic derivatives of
$\mu_*\Delta_{\rm pair}$ contain the independently prescribed
effective-mass and pairing-gap responses. For the constant
$\mu_*$ and $\Delta_{\rm pair}$ used here, the five responses reduce to
\begin{gather}
 \mathcal L^{(n)}=-\frac1{3n},\qquad
 \mathcal L^{(N)}=0,\notag\\
 \mathcal L^{(nn)}=\frac1{3n^2},\qquad
 \mathcal L^{(nN)}=\mathcal L^{(NN)}=0.
 \label{eq:tension_constant_bcs_derivatives}
\end{gather}
Thus constant microscopic inputs still produce nonzero neutron
density derivatives. Composition dependence through $H_0$ also
remains when $\mathcal A\ne0$.

To express the induced momentum and stress shifts covariantly,
consider the current invariants $I_{xy}=-j_x\cdot j_y$ before
imposing entropy locking. Here $x,y$ label the currents
$(N_s^\mu,N_c^\mu,su^\mu)$ rather than the density arguments above.
Relative to the unit normal $o^\mu$, write
$j_x^\mu=\mathcal D_xo^\mu+(0,\mathcal J_x^i)$.
\endgroup
\begin{align}
 K^\Omega_{xx}&=2\frac{\partial F}{\partial I_{xx}},\qquad
 K^\Omega_{xy}=\frac{\partial F}{\partial I_{xy}}\quad(x\ne y),
 \label{eq:rw_material_coefficient_matrix}\\
 \delta H_x&=\sum_yK^\Omega_{xy}\mathcal D_y,\qquad
 p_{xi}=\sum_yK^\Omega_{xy}\mathcal J_{yi},\notag\\
 Z^{\mu\nu}&=\sum_xj_x^\mu\delta\mu_x^\nu,\qquad
 Z=g_{\mu\nu}Z^{\mu\nu},
 \label{eq:tension_general_density_response}
\end{align}
\begingroup
Here $\delta\mu^x_\nu=\sum_yK^\Omega_{xy}j_{y\nu}$.
The prefix $\delta$ denotes the line-induced shift, and $p_{xi}$
is its spatial component. It does not denote a characteristic
perturbation in these equations. For $F(n,N,\mathcal N)$, only
$K^\Omega_{nn}=(\partial F/\partial n)/n$ and
$K^\Omega_{cc}=(\partial F/\partial N)/N$ are nonzero. Consequently,
$Z^{\mu\nu}-Zg^{\mu\nu}=C_\Omega^{\mu\nu}$, and the general
invariant stress reduces exactly to
Eq.~\eqref{eq:tension_draft_stress}. The absence of off-diagonal
$K^\Omega_{xy}$ does not remove $F_{nN}$ or $F_{x\mathcal N}$:
differentiating the diagonal coefficients retains the mixed
responses required by the characteristic analysis.
\endgroup

%% file: appendix_tension_electric_adm.tex
\allowdisplaybreaks[1]
\section{Primitive vortex evolution and energy exchange}
\label{app:tension_electric_adm}

The conservative electric equation evolves
$\mathfrak E^i=\sqrt\gamma a_\Omega E^i$. Its constitutive weight
changes with both the vortex density and the material state.
Recovering $\partial_tE_i$ therefore requires the
material time derivatives as well. We derive this coupled time
system and its primitive inverse for the characteristic analysis in
Sec.~\ref{app:rw_derivation}. We then obtain the vortex-energy
identity used to track exchange with matter in
Sec.~\ref{sec:tension_electric_and_drift}.

We consider smooth states with $E_iB^i=0$ and
$Q_\Omega^2=B^2-E^2>0$. Spatial indices, contractions and curls use
the Eulerian three-metric. We use
$\mathcal D_t=\partial_t-\mathcal L_\beta$, so that
$\mathcal D_t\gamma_{ij}=-2\alpha K_{ij}$, and write $v^i=v_L^i$.
The provisional update sets $\mathcal I_\parallel=0$ as in
Sec.~\ref{sec:numerical_electric}. Its constraint projection is
specified below.

\subsection{Inverting the constitutive electric response}

Expanding the weighted current of
Eq.~\eqref{eq:tension_vortex_response_current} gives
\begin{equation}
 \alpha(\mathcal J_{\Omega\perp})^i
 =\mathcal D_t(a_\Omega E)^i
       -\alpha K a_\Omega E^i
       -\epsilon^{ijk}\nabla_j(\alpha a_\Omega B_k).
 \label{eq:tension_weighted_current_adm}
\end{equation}
The force law supplies the current through
Eq.~\eqref{eq:tension_evolution_drift_current}. To extract the
electric derivative, differentiate
$a_\Omega=F_{\mathcal N}/(\kappa_v^2\mathcal N)$, treating
$\mathcal N$ and the material invariants $I_A$ as independent
constitutive arguments. Define
$c_\Omega^2=\mathcal NF_{\mathcal N\mathcal N}/F_{\mathcal N}$,
$r_\Omega=(c_\Omega^2-1)/Q_\Omega^2$ and
$b_A=(\partial\ln a_\Omega/\partial I_A)_{\mathcal N}
=F_{\mathcal NA}/F_{\mathcal N}$.
Induction and the chain rule then give
\begin{align}
 \mathcal D_tB^i&=-\epsilon^{ijk}\nabla_j(\alpha E_k)+\alpha KB^i,
 \label{eq:tension_electric_induction_adm}\\
 \mathcal D_t\ln a_\Omega
 &=\tfrac12r_\Omega\mathcal D_tQ_\Omega^2+\chi_\Omega,\qquad
 \chi_\Omega=\sum_Ab_A\mathcal D_tI_A,
 \label{eq:tension_electric_eos_derivative}\\
 \tfrac12\mathcal D_tQ_\Omega^2
 &=B_i\mathcal D_tB^i-E^i\mathcal D_tE_i
                  -\alpha K_{ij}(B^iB^j+E^iE^j).
 \label{eq:tension_electric_metric_chain}
\end{align}
The scalar $\chi_\Omega$ is the change in the electric weight due
to the evolving material state. Induction fixes the magnetic
derivative in the last equation. Lowering the electric index in
Eq.~\eqref{eq:tension_weighted_current_adm} also differentiates the
metric and supplies $-2\alpha K_{ij}E^j$.

With $d_E=1-r_\Omega E^2$, collect the resulting spatial and force
terms into $H_i$ and define its electric response by
\begin{align}
 H_i={}&\frac{\epsilon_i{}^{jk}\nabla_j(\alpha a_\Omega B_k)}{a_\Omega}
    -\frac{\alpha \nabla_j(a_\Omega E^j)}{a_\Omega}v_i\notag\\
 &+\frac{\alpha(B\times f_\Omega)_i}{a_\Omega B^2}
       +\alpha K E_i-2\alpha K_{ij}E^j\notag\\
 &+r_\Omega E_i\bigl[B^j\epsilon_j{}^{kl}\nabla_k(\alpha E_l)-\alpha K B^2\notag\\
 &\hspace{24mm}+\alpha K_{jk}(B^jB^k+E^jE^k)\bigr],\notag\\
 G_i={}&H_i+\frac{r_\Omega E_iE^jH_j}{d_E},\qquad
 e_i=E_i/d_E,
 \label{eq:tension_electric_primitive_rhs}
\end{align}
The remaining electric equation is
\begin{equation}
 (\delta_i{}^j-r_\Omega E_iE^j)\mathcal D_tE_j\big|_{\rm trial}
        =H_i-E_i\chi_\Omega,
 \label{eq:tension_electric_transverse_time}
\end{equation}
The operator on the left has the rank-one inverse
$\delta_i{}^j+r_\Omega E_iE^j/d_E$, giving
$\mathcal D_tE_i|_{\rm trial}=G_i-e_i\chi_\Omega$.
For the adopted line energy, Eq.~\eqref{eq:rw_line_derivatives}
gives $0<c_\Omega^2<1$, hence $r_\Omega<0$ and $d_E\geq1$.
This electric inversion is regular throughout the stated
$Q_\Omega^2>0$ domain. Its remaining unknown is the material
response $\chi_\Omega$.

\subsection{Closing the material time derivatives}

The material primitive vector $P$ has the nine components in
Eq.~\eqref{eq:summary_density_primitives}. At fixed spatial metric,
the number densities, total stress and canonical momentum define
the nine-component map $\mathcal U(P,E,B,\gamma)$ below. All
Jacobians hold the other arguments fixed. Writing their balance
laws as $\mathcal R_{\mathcal U}=\mathcal D_t\mathcal U$, set
\begin{align}
 \mathcal U&=(D,D_R,E_{\rm tot},S_{{\rm tot},i},\xi_i),\notag\\
 \mathcal A&=\partial_P\mathcal U,\qquad
 \mathcal B_E=\partial_E\mathcal U,\qquad
 \mathcal B_B=\partial_B\mathcal U,\notag\\
 R&=\mathcal R_{\mathcal U}-\mathcal B_B\mathcal D_tB
                    -(\partial_\gamma\mathcal U):\mathcal D_t\gamma,
 \label{eq:rw_explicit_balance_rhs}
\end{align}
The spatial right sides $\mathcal R_{\mathcal U}$ come from
the balance laws
\eqref{eq:valencia_charge_unified}--\eqref{eq:valencia_energy_unified}
and \eqref{eq:numerical_xi_continuum}, after removing density
weights and using $E_{\rm tot}=\tau_{\rm tot}+D$.
The subtraction in $R$ removes the already known magnetic and
metric time derivatives. The material chain rule is then
$\mathcal A\mathcal D_tP+\mathcal B_E\mathcal D_tE=R$.
Substitute $\mathcal D_tE=G-e\chi_\Omega$ and define
\begin{align*}
 p_0&=\mathcal A^{-1}(R-\mathcal B_EG),\qquad
 \upsilon=\mathcal A^{-1}\mathcal B_Ee,\\
 \zeta&=\sum_Ab_A\partial_PI_A,\qquad
 \zeta_\gamma=\sum_Ab_A(\partial_\gamma I_A):\mathcal D_t\gamma.
\end{align*}
The material derivative becomes
$\mathcal D_tP=p_0+\upsilon\chi_\Omega$. Since the invariants
depend on $P$ and $\gamma$, their chain rule also gives
$\chi_\Omega=\zeta\mathcal D_tP+\zeta_\gamma$.
Combining these two relations closes the coupling through one
scalar response:
\begin{equation}
 \chi_\Omega=\frac{\zeta p_0+\zeta_\gamma}{1-\zeta\upsilon}.
 \label{eq:rw_explicit_material_time_response}
\end{equation}
The provisional electric derivative is therefore
\begin{equation}
 \left.\partial_tE_i\right|_{\rm trial}
 =\beta^j\nabla_jE_i+E_j\nabla_i\beta^j+G_i-e_i\chi_\Omega,
 \label{eq:tension_main_adm_electric}
\end{equation}
There is no remaining implicit time derivative when
$\det\mathcal A\ne0$, $d_E\ne0$ and $1-\zeta\upsilon\ne0$.
These conditions specify the domain of this elimination.
If $\mathcal A$ is singular while the full time matrix $M_t$ of
Eq.~\eqref{eq:rw_full_time_matrix} is regular, solve the material
chain rule and
Eqs.~\eqref{eq:tension_electric_induction_adm} and
\eqref{eq:tension_electric_transverse_time} jointly for
$\mathcal D_t(P,E,B)$. A singular full time matrix does not define
a unique primitive time derivative. The energy-based material map
also requires $T_c>0$.
For a line energy independent of the material invariants,
$b_A=0$ gives $\chi_\Omega=0$ and recovers the electric response
at fixed constitutive material parameters.

Before inversion, the material chain rule, electric response and
induction equation give the time matrix
\eqref{eq:rw_full_time_matrix} used in
Sec.~\ref{app:rw_derivation}.

\subsection{Energy exchanged with matter}

The line energy changes both when vortices move and when the
material state changes. To distinguish these contributions, use
the two-form stress
$\bar T_\Omega=T_\Omega-C_\Omega$ from
Eq.~\eqref{eq:tension_main_stress_divergence}. Its Eulerian
projections are $\bar{\mathcal{E}}_\Omega=F+a_\Omega E^2$,
$\bar m=a_\Omega E\times B$ and
$\bar P_{\Omega ij}=(a_\Omega B^2-F)\gamma_{ij}
-a_\Omega(E_iE_j+B_iB_j)$.
Projecting its divergence along the Eulerian observer and using
$u_L^\nu f_{\Omega\nu}=0$ to express the force work gives
\begin{align}
 \partial_t(\sqrt\gamma\bar{\mathcal{E}}_\Omega)
 &+\partial_j[\sqrt\gamma(\alpha\bar m^j-\beta^j\bar{\mathcal{E}}_\Omega)]\notag\\
 &=\sqrt\gamma\bigl[\alpha v^if_{\Omega i}
       +\alpha\bar P_\Omega^{ij}K_{ij}\notag\\
 &\hspace{15mm}-\bar m^i\partial_i\alpha
       +\sum_A\frac{\partial F}{\partial I_A}\mathcal D_tI_A\bigr].
 \label{eq:tension_main_adm_energy}
\end{align}
With $F_A=\partial F/\partial I_A$ at fixed $\mathcal N$, the term
$\sum_AF_A\mathcal D_tI_A$ is reversible exchange as the material
state changes. Its counterpart in
Eq.~\eqref{eq:tension_main_matter_exchange} cancels it in the total
energy balance. The full conserved stress includes $C_\Omega$
through the moments of Eq.~\eqref{eq:rw_primitive_stress}.
Equation~\eqref{eq:tension_main_adm_energy} therefore checks the
exchange terms, while the numerical method evolves total energy
with Eq.~\eqref{eq:valencia_energy_unified}.

Finally, the provisional electric derivative is followed by the
projection of Eq.~\eqref{eq:tension_electric_algebraic_projection}
and joint primitive recovery. This removes the component parallel
to $B$, as in the algebraic force-free projection written in
Ref.~\cite[Sec.~2.3, Eq.~(27)]{Alic:2012df}.
Linearization at $E_iB^i=0$ gives
$B^i\delta E_i+E_i\delta B^i=0$. Applying this constraint to the
primitive evolution is the projection used in
Sec.~\ref{app:rw_derivation} and in the characteristic analysis of
Sec.~\ref{app:regulated_characteristics}.

%% file: appendix_imex_tableau_tension.tex
\renewcommand{\arraystretch}{1.3}
\section{Four-stage IMEX Runge--Kutta coefficients}
\label{app:imex_tableau}

The stage equations \eqref{eq:imex_stage_predictor} use the
IMEX-SSP3(4,3,3) pair of Ref.
\cite[Table~6]{2010arXiv1009.2757P}.
We retain the convention of Sec.~\ref{sec:tension_imex}:
$a_{kl}$ multiplies the explicit operator $\mathcal H$, whereas
$\widetilde a_{kl}$ multiplies the implicit operator $\mathcal I$.
The explicit Butcher tableau is
\begin{equation}
 \begin{array}{c|cccc}
 c_k & \multicolumn{4}{c}{a_{kl}}\\
 \hline
 0 & 0 & 0 & 0 & 0\\
 0 & 0 & 0 & 0 & 0\\
 1 & 0 & 1 & 0 & 0\\
 \tfrac12 & 0 & \tfrac14 & \tfrac14 & 0\\
 \hline
 b_l & 0 & \tfrac16 & \tfrac16 & \tfrac23
 \end{array}
 \label{eq:imex_explicit_tableau}
\end{equation}
The corresponding diagonally implicit tableau is
\begin{equation}
 \setlength{\arraycolsep}{3pt}
 \begin{array}{c|cccc}
 \widetilde c_k & \multicolumn{4}{c}{\widetilde a_{kl}}\\
 \hline
 \alpha_{\rm I} & \alpha_{\rm I} & 0 & 0 & 0\\
 0 & -\alpha_{\rm I} & \alpha_{\rm I} & 0 & 0\\
 1 & 0 & 1-\alpha_{\rm I} & \alpha_{\rm I} & 0\\
 \tfrac12 & \beta_{\rm I} & \eta_{\rm I}
    & \tfrac12-\beta_{\rm I}-\eta_{\rm I}-\alpha_{\rm I}
    & \alpha_{\rm I}\\
 \hline
 \widetilde b_l & 0 & \tfrac16 & \tfrac16 & \tfrac23
 \end{array}
 \label{eq:imex_implicit_tableau}
\end{equation}
The numerical coefficients used in the implementation are
\begin{align}
 \alpha_{\rm I}&=0.24169426078821,\notag\\
 \beta_{\rm I}&=0.06042356519705,\notag\\
 \eta_{\rm I}&=0.12915286960590.
 \label{eq:imex_tableau_parameters}
\end{align}
These constants are distinct from the lapse $\alpha$ and shift
$\beta^i$. In particular, all four implicit stages have
$\widetilde a_{kk}=\alpha_{\rm I}$, so the local stage size in
Eq.~\eqref{eq:imex_geometry} is
$\Delta t_{\rm I}=\alpha\alpha_{\rm I}\Delta t$.

The abscissae are the row sums,
$c_k=\sum_l a_{kl}$ and
$\widetilde c_k=\sum_l\widetilde a_{kl}$.
Explicit and implicit terms with prescribed time dependence are
evaluated at $t^n+c_k\Delta t$ and
$t^n+\widetilde c_k\Delta t$, respectively.
There are four implicit solves but only three explicit
right-hand-side evaluations: the first column of $a_{kl}$ and
the first weight $b_1$ vanish, so $\mathcal H(U^{(1)})$ is not
needed. The explicit evaluations occur at stages $2,3,4$, at
times $t^n$, $t^n+\Delta t$ and $t^n+\Delta t/2$.

After the stages, the Runge--Kutta update is
\begin{align}
 U^{n+1}=U^n+\Delta t\sum_{k=1}^{4}\bigl[
 b_k\mathcal H(U^{(k)})
 +\widetilde b_k\mathcal I(U^{(k)})\bigr].
 \label{eq:imex_final_combination}
\end{align}
The final state is this weighted combination, rather than the
fourth stage. The implicit sources entering the stage predictors
and final sum are obtained from the converged nonlinear solve
through Eq.~\eqref{eq:imex_source_replay}. Additionally,
finite algebraic
projection corrections are applied separately afterward and are not
included in those sources.

%% file: appendix_riemann_analytic_tension.tex
\section{Riemann solutions for first and second sound}
\label{app:riemann_analytic}

We construct the longitudinal Riemann solution for a finite-temperature
relativistic superfluid without line tension with generating pressure $P(T,\mu,e)$.
The equation of state enters through its thermodynamic responses.
\revadd{We use the primitive state $\boldsymbol q=(e,T,w,k)$, where
$e=\varepsilon_\xi$, $w=z^x$ and $k=\xi_x$. The states
$\boldsymbol q_1,\boldsymbol q_2,\boldsymbol q_3$ are the
intermediate plateaus in spatial order, and $\boldsymbol q_a$ is
the sonic attachment state of the compound wave. The shock entropy
flux is $\mathcal J_s=[sw]-V[sW]$, with shock speed $V$ and
$[f]=f_R-f_L$.}

\begin{table*}[tbp]
 \centering
 \begin{tabular}{llrrrr}
 \hline
 Problem & State & $e$ & $T$ & $w$ & $k$\\
 \hline
 RP1 & $\boldsymbol q_1$ & 1.135596910 & 0.027097045 & 0.060068151 & 0.039928837\\
     & $\boldsymbol q_2$ & 1.134449695 & 0.029854877 & 0.005298649 & 0.046200685\\
     & $\boldsymbol q_a$ & 1.133091705 & 0.032868792 & 0.055750983 & 0.038147469\\
     & $\boldsymbol q_3$ & 1.132773869 & 0.033220793 & 0.062247163 & 0.036355673\\
 RP2 & $\boldsymbol q_1$ & 1.120040622 & 0.030010197 & -0.000192834 & -0.000119301\\
     & $\boldsymbol q_2$ & 1.119583444 & 0.030710504 & -0.012494691 & 0.002665155\\
     & $\boldsymbol q_3$ & 1.119161465 & 0.031495618 & -0.000078526 & -0.000048812\\
 \hline
 \end{tabular}
 \caption{Intermediate Riemann states in spatial order, with
 $\boldsymbol q_a$ the RP1 sonic attachment. Values follow from
 Eqs.~\eqref{eq:rp_four_wave_matching} and
 \eqref{eq:rp_compound_attachment}, independently of the finite-volume
 plateaus, and are rounded for display.}
 \label{tab:rp_exact_states}
\end{table*}

\begin{table*}[tbp]
 \centering
 \begin{tabular}{lllr}
 \hline
 Problem & Family and wave & Speed or fan interval & $\mathcal J_s$ (shock)\\
 \hline
 RP1 & $1$: first-sound rarefaction & $[-0.367740037,-0.318251532]$ & ---\\
     & $2$: second-sound shock & $-0.144868317$ & $2.78\times10^{-8}$\\
     & $3$: second-sound rarefaction & $[0.194875372,0.210151164]$ & ---\\
     & $3$: attached second-sound shock & $0.210151164$ & $4.39\times10^{-12}$\\
     & $4$: first-sound shock & $0.367401386$ & $4.10\times10^{-7}$\\
 RP2 & $1$: first-sound shock & $-0.340557884$ & $1.42\times10^{-14}$\\
     & $2$: second-sound shock & $-0.163052629$ & $1.74\times10^{-10}$\\
     & $3$: second-sound rarefaction & $[0.155381626,0.157484525]$ & ---\\
     & $4$: first-sound rarefaction & $[0.342216775,0.342285920]$ & ---\\
 \hline
 \end{tabular}
 \caption{Wave patterns of the four acoustic families. RP1 contains
 a compound third-family wave. Fan intervals list the spatial left and
 right edges in $x/t$; shock entropy fluxes are defined by
 Eq.~\eqref{eq:rp_shock_entropy}.}
 \label{tab:rp_exact_waves}
\end{table*}

\begin{figure*}[tbp]
 \centering
 \includegraphics[width=\textwidth]{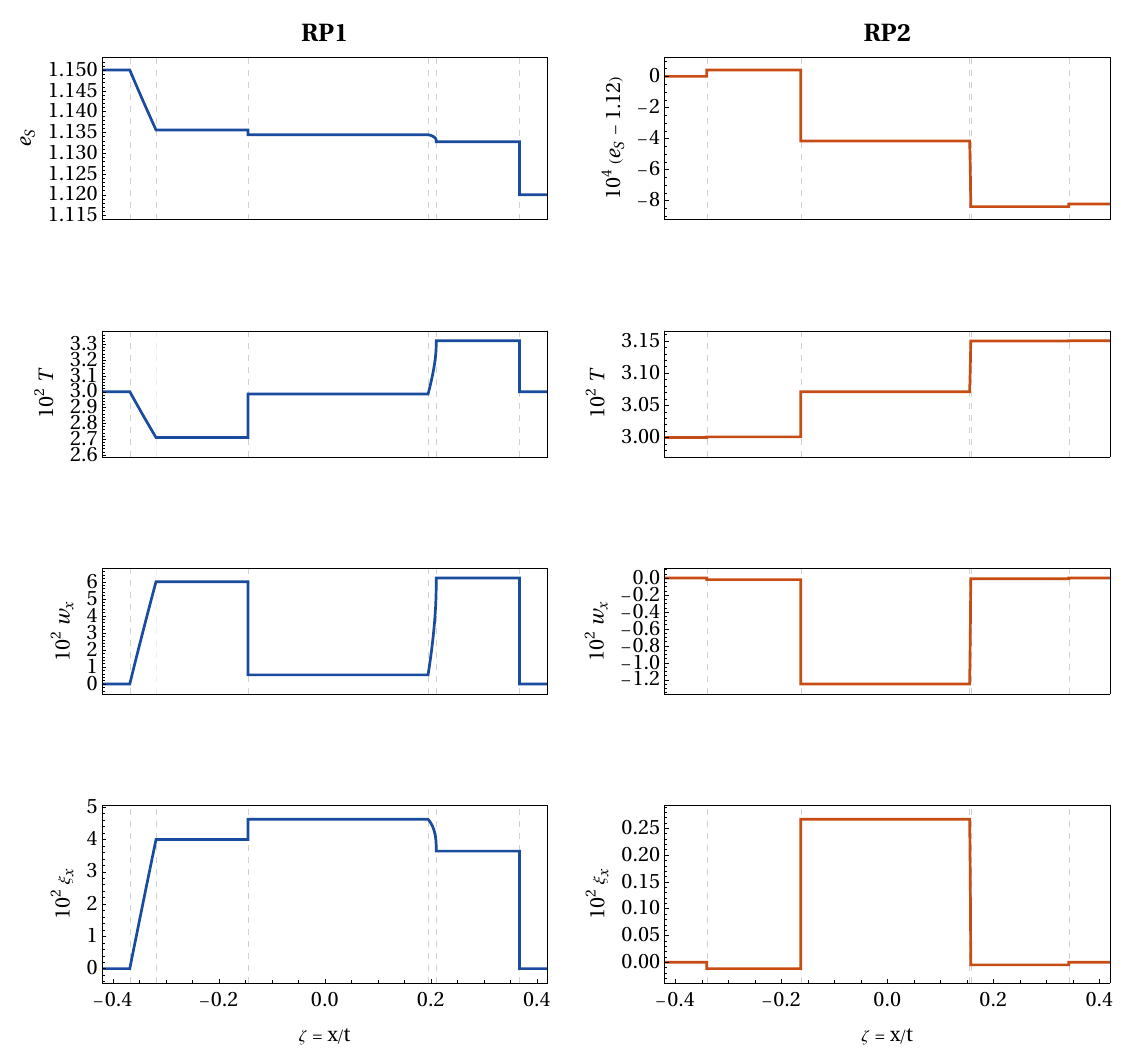}
 \caption{Self-similar solutions for discontinuous RP1 (left) and
 RP2 (right) data. Rows show $e_{\rm S}$, $T$, $w_x$, and $\xi_x$
 with the indicated scales. Vertical guides mark fan edges and shock
 speeds; jumps connect the one-sided shock states. The RP1 second-sound
 rarefaction is attached to a sonic shock. {These
 references are overlaid on the numerical profiles in
 Figs.~\ref{fig:riemann-first} and \ref{fig:riemann-second}{, which evolve the
 same discontinuous data}.}}
 \label{fig:rp_exact_profiles}
\end{figure*}

\subsection{Basic setup}

We consider source-free flow in flat spacetime, depending only on
$(t,x)$, with vanishing transverse velocities and canonical momenta.
We use the primitives
\begin{equation}
 \boldsymbol q=(e,T,w,k),\qquad
 e=\varepsilon_\xi,\quad w=z^x,\quad k=\xi_x,
 \label{eq:rp_primitives}
\end{equation}
with kinematic scalars
\begin{equation}
 W=\sqrt{1+w^2},\quad \Xi=\sqrt{e^2+k^2},\quad
 \mu=W\Xi-wk.
 \label{eq:rp_kinematics}
\end{equation}
The normal-excitation and phase velocities are $\revadd{v_c=w/W}$ and
{$v_{\rm phase}=k/\Xi$}, respectively. For an arbitrary
pressure, the thermodynamic responses are
\begin{align}
 s&=\frac{\partial P}{\partial T},&
 \rho&=\frac{\partial P}{\partial\mu},\notag\\
 Y&=\frac{1}{e}\frac{\partial P}{\partial e},&
 C&=Ts+\mu\rho.
 \label{eq:rp_general_responses}
\end{align}
The derivatives hold the other arguments $(T,\mu,e)$ fixed. We work
on the finite-temperature branch with $T>0$, $e>0$, $s>0$, and $Y>0$.

Under these conditions, the canonical momentum evolves according to
$\partial_t k+\partial_x\Xi=0$. With $Y=f^2$, the four conservative
variables and their fluxes are
\begin{align}
 \boldsymbol U&=(D,E,S,k)^{\mathsf T},\notag\\
 D&=\rho W+Y\Xi,\notag\\
 E&=CW^2-P+Y\Xi^2,\notag\\
 S&=CWw+Y\Xi k,
 \label{eq:rp_conserved}\\
 \boldsymbol F&=(J,S,\Pi,\Xi)^{\mathsf T},\notag\\
 J&=\rho w+Yk,\qquad
 \Pi=Cw^2+P+Yk^2,\notag\\
 \partial_t\boldsymbol U+\partial_x\boldsymbol F&=0.
 \label{eq:rp_flux}
\end{align}
Here $S=T_{\rm mat}^{0x}$ and $\Pi=T_{\rm mat}^{xx}$. {These equal the total projections in this sector.}
At $t=0$, the fluid occupies constant states $\boldsymbol q_L$ for
$x<0$ and $\boldsymbol q_R$ for $x>0$.

\subsection{Characteristic system}

Differentiating Eqs.~\eqref{eq:rp_conserved} and \eqref{eq:rp_flux},
using Eq.~\eqref{eq:rp_general_responses} and the full kinematic
dependence of $\mu$, gives
\begin{equation}
 A=\frac{\partial\boldsymbol U}{\partial\boldsymbol q},\qquad
 B=\frac{\partial\boldsymbol F}{\partial\boldsymbol q},\qquad
 \det(B-\lambda A)=0.
 \label{eq:rp_characteristic_pencil}
\end{equation}
The construction requires an invertible $A$ and four real, distinct,
subluminal acoustic roots along the wave curves, in addition to the
thermodynamic domain of the chosen equation of state.
For smooth flow, the conservation residuals $\mathcal R_D,\mathcal R_E,
\mathcal R_S,\mathcal R_k$ satisfy
\begin{equation}
 W\mathcal R_E-w\mathcal R_S-\mu\mathcal R_D
 =T\mathcal R_s+Y(Wk-w\Xi)\mathcal R_k,
 \label{eq:rp_entropy_identity}
\end{equation}
where $\mathcal R_s=\partial_t(sW)+\partial_x(sw)$. Smooth flow thus
conserves entropy at $T>0$. At shocks, entropy obeys an inequality
\cite{doi:10.1007/s10255-003-0074-6}, and energy remains an independent conservation law.

At coflow, $w=k=0$ and $\mu=e$. Define
$\bar P(T,e)=P(T,e,e)$, its Hessian $H=\partial^2\bar P/\partial(T,e)^2$,
and the responses $(s_0,\rho_0,Y_0,C_0)$ from
Eq.~\eqref{eq:rp_general_responses}, evaluated at coflow. The particle density at rest is
$\bar P_e=\rho_0+eY_0$, whereas $\rho_0$ is the native normal response.

For $\boldsymbol\theta=(\delta T,\delta e)^{\mathsf T}$ and
$\boldsymbol v=(\delta w,\delta k)^{\mathsf T}$, the linearized system
is symmetric after replacing the smooth energy row by entropy,
subtracting $eY_0$ times the canonical row from momentum, and
multiplying the canonical row by $Y_0$:
\begin{align}
 H\partial_t\boldsymbol\theta
       +G\partial_x\boldsymbol v&=0,\notag\\
 K\partial_t\boldsymbol v
       +G^{\mathsf T}\partial_x\boldsymbol\theta&=0,\notag\\
 G&=\begin{pmatrix}s_0&0\\\rho_0&Y_0\end{pmatrix},\qquad
 K=\operatorname{diag}(C_0,Y_0).
 \label{eq:rp_coflow_blocks}
\end{align}
These operations require smooth flow. A mode of speed $\lambda$
satisfies
\begin{align}
 (\lambda^2H-M)\boldsymbol\theta&=0,\notag\\
 M=GK^{-1}G^{\mathsf T}
 &=\begin{pmatrix}
 s_0^2/C_0&s_0\rho_0/C_0\\
 s_0\rho_0/C_0&\rho_0^2/C_0+Y_0
 \end{pmatrix}.
 \label{eq:rp_sound_matrix}
\end{align}
With $h=\det H$, $d=\det M=s_0^2Y_0/C_0$ and
$b=H_{22}M_{11}+H_{11}M_{22}-2H_{12}M_{12}$, the sound speeds are
\begin{align}
 c_{1,2}^2&=\frac{b\pm\sqrt{b^2-4hd}}{2h},\notag\\
 (\lambda_1,\lambda_2,\lambda_3,\lambda_4)
 &=(-c_1,-c_2,c_2,c_1).
 \label{eq:rp_sound_roots}
\end{align}
Both squared speeds are positive for $H>0$, $C_0>0$, and $Y_0>0$;
we further require $c_2<c_1<1$. Since $d$ is positive, there is no
zero-speed longitudinal contact at coflow. Transverse modes are
excluded by the one-dimensional reduction.

In $(T,e,w,k)$ ordering, an explicit polarization is
\begin{align}
 \boldsymbol\theta_\lambda
 &=\begin{pmatrix}M_{12}-\lambda^2H_{12}\\
                    \lambda^2H_{11}-M_{11}\end{pmatrix},\notag\\
 \boldsymbol r_\lambda
 &=\begin{pmatrix}\boldsymbol\theta_\lambda\\
            K^{-1}G^{\mathsf T}\boldsymbol\theta_\lambda/\lambda
   \end{pmatrix}.
 \label{eq:rp_sound_eigenvector}
\end{align}
If the thermal vector vanishes, use a null vector of the other row of
$\lambda^2H-M$. Interchanging the first two entries gives
$\boldsymbol r_\lambda$ in ordering~\eqref{eq:rp_primitives}.
Suppose both initial states are infinitesimal perturbations of the
same uniform background. Evaluating the characteristic speeds and
eigenvectors at that state, define
$R=(\boldsymbol r_{\lambda_1},\ldots,
\boldsymbol r_{\lambda_4})$ and
$\boldsymbol\beta=R^{-1}(\boldsymbol q_R-\boldsymbol q_L)$.
The linear solution is
\begin{equation}
 \boldsymbol q(x,t)=\boldsymbol q_L+
 \sum_{j=1}^4\beta_j\boldsymbol r_{\lambda_j}
        \Theta(x/t-\lambda_j).
 \label{eq:rp_linear_solution}
\end{equation}
{Here $\Theta$ is the Heaviside step function.}
\subsection{Rarefaction curves}

For the initial discontinuity, the similarity coordinate is $\zeta=x/t$.
Within a rarefaction, $(B-\zeta A)d\boldsymbol q/d\zeta=0$, hence
$\zeta=\lambda_j(\boldsymbol q)$ with
$(B-\lambda_jA)\boldsymbol r_j=0$. Choosing $q_m$ such that
$r_{j,m}\ne0$ and strength $a=q_m-q_{m,-}$ gives the wave curve
\revadd{$\mathcal R_j$, with the starting state $\boldsymbol q_-$
held fixed in the differential equation,}
\begin{equation}
 \frac{d\revadd{\mathcal R_j}}{da}
 =\frac{\boldsymbol r_j(\revadd{\mathcal R_j})}{r_{j,m}(\revadd{\mathcal R_j})},
 \qquad \revadd{\mathcal R_j(0)}=\boldsymbol q_-.
 \label{eq:rp_rarefaction_quadrature}
\end{equation}
The speed must increase monotonically from the spatial left to the
right edge. Inverting $\lambda_j(\revadd{\mathcal R_j(a)})=\zeta$ then gives
the fan profile; a turning point terminates the elementary fan.
Away from coflow, Eq.~\eqref{eq:rp_characteristic_pencil} is evaluated
on the full two-current state, retaining the relative flow.

\subsection{Shock curves}

For a front at $x=Vt$, define $[X]=X_+-X_-$, where $-$ and $+$
denote the spatial left and right states. The four Rankine--Hugoniot
conditions at the shock are
\begin{align}
 [\rho w+Yk]&=V[\rho W+Y\Xi],\notag\\
 [CWw+Y\Xi k]&=V[CW^2-P+Y\Xi^2],\notag\\
 [Cw^2+P+Yk^2]&=V[CWw+Y\Xi k],\notag\\
 [\Xi]&=V[k].
 \label{eq:rp_shock_equations}
\end{align}
At prescribed $q_{m,+}=q_{m,-}+a_j$, these determine the other three
right-state primitives and $V$. We follow the branch tangent to
$\boldsymbol r_j/r_{j,m}$ as $a_j\to0$, with
$V\to\lambda_j(\boldsymbol q_-)$.
\revadd{We denote this Hugoniot branch by $\mathcal H_j$, with shock
strength $a_j$ and starting state $\boldsymbol q_-$.}

In the shock frame, with $\gamma_V=(1-V^2)^{-1/2}$,
\begin{align}
 W'&=\gamma_V(W-Vw),& w'&=\gamma_V(w-VW),\notag\\
 \Xi'&=\gamma_V(\Xi-Vk),& k'&=\gamma_V(k-V\Xi).
 \label{eq:rp_shock_boost}
\end{align}
The four common invariants are
\begin{align}
 \mathcal Q &=\Xi',& j&=\rho w'+Yk',\notag\\
 f&=CW'w'+Y\mathcal Q k',& m&=Cw'^2+P+Yk'^2.
 \label{eq:rp_shock_invariants}
\end{align}
They satisfy $k'^2=\mathcal Q^2-e^2$ and $\mu=W'\mathcal Q-w'k'$.
The canonical-momentum balance distinguishes these conditions from
the single-fluid Taub adiabat \cite{Taub:1948zz}.

\subsection{Shock admissibility}

Admissible shocks have subluminal $V$, satisfy the state bounds,
and produce nonnegative physical entropy,
\begin{equation}
 \mathcal J_s=[sw]-V[sW]\geq0,
 \qquad [sw']=\gamma_V\mathcal J_s.
 \label{eq:rp_shock_entropy}
\end{equation}
{An ordinary compressive shock of family $j$ must also
satisfy the Lax inequalities \cite{Lax:1957hec}}
\begin{align}
 \lambda_{j-1}(\boldsymbol q_-) &< V
                         <\lambda_j(\boldsymbol q_-),\notag\\
 \lambda_j(\boldsymbol q_+) &< V
                         <\lambda_{j+1}(\boldsymbol q_+),
 \label{eq:rp_lax_conditions}
\end{align}
where $\lambda_0=-\infty$ and $\lambda_5=+\infty$.
Neither a pressure nor a temperature jump alone selects the wave type.
When genuine nonlinearity is lost, we additionally impose the Liu
condition on the connected Hugoniot curve
\cite{1975JDE....18..218L,doi:10.1007/s10255-003-0074-6}. At fixed spatial-left state,
the terminal shock is no faster than any intermediate shock on the
same branch,
\begin{equation}
 V(\boldsymbol q_-,\mathcal H_j(\theta;\boldsymbol q_-))
 \geq V(\boldsymbol q_-,\boldsymbol q_+),
 \label{eq:rp_liu_condition}
\end{equation}
for $\theta$ between zero and the terminal strength. At zero
strength, the speed tends to $\lambda_j(\boldsymbol q_-)$. When a
rarefaction attaches to a shock, its edge travels with the shock.
The shock speed therefore equals the characteristic speed in their
common state, saturating the Lax bound on that side.

\subsection{Wave matching}

Let $\mathcal W_j(a;\boldsymbol q)$ be the admissible wave curve
of family $j$, allowing elementary shocks, rarefactions, and compound
waves when required. Starting from
$\boldsymbol q_0=\boldsymbol q_L$, successive waves satisfy
\begin{equation}
 \boldsymbol q_j=\mathcal W_j(a_j;\boldsymbol q_{j-1}),
 \quad j=1,\ldots,4,\qquad
 \boldsymbol q_4=\boldsymbol q_R.
 \label{eq:rp_four_wave_matching}
\end{equation}
For four elementary waves, the endpoint condition gives four equations
for their strengths. Each curve is parameterized by a primitive with
nonzero tangent component. The linear decomposition supplies an initial
guess. The solution is constant between waves,
with shocks at $x/t=V$ and fans given by
Eq.~\eqref{eq:rp_rarefaction_quadrature}. Wave supports must be ordered
and disjoint, except at a sonic attachment.

If $\lambda_j$ turns along a rarefaction curve, continuing that curve
would make the self-similar profile multivalued. An admissible
continuation may instead require a compound wave.

For a rarefaction followed by a shock in family $j$, the corresponding
step in Eq.~\eqref{eq:rp_four_wave_matching} becomes
\begin{align}
 \boldsymbol q_a&=\mathcal R_j(a_j;\boldsymbol q_{j-1}),\notag\\
 \boldsymbol q_j&=\mathcal H_j(b_j;\boldsymbol q_a),\notag\\
 \revadd{V}(\boldsymbol q_a,\boldsymbol q_j)
   &=\lambda_j(\boldsymbol q_a)
     >\lambda_j(\boldsymbol q_j).
 \label{eq:rp_compound_attachment}
\end{align}
The common state $\boldsymbol q_a$ is fixed by the finite shock
conditions and equality of the shock and fan-edge speeds; it has no
intervening plateau. The additional wave strength is fixed by this
sonic condition. The retained fan must be monotone, and the shock must
satisfy the entropy and Liu conditions. A turning point alone does
not determine the attachment state. The analogous construction applies
when the shock precedes the rarefaction, with sonic matching on the
other side of the shock.

\subsection{Reference solution}

For the reference solution used in the main text, we specialize to the CP pressure
with $\Gamma=2$. Set $\Delta=e^3-\mu^2$ and $p=P_{\rm ph}$.
The pressure and logarithmic phonon-pressure derivatives are
\begingroup\colorlet{revisionblue}{mainrevisionblue}\color{revisionblue}
\begin{align}
 P&=\frac{(e-1)^2}{4 {\kappa_p}}+p,
 &p&=\frac{27T^4\mu^4e^{3/2}\sqrt{e-1}}{256\Delta^2},
 \label{eq:rp_pressure}\\
 L_T&=\frac4T,\qquad
 L_\mu=\frac4\mu+\frac{4\mu}{\Delta},\notag\\
 L_e&=\frac{3}{2e}+\frac{1}{2(e-1)}-\frac{6e^2}{\Delta}.
 \label{eq:rp_logderivatives}
\end{align}
\endgroup
Here $L_a=\partial_a\log p$ is evaluated at independent $(T,\mu,e)$,
before imposing Eq.~\eqref{eq:rp_kinematics}. The responses are
\begingroup\colorlet{revisionblue}{mainrevisionblue}\color{revisionblue}
\begin{align}
 s&=\frac{4p}{T},\qquad \rho=pL_\mu,\notag\\
 Y&=\frac{e-1}{2 {\kappa_p}e}+\frac{pL_e}{e},\notag\\
 C&=4p\frac{2e^3-\mu^2}{\Delta}.
 \label{eq:rp_responses}
\end{align}
\endgroup
The second derivatives are
\begingroup\colorlet{revisionblue}{mainrevisionblue}\color{revisionblue}
\begin{equation}
 P_{ab}=p(L_aL_b+\partial_bL_a)
       +\frac{\delta_{ae}\delta_{be}}{2 {\kappa_p}},
 \quad a,b\in\{T,\mu,e\}.
 \label{eq:rp_hessian_native}
\end{equation}
\endgroup
The thermodynamic branch requires $T>0$, $e>1$, $e^2\leq\mu^2<e^3$,
and $Y>0$.

For this EOS, the thermal amplitude can be eliminated algebraically.
Define the cold pressure, stiffness, and response factors
\begingroup\colorlet{revisionblue}{mainrevisionblue}\color{revisionblue}
\begin{align}
 P_{\rm c}&=\frac{(e-1)^2}{4{\kappa_p}},&
 Y_{\rm c}&=\frac{e-1}{2{\kappa_p}e},\notag\\
 R&=\frac{4e^3}{\mu\Delta},&
 Z&=-\frac{L_e}{e},&
 a_{\rm h}&=\mu\left(2-\frac{\mu^2}{e^3}\right).
 \label{eq:rp_shock_amplitude_factors}
\end{align}
\endgroup
Thus $\rho=Rp$, $Y=Y_{\rm c}-Zp$, and $C=a_{\rm h}Rp$.
At fixed $\mathcal Q $, introduce the signed roots
\begin{align}
 \eta=k'&=\sigma_\xi\sqrt{\mathcal Q ^2-e^2},\qquad
 \chi=\sigma_r\sqrt{\mu^2-e^2},\notag\\
 &\sigma_\xi,\sigma_r\in\{-1,+1\}.
 \label{eq:rp_shock_signed_roots}
\end{align}
The normal four-velocity is
\begin{equation}
 W'=\frac{\mu \mathcal Q +\eta\chi}{e^2},\qquad
 w'=\frac{\mu\eta+\mathcal Q \chi}{e^2}.
 \label{eq:rp_shock_velocity_elimination}
\end{equation}
These expressions satisfy $W'^2-w'^2=1$ and $\mathcal Q W'-\eta w'=\mu$.
Each sign is continued along the branch and changes only when its
root vanishes.

Set
\begin{align}
 \mathcal B_N&=Rw'-Z\eta,\notag\\
 \mathcal B_E&=a_{\rm h}RW'w'-Z\mathcal Q \eta,\notag\\
 \mathcal B_M&=a_{\rm h}Rw'^2+1-Z\eta^2.
 \label{eq:rp_shock_amplitude_coefficients}
\end{align}
The invariants are affine in $p$:
\begin{equation}
 \begin{pmatrix}
 j-Y_{\rm c}\eta\\ f-Y_{\rm c}\mathcal Q\eta\\
 m-P_{\rm c}-Y_{\rm c}\eta^2
 \end{pmatrix}
 =p\begin{pmatrix}\mathcal B_N\\\mathcal B_E\\\mathcal B_M\end{pmatrix}.
 \label{eq:rp_shock_amplitude_equations}
\end{equation}
For $\mathcal B_M\ne0$, elimination gives
\begin{align}
 (j-Y_{\rm c}\eta)\mathcal B_M
  &=(m-P_{\rm c}-Y_{\rm c}\eta^2)\mathcal B_N,\notag\\
 (f-Y_{\rm c}\mathcal Q \eta)\mathcal B_M
  &=(m-P_{\rm c}-Y_{\rm c}\eta^2)\mathcal B_E
 \label{eq:rp_shock_two_scalar_equations}
\end{align}
which determine $(e,\mu)$ at fixed upstream invariants and front
speed. The positive thermal amplitude and temperature follow from
\begin{equation}
 p=\frac{m-P_{\rm c}-Y_{\rm c}\eta^2}{\mathcal B_M},\qquad
 T=\left(\frac{256p\Delta^2}
 {27\mu^4e^{3/2}\sqrt{e-1}}\right)^{1/4}.
 \label{eq:rp_shock_temperature_recovery}
\end{equation}
The inverse boost gives $w=\gamma_V(w'+VW')$ and
$k=\gamma_V(\eta+V\mathcal Q)$. At $\mathcal B_M=0$, use the undivided
system~\eqref{eq:rp_shock_amplitude_equations}. We discard the trivial
upstream root and check all candidates against the original jump
conditions and state bounds.

We use $\revadd{\kappa_p=58}$ and parameterize first-sound curves by
$e$ and second-sound curves by $T$. At $(e,T)=(1.12,0.03)$, the coflow
speeds and second-sound polarization are
\begin{align}
 c_1&=0.340475423036097,\notag\\
 c_2&=0.161973446292176,\notag\\
 r_{2,e}/r_{2,T}&=-0.547885080143339.
 \label{eq:rp_background_numbers}
\end{align}
Here $r_2$ denotes the second-sound polarization of
Eq.~\eqref{eq:rp2-ic}, with either propagation sign.

The $\pm c_2$ eigenvectors can share a thermal vector with opposite
velocity perturbations. Their equal-amplitude sum gives the RP2
thermodynamic jump with zero velocities. A finite jump along this
tangent need not follow the nonlinear second-sound curves and excites
first sound at higher order.

For this specialization, RP1 has a third-family rarefaction attached
to a shock, as in Eq.~\eqref{eq:rp_compound_attachment} with $j=3$.
Its endpoint and sonic conditions give five equations for
$(a_1,a_2,a_3,b_3,a_4)$. RP2 has four elementary waves.

RP1 uses the endpoint states of Eq.~\eqref{eq:rp1-ic}.
The archived RP2 initialization has a one-sided $5\%$ temperature
increase, with asymptotic states
\begin{align}
 \boldsymbol q_L&=(1.12,0.03,0,0),\notag\\
 \boldsymbol q_R&=(1.119178172379455,0.0315,0,0).
 \label{eq:rp2_exact_endpoints}
\end{align}
We retain the archived eigenvector ratio; using
Eq.~\eqref{eq:rp_background_numbers} changes $e_R$ by
$3.3\times10^{-13}$.

Tables~\ref{tab:rp_exact_states} and \ref{tab:rp_exact_waves} list the
matched states and wave patterns. The RP2 central plateau has
oppositely directed $w$ and $k$ and a small first-sound contribution.
All shocks conserve the four evolved quantities and produce
nonnegative entropy.

The reference assumes a step on the infinite line, or a local periodic
problem before neighboring waves interact. {The simulations impose
the same discontinuous data, so the comparison is direct: the captured
fronts converge to this reference in $L_1$ at first order, with front
speeds reproduced to $0.3\%$ or better at $N_x=2048$.}
Overlays use the numerical
output times, retain both shock traces, and convert to the plotted
velocities, $sW/D$, and $D$, without fitting fronts or plateaus.

%% file: appendix_kelvin_native_tension.tex
\begingroup
\section{Kelvin waves}
\label{app:kelvin_native}
\label{app:kelvin_tension_prediction}

We formulate the Kelvin modes used in
Secs.~\ref{sec:results-kelvin} and \ref{sec:results-kelvin-tension}.
Without line stress, bending is restored by the flow induced by the
displaced column. With line stress, the line energy supplies an
additional restoring force and inertia. We work in Minkowski spacetime,
set $c=1$, and use orthonormal cylindrical components with amplitudes
$q = q(r)e^{i(\theta+kz-\omega t)}$. Thus $m=1$,
$\operatorname{Re}\omega$ is the signed frequency and
$-\operatorname{Im}\omega$ is the decay rate.

\begin{table*}[tbp]
\centering
{
\begin{tabular}{ccccc}
\hline\hline
$\mathcal R$ & $\operatorname{Re}\omega_n$ & $-\operatorname{Im}\omega_n$ & $\operatorname{Re}\omega_p$ & $-\operatorname{Im}\omega_p$ \\
\hline
$0$ & $-0.0816580531$ & $0$ & $-0.0817655374$ & $0$ \\
$10^{-5}$ & $-0.0816582672$ & $2.785430\times10^{-6}$ & $-0.0817653231$ & $8.247372\times10^{-6}$ \\
$3\times10^{-5}$ & $-0.0816599687$ & $8.087460\times10^{-6}$ & $-0.0817636198$ & $2.501094\times10^{-5}$ \\
$10^{-4}$ & $-0.0816745609$ & $1.590843\times10^{-5}$ & $-0.0817490068$ & $9.441959\times10^{-5}$ \\
\hline\hline
\end{tabular}}
\caption{Force-equilibrium $m=1$, $ka=1$ co-rotating column
frequencies in units $c/a$, with $\mathcal Z=\mathcal R'=0$.
Subscripts track the neutron and proton branches from zero drag;
positive $-\operatorname{Im}\omega$ denotes decay.}
\label{tab:kelvin-rel-force}
\end{table*}

\subsection{Column without line stress}
\label{app:kelvin_force_equilibrium}

We begin with the cold, unentrained NC model with
$\mu_x=m_x+2K_xn_x$ and $c_x^2=(\mu_x-m_x)/\mu_x$ for
$x\in\{n,p\}$. The canonical momentum covectors are
$\pi^x_\mu=\mu_xu^x_\mu$, with $\pi^n_\mu=\xi_\mu$.
In the absence of lift and drag, the lines move with the neutrons,
and their spatial canonical momentum obeys
\begin{equation}
 \partial_t\boldsymbol\xi
 =\boldsymbol v_n\times(\nabla\times\boldsymbol\xi)-\nabla\Phi_n,
 \label{eq:native_kelvin_equations}
\end{equation}
with $\Phi_n=-\xi_0=\mu_nW$. For a stationary co-rotating
column, both fluids have the Kaufmann--Scully azimuthal speed
\cite[Sec.~1.1, Eq.~(1.8)]{2018arXiv180505064G}
\begin{equation}
 V=\frac{\Omega_ca^2r}{r^2+a^2},\qquad
 \frac{d\ln\mu_x}{dr}=\frac{W^2V^2}{r}.
 \label{eq:native_kelvin_equilibrium}
\end{equation}
Here $W=(1-V^2)^{-1/2}$. Fixing $\mu_{x,\infty}$ gives
\begin{align}
 n_x(r)&=\frac{\mu_{x,\infty}e^{-I(r)}-m_x}{2K_x},\notag\\
 I(r)&=\int_r^\infty\frac{V(s)^2}{s[1-V(s)^2]}\,ds\notag\\
 &=\frac{M}{\sqrt{4-M^2}}\left[\frac\pi2-
 \arctan\frac{2r^2/a^2+2-M^2}{M\sqrt{4-M^2}}\right],
 \label{eq:kelvin_density_profile}
\end{align}
where $M=\Omega_ca$. We require $0<M<2$ and
$\mu_{x,\infty}e^{-I(0)}>m_x$ so that the speed is subluminal
and the densities are positive everywhere. In the free-neutron
benchmark the proton fluid instead rests with $\mu_p=3$ and
$n_p=(3-m_p)/(2K_p)$. Its perturbation vanishes.

Express the orthonormal momentum amplitudes as
$\delta\pi_r^x=iR_x$, $\delta\pi_\theta^x=P_x$ and
$\delta\pi_z^x=Q_x$, and set
$H_x=\delta\Phi_x-VP_x=\delta\mu_x/W$, with
$\Phi_x=\mu_xW$. With primes denoting $d/dr$, define
$\varpi=\omega-V/r$, $\zeta=W^2(V'+V/r)$,
$B_n=\Phi_n\zeta$, $\eta=n_n/n_p$ and
$L_x=[1+(c_x^{-2}-1)W^2V^2]/r$.
For the reference solutions used to validate the code, we set $\mathcal Z=\mathcal R'=0$ and only retain
carrier drag $\mathcal R$. Linearizing
Eq.~\eqref{eq:moments_vortex_force} about the common circular flow gives
\begin{align}
 \Delta_r&=R_n/\Phi_n-R_p/\Phi_p,\notag\\
 \Delta_\theta&=\frac{P_n/W^2-VH_n}{\Phi_n}
              -\frac{P_p/W^2-VH_p}{\Phi_p},\notag\\
 \mathcal V_r&=\frac{\mathcal R^2\Delta_r
                       +i\mathcal RW\Delta_\theta}{1+\mathcal R^2},\notag\\
 \mathcal V_\theta&=\frac{\mathcal R^2\Delta_\theta
                       +i\mathcal R\Delta_r/W}{1+\mathcal R^2}.
 \label{eq:kelvin_drag_amplitudes}
\end{align}
The line-velocity departures are
$\delta v_L^r-\delta v_n^r=-i\mathcal V_r$ and
$\delta v_L^\theta-\delta v_n^\theta=-\mathcal V_\theta$.
Thus $\delta f_r=-n_nWB_n\mathcal V_\theta$ and
$\delta f_\theta=in_nWB_n\mathcal V_r$ act on the neutrons,
with equal and opposite forces on the protons and $\delta f_z=0$.
Number conservation and the three momentum components reduce to
\begin{align}
 \varpi P_x&=\zeta R_x+H_x/r-\epsilon_xB_n\mathcal V_r,\notag\\
 Q_x&=kH_x/\varpi,\notag\\
 H_x'&=2VP_x/r-\zeta VH_x-\varpi R_x
                    -\epsilon_xB_n\mathcal V_\theta,\notag\\
 R_x'&=-L_xR_x+(\omega V-1/r)P_x\notag\\
 &\quad+[\varpi W^2(c_x^{-2}-V^2)+V/r-k^2/\varpi]H_x.
 \label{eq:native_kelvin_radial_system}
\end{align}
Here $\epsilon_n=1$ and $\epsilon_p=-\eta$.
The two angular equations form a linear $2\times2$ system for
$(P_n,P_p)$ at each radius. After this solve, the four ODEs for
$(R_n,H_n,R_p,H_p)$ close the problem. The axial momenta are
algebraic. No independent vorticity amplitude is needed, since
$\delta\boldsymbol B_n=\nabla\times\delta\boldsymbol\pi^n$.
At $\mathcal R=0$ both exchange amplitudes vanish and the two
fluids decouple. The neutron block then also describes the
free-neutron reference solution with the proton fluid at rest.

Regularity requires $P_x(0)=-R_x(0)$ and
$H_x=h_xr+O(r^3)$, where
\begin{align}
 h_x&=-(\omega+\Omega_c)R_x(0)
        +\epsilon_xB_n(0)g\Delta_r(0),\notag\\
 g&=\frac{\mathcal R^2-i\mathcal R}{1+\mathcal R^2}.
 \label{eq:kelvin_axis_data}
\end{align}
Two independent choices of $(R_n(0),R_p(0))$ span the regular
inner solutions. At large radius, $V,B_n\to0$ and
$\mu_x\to\mu_{x,\infty}$. In this uniform, nonrotating limit,
eliminating $R_x$ using
$R_x=-H_x'/\omega$ gives
\begin{equation}
 H_x''+\frac{H_x'}r-\left(q_x^2+\frac1{r^2}\right)H_x=0,
 \qquad q_x^2=k^2-\frac{\omega^2}{c_{x,\infty}^2}.
 \label{eq:kelvin_exterior_bessel}
\end{equation}
This is the modified Bessel equation of order $m=1$. Its
$I_1(q_xr)$ solution grows, whereas $K_1(q_xr)$ decays as
$e^{-q_xr}/\sqrt{q_xr}$ on the branch $\operatorname{Re}q_x>0$.
Localization therefore selects $H_x\propto K_1(q_xr)$.
The radial decay constant is fixed by the frequency. For a real localized mode,
$|\omega|<c_{x,\infty}|k|$.

We obtain the spectrum by two-sided radial shooting. For each trial
complex $\omega$, eliminate $P_x,Q_x$, integrate the two regular
axis solutions outward, and integrate the two decaying exterior
solutions inward from a large radius $r_b$. The exterior boundary conditions obey
$H_x'/H_x=q_xK_1'(q_xr_b)/K_1(q_xr_b)$, where the prime on
$K_1$ denotes its argument derivative. At an intermediate radius,
the four amplitudes must agree.\\
\begingroup
Let the columns of $Y_{\rm in}$ and $Y_{\rm out}$ be the two
regular inner and two decaying outer solutions evaluated at that
radius. A mode has coefficient pairs $a$ and $b$ satisfying
$Y_{\rm in}a=Y_{\rm out}b$. These four homogeneous matching
conditions admit a nonzero coefficient vector only when the
$4\times4$ matrix $(Y_{\rm in},-Y_{\rm out})$ is singular. We therefore solve
\begin{equation}
 \det\bigl(Y_{\rm in},-Y_{\rm out}\bigr)=0
 \label{eq:kelvin_matching_dispersion}
\end{equation}
for $\operatorname{Re}\omega$ and $\operatorname{Im}\omega$.
At a root, a nonzero null vector $(a_1,a_2,b_1,b_2)$ gives
the inner mode $a_1Y_{{\rm in},1}(r)+a_2Y_{{\rm in},2}(r)$ and
the outer mode $b_1Y_{{\rm out},1}(r)+b_2Y_{{\rm out},2}(r)$.
They agree at the matching radius and form one radial eigenfunction.
Only its overall amplitude remains arbitrary. We continue each
zero-drag root in $\mathcal R$, increasing $r_b$ and tightening the
radial integration until the frequency converges.
\endgroup
To compute the values in Table~\ref{tab:kelvin-rel-force}, we use an adaptive eighth-order Runge--Kutta integration with relative
tolerance $2\times10^{-11}$ and absolute tolerance $2\times10^{-13}$,
starting at $r=10^{-5}a$, matching at $2a$ and imposing exterior
boundary conditions at $16a$.
Repeating with exterior radius $32a$, axis radius $2\times10^{-6}a$,
matching radius $3a$ and tenfold tighter tolerances changes each
complex frequency by less than $3\times10^{-12}c/a$.

The resulting mode is specified, up to an overall amplitude, by
$q(r)=(R_n,H_n,R_p,H_p)$ and the algebraic $P_x,Q_x$.
The physical perturbations are their real Fourier parts, with
$\delta\mu_x=WH_x$, $\delta n_x=WH_x/(2K_x)$ and
\begin{equation}
 (\delta v_x^r,\delta v_x^\theta,\delta v_x^z)
 =\frac{(iR_x,P_x/W^2-VH_x,Q_x)}{\Phi_x}.
 \label{eq:kelvin_mode_reconstruction}
\end{equation}
For the free-neutron test at $ka=1$, this gives
$\omega=-0.08165805313\,c/a$ and $q_na\simeq0.993310$.
The proton amplitudes vanish. Equation~\eqref{eq:native_kelvin_seed}
and its stated centroid normalization set the seed amplitude.

Table~\ref{tab:kelvin-rel-force} instead uses the co-rotating
two-fluid equilibrium with $a=1$, $\Omega_c=0.4$,
$m_n=m_p=1$, $K_n=0.024$, $K_p=0.216$ and
$(\mu_{n,\infty},\mu_{p,\infty})=(2,3)$. The maximum speed is $0.2$.
At small carrier drag the decay rates are linear in $\mathcal R$,
with slopes $0.279630$ and $0.823650$ in units $c/a$ at $ka=1$;
the real-frequency shifts are quadratic.
\endgroup

\input{appendix_kelvin_tension_prediction}

%% file: appendix_kelvin_tension_prediction.tex
\subsection{Column with line stress}
\label{app:kelvin_tension}
\label{sec:kt_column}
\label{sec:kt_explicit_operator}
\label{sec:kt_warm_benchmark}

The reference mode for Sec.~\ref{sec:results-kelvin-tension}
includes the restoring force and inertia supplied by the line energy
$F(n,N,\mathcal N)$. We first isolate this response with prescribed
material currents, then construct the mode of a rotating column in
which both fluids respond.

For a uniform array $B_0=\kappa_v\mathcal N_*>0$, a transverse
displacement $\boldsymbol\eta(z,t)$ bends the lines together.
\revadd{The reference neutron and carrier densities are $n_*$ and
$N_*$, and $v_n,v_c$ denote their transverse velocity perturbations.}
The tension and inertia per line are both
$\mathcal T_*=F_{\mathcal N}>0$. Define
$J\boldsymbol\eta=\hat z\times\boldsymbol\eta$,
$\nu_*=n_*+\mathcal ZN_*$, $G=\kappa_v\nu_*>0$ and
$D=\kappa_v(\mathcal R\nu_*+\mathcal R'n_*)$.
The balance of line stress, lift and drag gives
\begin{align}
 \mathcal T_*(\boldsymbol\eta_{tt}-\boldsymbol\eta_{zz})
 +(D\mathsf I-GJ)\boldsymbol\eta_t
 &=-\kappa_vJ(n_*v_n+\mathcal ZN_*v_c)\notag\\
 &\quad+\kappa_v(\mathcal R\nu_*v_c+\mathcal R'n_*v_n).
 \label{eq:kt_displacement}
\end{align}
For resting material currents,
$w=\eta_x+i\eta_y\propto e^{i(kz-\omega t)}$ reduces this balance
to a quadratic. Its zero-drag slow root is
\begin{align}
 \mathcal T_*\omega^2+(G+iD)\omega-\mathcal T_*k^2&=0,
 \label{eq:kt_dispersion}\\
 \omega_K=\frac{\sqrt{G^2+4\mathcal T_*^2k^2}-G}{2\mathcal T_*}
 &\simeq\frac{\mathcal T_*k^2}{G}\quad(D=0).
 \label{eq:kt_nondissipative}
\end{align}
The $k^2$ approximation requires $\mathcal T_*|k|\ll G$,
line inertia lowers the frequency relative to this law.
Opposite polarization reverses the sign of $G$ in the quadratic.
Magnus lift does no work,
and $\mathcal R'=0$, $\mathcal R\geq0$ damps the mode.
The group speed approaches unity in the formal large-$k$ limit,
subject to continuum averaging.
Our matching retains the neutron logarithm of
Ref.~\cite[Eq.~(150)]{Rau:2020wnv}, with fixed outer cutoff
and no magnetic term or microscopic wavelength-dependent logarithm.
A canonical column also requires the material response, since its
vorticity is the curl of the neutron momentum.

\begin{figure*}[t]
\centering
\includegraphics[width=\textwidth]{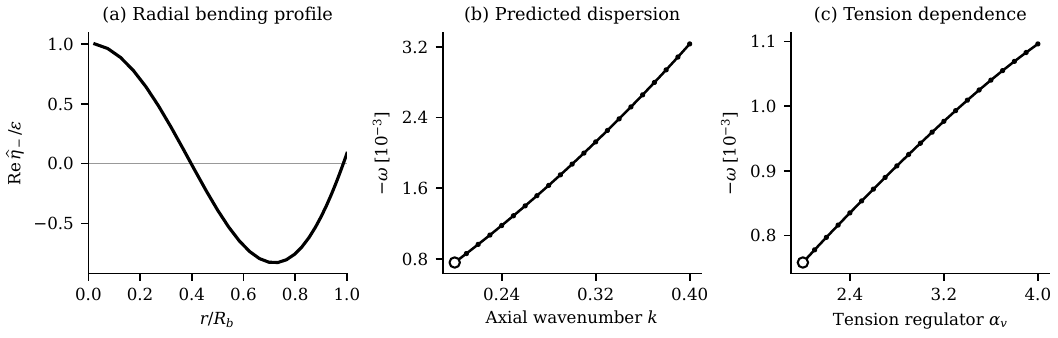}
\caption{Warm-column bending modes for
Sec.~\ref{sec:kt_warm_benchmark}.
\revadd{Here $\widehat\eta_-=(b_r-ib_\theta)/(ikB)$ is the complex
radial displacement amplitude, $b=\delta B$ and $\epsilon$ is its
maximum initial magnitude.}
(a) Radial displacement $\operatorname{Re}\widehat\eta_-/\epsilon$
with \eqref{eq:kt_warm_initial_data}. (b) Bending branch continued in
$k$ at $\revadd{\alpha_{\rm v}=2}$.
(c) Tension response at $k=0.2$, with equilibrium recomputed at fixed
line rotation. Circles mark the baseline. These are eigenmodes of
the coupled column, distinct from the evolution measurements in
Sec.~\ref{sec:results-kelvin-tension}.}
\label{fig:kt_warm_theory}
\end{figure*}

\input{appendix_kelvin_coupled_operator}
\input{appendix_kelvin_thermal_operator}

%% file: appendix_kelvin_coupled_operator.tex
For the column, set $n=\nu_n$, $N=n_p$ and
$\mathcal A=\mathcal R=\mathcal R'=0$, retaining constant
$\mathcal Z\geq0$. \revadd{The carrier rests in the background and advects
entropy, with $e_{\rm th}=C_{\rm th}s^{4/3}$,
$T_c=4C_{\rm th}s^{1/3}/3$ and constant $\varsigma=s/N$.}
The density dependence of $F$ modifies the material enthalpies:
$h_n=\mu_n+F_n$, $\revadd{h_c=\mu_p+F_N}$ and
$\revadd{\widehat h_c=h_c+\varsigma T_c}$.
For neutron speed $v$, put $p=h_nWv$ and $\Xi=h_nW$.
\revadd{For line angular velocity $\Omega_L$ and cylindrical radius
$r$, rigid line rotation has speed $u=\Omega_Lr$.} This gives $E_r=-uB$,
$\mathcal N=B\sqrt{1-u^2}/\kappa_v$ and $H_\Omega=a_\Omega B$.
Stationarity then requires
\begin{align}
 p'+p/r&=B,\quad\Xi'=-E_r,\quad\widehat h_c'=-\mathcal ZE_r,\notag\\
 H_\Omega'&=\frac{nW(v-u)-\mathcal ZNu+2\Omega_LuH_\Omega}{1-u^2}.
 \label{eq:kt_warm_equilibrium}
\end{align}
Axis regularity requires even densities and $B$, with
$p=B(0)r/2+O(r^3)$. The solution must retain $B>0$, $|u|<1$,
regular primitive recovery and the characteristic conditions of
Appendix~\ref{app:regulated_characteristics}.

The perturbation has eleven amplitudes,
$\revadd{q=(\delta n,\delta N,\delta s,\delta\xi,\delta\pi_c,
\mathcal E_r,\mathcal E_\theta)}$, where
$\revadd{\delta\pi_c=\widehat h_c\delta v_c}$ and both momenta are spatial
vectors. Canonical curl fixes $b=\delta B=\nabla\times\delta\xi$;
simplicity fixes $e_z=-E_rb_r/B$, with $e=\delta E$.
The independent electric amplitudes must include the stiffness variation:
\begin{align}
 \mathcal E&=a_\Omega e+E_r\delta a_\Omega\hat r,\notag\\
 \mathcal H&=a_\Omega b+B\delta a_\Omega\hat z,\qquad
 \delta\mathcal N=\frac{Bb_z-E_re_r}{\kappa_v^2\mathcal N}.
 \label{eq:kt_weighted_variables}
\end{align}
Here $a_\Omega=F_{\mathcal N}/(\kappa_v^2\mathcal N)$.
Vary $a_\Omega,h_n,h_c$ with all material and vortex-density
responses in Appendix~\ref{app:tension_constitutive_derivatives}.
These relations recover the physical $e$ from $\mathcal E$;
setting $\delta a_\Omega=0$ would omit part of the line response.
With $D_n=nW$, the number and entropy currents vary as
\begingroup\colorlet{revisionblue}{mainrevisionblue}\color{revisionblue}
\begin{align}
 \delta\Xi&=\delta h_n/W+v\delta\xi_\theta,\notag\\
 \delta D_n&=W\delta n+\frac n{h_n}(\delta\Xi-W\delta h_n),\notag\\
 \delta J_n&=\frac n{h_n}\delta\xi
       +Wv\left(\delta n-\frac n{h_n}\delta h_n\right)\hat\theta,\notag\\
 (\delta J_c,\delta J_S)&=(N,s)\delta\pi_c/\widehat h_c.
 \label{eq:kt_inverted_currents}
\end{align}
\endgroup
The line equation couples these currents through
$K^\mu=N_s^\mu+\mathcal ZN_c^\mu-\mathcal J_\Omega^\mu$.
Its background components are
$K^0=nW+\mathcal ZN+(r a_\Omega E_r)'/r$ and
$K^\theta=nWv-H_\Omega'=-K^0E_r/B$; its perturbations are
$\delta K^0=\delta D_n+\mathcal Z\delta N+\nabla\cdot\mathcal E$
and $\delta K=\delta J_n+\mathcal Z\delta J_c+i\omega\mathcal E
+\nabla\times\mathcal H$.
Number and entropy conservation, the material momentum balances
and the two transverse line equations therefore give
\begingroup\colorlet{revisionblue}{mainrevisionblue}\color{revisionblue}
\begin{align}
 -i\omega(\delta D_n,\delta N,\delta s)
       +\nabla\cdot(\delta J_n,\delta J_c,\delta J_S)&=0,\notag\\
 -i\omega\delta\xi+\nabla\delta\Xi+e&=0,\notag\\
 -i\omega\delta\pi_c+\nabla\delta h_c+\varsigma\nabla\delta T_c
       +\frac{T_c'}N(\delta s-\varsigma\delta N)\hat r\qquad&\notag\\
       +\mathcal Z(e+\delta v_c\times B\hat z)&=0,\notag\\
 B\delta K_\perp+K^0Je_\perp
       +(E_r\delta K^0+K^\theta b_z)\hat\theta&=0.
 \label{eq:kt_coupled_rows}
\end{align}
\endgroup
\revadd{Here $\delta T_c=T_c\delta s/(3s)$}, and derivatives are orthonormal
cylindrical operators with $\partial_\theta=i$, $\partial_z=ik$.
The last row acts in the $(r,\theta)$ plane.
\revadd{Conservation implies $\omega(\delta s-\varsigma\delta N)=0$:
the finite-frequency mode has $\delta s=\varsigma\delta N$, while
the entropy contact remains independent.}

Regular $m=1$ fields have scalar/axial amplitudes $O(r)$ and
transverse components $X_r+iX_\theta=O(r^2)$,
$X_r-iX_\theta=O(1)$ at the axis.
At the reflecting wall $r=R_b$, impose
$\revadd{\delta\xi_r=\delta\pi_{c,r}=\mathcal E_\theta=0}$.
Induction then fixes $b_r=0$ except at $\omega=\Omega_L$,
where the initial boundary flux must be retained.
With periodic $z$, radial collocation after constraint elimination
gives $\mathsf A q=\omega\mathsf Mq$.
{Identify the bending branch by following its frequency and
displacement continuously from zero additional lift,
$\mathcal Z=0$, and then as $k$ and $\revadd{\alpha_{\rm v}}$ vary. This keeps
the same bending mode among the fluid and line modes, whose
frequency ordering can change near crossings.
Figure~\ref{fig:kt_warm_theory}(b,c) shows the resulting dispersion
and tension dependence. Require residual and radial convergence.
Fluid backreaction and the wall now
select the frequency; the local quadratic alone is insufficient.
The closure contains no lattice shear and hence no Tkachenko branch.}

%% file: appendix_kelvin_thermal_operator.tex
{The benchmark is a warm, rotating neutron column inside a
reflecting cylindrical wall, with periodic $z$ and a resting carrier.
The vortex lines rotate rigidly, while the radial profiles satisfy
the canonical-curl relation and force balance in
Eq.~\eqref{eq:kt_warm_equilibrium}. We calculate the normal mode
and evolve a bending perturbation about this same stationary state to
separate bending from background readjustment. Comparing normal
modes with time evolution follows the approach of
Ref.~\cite{2008PhRvL.101b0402S}, which treats a trapped
Bose--Einstein condensate.}
Take constant positive microscopic inputs in
Eq.~\eqref{eq:tension_bcs_core_law}, $n_*=1$ and
$L_\Omega=10-\tfrac13\ln(n/n_*)$, retaining the density responses.
This prescription is not a calibrated pairing model.
Use $\revadd{m_n=m_p=1}$, $\revadd{\Gamma_n=\Gamma_p=2}$, $K_n=0.05$,
$\revadd{K_p=0.08}$, $\kappa_v=0.02$, $\mathcal N_0=0.1$,
$\revadd{\alpha_{\rm v}=2}$, $C_{\rm th}=0.1$, $\mathcal Z=0.5$, $R_b=10$ and
$k=0.2$, with central $(n,N,s,B)=(1,0.1,0.02,0.002)$.
Fix $\Omega_L=n(0)B(0)/[2h_n(0)(n(0)+\mathcal ZN(0))]$.
The bending benchmark uses the reference
\begin{equation}
 \omega=-7.5818344185\times10^{-4},\qquad
 2\pi/|\omega|\simeq8287.16.
 \label{eq:kt_warm_reference_frequency}
\end{equation}
For $L_z=10\pi$, initialize all amplitudes with
\begin{equation}
 \begin{aligned}
 \delta q(r,\theta,z,0)&=\operatorname{Re}[\widehat q(r)e^{i(\theta+kz)}],\\
 \max_r|\widehat\eta_-|&=\epsilon=10^{-3},\qquad
 \widehat\eta_-=\frac{b_r-ib_\theta}{ikB},
 \end{aligned}
 \label{eq:kt_warm_initial_data}
\end{equation}
and choose the phase so $\widehat\eta_-$ is positive near the axis.
\revadd{Include $\delta T_c/T_c=\delta N/(3N)$}, the full $\delta(a_\Omega E)$
and the density-response stress once in the conserved initial data.
At fixed $\Omega_L$ and $k$, recomputing equilibrium with
$\revadd{\alpha_{\rm v}=4}$ gives the reference $\omega=-1.0960939235\times10^{-3}$,
a $44.6\%$ increase in $|\omega|$
(Fig.~\ref{fig:kt_warm_theory}).

%% file: appendix_cowling_modes_tension.tex
\section{Radial two-fluid neutron star oscillation modes in the relativistic Cowling approximation}
\label{app:model_b_radial_cowling}

This appendix derives the reference frequencies for nonrotating radial two-fluid neutron star oscillations without vortex line tension in
Eq.~\eqref{eq:results_star_eigenvalues} from a global radial
boundary-value problem. We linearize the cold NC continuum equations
evolved in Sec.~\ref{sec:results-star-breathing} about the same stellar
equilibrium. The relativistic Cowling approximation fixes the spacetime
while allowing fluid perturbations
\cite{Cowling:1941nqk,1983ApJ...268..837M}. Here the two constituent
currents may move independently. We do not include drag. {In the generalized-Iordanskii parametrization, we set
$\mathcal Z=\mathcal R=\mathcal R'=0$, corresponding to free neutron-current
transport.}
The frequencies depend on the stellar stratification and boundary
conditions as well as the local acoustic speeds.

\subsection{Background and perturbation variables}

{In the spherical limit of a quasi-isotropic nonrotating stellar equilibrium,
the metric and spatial volume factor are}
\begin{align}
 ds^2&=-\alpha(r)^2dt^2
      +A(r)^2\left[dr^2+r^2d\Omega^2\right],
 \notag\\
 \sqrt\gamma&=A^3r^2\sin\theta .
 \label{eq:app_cowling_background}
\end{align}
\revadd{Here $A(r)$ is the spatial conformal factor,
$d\Omega^2=d\theta^2+\sin^2\theta\,d\phi^2$ is the unit-sphere
metric, and a prime denotes $d/dr$.}
{ The shift and extrinsic curvature
vanish, and we impose $\delta\alpha=\delta A=0$. Both background currents
are static, so their canonical Euler equations integrate to}
\revadd{constants $C_I$ satisfying}
\begin{equation}
 \alpha\mu_I=C_I,
 \qquad
 \mu_I'=-\mu_I\frac{\alpha'}{\alpha},
 \qquad I\in\{n,p\}.
 \label{eq:app_cowling_first_integrals}
\end{equation}
{We use $I$ for the constituent label to distinguish it from the
thermodynamic scalar $X=\xi_\mu\xi^\mu$, write
$\bar n_n\equiv\nu_n$ and $\bar n_p\equiv n_p$, and define the diagonal
susceptibility}
\begin{equation}
 \mathsf S_I\equiv\frac{d\bar n_I}{d\mu_I}.
 \label{eq:app_cowling_susceptibility}
\end{equation}

{The stellar equation of state in
Eqs.~\eqref{eq:main_model_b_master} and \eqref{eq:results_star_eos} has
$m_n=m_p=1$ and gives}
\begin{align}
 \bar n_n&=\frac{[C_n/\alpha-1]_+}{2K_n},
 &\mathsf S_n&=\frac{1}{2K_n},
 \notag\\
 \bar n_p&=\left[\frac{3[C_p/\alpha-1]_+}{4K_p}\right]^{1/3},
 &\mathsf S_p&=\frac{1}{4K_p\bar n_p^2}.
 \label{eq:app_cowling_gp4_thermo}
\end{align}
Here $[x]_+\equiv\max(x,0)$. We use
$K_p=0.7183075535$ as in Eq.~\eqref{eq:main_model_b_master}.

{We take Eulerian perturbations proportional to $e^{i\omega t}$ and
write the contravariant radial Eulerian three-velocity as}
\begin{equation}
 \delta v_I^r=i\omega\mathcal U_I,
 \qquad
 \zeta_I=\alpha\mathcal U_I,
 \qquad
 y_I=\alpha\,\delta\mu_I .
 \label{eq:app_cowling_variables}
\end{equation}
The coordinate displacement is $\zeta_I$, since
$dr/dt=\alpha v_I^r$ at zero shift. On the static,
unentrained background, $\Xi=\varepsilon_\xi=\mu_n$, and the primitive
perturbations satisfy
$\delta\mu=\delta\varepsilon_\xi=\delta\mu_n$ and
$\delta\chi=\delta\mu_p-\delta\mu_n$. Thus
$y_p=\alpha\,\delta\mu_p$, whereas $\alpha\,\delta\chi=y_p-y_n$.

\subsection{Linearized first-order eigenproblem}

\revadd{Write $\mu^n_\mu\equiv\xi_\mu$ for the neutron canonical momentum.}
{Equation~\eqref{eq:moments_carter_map} gives the radial canonical
momenta in terms of the Eulerian velocities:}
\begin{align}
 \begin{pmatrix}\delta\mu^n_r\\ \delta\mu^p_r\end{pmatrix}
 &=A^2\mathsf M
 \begin{pmatrix}\delta v_n^r\\ \delta v_p^r\end{pmatrix},
 \notag\\
 \mathsf M&=
 \begin{pmatrix}
  \mu_n-\mathcal A n_p&\mathcal A n_p\\
  \mathcal A\nu_n&\mu_p-\mathcal A\nu_n
 \end{pmatrix}.
 \label{eq:app_cowling_momenta}
\end{align}

{Linearizing species conservation and the radial canonical Euler
equations gives}
\begin{align}
 i\omega\,\delta\bar n_I
 &=-\frac{1}{A^3r^2}\frac{d}{dr}
   \left(A^3r^2\alpha\bar n_I\delta v_I^r\right),
 \notag\\
 i\omega\,\delta\mu^I_r
 &=-\frac{d}{dr}\left(\alpha\,\delta\mu_I\right).
 \label{eq:app_cowling_linear_equations}
\end{align}
{Substituting Eq.~\eqref{eq:app_cowling_variables} yields the
first-order system,}
\begin{align}
 \mathcal U_I'
 &=-\left[\ln\!\left(A^3r^2\alpha\bar n_I\right)\right]'
    \mathcal U_I
   -\frac{\mathsf S_I}{\alpha^2\bar n_I}\,y_I,
 \label{eq:app_cowling_continuity}\\
 \begin{pmatrix}y_n'\\y_p'\end{pmatrix}
 &=\omega^2A^2\mathsf M
   \begin{pmatrix}\mathcal U_n\\\mathcal U_p\end{pmatrix}.
 \label{eq:app_cowling_first_order}
\end{align}
The sign in Eq.~\eqref{eq:app_cowling_first_order} follows from the two
factors of $i\omega$. For the separable cold stellar equation of state with
$\mathcal A=0$, both the inertial and susceptibility matrices are diagonal,
and the neutron and proton acoustic families decouple exactly. In the local
WKB limit, Eqs.~\eqref{eq:app_cowling_continuity} and
\eqref{eq:app_cowling_first_order} reduce to
$c_{{\rm ac},I}^{\,2}\mathsf M\bm{\mathcal U}=K\bm{\mathcal U}$ for each
proper acoustic phase speed $c_{{\rm ac},I}$, where
$K=\operatorname{diag}(\nu_n\,d\mu_n/d\nu_n,
n_p\,d\mu_p/dn_p)$ is the local compressibility matrix. The largest local
acoustic speed therefore satisfies
$c_{\rm ac}^2=\lambda_{\max}(\mathsf M^{-1}K)$.

{The same system follows directly from the evolved variables.
To first order on this background,
{the fluid EOS and its Eulerian stress projections give}}
\begin{align}
 \delta D_B&=\delta\nu_n+\delta n_p,
 &\delta D_R&=\delta n_p,
 \notag\\
 \delta S_r&=A^2\nu_n\mu_n\delta v_n^r
              +A^2n_p\mu_p\delta v_p^r,
 \notag\\
 \delta E&=\mu_n\delta\nu_n+\mu_p\delta n_p,
 \notag\\
 \delta\Psi&=\nu_n\delta\mu_n+n_p\delta\mu_p.
 \label{eq:app_cowling_valencia_map}
\end{align}
Subtracting carrier conservation from total-baryon conservation recovers the
neutron continuity equation, and the neutron canonical equation supplies its
Euler equation. Subtracting the neutron contribution from radial
total-momentum conservation then gives the proton Euler equation. The
background-gradient terms cancel by
Eq.~\eqref{eq:app_cowling_first_integrals}, so the eigenproblem is a linear
projection of the evolved Valencia-plus-canonical system. Table~\ref{tab:app_cowling_modes}
compares two numerical solutions with the endpoint conditions specified
below.

\begin{table*}[t]
 \caption{{Lowest eigenfrequencies for radial neutron star oscillations of the
 nonrotating {$M=1.47137\,M_\odot$} superfluid neutron star with the
 equation of state in {Eq.~\eqref{eq:results_star_eos}}.
 Frequencies are
 coordinate-time angular frequencies in units of $t_u^{-1}$. The last column provides an error estimate by comparing the solutions from two different methods, shooting and spectral collocation (with $N_C$ collocation points).}}
 \label{tab:app_cowling_modes}
 \begin{ruledtabular}
 \begin{tabular}{lccc}
 Mode & Shooting & $N_C=220$ & Relative difference\\
 \hline
 $n:F$   & 0.74438946 & 0.74438937 & $1.25\times10^{-7}$\\
 $p:F$   & 1.13244510 & 1.13242078 & $2.15\times10^{-5}$\\
 $n:p_1$ & 1.25676622 & 1.25676572 & $4.00\times10^{-7}$\\
 $n:p_2$ & 1.74989668 & 1.74989501 & $9.54\times10^{-7}$\\
 $p:p_1$ & 1.97741085 & 1.97732007 & $4.59\times10^{-5}$
 \end{tabular}
 \end{ruledtabular}
\end{table*}

\subsection{Boundary conditions}
To solve the system, we need to specify boundary conditions at the center of the star and at its surface.
{Regularity at the center leaves two independent amplitudes, with}
\begin{align}
 y_I(r)&=y_{I,c}+O(r^2),
 \notag\\
 \mathcal U_I(r)&=-\frac{\mathsf S_{I,c}y_{I,c}}
 {3\alpha_c^2\bar n_{I,c}}\,r+O(r^3).
 \label{eq:app_cowling_center}
\end{align}
{At the common surface $r_s$, each freely moving constituent
satisfies}
\begin{equation}
 \Delta_{\!L}\mu_I
 =\delta\mu_I+\zeta_I\mu_I'
 =\frac{y_I}{\alpha}+\alpha\mathcal U_I\mu_I'=0.
 \label{eq:app_cowling_surface}
\end{equation}
This vanishing Lagrangian chemical-potential perturbation automatically selects the correct regular solution.

{We regularize the continuity equation before
discretizing it. Defining}
\begin{equation}
 b_I\equiv\frac{\bar n_I}{\mathsf S_I}
      =(\Gamma_I-1)(\mu_I-m_I),
 \label{eq:app_cowling_bi}
\end{equation}
{recasts Eq.~\eqref{eq:app_cowling_continuity} as}
\begin{align}
 0={}&\alpha b_I\mathcal U_I'
 +\left(\alpha'b_I-\mu_I\alpha'\right)\mathcal U_I
 \notag\\
 &+\alpha b_I\left[\ln(A^3r^2)\right]'\mathcal U_I
 +\frac{y_I}{\alpha}.
 \label{eq:app_cowling_regularized}
\end{align}
At $r_s$, $b_I\rightarrow0$. Using
Eq.~\eqref{eq:app_cowling_first_integrals}, the regularized row then reduces
exactly to Eq.~\eqref{eq:app_cowling_surface}. For the stellar proton
equation of state, $b_p=3(\mu_p-m_p)$ supplies the factor of three in the
implemented proton row.

\subsection{Numerical solution}

We first solve the system using a shooting method. The imported $\alpha$ and $A^2$
profiles are represented by not-a-knot cubic splines through 400 uniformly
spaced \textsc{Sfluide} samples. We reconstruct the remaining coefficients
from Eqs.~\eqref{eq:app_cowling_first_integrals} and
\eqref{eq:app_cowling_gp4_thermo}, without interpolating either density
profile directly. Two regular basis solutions start at $r_0=10^{-4}r_s$.
Each reconstructed coefficient is then tabulated on the uniform grid
\begin{equation}
 \begin{aligned}
 r_j&=r_0+\frac{j}{6000}(r_1-r_0),\qquad j=0,\ldots,6000,\\
 r_1&=(1-10^{-4})r_s.
 \end{aligned}
 \label{eq:app_cowling_shooting_grid}
\end{equation}
and represented by a second not-a-knot cubic spline. For fixed $\omega$, let
$\bm q=(\mathcal U_n,\mathcal U_p,y_n,y_p)^T$, so that
Eqs.~\eqref{eq:app_cowling_continuity} and
\eqref{eq:app_cowling_first_order} read
$\bm q'=\mathsf A_\omega(r)\bm q$. An adaptive Dormand-Prince (DOP853) step from $r_m$ to
$r_m+h_m$ applies the 12-stage explicit Runge--Kutta update
\begin{align}
 \bm q_{m,a}&=\bm q_m+h_m\sum_{b<a}a_{ab}\bm k_b,
 \notag\\
 \bm k_a&=\mathsf A_\omega(r_m+c_ah_m)\bm q_{m,a},
 \qquad a=1,\ldots,12,
 \notag\\
 \bm q_{m+1}&=\bm q_m+h_m\sum_{a=1}^{12}b_a\bm k_a .
 \label{eq:app_cowling_dop853_update}
\end{align}
The method has order $8(5,3)$ \cite{doi:10.1007/978-3-540-78862-1}; its embedded
error estimates control the step size with relative and absolute tolerances
$10^{-9}$ and $10^{-13}$.
Evaluating
Eq.~\eqref{eq:app_cowling_surface} for both constituents on both basis
solutions gives a $2\times2$ residual matrix $\mathsf R(\omega)$, with modes
at
\begin{equation}
 \det\mathsf R(\omega)=0.
 \label{eq:app_cowling_shooting_det}
\end{equation}
We scan 141 uniformly spaced frequencies over
$0.2\leq\omega t_u\leq3$. Since the stellar system decouples exactly, we
bracket the two diagonal residuals separately and refine their roots with
Brent's method to an absolute frequency tolerance $10^{-10}$. This retains
coincident roots that a scan of the residual product could miss. Reducing
the surface offset from $10^{-3}r_s$ to $10^{-5}r_s$ moves the most sensitive
tabulated root, the proton fundamental, from $1.13293918$ to $1.13242530$,
toward the exact-surface value $1.13242078$.

As an independent check, we solve
Eqs.~\eqref{eq:app_cowling_regularized} and
\eqref{eq:app_cowling_first_order} by spectral collocation on a
Chebyshev--Gauss--Lobatto grid with
$N_C$ nodes on $[0,r_s]$ \cite{doi:10.1137/1.9780898719598}. The center rows impose
Eq.~\eqref{eq:app_cowling_center}, and the unmodified regularized surface rows
impose Eq.~\eqref{eq:app_cowling_surface}. For
$\bm U_{\rm eig}=(\bm{\mathcal U}_n,\bm{\mathcal U}_p,
\bm y_n,\bm y_p)^T$, this gives the $4N_C$-dimensional generalized
eigenproblem\revadd{, where $\mathsf L$ contains the spatial and
boundary equations and $\mathsf W$ multiplies $\omega^2$,}
\begin{equation}
 \mathsf L\bm U_{\rm eig}=\omega^2\mathsf W\bm U_{\rm eig}.
 \label{eq:app_cowling_collocation}
\end{equation}
Table~\ref{tab:app_cowling_modes} uses $N_C=220$. The largest relative difference
between shooting and collocation among the first five roots is
$4.6\times10^{-5}$. We adopt the exact-surface collocation roots as the
rounded reference frequencies. Shooting checks the discretization and
regulated endpoint independently. With $\bar\omega\equiv\omega t_u$, the
dimensional cyclic frequency is
\begin{equation}
 f_{\rm phys}=\frac{\bar\omega}{2\pi t_u},
 \qquad t_u=\frac{10\,\mathrm{km}}{c}=33.3564\,\mu\mathrm{s}.
 \label{eq:app_cowling_frequency_conversion}
\end{equation}
{For the nonrotating {$M=1.47137\,M_\odot$} star with the equation
of state in {Eq.~\eqref{eq:results_star_eos}},
the neutron fundamental, $\bar\omega_{n,F}=0.74438937$, therefore corresponds
to $3.5517\,\mathrm{kHz}$.}